 \documentclass[a4papper,10pt]{article}
\usepackage{amssymb}
\usepackage{graphicx}
\usepackage{ifpdf}
\begin{document}
\def\ctr#1{\hfil $\,\,\,#1\,\,\,$ \hfil}
\def\tstrut{\vrule height 2.7ex depth 1.0ex width 0pt}
\def\mystrut{\vrule height 3.7ex depth 1.6ex width 0pt}
\def \inparg{\leftskip = 40pt\rightskip = 40pt}
\def \outparg{\leftskip = 0 pt\rightskip = 0pt}
\def\lf{16\pi^2}
\def\beqn{\begin{eqnarray}}
\def\eeqn{\end{eqnarray}}
\def\llf{(16\pi^2)^2}
\def\lllf{(16\pi^2)^3}
\def\llllf{(16\pi^2)^4}
\def\tr{\, {\rm tr}}
\def\Ttil{\widetilde{T}}
\def\Btil{\widetilde{B}}
\def\Etil{\widetilde{E}}
\def\ga{\gamma}
\def\b{\textbf}
\def\frak{\frac}
\def\ttil{\widetilde{t}}
\def\util{\widetilde{u}}
 \def\btil{\widetilde{b}}
\def\mutil{\widetilde{\mu}}
\def\dtil{\widetilde{d}}
\def\ctil{\widetilde{c}}
\def\stil{\widetilde{s}}
\def\tautil{\widetilde{\tau}}
\def\etil{\widetilde{e}}
\def\nutil{\widetilde{\nu}}
\def\gtil{\widetilde{g}}
\def\bigpound{ hbox{\bigit\$}} 
\def \q{\quad}
\def\lf{16\pi^2}
\def\llf{(16\pi^2)^2}
\def\lllf{(16\pi^2)^3}
\def \qq{\qquad}
\def \b{\bigskip}
\def\GeV{{\rm GeV}}
 \def\ha{{1\over2}}

\def\half{{\textstyle{1\over2}}} 
\def\frak#1#2{{\textstyle{{#1}\over{#2}}}}
\def\frakk#1#2{{{#1}\over{#2}}}
\def\pa{\partial}
\def\Scal{{\cal S}}

\def\cosec{\hbox{cosec }}
\def\semi{;\hfil\break}
\def\NSVZ{{\rm NSVZ}}
\def\DRED{{\rm DRED}}
\def\DREDp{{\rm
DRED}'}
\def\ga{\gamma}
\def\de{\delta}
\def\ep{\epsilon}
\def \la{\lambda}
\def \La{\Lambda}
\def \th{\theta}
\def\al{\alpha}

\def \bfa{{\bf a}}
\def \bfb{{\bf b}}
\def \bfc{{\bf c}}
\def \bfi{{\bf i}}
\def \bfj{{\bf j}}
\def \bfk{{\bf k}}
\def \bfn{{\bf n}}
\def \bfp{{\bf p}}
\def \bfr{{\bf r}}

\def \bfA{{\bf A}}
\def \bfB{{\bf B}}
\def \bfE{{\bf E}}
\def \bfF{{\bf F}}
\def \bfJ{{\bf J}}
\def \bfL{{\bf L}}
\def \bfR{{\bf R}}
\def \bfT{{\bf T}}
\def \bfV{{\bf V}}

\def\bfAdot{\dot\bfA}

\def\alphadot{\dot\alpha}
\def\betadot{\dot\beta}
\def\thdot{\dot\theta}
\def\thddot{\ddot\theta}
\def\omegadot{\dot\omega}
\def\omegaddot{\ddot\omega}
\def\onedot{\dot 1}
\def\twodot{\dot 2}
\def\pdot{\dot p}
\def\pddot{\ddot p}
\def\qdot{\dot q}
\def\qddot{\ddot q}
\def\rdot{\dot r}
\def\rddot{\ddot r}
\def\bfrdot{{\bf{\dot r}}}
\def\bfrddot{{\bf{\ddot r}}}
\def\xdot{\dot x}
\def\xddot{\ddot x}
\def\ydot{\dot y}
\def\yddot{\ddot y}
\def\zdot{\dot z}
\def\zddot{\ddot z}

\def\psib{\overline{\psi}}
\def\betab{\overline{\beta}}
\def\epb{\overline{\epsilon}}
\def\lab{\overline{\lambda}}
\def\thb{\overline{\theta}}
\def\chib{\overline{\chi}}
\def\taub{{\overline{\tau}}}
\def\phib{\overline{\phi}}
\def\Phib{\overline\Phi}
\def\Pib{\overline\Pi}
\def\sigmat{\sigma^{\mu}_{\alpha\alphadot}}
\def\sigmab{\overline{\sigma}}
\def\sigmad{\sigmab^{\mu\alphadot\alpha}}
\def\xib{\overline{\xi}}
\def\nub{\overline{\nu}}
\def\TeV{{\rm TeV}}
\def\GeV{{\rm GeV}}

\def\tautilde{\tilde\tau}
\def\chitilde{\tilde\chi}
\def\nutilde{\tilde\nu}
\def\gatilde{\tilde\ga}

\def\btilde{\tilde b}
\def\dtilde{\tilde d}
\def\etilde{\tilde e}
\def\gtilde{\tilde g}
\def\mtilde{\tilde m}
\def\ttilde{\tilde t}
\def\utilde{\tilde u}
\def\ytilde{\tilde y}

\def\Atilde{\tilde A}
\def\Btilde{\tilde B}
\def\Ctilde{\tilde C}
\def\Dtilde{\tilde D}
\def\Etilde{\tilde E}
\def\Ttilde{\tilde T}
\def\Xtilde{\tilde X}
\def\Ytilde{\tilde Y}
\def\bbar{{\overline{b}}}
\def\dbar{{\overline{d}}}
\def\ebar{{\overline{e}}}
\def\fbar{{\overline{f}}}
\def\gbar{{\overline{g}}}
\def\jbar{{\overline{j}}}
\def\mbar{{\overline{m}}}
\def\qbar{{\overline{q}}}
\def\tbar{{\overline{t}}}
\def\ubar{{\overline{u}}}
\def\ybar{{\overline{y}}}
\def\Bbar{{\overline{B}}}
\def\Dbar{{\overline{D}}}
\def\Ebar{{\overline{E}}}
\def\Hbar{{\overline{H}}}
\def\Jbar{{\overline{J}}}
\def\Qbar{{\overline{Q}}}
\def\Qb{\overline{Q}}
\def\Ubar{{\overline{U}}}
\def\Wbar{{\overline{W}}}
\def\Zbar{{\overline{Z}}}
\def\fivebar{{\overline{5}}}
\def\tenbar{{\overline{5}}}
\def\threebar{{\overline{3}}}
\def\phibar{{\overline{\phi}}}
\def\taubar{{\overline{\tau}}}

\def\msbar{{\overline{\rm MS}}}
\def\drbar{{\overline{\rm DR}}}
\def\gahat{\hat{\gamma}}
\def\lahat{\hat{\lambda}}
\def\ephat{\hat{\epsilon}}
\def\ghat{\hat{g}}
\def\Bhat{\hat{B}}
\def\Khat{\hat{K}}

\def\lt{\lambda_{{}_T}}
\def\mt{m_{{}_T}}
\def\leff{{\cal L}_{eff}}
\def\tc{T_c}
\def\vev#1{\mathopen\langle #1\mathclose\rangle }
\def\Dslash{D\!\!\!\! /}
\def\dslash{\pa \!\!\! /}
\def\kslash{k\!\!\! /}
\def\pslash{p\!\!\! /}

\def\sy{supersymmetry}
\def\sic{supersymmetric}
\def\sa{supergravity}
\def\ssm{supersymmetric standard model}
\def\sm{standard model}
\def\ssb{spontaneous symmetry breaking}
\def\smgroup{$SU_3\otimes\ SU_2\otimes\ U_1$}
\def\stw{\sin^2\th_W}

\def\app{{Acta Phys.\ Pol.\ }{\bf B}}
\def\anp{Ann.\ Phys.\ }
\def\cmp{Comm.\ Math.\ Phys.\ }
\def\fortphys{{Fort.\ Phys.\ }{\bf A}}
\def\ijmpa{{Int.\ J.\ Mod.\ Phys.\ }{\bf A}}
\def\jetp{JETP\ }
\def\jetpl{JETP Lett.\ }
\def\jmp{J.\ Math.\ Phys.\ }
\def\mpla{{Mod.\ Phys.\ Lett.\ }{\bf A}}
\def\nc{Nuovo Cimento\ }
\def\npb{{Nucl.\ Phys.\ }{\bf B}}
\def\physrep{Phys.\ Reports\ }
\def\plb{{Phys.\ Lett.\ }{\bf B}}
\def\pnas{Proc.\ Natl.\ Acad.\ Sci.\ (U.S.)\ }
\def\pr{Phys.\ Rev.\ }
\def\prd{{Phys.\ Rev.\ }{\bf D}}
\def\prl{Phys.\ Rev.\ Lett.\ }
\def\ptp{Prog.\ Th.\ Phys.\ }
\def\sjnp{Sov.\ J.\ Nucl.\ Phys.\ }
\def\tmp{Theor.\ Math.\ Phys.\ }
\def\pw{Part.\ World\ }
\def\zpc{Z.\ Phys.\ {\bf C}}

\def\dfx{{{df}\over{dx}}}

\def\dyx{{{dy}\over{dx}}}
\def\ddyx{{{d^2 y}\over{dx^2}}}
\def\dddyx{{{d^3 y}\over{dx^3}}}
\def\pux{{{\pa u}\over{\pa x}}}
\def\puy{{{\pa u}\over{\pa y}}}
\def\pvx{{{\pa v}\over{\pa x}}}
\def\pvy{{{\pa v}\over{\pa y}}}
\def\pzx{{{\pa z}\over{\pa x}}}
\def\pzy{{{\pa z}\over{\pa y}}}

\def\msbar{{\overline{\rm MS}}}
\def\drbar{{\overline{\rm DR}}}

\catcode`\@=11
\def\and{\char`\&}
\def\TeV{{\rm TeV}}
\def\GeV{{\rm GeV}}
\def\eV{{\rm eV}}

\vskip .3in
\centerline{\large{\textbf{The Full  Two-Loop R-parity Violating  Renormalization Group Equations   }}}
\medskip
\centerline{\large{\textbf{for All Minimal Supersymmetric Standard Model Couplings }}}
\vskip .3in
\centerline{\bf  A.F.~Kord and A.~Yazdanian}
\bigskip
\centerline{\small{\it {Department of Physics,Sabzevar Tarbiat Moallem University ,
      P.O.Box 397, Sabzevar, Iran}}}
\small{\it{farzaneh@sttu.ac.ir}}
\vskip .3in

\begin{abstract}
We present the full two-loop $\beta$-functions for the minimal supersymmetric standard
model couplings, extended to include R-parity violating couplings through explicit R-parity
violation.
\end{abstract}

\section{Introduction}
The supersymmetric extension ~\cite{a1,a2} of the Standard Model(SM) which introduces new scalar particles for SM fermions and new fermionic particles for SM bosons is a model which baryon ($B$) and lepton ($L$) number are not automatically conserved in the renormalisable Lagrangian. In the SM, the baryon and lepton number are conserved
automatically because it is impossible to write down renormalizable gauge-invariant
interactions that violate baryon and lepton number. However, in supersymmetric
extensions of the SM new interactions
are allowed to violate baryon or lepton number, so a discrete symmetry is often imposed to prohibit the unwanted interactions that violate $B$ and/or $L$ and lead to unacceptably large
effects~\cite{a3}. The most common discrete symmetry is $R$-parity which is defined~\cite{a4}:

\begin{eqnarray}
&&R_p=(-1)^{2S+3B+L},
\end{eqnarray}
where $S$ is the spin of the particle. All SM particles including the two scalar Higgs doublets are  even under this
transformation, and their superpartners are odd, which forces SUSY particles to always be made in pairs
and forbids  decay of  the Lightest Supersymmetric Particle (LSP).

Alternatively, since R-conservation is not theoretically
motivated by any known principle one can allow the $B$ and $L$ violating interactions to remain in the susy lagrangian, and constrain the couplings to be consistent with present experimental data. Both present of possible baryon and lepton number violating couplings may lead to unacceptably large effects such as proton decay. However; one can consider scenario which allow  $B$ or $L$ violating interactions~\cite{a4}.
Therefore, the
possibility of R-nonconservation deserves consideration. If we require a supersymmetric
Standard Model which is only minimal in particle content the superpotential is modified to
allow for additional R-parity violating interactions. R-parity violating interactions come
from either superpotential terms which violate baryon number, separate terms which violate
lepton-number, or from soft supersymmetry breaking terms. There are different kinds
of such terms, of dimensions 4, 3, or 2, with a potentially rich flavour structure.

The rich phenomenology implied by $R$-parity violation has
now gained full attention in the search for supersymmetry. The R-parity violating terms
in supersymmetry have theoretical and phenomenological implications in supersymmetric
theories~\cite{a6},such as:generating majorana neutrino masses, mediating various flavour and lepton number
violating processes,  modifying the signatures of supersymmetric particles at colliders, and possibility of decay of the lightest supersymmetric particle(LSP). It can also modify the Higgs sector \cite{a7,a8}.

All parameters of the MSSM lagrangian such as gauge couplings and Yukawa
couplings are scale dependent when one renormalizes the theory and replaces
the  bare parameters of the theory with running parameters. They are energy,
or equivalently distance, dependent through virtual corrections. One
can relate the values of these parameters at different scales using the Renormalisation
Group Equations(RGEs), which involve the $\beta$-functions.
We know low-energy supersymmetry is a perturbative theory, which relates
observed phenomena at low energies to physics at very high-energy scales.
For example, it is remarkable that in the MSSM, the three gauge couplings
appear to unify at  scale $2-3\times 10^{16}\GeV$ (the GUT scale). With the parameters
of the model specified at the input GUT scale by some candidate
organizing principle, one can run the couplings and masses down to low energies
using the renormalization group equations and make predictions about
the sparticle masses and other low-energy phenomena. Therefore the knowledge
of the $\beta$-functions and anomalous dimensions of the theory is vital.
The gauge $\beta$ functions  and the matter multiplet anomalous dimensions  have been
calculated up to four and three loop respectively~\cite{a9, a10} in the R-parity conserving
(RPC) case. The full three loop $\beta$-functions including soft $\beta$-functions have been
presented in~\cite{a11}. The two loop gauge $\beta$  functions and anomalous dimensions in the
R-parity violating (RPV) case have been calculated in~\cite{a12}. In particular, Ref ~\cite{a13} contained a
complete set of one-loop $\beta$-functions for RPV parameters. The full two loop $\beta$-functions for the baryon number violating (BNV) couplings have been presented in ~\cite{a14}. In this paper, we shall write down explicitly all possible R-parity violating
terms in the framework of the Minimal Supersymmetric Standard Model, assuming the most general breaking of R-parity, and then will complete results of two loop $\beta$-functions for all $R$-parity violating parameters of the MSSM.
\section{Beta Functions}
For a general N = 1 supersymmetric gauge theory with superpotential
\begin{eqnarray}
&&W=  \frac{1}{2}\mu ^{ij}\Phi _i \Phi _j +\frac{1}{6}Y^{ijk}\Phi _i \Phi _j \Phi _k+
 L ^i \Phi _i,
\end{eqnarray}
the standard soft supersymmetry-breaking scalar terms are as follows
\begin{equation}
\mathcal{-L}_{soft}=   \frac{1}{2}b ^{ij}\phi _i \phi _j +\frac{1}{6}h^{ijk}\phi _i \phi _j \phi _k+\frac{1}{2}{(m^2)}^j_i\phi ^i\phi_j+\frac{1}{2} M\bar{\lambda}\lambda +h.c.
\end{equation}
Here ${\lambda}$ is  the gaugino and $M$ is its mass. The renormalization group equations for the gauge coupling and  the superpotential parameters are given by:
 \begin{equation}
\beta _g=\mu \frac{dg}{d\mu}, \ \ \beta _{ijk}=\mu \frac{dY_{ijk}}{d\mu},\ \ \beta _{ij}=\mu \frac{d\mu _{ij}}{d\mu} \ \       \mbox{and} \ \   \beta _i=\mu \frac {dL _i}{d\mu},
\end{equation}
where $ \mu$ is the renormalization mass scale. The $\beta$-functions $\beta _{ijk}, \beta _{ij}$and $\beta _i$ are determined by the anomalous dimension $(\gamma )$ of the chiral supermultiplet $\Phi _i$ as follows:
 \begin{eqnarray}
&& \beta ^i=L^j \gamma ^i_j     \nonumber\\
&&\beta ^{ij}=\mu ^{ik}\gamma _k^j +\mu ^{kj}\gamma _k^i \nonumber\\
&&\beta ^{ijk}=Y^{ijl}\gamma _l^k+Y^{ilk}\gamma _l^j+Y^{ljk}\gamma _l^i.
\end{eqnarray}
The one loop results for the gauge coupling $\beta _g$ and $\gamma $ are given by:
 \begin{equation}
16\pi ^2 \beta _g^{(1)}=gQ, \ \ \  16\pi ^2(\gamma ^{(1)})^i_j=P^i_j,
\end{equation}
where  $Q$ and $P^i_j$ are given by:
\begin{eqnarray}
&& Q=T(R)-3C(G)   \nonumber\\
 && P^i_j=\frac{1}{2}Y^{ikl}Y_{jkl}-2g^2C(R)^i_j,
\end{eqnarray}
where $Y_{ijk}=(Y^{ijk})^*$ and $T(R)$ is the Dynkin index summed over all chiral multiplets. The Casimir invariants $C(G)$ and $C(R)^i_j$ are defined by:
 \begin{equation}
T(R)\delta _{ab}=Tr(R_aR_b), \ \ \ C(G)\delta _{ab}=f_{acd}f_{bcd}, \ \ \ \mbox{and}  \ C(R)^i_j=(R_aR_a)^i_j.
\end{equation}

The two loop $\beta$-functions for the dimensionless couplings are given by:
\begin{eqnarray}
&&(16\pi ^2)^2\beta _g^{(2)}=2g^5C(G)Q-2g^3r^{-1}{C(R)}^i_jP^j_i, \\
&&(16\pi ^2)^2\gamma ^{(2)i}_j=[-Y_{jmn}Y^{mpi}-2g^2C(R)^p_j\delta ^i_n]P^n_p+2g^4C(R)^i_jQ,
\end{eqnarray}
where $r$ is the number of generators of the symmetry group, and $Q$ and $P^i_j$ are defined above.

The complete exact results for the soft â-functions are given by ~\cite{a15,a16, a17}:
\begin{eqnarray}
&&{\beta}^{ij}_b={\gamma}^i_{\ l}b^{lj}+{\gamma}^j_{\ l}b^{il}-2{\gamma}^i_{1\ l}{\mu}^{lj}
-2{\gamma}^j_{1\ l}{\mu}^{il}
\nonumber\\&&
{\beta}^{ijk}_h={\gamma}^i_{\ l}h^{ljk}+{\gamma}^j_{\ l}h^{ilk}+{\gamma}^k_{\ l}h^{ijl}
-2({\gamma}^i_{1\ l}Y^{ljk}+{\gamma}^j_{1\ l}Y^{ilk}+{\gamma}^k_{1\ l}Y^{ijl})\nonumber\\&&
{\beta}_{M_a}=2\Theta (\frac{{\beta}_{g_a}}{g_a})
\nonumber\\&&
({\beta}_{m^2})^i_{\ j}=\left[\triangle +X\frac{\partial}{\partial g_a}\right]{\gamma}^i_{\ j},
\end{eqnarray}
where
\begin{eqnarray}
&&\Theta=\left( M_ag^2_a\frac{\partial}{\partial g^2_a} -h^{lmn}\frac{\partial}{\partial Y^{lmn}}    \right) \\&& (\gamma _1)^i_{\ j}=\Theta \gamma^i_{\ j}\\&&
\triangle=2\Theta {\Theta}^{\ast}+2M_aM_a^{\ast}g^2_a\frac{\partial}{{\partial g^2_a}}+{\widetilde{Y}}_{lmn}\frac{\partial}{\partial Y_{lmn}}+{\widetilde{Y}}^{lmn}\frac{\partial}{\partial Y^{lmn}}.
\end{eqnarray}
Here $M$ is the gaugino mass, $\widetilde{Y}^{ijk}=(m^2)^i_{\ l}Y^{jkl}+(m^2)^j_{\ l}Y^{ikl}+(m^2)^k_{\ l}Y^{ijl}$,
and
\begin{eqnarray}
&&X^{\NSVZ}=-2{g^3\over{16\pi^2}}
{S\over{\left[1-2g^2 C(G)(16\pi^2)^{-1}\right]}},
\end{eqnarray}
where
\begin{eqnarray}
S =  r^{-1}\tr [m^2C(R)] -MM^* C(G),
\end{eqnarray}
$C(R),C(G)$ being the quadratic Casimirs for the matter and adjoint
representations respectively and $X$  function above is given in the NSVZ scheme ~\cite{a18}.
Eq.~(11)holds in a class of renormalisation schemes that includes $\DREDp$~\cite{a19}, which we will use
throughout. There is no corresponding exact form for X in the $\DREDp$~\cite{a19}; however
we only require here the leading contribution which is the same in both schemes; the subleading
 $\DREDp$ contribution is given in Ref~\cite{a20} . These formulae can readily be specialised to the case of the RPV MSSM and their implementation can be automated.
\section{The R-parity violating MSSM scenario}
The unbrokrn $N=1$ theory is defined by the superpotential
\begin{eqnarray}
&&W=W_1+W_2,
\end{eqnarray}
where
\begin{eqnarray}
W_1={\epsilon}_{ab}\left[(Y_U)_{ij}Q^{ax}_iH^b_2U^c_{jx}+(Y_D)_{ij}Q^{ax}_iH^b_1D^c_{jx}
+(Y_E)_{ij}L^a_iH^b_1E^c_j -\mu H^a_1H^b_2\right],
\end{eqnarray}
and
\begin{eqnarray}
W_{ 2}&=&{\epsilon}_{ab}\left[\frac{1}{2}{(\Lambda_{E^k})}_{ij}L^a_iL^b_jE^c_k
+{(\Lambda_{D^k})}_{ij}L^a_iQ^{xb}_jD^c_{kx}\right]+\frac{1}{2}{\epsilon}_{xyz}{(\Lambda_{U^i})}_{jk}U^{cx}_iD^{cy}_jD^{cz}_k\nonumber\\  &&-{\epsilon}_{ab}{\kappa}^iL^a_iH^b_2).
\end{eqnarray}

In above equations, there is a summation over the generation indices $(i,j,k=1,2,3)$ and gauge indices $(x,y,z=1,2,3$ and $a,b=1,2)$. ${\epsilon}_{ab}$ and ${\epsilon}_{xyz}$ are  totally anti-symmetric tensors.

The soft SUSY breaking Lagrangian is given by:
\begin{eqnarray}
-L_{soft}&=& (m^{2}_{\widetilde{Q}})_{ij}\widetilde{Q}_i^\dagger{\widetilde{Q}}_j +(m^{2}_{\widetilde{L}})_{ij}\widetilde{L}_i^\dagger{\widetilde{L}}_j +(m^{2}_{\widetilde{U}^c})_{ij}\widetilde{U}_i^{ c}{\widetilde{U}^{\dagger c}}_j
+(m^{2}_{\widetilde{D}^c})_{ij}\widetilde{D}_i^{ c}{\widetilde{D}^{\dagger c}}_j \nonumber\\  &&
+(m^{2}_{\widetilde{E}^c})_{ij}\widetilde{E}_i^{ c}{\widetilde{E}^{\dagger c}}_j+      m^{2}_{H_2}\hat{H}^{\dagger}_2\hat{H}_2   +
m^{2}_{H_1}\hat{H}^{\dagger}_1\hat{H}_1 +\Big[
(h_U)_{ij}\hat{H}_2\widetilde{Q}_i\widetilde{U}^c_j  \nonumber\\&&
+(h_D)_{ij}\hat{H}_1\widetilde{Q}_i\widetilde{D}^c_j+
(h_E)_{ij}\hat{H}_1\widetilde{L}_i\widetilde{E}^c_j -B\hat{H}_1\hat{H}_2+h.c.\Big]\nonumber\\&&
 +
\Big[\frac{1}{2}M_1\widetilde{B}\widetilde{B}
+\frac{1}{2}M_2\widetilde{W}\widetilde{W} + \frac{1}{2}M_3\widetilde{g}\widetilde{g}+h.c.\Big]-L_{2},
\end{eqnarray}
where we have suppressed $SU(2)$ indices. Here  $\widetilde{B}, \widetilde{W}$ and $\widetilde{g}$ are the gaugino fields, $\widetilde{Q},\widetilde{U}, \widetilde{D}$ and $\widetilde{L},\widetilde{E}$ are the squark and slepton fields, respectively, and $\hat{H}_{1,2} $ are the $SU(2)$ doublet Higgs fields. $L_{2} $ includes terms which violate R-parity and it is given by:
\begin{eqnarray}
-L_{2}&=&{\epsilon}_{ab}\left[\frac{1}{2}(h_{E^k})_{ij}\widetilde{L}^a_i\widetilde{L}^b_j\widetilde{E}^c_k+(h_{D^k})_{ij}\widetilde{L}^a_i\widetilde{Q}^{xb}_j\widetilde{D}^c_{kx}+h.c\right]
\nonumber\\&&+\left[\frac{1}{2}{\epsilon}_{xyz}(h_{U^i})_{jk}\widetilde{U}^{cx}_i\widetilde{D}^{cy}_j\widetilde{D}^{cz}_{k} -
D_i\hat{H}^a_{2}\widetilde{L}_{ia}+h.c\right] \nonumber\\&&+  \widetilde{L}^{\dagger}_{ia}(m^2_{\tilde{L}_iH_1})\hat{H}^{ a}_1+
\hat{H}^{\dagger a}_1(m^2_{H_1\tilde{L}_i})\widetilde{L}_{ia}
\end{eqnarray}
\section{Renormalization group equations for the R-parity violating MSSM scenario}
In In this paper we present the full two loop beta functions in the $R$-parity violating
(RPV) MSSM scenario. Firstly, We  demonstrate the results for the Yukawa coupling $\beta$-functions. In terms of the anomalous dimensions, the Yukawa $\beta$-functions are given by\cite{a12}:

\begin{equation}
 \frac{d}{dt}(Y_E)_{ij}=(Y_E)_{ik}{\gamma}^{E_j}_{E_k} +(Y_E)_{ij}{\gamma}_{H_1}^{H_1}
+(Y_E)_{kj}{\gamma}^{L_i}_{L_k}-{(\Lambda_{E^j})}_{ki}{\gamma}^{H_1}_{L_k}
\end{equation}
\begin{equation}
\frac{d}{dt}(Y_D)_{ij}=(Y_D)_{ik}{\gamma}^{D_j}_{D_k}+(Y_D)_{ij}{\gamma}_{H_1}^{H_1}
+(Y_D)_{kj}{\gamma}^{Q_i}_{Q_k}-{{(\Lambda_{D^j})}}_{ki}{\gamma}^{H_1}_{L_k}
\end{equation}
\begin{equation}
\frac{d}{dt}(Y_U)_{ij}=(Y_U)_{ik}{\gamma}^{U_j}_{U_k} +(Y_U)_{ij}{\gamma}^{H_2}_{H_2}
+(Y_U)_{kj}{\gamma}^{Q_i}_{Q_k}
\end{equation}
\begin{equation}
\frac{d}{dt}{(\Lambda_{E^k})}_{ij}={(\Lambda_{E^l})}_{ij}{\gamma}^{E_k}_{E_l}+{(\Lambda_{E^k})}_{il}{\gamma}^{L_j}_{L_l}+(Y_E)_{ik}{\gamma}^{L_j}_{H_1}-{(\Lambda_{E^k})}_{jl}{\gamma}^{L_i}_{L_l}
-(Y_E)_{jk}{\gamma}^{L_i}_{H_1}
\end{equation}
\begin{equation}
\frac{d}{dt}{{(\Lambda_{D^k})}}_{ij}={{(\Lambda_{D^l})}}_{ij}{\gamma}^{D_k}_{D_l}+
{{(\Lambda_{D^k})}}_{il}{\gamma}^{Q_j}_{Q_l}+
{{(\Lambda_{D^k})}}_{lj}{\gamma}^{L_i}_{L_l}
-(Y_D)_{jk}{\gamma}^{L_i}_{H_1}
\end{equation}
\begin{equation}
\frac{d}{dt}{{(\Lambda_{U^i})}{}}_{jk}={{(\Lambda_{U^i})}{}}_{jl}{\gamma}^{D_k}_{D_l}+{{(\Lambda_{U^i})}{}}_{lk}{\gamma}^{D_j}_{D_l}+
{{(\Lambda_{U^l})}{}}_{jk}{\gamma}^{U_i}_{U_l}
\end{equation}
Here $t=\ln(Q)$ and ${\gamma}^{F_i}_{F_j}$ are the anomalous dimensions, and they are given by
\begin{equation}
{\gamma}^{F_i}_{F_j}=\frac{1}{({16{\pi}^2})}{{\gamma}^{(1)}}^{F_i}_{F_j}+\frac{1}{({16{\pi}^2})^2}{{\gamma}^{(2)}}^{F_i}_{F_j}+\cdots
\end{equation}
where ${\gamma}^{(1)}, {\gamma}^{(2)}$ and $\cdots$ are  $1-$loop, $2-$loop, $\cdots$
anomalous dimensions. The two-loop anomalous dimensions  and  gauge $\beta$-functions   of  the R-parity violating MSSM   have been given in  Ref~\cite{a12}. Using the results of Ref~\cite{a12} the two-loop Yukawa $\beta$-functions for the $R$-parity violating MSSM are given by:
\begin{eqnarray}
(16\pi^2)^2\beta^{(2)}_{(Y_E)_{ij}}&=&(16\pi^2)^2\beta^{(2)}_{(Y_E)_{RPC}}
\nonumber\\&&+(3g_2^2-\frac{3}{5}g_1^2)(Y_E)_{ik}Tr(\Lambda_{E^j}\Lambda_{E^k}^{\dagger})-2(E\Lambda_{E^l}\Lambda_{E^l}^{\dagger}Y_E+3E\Lambda_{D^l}\Lambda_{D^l}^{\dagger}Y_E)_{ij}
\nonumber\\&&+2(3E\Lambda_{E^j}\Lambda_{D^m}^*Y_D-E\Lambda_{E^j}\Lambda_{E^m}^{\dagger}Y_E)_{im}\nonumber\\&&-2(Y_E)_{ik}Tr[\Lambda_{E^j}\Lambda_{E^k}^{\dagger}(E+\Lambda_{E^l}\Lambda_{E^l}^{\dagger}+3\Lambda_{D^l}\Lambda_{D^l}^{\dagger})]
\nonumber\\&&+2(Y_E)_{ik}(3Y_D^{\dagger}\Lambda_{D^m}^t\Lambda_{E^k}^{\dagger}Y_E-Y_E^{\dagger}\Lambda_{E^m}\Lambda_{E^k}^{\dagger}Y_E)_{mj}-(Y_E)_{ij}[(\tilde{E})_{nl}Tr(\Lambda_{E^l}^{\dagger}\Lambda_{E^n})]
\nonumber\\&&-(Y_E)_{ij}[6(\tilde{D})_{lp}Tr(\Lambda_{D^p}^{\dagger}\Lambda_{D^l})+Tr(3E\Lambda_{D^q}\Lambda_{D^q}^{\dagger}+E\Lambda_{E^q}\Lambda_{E^q}^{\dagger}+6\tilde{D}\Lambda_{U^q}^{\dagger}\Lambda_{U^q}
\nonumber\\&&+3D\Lambda_{D^q}^t\Lambda_{D^q}^*)]+(16g_3^2-\frac{2}{5}g_1^2)(\Lambda_{D^q}\Lambda_{D^q}^{\dagger}Y_E)_{ij}+\frac{6}{5}g_1^2(\Lambda_{E^q}\Lambda_{E^q}^{\dagger}Y_E)_{ij}
\nonumber\\&&-(\tilde{E})_{mj}(Y_E)_{il}Tr(\Lambda_{E^l}^{\dagger}\Lambda_{E^m})-2(\tilde{E})_{lm}(\Lambda_{E^l}\Lambda_{E^m}^{\dagger}Y_E)_{ij}-(\Lambda_{E^l}\Lambda_{E^m}^{\dagger}Y_E)_{ij}Tr(\Lambda_{E^l}^{\dagger}\Lambda_{E^m})
\nonumber\\&&-(Y_E)_{im}(3Y_E^t\Lambda_{E^m}^{\dagger}\Lambda_{D^p}Y_D^*+Y_E^t\Lambda_{E^m}^{\dagger}\Lambda_{E^p}Y_E^*)_{jp}+(\tilde{E})_{mj}(3\Lambda_{E^m}\Lambda_{D^P}^*Y_D
\nonumber\\&&+\Lambda_{E^m}\Lambda_{E^p}^*Y_E)_{ip}-(Y_E^t\Lambda_{E^m}^{\dagger}E\Lambda_{E^m}+3Y_E^t\Lambda_{E^m}^{\dagger}\Lambda_{D^p}\Lambda_{D^p}^{\dagger}\Lambda_{E^m}
\nonumber\\&&+Y_E^t\Lambda_{E^m}^{\dagger}\Lambda_{E^p}\Lambda_{E^p}^{\dagger}\Lambda_{E^m})_{ji}-6(\Lambda_{E^j}\Lambda_{D^n}^*Y_D\Lambda_{U^m}^{\dagger}\Lambda_{U^m})_{in}
\nonumber\\&&-6(\Lambda_{D^l}\Lambda_{D^m}^{\dagger}Y_E)_{ij}[(\tilde{D})_{lm}+Tr(\Lambda_{D^l}^{\dagger}\Lambda_{D^m})+(\Lambda_{U^q}\Lambda_{U^q}^{\dagger})_{ml}]
\nonumber\\&&-3(Y_E^t\Lambda_{D^m}^*D\Lambda_{D^m}^t+Y_E^t\Lambda_{D^m}^*T\Lambda_{D^m}^t)_{ji}-3(\Lambda_{D^m}\Lambda_{D^p}^{\dagger}\Lambda_{D^p}\Lambda_{D^m}^{\dagger}Y_E)_{ij}
\nonumber\\&&+(\frac{2}{5}g_1^2-16g_3^2)(\Lambda_{E^j}\Lambda_{D^q}^*Y_D)_{iq}-\frac{6}{5}g_1^2(\Lambda_{E^j}\Lambda_{E^q}^*Y_E)_{iq}-(3\Lambda_{E^j}\Lambda_{E^q}^{\dagger}\Lambda_{D^p}\Lambda_{D^p}^{\dagger}Y_E)_{iq}
\nonumber\\&&-(\Lambda_{E^j}\Lambda_{E^q}^{\dagger}\Lambda_{E^p}\Lambda_{E^p}^{\dagger}Y_E+3\Lambda_{E^j}\Lambda_{E^q}^{\dagger}EY_E)_{iq}-(\Lambda_{E^j}\Lambda_{E^n}^{\dagger}Y_E)_{il}Tr(\Lambda_{E^l}^{\dagger}\Lambda_{E^n})
\nonumber\\&&-3(\Lambda_{E^j}E^*\Lambda_{D^m}^*Y_D-\Lambda_{E^j}E^*\Lambda_{E^m}^*Y_E)_{im}+9(\Lambda_{E^j}\Lambda_{D^n}^*DY_D)_{in}
\nonumber\\&&+6(\Lambda_{E^j}\Lambda_{D^n}^*Y_D)_{il}Tr(\Lambda_{D^l}\Lambda_{D^n}^{\dagger})+3(\Lambda_{E^j}\Lambda_{D^q}^*TY_D+3\Lambda_{E^j}\Lambda_{D^q}^*\Lambda_{D^m}^t\Lambda_{D^m}^*Y_D)_{iq}.
\end{eqnarray}
\begin{eqnarray}
(16\pi^2)^2\beta^{(2)}_{(Y_D)_{ij}}&=&(16\pi^2)^2\beta^{(2)}_{(Y_D)_{RPC}}+\frac{6}{5}g_1^2(\Lambda_{D^j}^t\Lambda_{E^q}^*Y_E)_{iq}-(\frac{2}{5}g_1^2-16g_3^2)(\Lambda_{D^j}^t\Lambda_{D^q}^*Y_D)_{iq}
\nonumber\\&&+(\Lambda_{D^j}^t\Lambda_{E^q}^{\dagger}[3\Lambda_{D^p}\Lambda_{D^p}^{\dagger}Y_E+\Lambda_{E^p}\Lambda_{E^p}^{\dagger}Y_E+3EY_E)_{iq}
\nonumber\\&&+(\Lambda_{D^j}^t\Lambda_{E^n}^{\dagger}Y_E)_{il}Tr(\Lambda_{E^l}^{\dagger}\Lambda_{E^n})+3(\Lambda_{D^j}^t\tilde{E}[\Lambda_{D^m}^*Y_D-\Lambda_{E^m}^*Y_E])_{im}
\nonumber\\&&-(\Lambda_{D^j}^t\Lambda_{D^n}^*[9DY_D+6Y_D\Lambda_{U^m}^{\dagger}\Lambda_{U^m}])_{in}+6Tr(\Lambda_{D^l}\Lambda_{D^n}^{\dagger})(\Lambda_{D^j}^t\Lambda_{D^n}^*Y_D)_{in}
\nonumber\\&&+3(\Lambda_{D^j}^t\Lambda_{D^q}^*([TY_D+\Lambda_{D^m}^t\Lambda_{D^m}^*Y_D])_{iq}+(\frac{16}{3}g_3^2+\frac{16}{15}g_1^2)(\Lambda_{U^q}\Lambda_{U^q}^{\dagger}Y_D^t)_{ji}
\nonumber\\&&+(6g_2^2+\frac{2}{5}g_1^2)(Y_D)_{ik}Tr(\Lambda_{D^j}\Lambda_{D^k}^{\dagger})-2(Y_D^t\Lambda_{D^p}^{\dagger}\Lambda_{D^p}D^t)_{ji}
\nonumber\\&&-2(Y_D)_{ik}Tr(\Lambda_{D^j}^t\Lambda_{D^k}^*[D+T]+\Lambda_{D^k}^{\dagger}\Lambda_{D^j}\Lambda_{D^q}^{\dagger}\Lambda_{D^q})
\nonumber\\&&-(Y_D)_{ik}Tr(6\Lambda_{D^q}\Lambda_{D^q}^{\dagger}\Lambda_{D^j}\Lambda_{D^k}^{\dagger}+2\Lambda_{E^q}\Lambda_{E^q}^{\dagger}\Lambda_{D^j}\Lambda_{D^k}^{\dagger}+2E\Lambda_{D^j}\Lambda_{D^k}^{\dagger})
\nonumber\\&&-4(Y_D\Lambda_{U^m}^{\dagger}\Lambda_{U^p}\Lambda_{U^p}^{\dagger}\Lambda_{U^m})_{ij}+4(Y_D\Lambda_{U^m}^{\dagger})_{in}(\Lambda_{U^m})_{jl}Tr(\Lambda_{D^l}\Lambda_{D^n}^{\dagger})
\nonumber\\&&-(6D\Lambda_{D^j}^t\Lambda_{D^p}^*Y_D+2D\Lambda_{D^j}^t\Lambda_{E^q}^*Y_E)_{ip}
-(Y_D)_{ik}[6Y_D^t\Lambda_{D^k}^{\dagger}\Lambda_{D^p}Y_D^*
\nonumber\\&&+2Y_D^t\Lambda_{D^j}^{\dagger}\Lambda_{E^p}Y_E^*]_{jp}-2(Y_D\Lambda_{U^n}^{\dagger}\Lambda_{U^l})_{ij}[Tr(\Lambda_{U^l}^{\dagger}\Lambda_{U^n})+2(\tilde{T})_{ln}]
\nonumber\\&&-(Y_D)_{ij}Tr(3E\Lambda_{D^q}\Lambda_{D^q}^{\dagger}+E\Lambda_{E^q}\Lambda_{E^q}^{\dagger}+6\tilde{D}\Lambda_{U^q}^{\dagger}\Lambda_{U^q}+3D\Lambda_{D^q}^t\Lambda_{D^q}^*
\nonumber\\&&-(Y_D)_{ij}\tilde{E}_{nl}Tr(\Lambda_{E^l}^{\dagger}\Lambda_{E^n})-6(Y_D)_{ij}\tilde{D}_{lp}Tr(\Lambda_{D^p}^{\dagger}\Lambda_{D^l})+\frac{2}{5}g_1^2(Y_D^t\Lambda_{D^q}^{\dagger}\Lambda_{D^q})_{ji}
\nonumber\\&&-2(Y_D^t\Lambda_{D^l}^{\dagger}\Lambda_{D^m})_{ji}[\tilde{D}_{ml}+(\Lambda_{U^q}\Lambda_{U^q}^{\dagger})_{lm}+Tr(\Lambda_{D^m}^{\dagger}\Lambda_{D^l})]
\nonumber\\&&-[Y_D^t\Lambda_{D^m}^{\dagger}(E\Lambda_{D^m}+3\Lambda_{D^q}\Lambda_{D^q}^{\dagger}\Lambda_{D^m}+\Lambda_{E^q}\Lambda_{E^q}^{\dagger}\Lambda_{D^m})]_{ji}
\nonumber\\&&-(Y_D)_{il}[3Y_D^t\Lambda_{D^l}^{\dagger}\Lambda_{D^m}Y_D^*+Y_D^t\Lambda_{D^l}^{\dagger}\Lambda_{E^m}Y_E^*]_{jm}-2(Y_D\Lambda_{U^q}^{\dagger}\Lambda_{U^q}\tilde{D})_{ij}
\nonumber\\&&-2(Y_D)_{im}\tilde{D}_{lj}Tr(\Lambda_{D^m}^{\dagger}\Lambda_{D^l})-\tilde{D}_{lj}(3\Lambda_{D^l}^t\Lambda_{D^m}^*Y_D+\Lambda_{D^l}^t\Lambda_{E^m}^*Y_E)_{im}
\nonumber\\&&-(Y_U)_{il}(Y_U^{\dagger}Y_D)_{pj}Tr(\Lambda_{U^l}^{\dagger}\Lambda_{U^p})-4(\Lambda_{U^m}\tilde{D}\Lambda_{U^m}^{\dagger}Y_D^t)_{ji}.
\end{eqnarray}
\begin{eqnarray}
(16\pi^2)^2\beta^{(2)}_{(Y_U)_{ij}}&=&(16\pi^2)^2\beta^{(2)}_{(Y_U)_{RPC}}+(\frac{8}{3}g_3^2-\frac{4}{15}g_1^2)(Y_U)_{ik}Tr(\Lambda_{U^j}\Lambda_{U^k}^{\dagger})-2(T\Lambda_{D^m}^t\Lambda_{D^m}^*Y_U)_{ij}
\nonumber\\&&-4(Y_U)_{ik}[(\Lambda_{U^k}^{\dagger}\Lambda_{U^j})_{lm}Tr(\Lambda_{D^m}^{\dagger}\Lambda_{D^l})+Tr(\Lambda_{U^k}^{\dagger}\Lambda_{U^j}\tilde{D})
\nonumber\\&&+Tr(\Lambda_{U^j}\Lambda_{U^k}^{\dagger}\Lambda_{U^p}\Lambda_{U^p}^{\dagger})]-(Y_U)_{il}(\tilde{T})_{pj}Tr(\Lambda_{U^l}^{\dagger}\Lambda_{U^p})
\nonumber\\&&-3(Y_U)_{ij}[Tr(\Lambda_{D^q}^t\Lambda_{D^q}^*)+\tilde{T}_{pm}Tr(\Lambda_{U^m}^{\dagger}\Lambda_{U^p})]+\frac{2}{5}g_1^2(Y_U^t\Lambda_{D^q}^{\dagger}\Lambda_{D^q})_{ji}
\nonumber\\&&-2(Y_U^{\dagger}\Lambda_{D^l}^{\dagger}\Lambda_{D^m})_{ji}[\tilde{D}_{ml}+(\Lambda_{U^q}\Lambda_{U^q}^{\dagger})_{lm}+Tr(\Lambda_{D^m}^{\dagger}\Lambda_{D^l})]
\nonumber\\&&-(Y_U^t\Lambda_{D^m}^{\dagger}E\Lambda_{D^m}+3Y_U^t\Lambda_{D^m}^{\dagger}\Lambda_{D^q}\Lambda_{D^q}^{\dagger}\Lambda_{D^m}+Y_U^t\Lambda_{D^m}^{\dagger}\Lambda_{E^q}\Lambda_{E^q}^{\dagger}\Lambda_{D^m})_{ji}
\nonumber\\&&-(Y_D)_{il}(3Y_U^t\Lambda_{D^l}^{\dagger}\Lambda_{D^m}Y_D^*+Y_U^t\Lambda_{D^l}^{\dagger}\Lambda_{E^m}Y_E^*)_{jm}
\nonumber\\&&-2(Y_D\Lambda_{U^q}^{\dagger}\Lambda_{U^q}Y_D^{\dagger}Y_U)_{ij}-2(Y_D)_{im}(Y_D^{\dagger}Y_U)_{lj}Tr(\Lambda_{D^m}^{\dagger}\Lambda_{D^l})
\nonumber\\&&-(Y_D^{\dagger}Y_U)_{lj}(3\Lambda_{D^l}^t\Lambda_{D^m}^*Y_D+\Lambda_{D^l}^t\Lambda_{E^m}^*Y_E)_{im}.
\end{eqnarray}

\begin{eqnarray}
(16\pi^2)^2\beta^{(2)}_{(\Lambda_{E^k})_{ij}}&=&(\Lambda_{E^l})_{ij}\{(6g_2^2-\frac{6}{5}g_1^2)(\tilde{E})_{lk}+(3g_2^2-\frac{3}{5}g_1^2)Tr(\Lambda_{E^k}\Lambda_{E^l}^{\dagger})
\nonumber\\&&-2(Y_E^{\dagger}EY_E+Y_E^{\dagger}\Lambda_{E^q}\Lambda_{E^q}^{\dagger}Y_E+3Y_E^{\dagger}\Lambda_{D^q}\Lambda_{D^q}^{\dagger}Y_E)_{lk}
\nonumber\\&&-2(\tilde{E})_{lk}Tr(E+3D)+2(3Y_E^{\dagger}\Lambda_{E^k}\Lambda_{D^m}^*Y_D
\nonumber\\&&-Y_E^{\dagger}\Lambda_{E^k}\Lambda_{E^m}^{\dagger}Y_E)_{lm}-2Tr[\Lambda_{E^k}\Lambda_{E^l}^{\dagger}(E+\Lambda_{E^q}\Lambda_{E^q}^{\dagger}
\nonumber\\&&+3\Lambda_{D^q}\Lambda_{D^q}^{\dagger})]+2(3Y_D^{\dagger}\Lambda_{D^m}^t\Lambda_{E^l}^{\dagger}Y_E-Y_E^{\dagger}\Lambda_{E^m}\Lambda_{E^l}^{\dagger}Y_E)_{mk}\}
\nonumber\\&&+(\Lambda_{E^k})_{il}\{(16g_3^2-\frac{2}{5}g_1^2)(\Lambda_{D^q}\Lambda_{D^q}^{\dagger})_{jl}-2(E^2)_{jl}
\nonumber\\&&+\frac{6}{5}g_1^2(E+\Lambda_{E^q}\Lambda_{E^q}^{\dagger})_{jl}+(Y_E^*)_{lm}(3\Lambda_{E^m}\Lambda_{D^p}^*Y_D+\Lambda_{E^m}\Lambda_{E^p}^*Y_E)_{jp}
\nonumber\\&&-(Y_E^{\dagger})_{ml}(Y_E)_{jn}Tr(\Lambda_{E^n}^{\dagger}\Lambda_{E^m})-2(\Lambda_{E^n}\Lambda_{E^m}^{\dagger})_{jl}(\tilde{E})_{nm}
\nonumber\\&&-(\Lambda_{E^n}\Lambda_{E^m}^{\dagger})_{jl}Tr(\Lambda_{E^n}^{\dagger}\Lambda_{E^m})-(E)_{jl}Tr(E+3D)
\nonumber\\&&-(Y_E)_{jm}(3\Lambda_{E^m}^{\dagger}\Lambda_{D^p}Y_D^*+\Lambda_{E^m}^{\dagger}\Lambda_{E^p}Y_E^*)_{lp}
\nonumber\\&&-(\Lambda_{E^m}^{\dagger}E\Lambda_{E^m}+3\Lambda_{E^m}^{\dagger}\Lambda_{D^p}\Lambda_{D^p}^{\dagger}\Lambda_{E^m}+\Lambda_{E^m}^{\dagger}\Lambda_{E^p}\Lambda_{E^p}^{\dagger}\Lambda_{E^m})_{lj}
\nonumber\\&&-6(\Lambda_{D^n}\Lambda_{D^m}^{\dagger})_{jl}[(\tilde{D})_{nm}+Tr(\Lambda_{D^n}^{\dagger}\Lambda_{D^m})+(\Lambda_{U^q}\Lambda_{U^q}^{\dagger})_{mn}]
\nonumber\\&&-3(\Lambda_{D^m}^*D\Lambda_{D^m}^t+\Lambda_{D^m}^*T\Lambda_{D^m}^t)_{lj}-3(\Lambda_{D^m}\Lambda_{D^p}^{\dagger}\Lambda_{D^p}\Lambda_{D^m}^{\dagger})_{jl}\}
\nonumber\\&&-(\Lambda_{E^k})_{jl}\{(16g_3^2-\frac{2}{5}g_1^2)(\Lambda_{D^q}\Lambda_{D^q}^{\dagger})_{il}+\frac{6}{5}g_1^2(E+\Lambda_{E^q}\Lambda_{E^q}^{\dagger})_{il}
\nonumber\\&&-2(E^2)_{il}-(Y_E^{\dagger})_{ml}(Y_E)_{in}Tr(\Lambda_{E^n}^{\dagger}\Lambda_{E^m})-2(\Lambda_{E^n}\Lambda_{E^m}^{\dagger})_{il}(\tilde{E})_{nm}
\nonumber\\&&-(\Lambda_{E^n}\Lambda_{E^m}^{\dagger})_{il}Tr(\Lambda_{E^n}^{\dagger}\Lambda_{E^m})-E_{il}Tr(E+3D)+(Y_E^*)_{lm}(3\Lambda_{E^m}\Lambda_{D^p}^*Y_D
\nonumber\\&&+\Lambda_{E^m}\Lambda_{E^p}^*Y_E)_{ip}-(\Lambda_{E^m}^{\dagger}E\Lambda_{E^m}+3\Lambda_{E^m}^{\dagger}\Lambda_{D^p}\Lambda_{D^p}^{\dagger}\Lambda_{E^m}
\nonumber\\&&+\Lambda_{E^m}^{\dagger}\Lambda_{E^p}\Lambda_{E^p}^{\dagger}\Lambda_{E^m})_{li}-(Y_E)_{im}(3\Lambda_{E^m}^{\dagger}\Lambda_{D^p}Y_D^*+\Lambda_{E^m}^{\dagger}\Lambda_{E^p}Y_E^*)_{lp}
\nonumber\\&&-6(\Lambda_{D^n}\Lambda_{D^m}^{\dagger})_{il}[(\tilde{D})_{nm}+Tr(\Lambda_{D^n}^{\dagger}\Lambda_{D^m})+(\Lambda_{U^q}\Lambda_{U^q}^{\dagger})_{mn}]
\nonumber\\&&-3(\Lambda_{D^m}^*D\Lambda_{D^m}^t+\Lambda_{D^m}^*T\Lambda_{D^m}^t)_{li}-3(\Lambda_{D^m}\Lambda_{D^p}^{\dagger}\Lambda_{D^p}\Lambda_{D^m}^{\dagger})_{il}\}
\nonumber\\&&+(Y_E)_{ik}\{(\frac{2}{5}g_1^2-16g_3^2)(\Lambda_{D^q}Y_D^*)_{jp}-\frac{6}{5}g_1^2(\Lambda_{E^q}Y_E^*)_{jp}
\nonumber\\&&-(3\Lambda_{E^q}^t\Lambda_{D^p}^*\Lambda_{D^p}^tY_E^*+\Lambda_{E^q}^t\Lambda_{E^p}^*\Lambda_{E^p}^tY_E^*+3\Lambda_{E^q}^tE^tY_E^*)_{jq}
\nonumber\\&&-(\Lambda_{E^n}^tY_E^*)_{jl}Tr(\Lambda_{E^l}^t\Lambda_{E^n}^*)-6(\Lambda_{D^n}Y_D^*\Lambda_{U^m}^t\Lambda_{U^m}^*)_{jn}
\nonumber\\&&-3(E\Lambda_{D^m}Y_D^*-E\Lambda_{E^m}Y_E^*)_{jm}+9(\Lambda_{D^n}D^tY_D^*)_{jn}
\nonumber\\&&+6(\Lambda_{D^n}Y_D^*)_{jl}Tr(\Lambda_{D^l}^*\Lambda_{D^n}^t)+3(Y_D^{\dagger}T\Lambda_{D^q}^t)_{qj}+3(\Lambda_{D^q}\Lambda_{D^m}^{\dagger}\Lambda_{D^m}Y_D^*)_{jq}\}
\nonumber\\&&-(Y_E)_{jk}\{(\frac{2}{5}g_1^2-16g_3^2)(\Lambda_{D^q}Y_D^*)_{iq}-\frac{6}{5}g_1^2(\Lambda_{E^q}Y_E^*)_{iq}
\nonumber\\&&-(3\Lambda_{E^q}^t\Lambda_{D^p}^*\Lambda_{D^p}^tY_E^*+\Lambda_{E^q}^t\Lambda_{E^p}^*\Lambda_{E^p}^tY_E^*+3\Lambda_{E^q}^tE^tY_E^*)_{iq}
\nonumber\\&&-(\Lambda_{E^n}^tY_E^*)_{il}Tr(\Lambda_{E^l}^t\Lambda_{E^n}^*)-3(E\Lambda_{D^m}Y_D^*-E\Lambda_{E^m}Y_E^*)_{im}
\nonumber\\&&+9(\Lambda_{D^n}D^tY_D^*)_{in}-6(\Lambda_{D^n}Y_D^*\Lambda_{U^m}^t\Lambda_{U^m}^*)_{in}+6(\Lambda_{D^n}Y_D^*)_{il}Tr(\Lambda_{D^l}^*\Lambda_{D^n}^t)
\nonumber\\&&+3(\Lambda_{D^q}Y_U^*Y_U^tY_D^*)_{iq}+3(\Lambda_{D^q}\Lambda_{D^m}^{\dagger}\Lambda_{D^m}Y_D^*)_{iq}\}
\nonumber\\&&+(\Lambda_{E^k})_{ij}[\frac{27}{2}g_1^4+\frac{15}{2}g_2^4+\frac{9}{5}g_2^2g_1^2]
\end{eqnarray}
\begin{eqnarray}
(16\pi^2)^2\beta^{(2)}_{(\Lambda_{D^k})_{ij}}&=&(\Lambda_{D^l})_{ij}\{(\frac{16}{3}g_3^2+\frac{16}{15}g_1^2)(\Lambda_{U^q}\Lambda_{U^q}^{\dagger})_{kl}+(6g_2^2+\frac{2}{5}g_1^2)[D_{kl}+Tr(\Lambda_{D^k}\Lambda_{D^l}^{\dagger})]
\nonumber\\&&-2(Y_D^{\dagger}DY_D+Y_D^{\dagger}TY_D)_{lk}-(6Y_D^{\dagger}\Lambda_{D^k}^t\Lambda_{D^p}^*Y_D)
\nonumber\\&&-2(\tilde{D})_{lk}Tr(E+3D)+4(\Lambda_{U^m}^{\dagger})_{ln}(\Lambda_{U^m})_{kh}Tr(\Lambda_{D^h}\Lambda_{D^n}^{\dagger})
\nonumber\\&&-(2Y_D^t\Lambda_{D^p}^{\dagger}\Lambda_{D^p}Y_D^*+4\Lambda_{U^m}\tilde{D}\Lambda_{U^m}^{\dagger})_{kl}-4(\Lambda_{U^m}^{\dagger}\Lambda_{U^p}\Lambda_{U^p}^{\dagger}\Lambda_{U^m})_{lk}
\nonumber\\&&-2Tr(\Lambda_{D^k}^t\Lambda_{D^l}^*(D+T)+\Lambda_{D^l}^{\dagger}\Lambda_{D^k}\Lambda_{D^q}^{\dagger}\Lambda_{D^q})
\nonumber\\&&-Tr(6\Lambda_{D^q}\Lambda_{D^q}^{\dagger}\Lambda_{D^k}\Lambda_{D^l}^{\dagger}+2\Lambda_{E^q}\Lambda_{E^q}^{\dagger}\Lambda_{D^k}\Lambda_{D^l}^{\dagger}+2E\Lambda_{D^k}\Lambda_{D^l}^{\dagger})
\nonumber\\&&-2(Y_D^{\dagger}\Lambda_{D^k}^t\Lambda_{E^p}^*Y_E)_{lp}-(6Y_D^t\Lambda_{D^l}^{\dagger}\Lambda_{D^p}Y_D^*+2Y_D^t\Lambda_{D^k}^{\dagger}\Lambda_{E^p}Y_E^*)_{kp}
\nonumber\\&&-2(\Lambda_{U^n}^{\dagger}\Lambda_{U^h})_{lk}[Tr(\Lambda_{U^h}^{\dagger}\Lambda_{U^n})+2(\tilde{T})_{hn}]\}
\nonumber\\&&+\frac{2}{5}g_1^2[(D\Lambda_{D^k}^t+2T\Lambda_{D^k}^t)_{ji}+(\Lambda_{D^k}\Lambda_{D^q}^{\dagger}\Lambda_{D^q})_{ij}]-2(D^2\Lambda_{D^k}^t)_{ji}
\nonumber\\&&-(D\Lambda_{D^k}^t)_{ji}Tr(E+3D)-2(T^2\Lambda_{D^k}^t)_{ji}-3(T\Lambda_{D^k}^t)_{ji}Tr(T)
\nonumber\\&&-2(\Lambda_{D^k}\Lambda_{D^p}^{\dagger}\Lambda_{D^m})_{ij}[(\tilde{D})_{mp}+(\Lambda_{U^q}\Lambda_{U^q}^{\dagger})_{pm}+Tr(\Lambda_{D^m}^{\dagger}\Lambda_{D^p})]
\nonumber\\&&-[\Lambda_{D^k}\Lambda_{D^m}^{\dagger}(E\Lambda_{D^m}+3\Lambda_{D^q}\Lambda_{D^q}^{\dagger}\Lambda_{D^m}+\Lambda_{E^q}\Lambda_{E^q}^{\dagger}\Lambda_{D^m})]_{ij}
\nonumber\\&&-(Y_D)_{jp}(3\Lambda_{D^k}\Lambda_{D^p}^{\dagger}\Lambda_{D^m}Y_D^*+\Lambda_{D^k}\Lambda_{D^p}^{\dagger}\Lambda_{E^m}Y_E^*)_{im}-2(Y_D\Lambda_{U^q}^{\dagger}\Lambda_{U^q}Y_D^{\dagger}\Lambda_{D^k}^t)_{ji}
\nonumber\\&&-2(Y_D)_{jm}(Y_D^{\dagger}\Lambda_{D^k}^t)_{pi}Tr(\Lambda_{D^m}^{\dagger}\Lambda_{D^p})-(Y_U)_{jp}(Y_U^{\dagger}\Lambda_{D^k}^t)_{ni}Tr(\Lambda_{U^p}^{\dagger}\Lambda_{U^n})
\nonumber\\&&-(\Lambda_{D^k}Y_D^*)_{ip}(3\Lambda_{D^p}^t\Lambda_{D^m}^*Y_D+\Lambda_{D^p}^t\Lambda_{E^m}^*Y_E)_{jm}
\nonumber\\&&+(Y_D)_{jk}\{-(\frac{2}{5}g_1^2-16g_3^2)(\Lambda_{D^q}Y_D^*)_{iq}+\frac{6}{5}g_1^2(\Lambda_{E^q}Y_E^*)_{iq}
\nonumber\\&&+(3\Lambda_{E^q}^t\Lambda_{D^p}^*\Lambda_{D^p}^tY_E^*+\Lambda_{E^q}^t\Lambda_{E^p}^*\Lambda_{E^p}^tY_E^*+3\Lambda_{E^q}^tE^*Y_E^*
\nonumber\\&&-3\Lambda_{D^q}Y_U^*Y_U^tY_D^*-3\Lambda_{D^q}\Lambda_{D^m}^{\dagger}\Lambda_{D^m}Y_D^*)_{iq}+(\Lambda_{E^n}^tY_E^*)_{il}Tr(\Lambda_{E^l}^t\Lambda_{E^n}^*)
\nonumber\\&&+3[E\Lambda_{D^m}Y_D^*-E\Lambda_{E^m}Y_E^*]_{im}-[9\Lambda_{D^n}D^tY_D^*
\nonumber\\&&-6\Lambda_{D^n}Y_D^*\Lambda_{U^m}^t\Lambda_{U^m}^*]_{in}-6(\Lambda_{D^n}Y_D^*)_{il}Tr(\Lambda_{D^l}^*\Lambda_{D^n}^t)\}
\nonumber\\&&+(\Lambda_{D^k})_{ij}[\frac{287}{90}g_1^4-\frac{16}{9}g_3^4+\frac{15}{2}g_2^4+g_1^2g_2^2+8g_3^2g_2^2+\frac{8}{9}g_3^2g_1^2]
\nonumber\\&&+(\Lambda_{D^q}\Lambda_{D^q}^{\dagger}\Lambda_{D^k})_{ij}(16g_3^2-\frac{2}{5}g_1^2)+\frac{6}{5}g_1^2(E\Lambda_{D^k}
\nonumber\\&&+\Lambda_{E^q}\Lambda_{E^q}^{\dagger}\Lambda_{D^k})_{ij}-(2E^2\Lambda_{D^k}-3\Lambda_{D^n}\Lambda_{D^p}^{\dagger}\Lambda_{D^p}\Lambda_{D^n}^{\dagger}\Lambda_{D^k})_{ij}
\nonumber\\&&-(Y_E^{\dagger}\Lambda_{D^k})_{nj}(Y_E)_{im}Tr(\Lambda_{E^m}^{\dagger}\Lambda_{E^n})-(\Lambda_{E^m}\Lambda_{E^n}^{\dagger}\Lambda_{D^k})_{ij}[(2\tilde{E})_{mn}
\nonumber\\&&+Tr(\Lambda_{E^m}^{\dagger}\Lambda_{E^n})]-(E\Lambda_{D^k})_{ij}Tr(E+3D)
\nonumber\\&&-(Y_E)_{in}(3\Lambda_{D^k}^t\Lambda_{E^n}^{\dagger}\Lambda_{D^p}Y_D^*+\Lambda_{D^k}^t\Lambda_{E^n}^{\dagger}\Lambda_{E^p}Y_E^*)_{jp}+(Y_E^{\dagger}\Lambda_{D^k})_{nj}(3\Lambda_{E^n}\Lambda_{D^p}^*Y_D
\nonumber\\&&+\Lambda_{E^n}\Lambda_{E^p}^*Y_E)_{ip}-(\Lambda_{D^k}^t\Lambda_{E^n}^{\dagger}E\Lambda_{E^n}+3\Lambda_{D^k}^t\Lambda_{E^n}^{\dagger}\Lambda_{D^p}\Lambda_{D^p}^{\dagger}\Lambda_{E^n}
\nonumber\\&&+\Lambda_{D^k}^t\Lambda_{E^n}^{\dagger}\Lambda_{E^p}\Lambda_{E^p}^{\dagger}\Lambda_{E^n})_{ji}-6(\Lambda_{D^m}\Lambda_{D^n}^{\dagger}\Lambda_{D^k})_{ij}[(\tilde{D})_{mn}+Tr(\Lambda_{D^m}^{\dagger}\Lambda_{D^n})
\nonumber\\&&+(\Lambda_{U^q}\Lambda_{U^q}^{\dagger})_{nm}]-3(\Lambda_{D^k}^t\Lambda_{D^n}^*D\Lambda_{D^n}^t+\Lambda_{D^k}^t\Lambda_{D^n}^*T\Lambda_{D^n}^t)_{ji}
\end{eqnarray}
\begin{eqnarray}
(16\pi^2)^2\beta^{(2)}_{(\Lambda_{U^i})_{jk}}&=&-(\frac{16}{3}g_3^2+\frac{16}{15}g_1^2)(\Lambda_{U^q}\Lambda_{U^q}^{\dagger}\Lambda_{U^i})_{kj}+(6g_2^2+\frac{2}{5}g_1^2)(\Lambda_{U^i})_{jl}Tr(\Lambda_{D^k}\Lambda_{D^l}^{\dagger})
\nonumber\\&&-(6g_2^2+\frac{2}{5}g_1^2)(D\Lambda_{U^i})_{kj}-2(\Lambda_{U^i}Y_D^{\dagger}DY_D)_{jk}-2(\Lambda_{U^i}Y_D^{\dagger}TY_D)_{jk}
\nonumber\\&&+2Tr(E+3D)[(\Lambda_{U^i}\tilde{D})_{kj}-(\Lambda_{U^i}\tilde{D})_{jk}]+2(Y_D^t\Lambda_{D^p}^{\dagger}\Lambda_{D^p}Y_D^*\Lambda_{U^i})_{kj}
\nonumber\\&&-2(\Lambda_{U^i})_{jl}Tr(\Lambda_{D^k}^t\Lambda_{D^l}^*(D+T)+\Lambda_{D^l}^{\dagger}\Lambda_{D^k}\Lambda_{D^q}^{\dagger}\Lambda_{D^q})
\nonumber\\&&-(\Lambda_{U^i})_{jl}Tr(6\Lambda_{D^q}\Lambda_{D^q}^{\dagger}\Lambda_{D^k}\Lambda_{D^l}^{\dagger}+2\Lambda_{E^q}\Lambda_{E^q}^{\dagger}\Lambda_{D^k}\Lambda_{D^l}^{\dagger}+2E\Lambda_{D^k}\Lambda_{D^l}^{\dagger})
\nonumber\\&&-(6\Lambda_{U^i}Y_D^{\dagger}\Lambda_{D^k}^t\Lambda_{D^p}^*Y_D+\Lambda_{U^i}Y_D^{\dagger}\Lambda_{D^k}^t\Lambda_{E^p}^*Y_E)_{jp}
\nonumber\\&&+4(\Lambda_{U^m}\tilde{D}\Lambda_{U^m}^{\dagger}\Lambda_{U^i})_{kj}-4(\Lambda_{U^i}\Lambda_{U^m}^{\dagger})_{kn}(\Lambda_{U^m})_{jh}Tr(\Lambda_{D^h}\Lambda_{D^n}^{\dagger})
\nonumber\\&&+4(\Lambda_{U^i}\Lambda_{U^m}^{\dagger})_{jn}(\Lambda_{U^m})_{kh}Tr(\Lambda_{D^h}\Lambda_{D^n}^{\dagger})-4(\Lambda_{U^i}\Lambda_{U^m}^{\dagger}\Lambda_{U^p}\Lambda_{U^p}^{\dagger}\Lambda_{U^m})_{jk}
\nonumber\\&&-2(\Lambda_{U^i}\Lambda_{U^n}^{\dagger}\Lambda_{U^h})_{jk}(Tr[\Lambda_{U^h}^{\dagger}\Lambda_{U^n}]+2\tilde{T}_{hn})-(\Lambda_{U^i})_{jl}(6Y_D^t\Lambda_{D^l}^{\dagger}\Lambda_{D^p}Y_D^*
\nonumber\\&&+2Y_D^t\Lambda_{D^k}^{\dagger}\Lambda_{E^p}Y_E^*)_{kp}+(\frac{16}{3}g_3^2+\frac{16}{15}g_1^2)(\Lambda_{U^q}\Lambda_{U^q}^{\dagger}\Lambda_{U^i})_{jk}
\nonumber\\&&+(6g_2^2+\frac{2}{5}g_1^2)[(D\Lambda_{U^i})_{jk}+Tr(\Lambda_{D^j}\Lambda_{D^l}^{\dagger})(\Lambda_{U^i})_{lk}]
\nonumber\\&&+2(\Lambda_{U^i}Y_D^{\dagger}DY_D)_{kj}+2(\Lambda_{U^i}Y_D^{\dagger}TY_D)_{kj}-2(Y_D^t\Lambda_{D^p}^{\dagger}\Lambda_{D^p}Y_D^*\Lambda_{U^i})_{jk}
\nonumber\\&&-2(\Lambda_{U^i})_{lk}Tr(\Lambda_{D^j}^t\Lambda_{D^l}^*(D+T)+\Lambda_{D^l}^{\dagger}\Lambda_{D^j}\Lambda_{D^q}^{\dagger}\Lambda_{D^q})
\nonumber\\&&-(\Lambda_{U^i})_{lk}Tr(6\Lambda_{D^q}\Lambda_{D^q}^{\dagger}\Lambda_{D^j}\Lambda_{D^l}^{\dagger}+2\Lambda_{E^q}\Lambda_{E^q}^{\dagger}\Lambda_{D^j}\Lambda_{D^l}^{\dagger}+2E\Lambda_{D^j}\Lambda_{D^l}^{\dagger})
\nonumber\\&&-4(\Lambda_{U^m}\tilde{D}\Lambda_{U^m}^{\dagger}\Lambda_{U^i})_{jk}+4(\Lambda_{U^i}\Lambda_{U^m}^{\dagger}\Lambda_{U^p}\Lambda_{U^p}^{\dagger}\Lambda_{U^m})_{kj}
\nonumber\\&&+2(\Lambda_{U^i}Y_D^{\dagger}\Lambda_{D^j}^t\Lambda_{E^p}^*Y_E)_{kp}-(\Lambda_{U^i})_{lk}(6Y_D^t\Lambda_{D^l}^{\dagger}\Lambda_{D^p}Y_D^*
\nonumber\\&&+2Y_D^t\Lambda_{D^j}^{\dagger}\Lambda_{E^p}Y_E^*)_{jp}+2(\Lambda_{U^i}\Lambda_{U^n}^{\dagger}\Lambda_{U^h})_{kj}[Tr(\Lambda_{U^h}^{\dagger}\Lambda_{U^n})+2(\tilde{T})_{hn}]
\nonumber\\&&+(6\Lambda_{U^i}Y_D^{\dagger}\Lambda_{D^j}^t\Lambda_{D^p}^*Y_D)_{kp}+(\Lambda_{U^l})_{jk}\{(\frac{8}{3}g_3^2-\frac{4}{15}g_1^2)Tr(\Lambda_{U^i}\Lambda_{U^l}^{\dagger})
\nonumber\\&&+(6g_2^2-\frac{2}{5}g_1^2)(\tilde{T})_{li}-2(Y_U^{\dagger}TY_U+Y_U^{\dagger}DY_U+Y_U^{\dagger}\Lambda_{D^m}^t\Lambda_{D^m}^*Y_U)_{li}
\nonumber\\&&-6\tilde{T}_{li}Tr(T)-4(\Lambda_{U^l}^{\dagger}\Lambda_{U^i})_{hm}Tr(\Lambda_{D^m}^{\dagger}\Lambda_{D^h})-4Tr(\Lambda_{U^i}\Lambda_{U^l}^{\dagger}\Lambda_{U^p}\Lambda_{U^p}^{\dagger})
\nonumber\\&&-4Tr(\Lambda_{U^l}^{\dagger}\Lambda_{U^i}\tilde{D})\}+(\Lambda_{U^i})_{jk}[-\frac{8}{3}g_3^4+\frac{28}{5}g_1^4+\frac{64}{15}g_3^2g_1^2]
\end{eqnarray}
The renormalisation equations for the bilinear terms can be written as:
\begin{equation}
\frac{d}{dt}{\mu}=\beta_{\mu}=\mu \{ {\gamma}^{H_1}_{H_1}+{\gamma}^{H_2}_{H_2} \}+{\kappa}^i{\gamma}^{H_1}_{L_i}
\end{equation}
\begin{equation}
\frac{d}{dt}{\kappa}^i=\beta_{k^i}={\kappa}^i {\gamma}^{H_2}_{H_2}+{\kappa}^p{\gamma}^{L_i}_{L_p} +{\mu}{\gamma}^{L_i}_{H_1}
\end{equation}

The two loop bilinear $\beta$-functions are given by:

\begin{eqnarray}
{(16\pi^2)^2}\beta^{(2)}_{\mu}&=&{(16\pi^2)^2}\beta^{(2)}_{\mu_{(RPC)}}+\mu[-Tr(3E\Lambda_{D^q}\Lambda_{D^q}^{\dagger}+E\Lambda_{E^q}\Lambda_{E^q}^{\dagger}+6\tilde{D}\Lambda_{U^q}^{\dagger}\Lambda_{U^q}+3D\Lambda_{D^q}^t\Lambda_{D^q}^*)
\nonumber\\&&-(\tilde{E})_{lm}Tr(\Lambda_{E^m}^{\dagger}\Lambda_{E^l})-6(\tilde{D})_{mp}Tr(\Lambda_{D^p}^{\dagger}\Lambda_{D^m})
\nonumber\\&&-Tr(3T\Lambda_{D^q}^t\Lambda_{D^q}^*)-3\tilde{T}_{lp}Tr(\Lambda_{U^p}^{\dagger}\Lambda_{U^l})]
\nonumber\\&&+\kappa^i[(\frac{2}{5}g_1^2-16g_3^2)(\Lambda_{D^q}^*Y_D)_{iq}-\frac{6}{5}g_1^2(\Lambda_{E^q}^*Y_E)_{iq}-(3\Lambda_{E^l}^{\dagger}\Lambda_{D^p}\Lambda_{D^p}^{\dagger}Y_E)_{il}
\nonumber\\&&-(\Lambda_{E^l}^{\dagger}\Lambda_{E^p}\Lambda_{E^p}^{\dagger}Y_E+3\Lambda_{E^l}^{\dagger}EY_E)_{il}-(\Lambda_{E^p}^{\dagger}Y_E)_{il}Tr(\Lambda_{E^l}^{\dagger}\Lambda_{E^p})
\nonumber\\&&-3(E^t\Lambda_{D^m}^*Y_D-E^t\Lambda_{E^m}^*Y_E)_{im}+9(\Lambda_{D^p}^*DY_D)_{ip}
\nonumber\\&&-6(\Lambda_{D^p}^*Y_D\Lambda_{U^m}^{\dagger}\Lambda_{U^m})_{ip}+6(\Lambda_{D^p}^*Y_D)_{il}Tr(\Lambda_{D^l}\Lambda_{D^p}^{\dagger})+3(\Lambda_{D^l}^*TY_D)_{il}
\nonumber\\&&+3(\Lambda_{D^l}^*\Lambda_{D^m}^t\Lambda_{D^m}^*Y_D)_{il}].
\end{eqnarray}
\begin{eqnarray}
{(16\pi^2)^2}\beta^{(2)}_{k^i}&=&\kappa^i\{\frac{15}{2}g_2^4+\frac{207}{50}g_1^4+\frac{9}{5}g_2^2g_1^2)+(16g_3^2+\frac{4}{5}g_1^2)Tr(T)
\nonumber\\&&-Tr(9TT+3TD)-3Tr(T\Lambda_{D^q}^t\Lambda_{D^q}^*)-3\tilde{T}_{lp}Tr(\Lambda_{U^p}^{\dagger}\Lambda_{U^l})\}
\nonumber\\&&+\kappa^p\{(16g_3^2-\frac{2}{5}g_1^2)(\Lambda_{D^q}\Lambda_{D^q}^{\dagger})_{ip}+\frac{6}{5}g_1^2(E+\Lambda_{E^q}\Lambda_{E^q}^{\dagger})_{ip}-2(E^2)_{ip}
\nonumber\\&&-(Y_E^{\dagger})_{kp}(Y_E)_{il}Tr(\Lambda_{E^l}^{\dagger}\Lambda_{E^k})-2(\Lambda_{E^l}\Lambda_{E^k}^{\dagger})_{ip}(\tilde{E})_{lk}-E_{ip}Tr(E+3D)
\nonumber\\&&-(\Lambda_{E^k}^{\dagger}E\Lambda_{E^k})_{pi}-(\Lambda_{E^l}\Lambda_{E^k}^{\dagger})_{ip}Tr(\Lambda_{E^l}^{\dagger}\Lambda_{E^k})-(Y_E)_{ik}(3\Lambda_{E^k}^{\dagger}\Lambda_{D^q}Y_D^*
\nonumber\\&&+\Lambda_{E^k}^{\dagger}\Lambda_{E^q}Y_E^*)_{pq}+(Y_E^{\dagger})_{kp}(3\Lambda_{E^k}\Lambda_{D^q}^*Y_D+\Lambda_{E^k}\Lambda_{E^q}^*Y_E)_{iq}
\nonumber\\&&-(3\Lambda_{E^k}^{\dagger}\Lambda_{D^q}\Lambda_{D^q}^{\dagger}\Lambda_{E^k}+\Lambda_{E^k}^{\dagger}\Lambda_{E^q}\Lambda_{E^q}^{\dagger}\Lambda_{E^k})_{pi}-6(\Lambda_{D^l}\Lambda_{D^k}^{\dagger})_{ip}[(\tilde{D})_{lk}
\nonumber\\&&+Tr(\Lambda_{D^l}^{\dagger}\Lambda_{D^k})+(\Lambda_{U^q}\Lambda_{U^q}^{\dagger})_{kl}]-3(\Lambda_{D^k}^*D\Lambda_{D^k}^t+\Lambda_{D^k}^*T\Lambda_{D^k}^t)_{pi}
\nonumber\\&&-3(\Lambda_{D^k}\Lambda_{D^q}^{\dagger}\Lambda_{D^q}\Lambda_{D^k}^{\dagger})_{ip}\}+\mu\{(\frac{2}{5}g_1^2-16g_3^2)(\Lambda_{D^q}Y_D^*)_{iq}-\frac{6}{5}g_1^2(\Lambda_{E^q}Y_E^*)_{iq}
\nonumber\\&&-(3\Lambda_{E^n}^t\Lambda_{D^p}^*\Lambda_{D^p}^tY_E^*+\Lambda_{E^n}^t\Lambda_{E^p}^*\Lambda_{E^p}^tY_E^*+3\Lambda_{E^n}^tE^tY_E^*)_{in}
\nonumber\\&&-3(E\Lambda_{D^m}Y_D^*-E\Lambda_{E^m}Y_E^*)_{im}+3(\Lambda_{D^n}\Lambda_{D^m}^{\dagger}\Lambda_{D^m}Y_D^*)_{in}
\nonumber\\&&-6(\Lambda_{D^q}Y_D^*\Lambda_{U^m}^t\Lambda_{U^m}^*)_{iq}-(\Lambda_{E^q}^t
Y_E^*)_{il}Tr(\Lambda_{E^l}^t\Lambda_{E^q}^*)
\nonumber\\&&+6(\Lambda_{D^q}Y_D^*)_{il}Tr(\Lambda_{D^l}^*\Lambda_{D^q}^t)+3(\Lambda_{D^n}Y_U^*Y_U^tY_D^*)_{in}+9(\Lambda_{D^q}D^tY_D^*)_{iq}\}.
\end{eqnarray}




\subsection{Soft Supersymmetry Breaking equations }
A precise study of the renormalization group equations must include R-parity violating terms because  they have direct and indirect effects on the mass-matrix, and  on other couplings of the MSSM lagrangian. R-parity violating couplings allow the generation of lepton-Higgs mixing that results in sneutrino vevs and neutrino masses.
By using the method that  has been  introduced in equations, the RGEs for the bilinear and trilinear soft SUSY breaking terms in R-parity violating scenario can be obtained. They are given by:

\begin{eqnarray}
{(16\pi^2)^2}\beta^{(2)}_{B}&=&{(16\pi^2)^2}\beta^{(2)}_{B_{(RPC)}}+B\{-Tr(3E\Lambda_{D^q}\Lambda_{D^q}^{\dagger}+E\Lambda_{E^q}\Lambda_{E^q}^{\dagger}+6\tilde{D}\Lambda_{U^q}^{\dagger}\Lambda_{U^q}
\nonumber\\&&+3D\Lambda_{D^q}^t\Lambda_{D^q}^*)-\tilde{E}_{lm}Tr(\Lambda_{E^m}^{\dagger}\Lambda_{E^l})-6(\tilde{D})_{mp}Tr(\Lambda_{D^p}^{\dagger}\Lambda_{D^m})
\nonumber\\&&-3Tr(T\Lambda_{D^q}^t\Lambda_{D^q}^*)-3(\tilde{T})_{lp}Tr(\Lambda_{U^p}^{\dagger}\Lambda_{U^l})\}
\nonumber\\&&+D^i\{(\frac{2}{5}g_1^2-16g_3^2)(\Lambda_{D^q}^*Y_D)_{iq}-\frac{6}{5}g_1^2(\Lambda_{E^q}^*Y_E)_{iq}-(3\Lambda_{E^l}^{\dagger}\Lambda_{D^p}\Lambda_{D^p}^{\dagger}Y_E
\nonumber\\&&+\Lambda_{E^l}^{\dagger}\Lambda_{E^p}\Lambda_{E^p}^{\dagger}Y_E+3\Lambda_{E^l}^{\dagger}EY_E)_{il}-(\Lambda_{E^p}^{\dagger}Y_E)_{il}Tr(\Lambda_{E^l}^{\dagger}\Lambda_{E^p})
\nonumber\\&&-3(E^t\Lambda_{D^m}^*Y_D-E^t\Lambda_{E^m}^*Y_E)_{im}+9(\Lambda_{D^p}^*DY_D)_{ip}
\nonumber\\&&-6(\Lambda_{D^p}^*Y_D\Lambda_{U^m}^{\dagger}\Lambda_{U^m})_{ip}+6(\Lambda_{D^p}^*Y_D)_{il}Tr(\Lambda_{D^l}\Lambda_{D^p}^{\dagger})+3(\Lambda_{D^l}^*TY_D)_{il}
\nonumber\\&&+3(\Lambda_{D^l}^*\Lambda_{D^m}^t\Lambda_{D^m}^*Y_D)_{il}\}-2\mu\{Tr(3h_EY_E^{\dagger}\Lambda_{D^q}\Lambda_{D^q}^{\dagger}+h_EY_E^{\dagger}\Lambda_{E^q}\Lambda_{E^q}^{\dagger})
\nonumber\\&&+Tr(6Y_D^{\dagger}h_D\Lambda_{U^q}^{\dagger}\Lambda_{U^q}+3h_DY_D^{\dagger}\Lambda_{D^q}^t\Lambda_{D^q}^*)+6(Y_D^{\dagger}h_D)_{np}Tr(\Lambda_{D^p}^{\dagger}\Lambda_{D^n})
\nonumber\\&&+Tr(Eh_{E^q}\Lambda_{E^q}^{\dagger})+(\tilde{E})_{ln}Tr(\Lambda_{E^n}^{\dagger}h_{E^l})+Tr(6\tilde{D}\Lambda_{U^q}^{\dagger}h_{U^q})
\nonumber\\&&+Tr(3Eh_{D^q}\Lambda_{D^q}^{\dagger}+3Dh_{D^q}^t\Lambda_{D^q}^*)+6(\tilde{D})_{np}Tr(\Lambda_{D^p}^{\dagger}h_{D^n})
\nonumber\\&&+Tr(3h_UY_U^{\dagger}\Lambda_{D^q}^t\Lambda_{D^q}^*)+(Y_E^{\dagger}h_E)_{ln}Tr(\Lambda_{E^n}^{\dagger}\Lambda_{E^l})+3(Y_U^{\dagger}h_U)_{lp}Tr(\Lambda_{U^p}^{\dagger}\Lambda_{U^l})
\nonumber\\&&+3(\tilde{T})_{lp}Tr(\Lambda_{U^p}^{\dagger}h_{U^l})+Tr(3Th_{D^q}^t\Lambda_{D^q}^*)\}
\nonumber\\&&-2\kappa^i\{(\frac{2}{5}M_1g_1^2-16M_3g_3^2)(\Lambda_{D^q}^*Y_D)_{iq}-6(\Lambda_{D^q}^*Y_D)_{il}Tr(h_{D^l}\Lambda_{D^q}^{\dagger})
\nonumber\\&&-\frac{6}{5}M_1g_1^2(\Lambda_{E^q}^*Y_E)_{iq}+(16g_3^2-\frac{2}{5}g_1^2)(\Lambda_{D^q}^*h_D)_{iq}+\frac{6}{5}g_1^2(\Lambda_{E^q}^*h_E)_{iq}
\nonumber\\&&+(3\Lambda_{E^l}^{\dagger}\Lambda_{D^p}\Lambda_{D^p}^{\dagger}h_E+\Lambda_{E^l}^{\dagger}\Lambda_{E^p}\Lambda_{E^p}^{\dagger}h_E+3\Lambda_{E^l}^{\dagger}h_E\tilde{E}+3\Lambda_{E^l}^{\dagger}Eh_E)_{il}
\nonumber\\&&+(\Lambda_{E^q}^{\dagger}h_E)_{il}Tr(\Lambda_{E^l}^{\dagger}\Lambda_{E^q})+3(Y_E^*h_E^t\Lambda_{D^m}^*Y_D-Y_E^*h_E^t\Lambda_{E^m}^*Y_E
\nonumber\\&&-E^t\Lambda_{E^m}^*h_E)_{im}-3(\Lambda_{D^l}^*h_UY_U^{\dagger}Y_D)_{il}-3(\Lambda_{D^l}^*h_{D^m}^t\Lambda_{D^m}^*Y_D)_{il}
\nonumber\\&&-9(\Lambda_{D^q}^*h_D\tilde{D}+\Lambda_{D^q}^*Dh_D)_{iq}+(\Lambda_{E^q}^{\dagger}Y_E)_{il}Tr(\Lambda_{E^l}^{\dagger}h_{E^q})
\nonumber\\&&+6(\Lambda_{D^q}^*h_D\Lambda_{U^m}^{\dagger}\Lambda_{U^m})_{iq}-6(\Lambda_{D^q}^*h_D)_{il}Tr(\Lambda_{D^l}\Lambda_{D^q}^{\dagger})
\nonumber\\&&-3(\Lambda_{D^l}^*\Lambda_{D^m}^t\Lambda_{D^m}^*h_D)_{il}+(\Lambda_{E^l}^{\dagger}h_{E^p}\Lambda_{E^p}^{\dagger}Y_E)_{il}+3(E^t\Lambda_{D^m}^*h_D)_{im}
\nonumber\\&&+6(\Lambda_{D^q}^*Y_D\Lambda_{U^m}^{\dagger}h_{U^m})_{iq}+3(\Lambda_{E^l}^{\dagger}h_{D^p}\Lambda_{D^p}^{\dagger}Y_E)_{il}-3(\Lambda_{D^l}^*Th_D)_{il}\}
\end{eqnarray}

\begin{eqnarray}
{(16\pi^2)^2}\beta^{(2)}_{D^i}&=&B\{(\frac{2}{5}g_1^2-16g_3^2)(\Lambda_{D^q}Y_D^*)_{iq}-\frac{6}{5}g_1^2(\Lambda_{E^q}Y_E^*)_{iq}-(3\Lambda_{E^n}^t\Lambda_{D^p}^*\Lambda_{D^p}^tY_E^*
\nonumber\\&&+\Lambda_{E^n}^t\Lambda_{E^p}^*\Lambda_{E^p}^tY_E^*+3\Lambda_{E^n}^tE^tY_E^*)_{in}-(\Lambda_{E^q}^tY_E^*)_{il}Tr(\Lambda_{E^l}^t\Lambda_{E^q}^*)
\nonumber\\&&-3(E\Lambda_{D^m}Y_D^*-E\Lambda_{E^m}Y_E^*)_{im}+9(\Lambda_{D^q}D^tY_D^*)_{iq}
\nonumber\\&&-6(\Lambda_{D^q}Y_D^*\Lambda_{U^m}^t\Lambda_{U^m}^*)_{iq}+6(\Lambda_{D^q}Y_D^*)_{il}Tr(\Lambda_{D^l}^*\Lambda_{D^q}^t)+3(Y_D^{\dagger}T\Lambda_{D^n}^t)_{ni}
\nonumber\\&&+3(\Lambda_{D^n}\Lambda_{D^m}^{\dagger}\Lambda_{D^m}Y_D^*)_{in}\}+D^l\{(16g_3^2-\frac{2}{5}g_1^2)(\Lambda_{D^q}\Lambda_{D^q}^{\dagger})_{il}
\nonumber\\&&+\frac{6}{5}g_1^2(E+\Lambda_{E^q}\Lambda_{E^q}^{\dagger})_{il}+(Y_E^{\dagger})_{ql}(3\Lambda_{E^q}\Lambda_{D^p}^*Y_D+\Lambda_{E^q}\Lambda_{E^p}^*Y_E)_{ip}
\nonumber\\&&-(Y_E^{\dagger})_{ql}(Y_E)_{in}Tr(\Lambda_{E^n}^{\dagger}\Lambda_{E^q})-2(\Lambda_{E^n}\Lambda_{E^q}^{\dagger})_{il}(\tilde{E})_{nq}
\nonumber\\&&-(\Lambda_{E^n}\Lambda_{E^q}^{\dagger})_{il}Tr(\Lambda_{E^n}^{\dagger}\Lambda_{E^q})-(E)_{il}Tr(E+3D)
\nonumber\\&&-(Y_E)_{iq}(3\Lambda_{E^q}^{\dagger}\Lambda_{D^p}Y_D^*+\Lambda_{E^q}^{\dagger}\Lambda_{E^p}Y_E^*)_{lp}-2(E^2)_{il}
\nonumber\\&&-(\Lambda_{E^q}^{\dagger}E\Lambda_{E^q}+3\Lambda_{E^q}^{\dagger}\Lambda_{D^p}\Lambda_{D^p}^{\dagger}\Lambda_{E^q}+\Lambda_{E^q}^{\dagger}\Lambda_{E^p}\Lambda_{E^p}^{\dagger}\Lambda_{E^q})_{li}
\nonumber\\&&-6(\Lambda_{D^n}\Lambda_{D^q}^{\dagger})_{il}[(\tilde{D})_{nq}+Tr(\Lambda_{D^n}^{\dagger}\Lambda_{D^q})+(\Lambda_{U^m}\Lambda_{U^m}^{\dagger})_{qn}]
\nonumber\\&&-3(\Lambda_{D^q}^*D\Lambda_{D^q}^t+\Lambda_{D^q}^*T\Lambda_{D^q}^t)_{li}-3(\Lambda_{D^q}\Lambda_{D^p}^{\dagger}\Lambda_{D^p}\Lambda_{D^q}^{\dagger})_{il}\}
\nonumber\\&&+D^i\{(\frac{15}{2}g_2^4+\frac{207}{50}g_1^4+\frac{9}{5}g_2^2g_1^2)+(16g_3^2+\frac{4}{5}g_1^2)Tr(T)
\nonumber\\&&-Tr(9T^2+3TD+3T\Lambda_{D^q}^t\Lambda_{D^q}^*)
\nonumber\\&&-3(\tilde{T})_{lq}Tr(\Lambda_{U^q}^{\dagger}\Lambda_{U^l})\}-2\kappa^i\{15M_2g_2^4+\frac{207}{25}M_1g_1^4
\nonumber\\&&+\frac{9}{5}(M_1+M_2)g_1^2g_2^2+(16M_3g_3^2+\frac{4}{5}M_1g_1^2)Tr(T)
\nonumber\\&&-(16g_3^2+\frac{4}{5}g_1^2)Tr(h_UY_U^{\dagger})+Tr(9h_UY_U^{\dagger}T+3h_UY_U^{\dagger}D
\nonumber\\&&+3h_UY_U^{\dagger}\Lambda_{D^q}^t\Lambda_{D^q}^*+9Th_UY_U^{\dagger})+3(Y_U^{\dagger}h_U)_{lq}Tr(\Lambda_{U^q}^{\dagger}\Lambda_{U^l})
\nonumber\\&&+Tr(3Th_DY_D^{\dagger})+3(\tilde{T})_{lq}Tr(\Lambda_{U^q}^{\dagger}h_{U^l})+Tr(3Th_{D^q}^t\Lambda_{D^q}^*)\}
\nonumber\\&&-2\mu\{(\frac{2}{5}M_1g_1^2-16M_3g_3^2)(\Lambda_{D^q}Y_D^*)_{iq}-\frac{6}{5}g_1^2(M_1\Lambda_{E^q}Y_E^*-h_{E^q}Y_E^*)_{iq}
\nonumber\\&&+(16g_3^2-\frac{2}{5}g_1^2)(h_{D^q}Y_D^*)_{iq}+(3\Lambda_{E^n}^tY_E^*h_E^tY_E^*)_{in}-3(\Lambda_{D^n}Y_U^*h_U^tY_D^*)_{in}
\nonumber\\&&+3(h_EY_E^{\dagger}\Lambda_{D^m}Y_D^*-h_EY_E^{\dagger}\Lambda_{E^m}Y_E^*)_{im}-9(\Lambda_{D^q}Y_D^*h_D^tY_D^*)_{iq}
\nonumber\\&&+(3h_{E^n}^t\Lambda_{D^p}^*\Lambda_{D^p}^tY_E^*+h_{E^n}^t\Lambda_{E^p}^*\Lambda_{E^p}^tY_E^*+\Lambda_{E^n}^t\Lambda_{E^p}^*h_{E^p}^tY_E^*
\nonumber\\&&+3h_{E^n}^tE^tY_E^*)_{in}+(h_{E^q}^tY_E^*)_{il}Tr(\Lambda_{E^l}^t\Lambda_{E^q}^*)+(\Lambda_{E^q}^tY_E^*)_{il}Tr(h_{E^l}^t\Lambda_{E^q}^*)
\nonumber\\&&-3(Eh_{E^m}Y_E^*)_{im}+6(\Lambda_{D^q}Y_D^*h_{U^m}^t\Lambda_{U^m}^*)_{iq}+3(\Lambda_{E^n}^t\Lambda_{D^p}^*h_{D^p}^tY_E^*)_{in}
\nonumber\\&&+3(Eh_{D^m}Y_D^*)_{im}-9(h_{D^q}D^tY_D^*)_{iq}+6(h_{D^q}Y_D^*\Lambda_{U^m}^t\Lambda_{U^m}^*)_{iq}
\nonumber\\&&-6(h_{D^q}Y_D^*)_{il}Tr(\Lambda_{D^l}^*\Lambda_{D^q}^t)-6(\Lambda_{D^q}Y_D^*)_{il}Tr(\Lambda_{D^l}^*h_{D^q}^t)
\nonumber\\&&-3(Y_D^{\dagger}Th_{D^n}^t)_{ni}-3(h_{D^n}\Lambda_{D^m}^{\dagger}\Lambda_{D^m}Y_D^*+\Lambda_{D^n}\Lambda_{D^m}^{\dagger}h_{D^m}Y_D^*)_{in}\}
\nonumber\\&&-2\kappa^l\{(16M_3g_3^2-\frac{2}{5}M_1g_1^2)(\Lambda_{D^q}\Lambda_{D^q}^{\dagger})_{il}+\frac{6}{5}M_1g_1^2(E+\Lambda_{E^q}\Lambda_{E^q}^{\dagger})_{il}
\nonumber\\&&-(16g_3^2-\frac{2}{5}g_1^2)(h_{D^q}\Lambda_{D^q}^{\dagger})_{il}-\frac{6}{5}g_1^2(h_EY_E^{\dagger}+h_{E^q}\Lambda_{E^q}^{\dagger})_{il}
\nonumber\\&&+2(h_EY_E^{\dagger}E+Eh_EY_E^{\dagger})_{il}+(h_EY_E^{\dagger})_{il}Tr(E+3D)
\nonumber\\&&+(Y_E^{\dagger})_{ql}(h_E)_{in}Tr(\Lambda_{E^n}^{\dagger}\Lambda_{E^q})+2(\Lambda_{E^n}\Lambda_{E^q}^{\dagger})_{il}(Y_E^{\dagger}h_E)_{nq}
\nonumber\\&&+(E)_{il}Tr(h_EY_E^{\dagger})+(h_E)_{iq}(3\Lambda_{E^q}^{\dagger}\Lambda_{D^p}Y_D^*+\Lambda_{E^q}^{\dagger}\Lambda_{E^p}Y_E^*)_{lp}
\nonumber\\&&-(Y_E^{\dagger})_{ql}(\Lambda_{E^q}\Lambda_{E^p}^*h_E)_{ip}+(\Lambda_{E^q}^{\dagger}h_EY_E^{\dagger}\Lambda_{E^q})_{li}+3(\Lambda_{D^q}^*h_UY_U^{\dagger}\Lambda_{D^q}^t)_{li}
\nonumber\\&&+(E)_{il}Tr(3h_DY_D^{\dagger})-3(Y_E^*)_{lq}(\Lambda_{E^q}\Lambda_{D^p}^*h_D)_{ip}
\nonumber\\&&+6(\Lambda_{D^n}\Lambda_{D^q}^{\dagger})_{il}(Y_D^{\dagger}h_D)_{nq}+3(\Lambda_{D^q}^*h_DY_D^{\dagger}\Lambda_{D^q}^t)_{li}
\nonumber\\&&+2(h_{E^n}\Lambda_{E^q}^{\dagger})_{il}(\tilde{E})_{nq}+(h_{E^n}\Lambda_{E^q}^{\dagger})_{il}Tr(\Lambda_{E^n}^{\dagger}\Lambda_{E^q})+(\Lambda_{E^n}\Lambda_{E^q}^{\dagger})_{il}Tr(\Lambda_{E^n}^{\dagger}h_{E^q})
\nonumber\\&&+(Y_E)_{iq}(\Lambda_{E^q}^{\dagger}h_{E^p}Y_E^*)_{lp}-(Y_E^{\dagger})_{ql}(3h_{E^q}\Lambda_{D^p}^*Y_D+h_{E^q}\Lambda_{E^p}^*Y_E)_{ip}
\nonumber\\&&+(\Lambda_{E^q}^{\dagger}Eh_{E^q}+3\Lambda_{E^q}^{\dagger}\Lambda_{D^p}\Lambda_{D^p}^{\dagger}h_{E^q}+\Lambda_{E^q}^{\dagger}h_{E^p}\Lambda_{E^p}^{\dagger}\Lambda_{E^q}+\Lambda_{E^q}^{\dagger}\Lambda_{E^p}\Lambda_{E^p}^{\dagger}h_{E^q})_{li}
\nonumber\\&&+6(\Lambda_{D^n}\Lambda_{D^q}^{\dagger})_{il}(h_{U^m}\Lambda_{U^m}^{\dagger})_{qn}+3(Y_E)_{iq}(\Lambda_{E^q}^{\dagger}h_{D^p}Y_D^*)_{lp}+3(\Lambda_{E^q}^{\dagger}h_{D^p}\Lambda_{D^p}^{\dagger}\Lambda_{E^q})_{li}
\nonumber\\&&+6(h_{D^n}\Lambda_{D^q}^{\dagger})_{il}[(\tilde{D})_{nq}+Tr(\Lambda_{D^n}^{\dagger}\Lambda_{D^q})+(\Lambda_{U^m}\Lambda_{U^m}^{\dagger})_{qn}]
\nonumber\\&&+6(\Lambda_{D^n}\Lambda_{D^q}^{\dagger})_{il}Tr(\Lambda_{D^n}^{\dagger}h_{D^q})+3(\Lambda_{D^q}^*Dh_{D^q}^t+\Lambda_{D^q}^*Th_{D^q}^t)_{li}
\nonumber\\&&+3(\Lambda_{D^q}\Lambda_{D^p}^{\dagger}h_{D^p}\Lambda_{D^q}^{\dagger}+h_{D^q}\Lambda_{D^p}^{\dagger}\Lambda_{D^p}\Lambda_{D^q}^{\dagger})_{il}+(Y_E^{\dagger})_{ql}(Y_E)_{in}Tr(\Lambda_{E^n}^{\dagger}h_{E^q})\}
\end{eqnarray}

\begin{eqnarray}
{(16\pi^2)^2}\beta^{(2)}_{(h_E)_{ij}}&=&{(16\pi^2)^2}\beta^{(2)}_{(h_E)_{(RPC)}}+(h_E)_{lj}\{(16g_3^2-\frac{2}{5}g_1^2)(\Lambda_{D^q}\Lambda_{D^q}^{\dagger})_{il}+\frac{6}{5}g_1^2(\Lambda_{E^q}\Lambda_{E^q}^{\dagger})_{il}
\nonumber\\&&-(Y_E^{\dagger})_{ml}(Y_E)_{iq}Tr(\Lambda_{E^q}^{\dagger}\Lambda_{E^m})-2(\Lambda_{E^n}\Lambda_{E^m}^{\dagger})_{il}(\tilde{E})_{nm}
\nonumber\\&&-(\Lambda_{E^n}\Lambda_{E^m}^{\dagger})_{il}Tr(\Lambda_{E^n}^{\dagger}\Lambda_{E^m})+(Y_E^*)_{lq}(3\Lambda_{E^q}\Lambda_{D^p}^*Y_D+\Lambda_{E^q}\Lambda_{E^p}^*Y_E)_{ip}
\nonumber\\&&-(Y_E)_{iq}(3\Lambda_{E^q}^{\dagger}\Lambda_{D^p}Y_D^*+\Lambda_{E^q}^{\dagger}\Lambda_{E^p}Y_E^*)_{lp}-(\Lambda_{E^q}^{\dagger}E\Lambda_{E^q}+3\Lambda_{E^q}^{\dagger}\Lambda_{D^p}\Lambda_{D^p}^{\dagger}\Lambda_{E^q}
\nonumber\\&&+\Lambda_{E^q}^{\dagger}\Lambda_{E^p}\Lambda_{E^p}^{\dagger}\Lambda_{E^q})_{li}-6(\Lambda_{D^n}\Lambda_{D^m}^{\dagger})_{il}[\tilde{D}_{nm}+Tr(\Lambda_{D^n}^{\dagger}\Lambda_{D^m})
\nonumber\\&&+(\Lambda_{U^q}\Lambda_{U^q}^{\dagger})_{mn}]-3(\Lambda_{D^q}^*D\Lambda_{D^q}^t+\Lambda_{D^q}^*T\Lambda_{D^q}^t)_{li}-3(\Lambda_{D^q}\Lambda_{D^p}^{\dagger}\Lambda_{D^p}\Lambda_{D^q}^{\dagger})_{il}\}
\nonumber\\&&-(h_E)_{ij}\{Tr(3E\Lambda_{D^q}\Lambda_{D^q}^{\dagger}+E\Lambda_{E^q}\Lambda_{E^q}^{\dagger}+6\tilde{D}\Lambda_{U^q}^{\dagger}\Lambda_{U^q}+3D\Lambda_{D^q}^t\Lambda_{D^q}^*)
\nonumber\\&&+\tilde{E}_{ml}Tr(\Lambda_{E^l}^{\dagger}\Lambda_{E^m})+6(\tilde{D})_{lq}Tr(\Lambda_{D^q}^{\dagger}\Lambda_{D^l})\}
\nonumber\\&&+(h_E)_{il}\{(3g_2^2-\frac{3}{5}g_1^2)Tr(\Lambda_{E^j}\Lambda_{E^l}^{\dagger})
\nonumber\\&&-2(Y_E^{\dagger}\Lambda_{E^n}\Lambda_{E^n}^{\dagger}Y_E+3Y_E^{\dagger}\Lambda_{D^n}\Lambda_{D^n}^{\dagger}Y_E)_{lj}+6(Y_E^{\dagger}\Lambda_{E^j}\Lambda_{D^m}^*Y_D)_{lm}
\nonumber\\&&-2(Y_E^{\dagger}\Lambda_{E^j}\Lambda_{E^m}^{\dagger}Y_E)_{lm}-2Tr(\Lambda_{E^j}\Lambda_{E^l}^{\dagger}E)
\nonumber\\&&-2Tr(\Lambda_{E^j}\Lambda_{E^l}^{\dagger}\Lambda_{E^n}\Lambda_{E^n}^{\dagger})-6Tr(\Lambda_{E^j}\Lambda_{E^l}^{\dagger}\Lambda_{D^n}\Lambda_{D^n}^{\dagger})+6(Y_D^{\dagger}\Lambda_{D^m}^t\Lambda_{E^l}^{\dagger}Y_E)_{mj}
\nonumber\\&&-2(Y_E^{\dagger}\Lambda_{E^m}\Lambda_{E^l}^{\dagger}Y_E)_{mj}\}+(h_{E^j})_{il}\{(\frac{2}{5}g_1^2-16g_3^2)(\Lambda_{D^q}^*Y_D)_{lq}-\frac{6}{5}g_1^2(\Lambda_{E^q}^*Y_E)_{lq}
\nonumber\\&&-3(\Lambda_{E^k}^{\dagger}\Lambda_{D^p}\Lambda_{D^p}^{\dagger}Y_E)_{lk}-(\Lambda_{E^k}^{\dagger}\Lambda_{E^p}\Lambda_{E^p}^{\dagger}Y_E)_{lk}-3(\Lambda_{E^k}^{\dagger}EY_E)_{lk}
\nonumber\\&&-(\Lambda_{E^q}^{\dagger}Y_E)_{ln}Tr(\Lambda_{E^n}^{\dagger}\Lambda_{E^q})-3(E^t\Lambda_{D^m}^*Y_D-E^t\Lambda_{E^m}^*Y_E)_{lm}+9(\Lambda_{D^q}^*DY_D)_{lq}
\nonumber\\&&-6(\Lambda_{D^q}^*Y_D\Lambda_{U^m}^{\dagger}\Lambda_{U^m})_{lq}+6(\Lambda_{D^q}^*Y_D)_{ln}Tr(\Lambda_{D^n}\Lambda_{D^q}^{\dagger})+3(\Lambda_{D^k}^*TY_D)_{lk}
\nonumber\\&&+3(\Lambda_{D^k}^*\Lambda_{D^m}^t\Lambda_{D^m}^*Y_D)_{lk}\}-2(Y_E)_{lj}\{(16M_3g_3^2-\frac{2}{5}M_1g_1^2)(\Lambda_{D^q}\Lambda_{D^q}^{\dagger})_{il}
\nonumber\\&&+\frac{6}{5}M_1g_1^2(\Lambda_{E^q}\Lambda_{E^q}^{\dagger})_{il}-(16g_2^3-\frac{2}{5}g_1)(h_{D^q}\Lambda_{D^q}^{\dagger})_{il}-\frac{6}{5}g_1^2(h_{E^q}\Lambda_{E^q}^{\dagger})_{il}
\nonumber\\&&+(Y_E^{\dagger})_{ql}(h_E)_{in}Tr(\Lambda_{E^n}^{\dagger}\Lambda_{E^q})+2(\Lambda_{E^n}\Lambda_{E^q}^{\dagger})_{il}(Y_E^{\dagger}h_E)_{nq}
\nonumber\\&&+(h_E)_{iq}(3\Lambda_{E^q}^{\dagger}\Lambda_{D^p}Y_D^*+\Lambda_{E^q}^{\dagger}\Lambda_{E^p}Y_E^*)_{lp}-(Y_E^*)_{lq}(\Lambda_{E^q}\Lambda_{E^p}^*h_E)_{ip}
\nonumber\\&&-3(Y_E^*)_{lq}(\Lambda_{E^q}\Lambda_{D^p}^*h_D)_{ip}+6(\Lambda_{D^n}\Lambda_{D^q}^{\dagger})_{il}(Y_D^{\dagger}h_D)_{nq}+3(\Lambda_{D^q}^*h_DY_D^{\dagger}\Lambda_{D^q}^t)_{li}
\nonumber\\&&+(\Lambda_{E^q}^{\dagger}h_EY_E^{\dagger}\Lambda_{E^q})_{li}+3(\Lambda_{D^q}^*h_UY_U^{\dagger}\Lambda_{D^q}^t)_{li}+(h_{E^n}\Lambda_{E^q}^{\dagger})_{il}Tr(\Lambda_{E^n}^{\dagger}\Lambda_{E^q})
\nonumber\\&&+(Y_E^{\dagger})_{ql}(Y_E)_{in}Tr(\Lambda_{E^n}^{\dagger}h_{E^q})+2(h_{E^n}\Lambda_{E^q}^{\dagger})_{il}(\tilde{E})_{nq}
\nonumber\\&&+(\Lambda_{E^n}\Lambda_{E^q}^{\dagger})_{il}Tr(\Lambda_{E^n}^{\dagger}h_{E^q})+(Y_E)_{iq}(\Lambda_{E^q}^{\dagger}h_{E^p}Y_E^*)_{lp}-(Y_E^*)_{lq}(3h_{E^q}\Lambda_{D^p}^*Y_D
\nonumber\\&&+h_{E^q}\Lambda_{E^p}^*Y_E)_{ip}+(\Lambda_{E^q}^{\dagger}Eh_{E^q})_{li}+3(\Lambda_{E^q}^{\dagger}\Lambda_{D^p}\Lambda_{D^p}^{\dagger}h_{E^q})_{li}
\nonumber\\&&+(\Lambda_{E^q}^{\dagger}h_{E^p}\Lambda_{E^p}^{\dagger}\Lambda_{E^q}+\Lambda_{E^q}^{\dagger}\Lambda_{E^p}\Lambda_{E^p}^{\dagger}h_{E^q})_{li}+6(\Lambda_{D^n}\Lambda_{D^q}^{\dagger})_{il}(h_{U^m}\Lambda_{U^m}^{\dagger})_{qn}
\nonumber\\&&+3(Y_E)_{iq}(\Lambda_{E^q}^{\dagger}h_{D^p}Y_D^*)_{lp}+3(\Lambda_{E^q}^{\dagger}h_{D^p}\Lambda_{D^p}^{\dagger}\Lambda_{E^q})_{li}+6(h_{D^n}\Lambda_{D^q}^{\dagger})_{il}[(\tilde{D})_{nq}
\nonumber\\&&+Tr(\Lambda_{D^n}^{\dagger}\Lambda_{D^q})+(\Lambda_{U^m}\Lambda_{U^m}^{\dagger})_{qn}]+6(\Lambda_{D^n}\Lambda_{D^q}^{\dagger})_{il}Tr(\Lambda_{D^n}^{\dagger}h_{D^q})+3(\Lambda_{D^q}^*Dh_{D^q}^t)_{li}
\nonumber\\&&+3(\Lambda_{D^q}^*Th_{D^q}^t)_{li}+3(\Lambda_{D^q}\Lambda_{D^p}^{\dagger}h_{D^p}\Lambda_{D^q}^{\dagger}+h_{D^q}\Lambda_{D^p}^{\dagger}\Lambda_{D^p}\Lambda_{D^q}^{\dagger})_{il}\}
\nonumber\\&&-2(Y_E)_{ij}\{Tr(3h_EY_E^{\dagger}\Lambda_{D^q}\Lambda_{D^q}^{\dagger}+h_EY_E^{\dagger}\Lambda_{E^q}\Lambda_{E^q}^{\dagger})+6(\tilde{D})_{lq}Tr(\Lambda_{D^q}^{\dagger}h_{D^l})
\nonumber\\&&+6Tr(Y_D^{\dagger}h_D\Lambda_{U^q}^{\dagger}\Lambda_{U^q})+3Tr(h_DY_D^{\dagger}\Lambda_{D^q}^t\Lambda_{D^q}^*)+(Y_E^{\dagger}h_E)_{mn}Tr(\Lambda_{E^n}^{\dagger}\Lambda_{E^m})
\nonumber\\&&+6(Y_D^{\dagger}h_D)_{lq}Tr(\Lambda_{D^q}^{\dagger}\Lambda_{D^l})+Tr(Eh_{E^q}\Lambda_{E^q}^{\dagger})+(\tilde{E})_{mn}Tr(\Lambda_{E^n}^{\dagger}h_{E^m})
\nonumber\\&&+6Tr(\tilde{D}\Lambda_{U^q}^{\dagger}h_{U^q})+3Tr(Eh_{D^q}\Lambda_{D^q}^{\dagger}+Dh_{D^q}^t\Lambda_{D^q}^*)\}
\nonumber\\&&-2(Y_E)_{il}\{(3M_2g_2^2-\frac{3}{5}M_1g_1^2)Tr(\Lambda_{E^j}\Lambda_{E^l}^{\dagger})+(\frac{3}{5}g_1^2-3g_2^2)Tr(h_{E^j}\Lambda_{E^l}^{\dagger})
\nonumber\\&&+6(Y_E^{\dagger}\Lambda_{D^q}\Lambda_{D^q}^{\dagger}h_E)_{lj}+2(Y_E^{\dagger}\Lambda_{E^j}\Lambda_{E^m}^{\dagger}h_E)_{lm}+2Tr(\Lambda_{E^j}\Lambda_{E^l}^{\dagger}h_EY_E^{\dagger})
\nonumber\\&&-6(Y_D^{\dagger}\Lambda_{D^m}^t\Lambda_{E^l}^{\dagger}h_E)_{mj}+2(Y_E^{\dagger}\Lambda_{E^m}\Lambda_{E^l}^{\dagger}h_E)_{mj}
\nonumber\\&&+2(Y_E^{\dagger}h_{E^q}\Lambda_{E^q}^{\dagger}Y_E)_{lj}-6(Y_E^{\dagger}h_{E^j}\Lambda_{D^m}^*Y_D)_{lm}+2(Y_E^{\dagger}h_{E^j}\Lambda_{E^m}^{\dagger}Y_E)_{lm}
\nonumber\\&&+2Tr[h_{E^j}\Lambda_{E^l}^{\dagger}(E+\Lambda_{E^q}\Lambda_{E^q}^{\dagger})]+6Tr(h_{E^j}\Lambda_{E^l}^{\dagger}\Lambda_{D^q}\Lambda_{D^q}^{\dagger})+2Tr(\Lambda_{E^j}\Lambda_{E^l}^{\dagger}h_{E^q}\Lambda_{E^q}^{\dagger})
\nonumber\\&&+2(Y_E^{\dagger}h_{E^m}\Lambda_{E^l}^{\dagger}Y_E)_{mj}+6(Y_E^{\dagger}h_{D^q}\Lambda_{D^q}^{\dagger}Y_E)_{lj}+6Tr(\Lambda_{E^j}\Lambda_{E^l}^{\dagger}h_{D^q}\Lambda_{D^q}^{\dagger})
\nonumber\\&&-6(Y_D^{\dagger}h_{D^m}^t\Lambda_{E^l}^{\dagger}Y_E)_{mj}+2(Y_E^{\dagger}\Lambda_{E^q}\Lambda_{E^q}^{\dagger}h_E)_{lj}-6(Y_E^{\dagger}\Lambda_{E^j}\Lambda_{D^m}^*h_D)_{lm}\}
\nonumber\\&&-2(\Lambda_{E^j})_{il}\{(\frac{2}{5}M_1g_1^2-16M_3g_3^2)(\Lambda_{D^q}^*Y_D)_{lq}-\frac{6}{5}M_1g_1^2(\Lambda_{E^q}^*Y_E)_{lq}
\nonumber\\&&+(16g_3^2-\frac{2}{5}g_1^2)(\Lambda_{D^q}^*h_D)_{lq}+\frac{6}{5}g_1^2(\Lambda_{E^q}^*h_E)_{lq}-6(\Lambda_{D^q}^*h_D)_{ln}Tr(\Lambda_{D^n}\Lambda_{D^q}^{\dagger})
\nonumber\\&&+3(\Lambda_{E^k}^{\dagger}\Lambda_{D^p}\Lambda_{D^p}^{\dagger}h_E+\Lambda_{E^k}^{\dagger}h_EY_E^{\dagger}Y_E+\Lambda_{E^k}^{\dagger}Eh_E)_{lk}
\nonumber\\&&+(\Lambda_{E^k}^{\dagger}\Lambda_{E^p}\Lambda_{E^p}^{\dagger}h_E)_{lk}+(\Lambda_{E^q}^{\dagger}h_E)_{ln}Tr(\Lambda_{E^n}^{\dagger}\Lambda_{E^q})+3(Y_E^*h_E^t\Lambda_{D^m}^*Y_D)_{lm}
\nonumber\\&&-3(Y_E^*h_E^t\Lambda_{E^m}^*Y_E+E^t\Lambda_{E^m}^*h_E)_{lm}-3(\Lambda_{D^k}^*h_UY_U^{\dagger}Y_D)_{lk}
\nonumber\\&&-9(\Lambda_{D^q}^*h_D\tilde{D}+\Lambda_{D^q}^*Dh_D)_{lq}+6(\Lambda_{D^q}^*h_D\Lambda_{U^m}^{\dagger}\Lambda_{U^m})_{lq}
\nonumber\\&&-3(\Lambda_{D^k}^*Th_D)_{lk}-3(\Lambda_{D^k}^*\Lambda_{D^m}^t\Lambda_{D^m}^*h_D)_{lk}+(\Lambda_{E^k}^{\dagger}h_{E^p}\Lambda_{E^p}^{\dagger}Y_E)_{lk}
\nonumber\\&&+(\Lambda_{E^q}^{\dagger}Y_E)_{ln}Tr(\Lambda_{E^n}^{\dagger}h_{E^q})+6(\Lambda_{D^q}^*Y_D\Lambda_{U^m}^{\dagger}h_{U^m})_{lq}+(\Lambda_{E^k}^{\dagger}h_{D^p}\Lambda_{D^p}^{\dagger}Y_E)_{lk}
\nonumber\\&&-6(\Lambda_{D^q}^*Y_D)_{ln}Tr(h_{D^n}\Lambda_{D^q}^{\dagger})-3(\Lambda_{D^k}^*h_{D^m}^t\Lambda_{D^m}^*Y_D)_{lk}+3(E^t\Lambda_{D^m}^*h_D)_{lm}\}
\end{eqnarray}
\begin{eqnarray}
{(16\pi^2)^2}\beta^{(2)}_{(h_D)_{ij}}&=&{(16\pi^2)^2}\beta^{(2)}_{(h_D)_{(RPC)}}+(h_D)_{lj}\{\frac{2}{5}g_1^2(\Lambda_{D^q}^{\dagger}\Lambda_{D^q})_{li}-2(\Lambda_{D^n}^{\dagger}\Lambda_{D^m})_{li}[(\tilde{D})_{mn}+(\Lambda_{U^q}\Lambda_{U^q}^{\dagger})_{nm}
\nonumber\\&&+Tr(\Lambda_{D^m}^{\dagger}\Lambda_{D^n})]-(\Lambda_{D^m}^{\dagger}E\Lambda_{D^m})_{li}-3(\Lambda_{D^m}^{\dagger}\Lambda_{D^q}\Lambda_{D^q}^{\dagger}\Lambda_{D^m})_{li}
\nonumber\\&&-(\Lambda_{D^m}^{\dagger}\Lambda_{E^q}\Lambda_{E^q}^{\dagger}\Lambda_{D^m})_{li}-3(Y_D)_{in}(\Lambda_{D^n}^{\dagger}\Lambda_{D^m}Y_D^*)_{lm}-(Y_D)_{in}(\Lambda_{D^n}^{\dagger}\Lambda_{E^m}Y_E^*)_{lm}
\nonumber\\&&-2(Y_D\Lambda_{U^q}^{\dagger}\Lambda_{U^q}Y_D^{\dagger})_{il}-2(Y_D)_{im}(Y_D^{\dagger})_{nl}Tr(\Lambda_{D^m}^{\dagger}\Lambda_{D^n})
\nonumber\\&&-(Y_D^*)_{ln}(3\Lambda_{D^n}^t\Lambda_{D^m}^*Y_D+\Lambda_{D^n}^t\Lambda_{E^m}^*Y_E)_{im}-(Y_U)_{in}(Y_U^{\dagger})_{ql}Tr(\Lambda_{U^n}^{\dagger}\Lambda_{U^q})\}
\nonumber\\&&-(h_D)_{ij}\{3Tr(D\Lambda_{D^q}^t\Lambda_{D^q}^*)+3Tr(E\Lambda_{D^q}\Lambda_{D^q}^{\dagger})+Tr(E\Lambda_{E^q}\Lambda_{E^q}^{\dagger})
\nonumber\\&&+6Tr(\tilde{D}\Lambda_{U^q}^{\dagger}\Lambda_{U^q})+(\tilde{E})_{mn}Tr(\Lambda_{E^n}^{\dagger}\Lambda_{E^m})+6(\tilde{D})_{nq}Tr(\Lambda_{D^q}^{\dagger}\Lambda_{D^n})\}
\nonumber\\&&+(h_D)_{il}\{(\frac{16}{3}g_3^2+\frac{16}{15}g_1^2)(\Lambda_{U^q}\Lambda_{U^q}^{\dagger})_{jl}+(6g_2^2+\frac{2}{5}g_1^2)Tr(\Lambda_{D^j}\Lambda_{D^l}^{\dagger})
\nonumber\\&&-2(Y_D^t\Lambda_{D^p}^{\dagger}\Lambda_{D^p}Y_D^*)_{jl}-2Tr[\Lambda_{D^j}^t\Lambda_{D^l}^*(D+T)]-2Tr(\Lambda_{D^l}^{\dagger}\Lambda_{D^j}\Lambda_{D^q}^{\dagger}\Lambda_{D^q})
\nonumber\\&&-6Tr(\Lambda_{D^q}\Lambda_{D^q}^{\dagger}\Lambda_{D^j}\Lambda_{D^l}^{\dagger})-2Tr(\Lambda_{E^q}\Lambda_{E^q}^{\dagger}\Lambda_{D^j}\Lambda_{D^l}^{\dagger})-2Tr(E\Lambda_{D^j}\Lambda_{D^l}^{\dagger})
\nonumber\\&&-4(\Lambda_{U^m}\tilde{D}\Lambda_{U^m}^{\dagger})_{jl}-4(\Lambda_{U^m}^{\dagger}\Lambda_{U^p}\Lambda_{U^p}^{\dagger}\Lambda_{U^m})_{lj}+4(\Lambda_{U^m}^{\dagger})_{lq}(\Lambda_{U^m})_{jn}Tr(\Lambda_{D^n}\Lambda_{D^q}^{\dagger})
\nonumber\\&&-6(Y_D^{\dagger}\Lambda_{D^j}^t\Lambda_{D^p}^*Y_D)_{lp}-2(Y_D^{\dagger}\Lambda_{D^j}^t\Lambda_{E^p}^*Y_E)_{lp}-6(Y_D^t\Lambda_{D^l}^{\dagger}\Lambda_{D^p}Y_D^*)_{jp}
\nonumber\\&&-2(Y_D^t\Lambda_{D^l}^{\dagger}\Lambda_{E^p}Y_E^*)_{jp}-2(\Lambda_{U^q}^{\dagger}\Lambda_{U^n})_{lj}[Tr(\Lambda_{U^n}^{\dagger}\Lambda_{U^q})+2(\tilde{T})_{nq}]\}
\nonumber\\&&-(h_{D^j})_{li}\{(\frac{2}{5}g_1^2-16g_3^2)(\Lambda_{D^q}^*Y_D)_{lq}-\frac{6}{5}g_1^2(\Lambda_{E^q}^*Y_E)_{lq}
\nonumber\\&&-3(\Lambda_{E^k}^{\dagger}\Lambda_{D^p}\Lambda_{D^p}^{\dagger}Y_E)_{lk}-3(\Lambda_{E^k}^{\dagger}EY_E)_{lk}-(\Lambda_{E^k}^{\dagger}\Lambda_{E^p}\Lambda_{E^p}^{\dagger}Y_E)_{lk}
\nonumber\\&&-(\Lambda_{E^q}^{\dagger}Y_E)_{ln}Tr(\Lambda_{E^n}^{\dagger}\Lambda_{E^q})-3(E^t\Lambda_{D^m}^*Y_D-E^t\Lambda_{E^m}^*Y_E)_{lm}
\nonumber\\&&+9(\Lambda_{D^q}^*DY_D)_{lq}-6(\Lambda_{D^q}^*Y_D\Lambda_{U^m}^{\dagger}\Lambda_{U^m})_{lq}+6(\Lambda_{D^q}^*Y_D)_{ln}Tr(\Lambda_{D^n}\Lambda_{D^q}^{\dagger})
\nonumber\\&&+3(\Lambda_{D^k}^*TY_D)_{lk}+3(\Lambda_{D^k}^*\Lambda_{D^m}^t\Lambda_{D^m}^*Y_D)_{lk}\}
\nonumber\\&&-2(Y_D)_{lj}\{\frac{2}{5}g_1^2[M_1(\Lambda_{D^q}^{\dagger}\Lambda_{D^q})_{li}-(\Lambda_{D^q}^{\dagger}h_{D^q})_{li}]+(\Lambda_{D^m}^{\dagger}h_EY_E^{\dagger}\Lambda_{D^m})_{li}
\nonumber\\&&+(Y_D^*)_{ln}(\Lambda_{D^n}^t\Lambda_{E^m}^*h_E)_{im}+(h_U)_{in}(Y_U^{\dagger})_{ql}Tr(\Lambda_{U^n}^{\dagger}\Lambda_{U^q})
\nonumber\\&&+2(\Lambda_{D^n}^{\dagger}\Lambda_{D^m})_{li}(Y_D^{\dagger}h_D)_{mn}+3(h_D)_{in}(\Lambda_{D^n}^{\dagger}\Lambda_{D^m}Y_D^*)_{lm}
\nonumber\\&&+(h_D)_{in}(\Lambda_{D^n}^{\dagger}\Lambda_{E^m}Y_E^*)_{lm}+2(h_D\Lambda_{U^q}^{\dagger}\Lambda_{U^q}Y_D^{\dagger})_{il}+2(h_D)_{im}(Y_D^{\dagger})_{nl}Tr(\Lambda_{D^m}^{\dagger}\Lambda_{D^n})
\nonumber\\&&+3(Y_D^*)_{ln}(\Lambda_{D^n}^t\Lambda_{D^m}^*h_D)_{im}+(\Lambda_{D^m}^{\dagger}h_{E^q}\Lambda_{E^q}^{\dagger}\Lambda_{D^m})_{li}+(Y_D)_{in}(\Lambda_{D^n}^{\dagger}h_{E^m}Y_E^*)_{lm}
\nonumber\\&&+2(\Lambda_{D^n}^{\dagger}\Lambda_{D^m})_{li}(h_{U^q}\Lambda_{U^q}^{\dagger})_{nm}+2(Y_D\Lambda_{U^q}^{\dagger}h_{U^q}Y_D^{\dagger})_{il}+(Y_U)_{in}(Y_U^{\dagger})_{ql}Tr(\Lambda_{U^n}^{\dagger}h_{U^q})
\nonumber\\&&+2(\Lambda_{D^n}^{\dagger}h_{D^m})_{li}[(\tilde{D})_{mn}+(\Lambda_{U^q}\Lambda_{U^q}^{\dagger})_{nm}+Tr(\Lambda_{D^m}^{\dagger}\Lambda_{D^n})]
\nonumber\\&&+2(\Lambda_{D^n}^{\dagger}\Lambda_{D^m})_{li}Tr(\Lambda_{D^m}^{\dagger}h_{D^n})+(\Lambda_{D^m}^{\dagger}Eh_{D^m})_{li}+(Y_D^*)_{ln}(h_{D^n}^t\Lambda_{E^m}^*Y_E)_{im}
\nonumber\\&&+3(\Lambda_{D^m}^{\dagger}h_{D^q}\Lambda_{D^q}^{\dagger}\Lambda_{D^m})_{li}+3(\Lambda_{D^m}^{\dagger}\Lambda_{D^q}\Lambda_{D^q}^{\dagger}h_{D^m})_{li}+(\Lambda_{D^m}^{\dagger}\Lambda_{E^q}\Lambda_{E^q}^{\dagger}h_{D^m})_{li}
\nonumber\\&&+3(Y_D)_{in}(\Lambda_{D^n}^{\dagger}h_{D^m}Y_D^*)_{lm}+2(Y_D)_{im}(Y_D^{\dagger})_{nl}Tr(\Lambda_{D^m}^{\dagger}h_{D^n})+3(Y_D^*)_{ln}(h_{D^n}^t\Lambda_{D^m}^*Y_D)_{im}\}
\nonumber\\&&-2(Y_D)_{ij}\{3Tr(h_EY_E^{\dagger}\Lambda_{D^q}\Lambda_{D^q}^{\dagger})+Tr(h_EY_E^{\dagger}\Lambda_{E^q}\Lambda_{E^q}^{\dagger})+(Y_E^{\dagger}h_E)_{mn}Tr(\Lambda_{E^m}\Lambda_{E^n}^{\dagger})
\nonumber\\&&+6Tr(Y_D^{\dagger}h_D\Lambda_{U^q}^{\dagger}\Lambda_{U^q})+3Tr(h_DY_D^{\dagger}\Lambda_{D^q}^t\Lambda_{D^q}^*)+6(Y_D^{\dagger}h_D)_{lq}Tr(\Lambda_{D^q}^{\dagger}\Lambda_{D^l})
\nonumber\\&&+Tr(Eh_{E^q}\Lambda_{E^q}^{\dagger})+(\tilde{E})_{mn}Tr(\Lambda_{E^n}^{\dagger}h_{E^m})+6Tr(\tilde{D}\Lambda_{U^q}^{\dagger}h_{U^q})
\nonumber\\&&+3Tr(Eh_{D^q}\Lambda_{D^q}^{\dagger})+3Tr(Dh_{D^q}^t\Lambda_{D^q}^*)+6(\tilde{D})_{lq}Tr(\Lambda_{D^q}^{\dagger}h_{D^l})\}
\nonumber\\&&-2(Y_D)_{il}\{(\frac{16}{3}M_3g_3^2+\frac{16}{15}M_1g_1^2)(\Lambda_{U^q}\Lambda_{U^q}^{\dagger})_{jl}
\nonumber\\&&-(\frac{16}{3}g_3^2+\frac{16}{15}g_1^2)(h_{U^q}\Lambda_{U^q}^{\dagger})_{jl}+(6M_2g_2^2+\frac{2}{5}M_1g_1^2)Tr(\Lambda_{D^j}\Lambda_{D^l}^{\dagger})
\nonumber\\&&-(6g_2^2+\frac{2}{5}g_1^2)Tr(h_{D^j}\Lambda_{D^l}^{\dagger})+2Tr(h_EY_E^{\dagger}\Lambda_{D^j}\Lambda_{D^l}^{\dagger})+2(Y_D^{\dagger}\Lambda_{D^j}^t\Lambda_{E^p}^*h_E)_{lp}
\nonumber\\&&+2Tr(\Lambda_{D^j}^t\Lambda_{D^l}^*h_UY_U^{\dagger})+4(\Lambda_{U^q}^{\dagger}\Lambda_{U^n})_{lj}(Y_U^{\dagger}h_U)_{nq}
\nonumber\\&&+2(h_D^t\Lambda_{D^p}^{\dagger}\Lambda_{D^p}Y_D^*)_{jl}+2Tr(\Lambda_{D^j}^t\Lambda_{D^l}^*h_DY_D^{\dagger})+4(\Lambda_{U^m}Y_D^{\dagger}h_D\Lambda_{U^m}^{\dagger})_{jl}
\nonumber\\&&+6(Y_D^{\dagger}\Lambda_{D^j}^t\Lambda_{D^p}^*h_D)_{lp}+6(h_D^t\Lambda_{D^l}^{\dagger}\Lambda_{D^p}Y_D^*)_{jp}+2(h_D^t\Lambda_{D^l}^{\dagger}\Lambda_{E^p}Y_E^*)_{jp}
\nonumber\\&&+2Tr(h_{E^q}\Lambda_{E^q}^{\dagger}\Lambda_{D^j}\Lambda_{D^l}^{\dagger})+2(Y_D^t\Lambda_{D^l}^{\dagger}h_{E^p}Y_E^*)_{jp}+4(h_{U^m}\tilde{D}\Lambda_{U^m}^{\dagger})_{jl}
\nonumber\\&&+4(\Lambda_{U^m}^{\dagger}h_{U^p}\Lambda_{U^p}^{\dagger}\Lambda_{U^m})_{lj}+4(\Lambda_{U^m}^{\dagger}\Lambda_{U^p}\Lambda_{U^p}^{\dagger}h_{U^m})_{lj}
\nonumber\\&&+2(\Lambda_{U^q}^{\dagger}h_{U^n})_{lj}Tr(\Lambda_{U^n}^{\dagger}\Lambda_{U^q})+4(\Lambda_{U^q}^{\dagger}h_{U^n})_{lj}(\tilde{T})_{nq}+2(\Lambda_{U^q}^{\dagger}\Lambda_{U^n})_{lj}Tr(\Lambda_{U^n}^{\dagger}h_{U^q})
\nonumber\\&&+2(Y_D^t\Lambda_{D^p}^{\dagger}h_{D^p}Y_D^*)_{jl}+2Tr(h_{D^j}^t\Lambda_{D^l}^*D)+2Tr(h_{D^j}^t\Lambda_{D^l}^*T)
\nonumber\\&&+6Tr(h_{D^q}\Lambda_{D^q}^{\dagger}\Lambda_{D^j}\Lambda_{D^l}^{\dagger})+6Tr(\Lambda_{D^q}\Lambda_{D^q}^{\dagger}h_{D^j}\Lambda_{D^l}^{\dagger})+2Tr(\Lambda_{E^q}\Lambda_{E^q}^{\dagger}h_{D^j}\Lambda_{D^l}^{\dagger}
\nonumber\\&&+Eh_{D^j}\Lambda_{D^l}^{\dagger})-4(\Lambda_{U^m}^{\dagger})_{lq}(\Lambda_{U^m})_{jn}Tr(h_{D^n}\Lambda_{D^q}^{\dagger})+6(Y_D^{\dagger}h_{D^j}^t\Lambda_{D^p}^*Y_D)_{lp}
\nonumber\\&&+2(Y_D^{\dagger}h_{D^j}^t\Lambda_{E^p}^*Y_E)_{lp}+6(Y_D^t\Lambda_{D^l}^{\dagger}h_{D^p}Y_D^*)_{jp}+2Tr(\Lambda_{D^l}^{\dagger}h_{D^j}\Lambda_{D^q}^{\dagger}\Lambda_{D^q})
\nonumber\\&&+2Tr(\Lambda_{D^l}^{\dagger}\Lambda_{D^j}\Lambda_{D^q}^{\dagger}h_{D^q})-4(\Lambda_{U^m}^{\dagger})_{lq}(h_{U^m})_{jn}Tr(\Lambda_{D^n}\Lambda_{D^q}^{\dagger})\}
\nonumber\\&&+2(\Lambda_{D^j})_{li}\{(\frac{2}{5}M_1g_1^2-16M_3g_3^2)(\Lambda_{D^q}^*Y_D)_{lq}-\frac{6}{5}M_1g_1^2(\Lambda_{E^q}^*Y_E)_{lq}
\nonumber\\&&+(16g_3^2-\frac{2}{5}g_1^2)(\Lambda_{D^q}^*h_D)_{lq}+\frac{6}{5}g_1^2(\Lambda_{E^q}^*h_E)_{lq}+3(\Lambda_{E^q}^{\dagger}\Lambda_{D^p}\Lambda_{D^p}^{\dagger}h_E)_{lq}
\nonumber\\&&+(\Lambda_{E^q}^{\dagger}\Lambda_{E^p}\Lambda_{E^p}^{\dagger}h_E)_{lq}+3(\Lambda_{E^q}^{\dagger}h_E\tilde{E}+\Lambda_{E^q}^{\dagger}Eh_E)_{lq}
\nonumber\\&&+(\Lambda_{E^q}^{\dagger}h_E)_{ln}Tr(\Lambda_{E^n}^{\dagger}\Lambda_{E^q})-3(\Lambda_{D^q}^*h_UY_U^{\dagger}Y_D+\Lambda_{D^q}^*h_{D^m}^t\Lambda_{D^m}^*Y_D)_{lq}
\nonumber\\&&+3(Y_E^*h_E^t\Lambda_{D^m}^*Y_D-Y_E^*h_E^t\Lambda_{E^m}^*Y_E-E^t\Lambda_{E^m}^*h_E)_{lm}
\nonumber\\&&+3(E^t\Lambda_{D^m}^*h_D)_{lm}-9(\Lambda_{D^q}^*h_D\tilde{D}+\Lambda_{D^q}^*Dh_D)_{lq}
\nonumber\\&&+6(\Lambda_{D^q}^*h_D\Lambda_{U^m}^{\dagger}\Lambda_{U^m})_{lq}-6(\Lambda_{D^q}^*h_D)_{ln}Tr(\Lambda_{D^n}\Lambda_{D^q}^{\dagger})-3(\Lambda_{D^q}^*Th_D)_{lq}
\nonumber\\&&-3(\Lambda_{D^q}^*\Lambda_{D^m}^t\Lambda_{D^m}^*h_D)_{lq}+(\Lambda_{E^q}^{\dagger}h_{E^p}\Lambda_{E^p}^{\dagger}Y_E)_{lq}+(\Lambda_{E^q}^{\dagger}Y_E)_{ln}Tr(\Lambda_{E^n}^{\dagger}h_{E^q})
\nonumber\\&&+6(\Lambda_{D^q}^*Y_D\Lambda_{U^m}^{\dagger}h_{U^m})_{lq}+3(\Lambda_{E^q}^{\dagger}h_{D^p}\Lambda_{D^p}^{\dagger}Y_E)_{lq}-6(\Lambda_{D^q}^*Y_D)_{ln}Tr(h_{D^n}\Lambda_{D^q}^{\dagger})\}
\end{eqnarray}
\begin{eqnarray}
{(16\pi^2)^2}\beta^{(2)}_{(h_U)_{ij}}&=&{(16\pi^2)^2}\beta^{(2)}_{(h_U)_{(RPC)}}
\nonumber\\&&+(h_U)_{il}\{(\frac{8}{3}g_3^2-\frac{4}{15}g_1^2)Tr(\Lambda_{U^j}\Lambda_{U^l}^{\dagger})-4Tr(\Lambda_{U^j}\Lambda_{U^l}^{\dagger}\Lambda_{U^p}\Lambda_{U^p}^{\dagger})
\nonumber\\&&-2(Y_U^{\dagger}\Lambda_{D^m}^t\Lambda_{D^m}^*Y_U)_{lj}-4(\Lambda_{U^l}^{\dagger}\Lambda_{U^j})_{nm}Tr(\Lambda_{D^m}^{\dagger}\Lambda_{D^n})
\nonumber\\&&-4Tr(\Lambda_{U^l}^{\dagger}\Lambda_{U^j}\tilde{D})\}+(h_U)_{lj}\{\frac{2}{5}g_1^2(\Lambda_{D^q}^{\dagger}\Lambda_{D^q})_{li}-(\Lambda_{D^m}^{\dagger}E\Lambda_{D^m})_{li}
\nonumber\\&&-2(\Lambda_{D^n}^{\dagger}\Lambda_{D^m})_{li}[(\tilde{D})_{mn}+(\Lambda_{U^q}\Lambda_{U^q}^{\dagger})_{nm}+Tr(\Lambda_{D^m}^{\dagger}\Lambda_{D^n})]
\nonumber\\&&-3(\Lambda_{D^m}^{\dagger}\Lambda_{D^q}\Lambda_{D^q}^{\dagger}\Lambda_{D^m})_{li}-(\Lambda_{D^m}^{\dagger}\Lambda_{E^q}\Lambda_{E^q}^{\dagger}\Lambda_{D^m})_{li}-(Y_D)_{in}(3\Lambda_{D^n}^{\dagger}\Lambda_{D^m}Y_D^*
\nonumber\\&&+\Lambda_{D^n}^{\dagger}\Lambda_{E^m}Y_E^*)_{lm}-2(Y_D\Lambda_{U^q}^{\dagger}\Lambda_{U^q}Y_D^{\dagger})_{il}-2(Y_D)_{im}(Y_D^{\dagger})_{nl}Tr(\Lambda_{D^m}^{\dagger}\Lambda_{D^n})
\nonumber\\&&-(Y_D^*)_{ln}(3\Lambda_{D^n}^t\Lambda_{D^m}^*Y_D+\Lambda_{D^n}^t\Lambda_{E^m}^*Y_E)_{im}-(Y_U)_{in}(Y_U^{\dagger})_{ql}Tr(\Lambda_{U^n}^{\dagger}\Lambda_{U^q})\}
\nonumber\\&&-(h_U)_{ij}\{3Tr(T\Lambda_{D^q}^t\Lambda_{D^q}^*)+3\tilde{T}_{lq}Tr(\Lambda_{U^q}^{\dagger}\Lambda_{U^l})\}
\nonumber\\&&-2(Y_U)_{il}\{(\frac{8}{3}M_3g_3^2-\frac{4}{15}M_1g_1^2)Tr(\Lambda_{U^j}\Lambda_{U^l}^{\dagger})+(\frac{4}{15}g_1^2-\frac{8}{3}g_3^2)Tr(h_{U^j}\Lambda_{U^l}^{\dagger})
\nonumber\\&&+2(Y_U^{\dagger}\Lambda_{D^m}^t\Lambda_{D^m}^*h_U)_{lj}+4Tr(\Lambda_{U^l}^{\dagger}\Lambda_{U^j}Y_D^{\dagger}h_D)+4(\Lambda_{U^l}^{\dagger}h_{U^j})_{nm}Tr(\Lambda_{D^m}^{\dagger}\Lambda_{D^n})
\nonumber\\&&+4Tr(h_{U^j}\Lambda_{U^l}^{\dagger}\Lambda_{U^p}\Lambda_{U^p}^{\dagger})+4Tr(\Lambda_{U^j}\Lambda_{U^l}^{\dagger}h_{U^p}\Lambda_{U^p}^{\dagger})+4Tr(\Lambda_{U^l}^{\dagger}h_{U^j}\tilde{D})
\nonumber\\&&+2(Y_U^{\dagger}h_{D^m}^t\Lambda_{D^m}^*Y_U)_{lj}+4(\Lambda_{U^l}^{\dagger}\Lambda_{U^j})_{nm}Tr(\Lambda_{D^m}^{\dagger}h_{D^n})\}
\nonumber\\&&-2(Y_U)_{lj}\{\frac{2}{5}M_1g_1^2(\Lambda_{D^q}^{\dagger}\Lambda_{D^q})_{li}-\frac{2}{5}g_1^2(\Lambda_{D^q}^{\dagger}h_{D^q})_{li}
\nonumber\\&&+(\Lambda_{D^m}^{\dagger}h_EY_E^{\dagger}\Lambda_{D^m})_{li}+(Y_D^*)_{ln}(\Lambda_{D^n}^t\Lambda_{E^m}^*h_E)_{im}
\nonumber\\&&+(h_U)_{in}(Y_U^{\dagger})_{ql}Tr(\Lambda_{U^n}^{\dagger}\Lambda_{U^q})+2(\Lambda_{D^n}^{\dagger}\Lambda_{D^m})_{li}(Y_D^{\dagger}h_D)_{mn}
\nonumber\\&&+(h_D)_{in}(3\Lambda_{D^n}^{\dagger}\Lambda_{D^m}Y_D^*+\Lambda_{D^n}^{\dagger}\Lambda_{E^m}Y_E^*)_{lm}+2(h_D\Lambda_{U^q}^{\dagger}\Lambda_{U^q}Y_D^{\dagger})_{il}
\nonumber\\&&+2(h_D)_{im}(Y_D^{\dagger})_{nl}Tr(\Lambda_{D^m}^{\dagger}\Lambda_{D^n})+3(Y_D^*)_{ln}(\Lambda_{D^n}^t\Lambda_{D^m}^*h_D)_{im}
\nonumber\\&&+(Y_D)_{in}(\Lambda_{D^n}^{\dagger}h_{E^m}Y_E^*)_{lm}+2(\Lambda_{D^n}^{\dagger}\Lambda_{D^m})_{li}(h_{U^q}\Lambda_{U^q}^{\dagger})_{nm}+2(Y_D\Lambda_{U^q}^{\dagger}h_{U^q}Y_D^{\dagger})_{il}
\nonumber\\&&+(Y_U)_{in}(Y_U^{\dagger})_{ql}Tr(\Lambda_{U^n}^{\dagger}h_{U^q})+(\Lambda_{D^m}^{\dagger}h_{E^q}\Lambda_{E^q}^{\dagger}\Lambda_{D^m})_{li}
\nonumber\\&&+2(\Lambda_{D^n}^{\dagger}h_{D^m})_{li}[(\tilde{D})_{mn}+(\Lambda_{U^q}\Lambda_{U^q}^{\dagger})_{nm}+Tr(\Lambda_{D^m}^{\dagger}\Lambda_{D^n})]
\nonumber\\&&+2(\Lambda_{D^n}^{\dagger}\Lambda_{D^m})_{li}Tr(\Lambda_{D^m}^{\dagger}h_{D^n})+(\Lambda_{D^m}^{\dagger}Eh_{D^m})_{li}+3(\Lambda_{D^m}^{\dagger}h_{D^q}\Lambda_{D^q}^{\dagger}\Lambda_{D^m})_{li}
\nonumber\\&&+3(\Lambda_{D^m}^{\dagger}\Lambda_{D^q}\Lambda_{D^q}^{\dagger}h_{D^m})_{li}+(\Lambda_{D^m}^{\dagger}\Lambda_{E^q}\Lambda_{E^q}^{\dagger}h_{D^m})_{li}+3(Y_D)_{in}(\Lambda_{D^n}^{\dagger}h_{D^m}Y_D^*)_{lm}
\nonumber\\&&+2(Y_D)_{im}(Y_D^{\dagger})_{nl}Tr(\Lambda_{D^m}^{\dagger}h_{D^n})+3(Y_D^*)_{ln}(h_{D^n}^t\Lambda_{D^m}^*Y_D)_{im}
\nonumber\\&&+(Y_D^*)_{ln}(h_{D^n}^t\Lambda_{E^m}^*Y_E)_{im}\}-6(Y_U)_{ij}\{(\tilde{T})_{lq}Tr(\Lambda_{U^q}^{\dagger}h_{U^l})+Tr(Th_{D^q}^t\Lambda_{D^q}^*)
\nonumber\\&&+Tr(h_UY_U^{\dagger}\Lambda_{D^q}^t\Lambda_{D^q}^*)+(Y_U^{\dagger}h_U)_{lq}Tr(\Lambda_{U^q}^{\dagger}\Lambda_{U^l})\}
\end{eqnarray}
\begin{eqnarray}
{(16\pi^2)^2}\beta^{(2)}_{(h_{E^k})_{ij}}&=&(h_{E^l})_{ij}\{(6g_2^2-\frac{6}{5}g_1^2)(\tilde{E})_{lk}+(3g_2^2-\frac{3}{5}g_1^2)Tr(\Lambda_{E^k}\Lambda_{E^l}^{\dagger})
\nonumber\\&&-2(\tilde{E}^2+Y_E^{\dagger}\Lambda_{E^q}\Lambda_{E^q}^{\dagger}Y_E+3Y_E^{\dagger}\Lambda_{D^q}\Lambda_{D^q}^{\dagger}Y_E)_{lk}
\nonumber\\&&-2(\tilde{E})_{lk}Tr(E+3D)+2(3Y_E^{\dagger}\Lambda_{E^k}\Lambda_{D^m}^*Y_D
\nonumber\\&&-Y_E^{\dagger}\Lambda_{E^k}\Lambda_{E^m}^{\dagger}Y_E)_{lm}-2Tr[\Lambda_{E^k}\Lambda_{E^l}^{\dagger}(E+\Lambda_{E^q}\Lambda_{E^q}^{\dagger}+3\Lambda_{D^q}\Lambda_{D^q}^{\dagger})]
\nonumber\\&&+2(3Y_D^{\dagger}\Lambda_{D^m}^t\Lambda_{E^l}^{\dagger}Y_E-Y_E^{\dagger}\Lambda_{E^m}\Lambda_{E^l}^{\dagger}Y_E)_{mk}\}
\nonumber\\&&+(h_E)_{ik}\{(\frac{2}{5}g_1^2-16g_3^2)(\Lambda_{D^q}Y_D^*)_{jq}+6(\Lambda_{D^q}Y_D^*)_{jl}Tr(\Lambda_{D^l}^*\Lambda_{D^q}^t)
\nonumber\\&&-(3\Lambda_{E^q}^t\Lambda_{D^p}^*\Lambda_{D^p}^tY_E^*+\Lambda_{E^q}^t\Lambda_{E^p}^*\Lambda_{E^p}^tY_E^*+3\Lambda_{E^q}^tE^tY_E^*)_{jq}
\nonumber\\&&-(Y_E^{\dagger}\Lambda_{E^q})_{lj}Tr(\Lambda_{E^q}^{\dagger}\Lambda_{E^l})-3(E\Lambda_{D^m}Y_D^*-E\Lambda_{E^m}Y_E^*)_{jm}
\nonumber\\&&+9(\Lambda_{D^q}D^tY_D^*)_{jq}-6(\Lambda_{D^q}Y_D^*\Lambda_{U^m}^t\Lambda_{U^m}^*)_{jq}-\frac{6}{5}g_1^2(\Lambda_{E^q}Y_E^*)_{jq}
\nonumber\\&&+3(\Lambda_{D^q}Y_U^*Y_U^tY_D^*)_{jq}+3(\Lambda_{D^q}\Lambda_{D^m}^{\dagger}\Lambda_{D^m}Y_D^*)_{jq}\}
\nonumber\\&&+(h_{E^k})_{lj}\{(16g_3^2-\frac{2}{5}g_1^2)(\Lambda_{D^q}\Lambda_{D^q}^{\dagger})_{il}+\frac{6}{5}g_1^2(E+\Lambda_{E^q}\Lambda_{E^q}^{\dagger})_{il}
\nonumber\\&&-2(E^2)_{il}-(Y_E^{\dagger})_{nl}(Y_E)_{im}Tr(\Lambda_{E^m}^{\dagger}\Lambda_{E^n})-2(\Lambda_{E^m}\Lambda_{E^n}^{\dagger})_{il}(\tilde{E})_{mn}
\nonumber\\&&-(\Lambda_{E^m}\Lambda_{E^n}^{\dagger})_{il}Tr(\Lambda_{E^m}^{\dagger}\Lambda_{E^n})-(E)_{il}Tr(E+3D)
\nonumber\\&&-(Y_E)_{in}(3\Lambda_{E^n}^{\dagger}\Lambda_{D^p}Y_D^*+\Lambda_{E^n}^{\dagger}\Lambda_{E^p}Y_E^*)_{lp}+(Y_E^*)_{ln}(3\Lambda_{E^n}\Lambda_{D^p}^*Y_D
\nonumber\\&&+\Lambda_{E^n}\Lambda_{E^p}^*Y_E)_{ip}-(\Lambda_{E^n}^{\dagger}E\Lambda_{E^n}+3\Lambda_{E^n}^{\dagger}\Lambda_{D^p}\Lambda_{D^p}^{\dagger}\Lambda_{E^n}+\Lambda_{E^n}^{\dagger}\Lambda_{E^p}\Lambda_{E^p}^{\dagger}\Lambda_{E^n})_{li}
\nonumber\\&&-6(\Lambda_{D^m}\Lambda_{D^n}^{\dagger})_{il}[(\tilde{D})_{mn}+Tr(\Lambda_{D^m}^{\dagger}\Lambda_{D^n})+(\Lambda_{U^q}\Lambda_{U^q}^{\dagger})_{nm}]
\nonumber\\&&-3(\Lambda_{D^n}^*D\Lambda_{D^n}^t+\Lambda_{D^n}^*T\Lambda_{D^n}^t)_{li}-3(\Lambda_{D^n}\Lambda_{D^p}^{\dagger}\Lambda_{D^p}\Lambda_{D^n}^{\dagger})_{il}\}
\nonumber\\&&-(h_E)_{jk}\{(\frac{2}{5}g_1^2-16g_3^2)(\Lambda_{D^q}Y_D^*)_{iq}-\frac{6}{5}g_1^2(\Lambda_{E^q}Y_E^*)_{iq}
\nonumber\\&&-(3\Lambda_{E^n}^t\Lambda_{D^p}^*\Lambda_{D^p}^tY_E^*+\Lambda_{E^n}^t\Lambda_{E^p}^*\Lambda_{E^p}^tY_E^*+3\Lambda_{E^n}^tE^tY_E^*)_{in}
\nonumber\\&&-3(E\Lambda_{D^m}Y_D^*-E\Lambda_{E^m}Y_E^*)_{im}+9(\Lambda_{D^q}D^tY_D^*)_{iq}
\nonumber\\&&-6(\Lambda_{D^q}Y_D^*\Lambda_{U^m}^t\Lambda_{U^m}^*)_{iq}+6(\Lambda_{D^q}Y_D^*)_{il}Tr(\Lambda_{D^l}^*\Lambda_{D^q}^t)
\nonumber\\&&+3(\Lambda_{D^n}Y_U^*Y_U^tY_D^*)_{in}+3(\Lambda_{D^n}\Lambda_{D^m}^{\dagger}\Lambda_{D^m}Y_D^*)_{in}-(\Lambda_{E^p}^tY_E^*)_{il}Tr(\Lambda_{E^l}^t\Lambda_{E^p}^*)\}
\nonumber\\&&+(h_{E^k})_{il}\{(16g_3^2-\frac{2}{5}g_1^2)(\Lambda_{D^q}\Lambda_{D^q}^{\dagger})_{jl}+\frac{6}{5}g_1^2(E+\Lambda_{E^q}\Lambda_{E^q}^{\dagger})_{jl}
\nonumber\\&&-2(E^2)_{jl}-(Y_E^{\dagger})_{nl}(Y_E)_{jm}Tr(\Lambda_{E^m}^{\dagger}\Lambda_{E^n})-2(\Lambda_{E^m}\Lambda_{E^n}^{\dagger})_{jl}(\tilde{E})_{mn}
\nonumber\\&&-(\Lambda_{E^m}\Lambda_{E^n}^{\dagger})_{jl}Tr(\Lambda_{E^m}^{\dagger}\Lambda_{E^n})-E_{jl}Tr(E+3D)
\nonumber\\&&-(Y_E)_{jn}(3\Lambda_{E^n}^{\dagger}\Lambda_{D^p}Y_D^*+\Lambda_{E^n}^{\dagger}\Lambda_{E^p}Y_E^*)_{lp}+(Y_E^*)_{ln}(3\Lambda_{E^n}\Lambda_{D^p}^*Y_D
\nonumber\\&&+\Lambda_{E^n}\Lambda_{E^p}^*Y_E)_{jp}-(\Lambda_{E^n}^{\dagger}E\Lambda_{E^n}+3\Lambda_{E^n}^{\dagger}\Lambda_{D^p}\Lambda_{D^p}^{\dagger}\Lambda_{E^n}+\Lambda_{E^n}^{\dagger}\Lambda_{E^p}\Lambda_{E^p}^{\dagger}\Lambda_{E^n})_{lj}
\nonumber\\&&-6(\Lambda_{D^m}\Lambda_{D^n}^{\dagger})_{jl}[(\tilde{D})_{mn}+Tr(\Lambda_{D^m}^{\dagger}\Lambda_{D^n})+(\Lambda_{U^q}\Lambda_{U^q}^{\dagger})_{nm}]
\nonumber\\&&-3(\Lambda_{D^n}^*D\Lambda_{D^n}^t+\Lambda_{D^n}^*T\Lambda_{D^n}^t)_{lj}-3(\Lambda_{D^n}\Lambda_{D^p}^{\dagger}\Lambda_{D^p}\Lambda_{D^n}^{\dagger})_{jl}\}
\nonumber\\&&-2(\Lambda_{E^l})_{ij}\{(6M_2g_2^2-\frac{6}{5}M_1g_1^2)(\tilde{E})_{lk}+(3M_2g_2^2-\frac{3}{5}M_1g_1^2)Tr(\Lambda_{E^k}\Lambda_{E^l}^{\dagger})
\nonumber\\&&-(6g_2^2-\frac{6}{5}g_1^2)(Y_E^{\dagger}h_E)_{lk}-(3g_2^2-\frac{3}{5}g_1^2)Tr(h_{E^k}\Lambda_{E^l}^{\dagger})+2(Y_E^{\dagger}h_E\tilde{E}
\nonumber\\&&+Y_E^{\dagger}Eh_E+Y_E^{\dagger}\Lambda_{E^q}\Lambda_{E^q}^{\dagger}h_E+3Y_E^{\dagger}\Lambda_{D^q}\Lambda_{D^q}^{\dagger}h_E)_{lk}
\nonumber\\&&+2(Y_E^{\dagger}h_E)_{lk}Tr(E+3D)+2(\tilde{E})_{lk}Tr(Y_E^{\dagger}h_E)
\nonumber\\&&+2(Y_E^{\dagger}\Lambda_{E^k}\Lambda_{E^m}^{\dagger}h_E)_{lm}+2Tr(\Lambda_{E^k}\Lambda_{E^l}^{\dagger}h_EY_E^{\dagger})
\nonumber\\&&-2(3Y_D^{\dagger}\Lambda_{D^m}^t\Lambda_{E^l}^{\dagger}h_E-Y_E^{\dagger}\Lambda_{E^m}\Lambda_{E^l}^{\dagger}h_E)_{mk}+6(\tilde{E})_{lk}Tr(Y_D^{\dagger}h_D)
\nonumber\\&&+2(Y_E^{\dagger}h_{E^k}\Lambda_{E^m}^{\dagger}Y_E)_{lm}-6(Y_E^{\dagger}\Lambda_{E^k}\Lambda_{D^m}^*h_D)_{lm}+2(Y_E^{\dagger}h_{E^q}\Lambda_{E^q}^{\dagger}Y_E)_{lk}
\nonumber\\&&-6(Y_E^{\dagger}h_{E^k}\Lambda_{D^m}^*Y_D)_{lm}+2Tr(h_{E^k}\Lambda_{E^l}^{\dagger}E)+6Tr(h_{E^k}\Lambda_{E^l}^{\dagger}\Lambda_{D^q}\Lambda_{D^q}^{\dagger})
\nonumber\\&&+2Tr(h_{E^k}\Lambda_{E^l}^{\dagger}\Lambda_{E^q}\Lambda_{E^q}^{\dagger})+2Tr(\Lambda_{E^k}\Lambda_{E^l}^{\dagger}h_{E^q}\Lambda_{E^q}^{\dagger})+2(Y_E^{\dagger}h_{E^m}\Lambda_{E^l}^{\dagger}Y_E)_{mk}
\nonumber\\&&+6(Y_E^{\dagger}h_{D^q}\Lambda_{D^q}^{\dagger}Y_E)_{lk}+6Tr(\Lambda_{E^k}\Lambda_{E^l}^{\dagger}h_{D^q}\Lambda_{D^q}^{\dagger})-6(Y_D^{\dagger}h_{D^m}^t\Lambda_{E^l}^{\dagger}Y_E)_{mk}\}
\nonumber\\&&-2(Y_E)_{ik}\{(\frac{2}{5}M_1g_1^2-16M_3g_3^2)(\Lambda_{D^q}Y_D^*)_{jq}-\frac{6}{5}M_1g_1^2(\Lambda_{E^q}Y_E^*)_{jq}
\nonumber\\&&-(\frac{2}{5}g_1^2-16g_3^2)(h_{D^q}Y_D^*)_{jq}+\frac{6}{5}g_1^2(h_{E^q}Y_E^*)_{jq}+(3\Lambda_{E^n}^tY_E^*h_E^tY_E^*)_{jn}
\nonumber\\&&+3(h_EY_E^{\dagger}\Lambda_{D^m}Y_D^*-h_EY_E^{\dagger}\Lambda_{E^m}Y_E^*)_{jm}-3(\Lambda_{D^n}Y_U^*h_U^tY_D^*)_{jn}
\nonumber\\&&-9(\Lambda_{D^q}Y_D^*h_D^tY_D^*)_{jq}+3(h_{E^n}^t\Lambda_{D^p}^*\Lambda_{D^p}^tY_E^*+h_{E^n}^tE^tY_E^*)_{jn}
\nonumber\\&&+(h_{E^n}^t\Lambda_{E^p}^*\Lambda_{E^p}^tY_E^*+\Lambda_{E^n}^t\Lambda_{E^p}^*h_{E^p}^tY_E^*)_{jn}+3(\Lambda_{E^n}^t\Lambda_{D^p}^*h_{D^p}^tY_E^*)_{jn}
\nonumber\\&&+(h_{E^q}^tY_E^*)_{jl}Tr(\Lambda_{E^l}^t\Lambda_{E^q}^*)+(Y_E^{\dagger}\Lambda_{E^q})_{lj}Tr(h_{E^l}^t\Lambda_{E^q}^*)-3(Eh_{E^m}Y_E^*)_{jm}
\nonumber\\&&+6(\Lambda_{D^q}Y_D^*h_{U^m}^t\Lambda_{U^m}^*)_{jq}+3(Eh_{D^m}Y_D^*)_{jm}-9(h_{D^q}D^tY_D^*)_{jq}
\nonumber\\&&+6(h_{D^q}Y_D^*\Lambda_{U^m}^t\Lambda_{U^m}^*)_{jq}-6(h_{D^q}Y_D^*)_{jl}Tr(\Lambda_{D^l}^*\Lambda_{D^q}^t)-6(\Lambda_{D^q}Y_D^*)_{jl}Tr(\Lambda_{D^l}^*h_{D^q}^t)
\nonumber\\&&-3(Y_D^{\dagger}Th_{D^n}^t)_{nj}-3(h_{D^n}\Lambda_{D^m}^{\dagger}\Lambda_{D^m}Y_D^*+\Lambda_{D^n}\Lambda_{D^m}^{\dagger}h_{D^m}Y_D^*)_{jn}\}
\nonumber\\&&+2(Y_E)_{jk}\{(\frac{2}{5}M_1g_1^2-16M_3g_3^2)(\Lambda_{D^q}Y_D^*)_{iq}-\frac{6}{5}M_1g_1^2(\Lambda_{E^q}Y_E^*)_{iq}
\nonumber\\&&-(\frac{2}{5}g_1^2-16g_3^2)(h_{D^q}Y_D^*)_{iq}+\frac{6}{5}g_1^2(h_{E^q}Y_E^*)_{iq}+(3\Lambda_{E^n}^tY_E^*h_E^tY_E^*)_{in}
\nonumber\\&&+3(h_EY_E^{\dagger}\Lambda_{D^m}Y_D^*-h_EY_E^{\dagger}\Lambda_{E^m}Y_E^*)_{im}-3(\Lambda_{D^n}Y_U^*h_U^tY_D^*)_{in}
\nonumber\\&&-9(\Lambda_{D^k}Y_D^*h_D^tY_D^*)_{ik}+3(h_{E^n}^t\Lambda_{D^p}^*\Lambda_{D^p}^tY_E^*+h_{E^n}^tE^tY_E^*)_{in}
\nonumber\\&&+(h_{E^n}^t\Lambda_{E^p}^*\Lambda_{E^p}^tY_E^*+\Lambda_{E^n}^t\Lambda_{E^p}^*h_{E^p}^tY_E^*)_{in}+(h_{E^k}^tY_E^*)_{il}Tr(\Lambda_{E^l}^t\Lambda_{E^k}^*)
\nonumber\\&&+(\Lambda_{E^k}^tY_E^*)_{il}Tr(h_{E^l}^t\Lambda_{E^k}^*)-3(Eh_{E^m}Y_E^*)_{im}+6(\Lambda_{D^q}Y_D^*h_{U^m}^t\Lambda_{U^m}^*)_{iq}
\nonumber\\&&+3(\Lambda_{E^n}^t\Lambda_{D^p}^*h_{D^p}^tY_E^*)_{in}+3(Eh_{D^m}Y_D^*)_{im}-9(h_{D^q}D^tY_D^*)_{iq}
\nonumber\\&&+6(h_{D^q}Y_D^*\Lambda_{U^m}^t\Lambda_{U^m}^*)_{iq}-6(h_{D^q}Y_D^*)_{il}Tr(\Lambda_{D^l}^*\Lambda_{D^q}^t)-6(\Lambda_{D^q}Y_D^*)_{il}Tr(\Lambda_{D^l}^*h_{D^q}^t)
\nonumber\\&&-3(h_{D^n}Y_U^*Y_U^tY_D^*)_{in}-3(h_{D^n}\Lambda_{D^m}^{\dagger}\Lambda_{D^m}Y_D^*+\Lambda_{D^n}\Lambda_{D^m}^{\dagger}h_{D^m}Y_D^*)_{in}\}
\nonumber\\&&-2(\Lambda_{E^k})_{il}\{(16M_3g_3^2-\frac{2}{5}M_1g_1^2)(\Lambda_{D^q}\Lambda_{D^q}^{\dagger})_{jl}+\frac{6}{5}M_1g_1^2(E+\Lambda_{E^q}\Lambda_{E^q}^{\dagger})_{jl}
\nonumber\\&&-(16g_3^2-\frac{2}{5}g_1^2)(h_{D^q}\Lambda_{D^q}^{\dagger})_{jl}-\frac{6}{5}g_1^2(h_EY_E^{\dagger}+h_{E^q}\Lambda_{E^q}^{\dagger})_{jl}
\nonumber\\&&+2(h_EY_E^{\dagger}E+Eh_EY_E^{\dagger})_{jl}+(Y_E^{\dagger})_{ql}(h_E)_{jn}Tr(\Lambda_{E^n}^{\dagger}\Lambda_{E^q})
\nonumber\\&&+2(\Lambda_{E^n}\Lambda_{E^q}^{\dagger})_{jl}(Y_E^{\dagger}h_E)_{nq}+(h_EY_E^{\dagger})_{jl}Tr(E+3D)
\nonumber\\&&+(E)_{jl}Tr(h_EY_E^{\dagger})+(h_E)_{jq}(3\Lambda_{E^q}^{\dagger}\Lambda_{D^p}Y_D^*+\Lambda_{E^q}^{\dagger}\Lambda_{E^p}Y_E^*)_{lp}
\nonumber\\&&-(Y_E^*)_{lq}(\Lambda_{E^q}\Lambda_{E^p}^*h_E)_{jp}+(\Lambda_{E^q}^{\dagger}h_EY_E^{\dagger}\Lambda_{E^q})_{lj}+3(\Lambda_{D^q}^*h_UY_U^{\dagger}\Lambda_{D^q}^t)_{lj}
\nonumber\\&&+3(E)_{jl}Tr(h_DY_D^{\dagger})-3(Y_E^{\dagger})_{ql}(\Lambda_{E^q}\Lambda_{D^p}^*h_D)_{jp}+6(\Lambda_{D^n}\Lambda_{D^q}^{\dagger})_{jl}(Y_D^{\dagger}h_D)_{nq}
\nonumber\\&&+3(\Lambda_{D^q}^*h_DY_D^{\dagger}\Lambda_{D^q}^t)_{lj}+(Y_E^{\dagger})_{ql}(Y_E)_{jn}Tr(\Lambda_{E^n}^{\dagger}h_{E^q})+2(h_{E^n}\Lambda_{E^q}^{\dagger})_{jl}(\tilde{E})_{nq}
\nonumber\\&&+(h_{E^n}\Lambda_{E^q}^{\dagger})_{jl}Tr(\Lambda_{E^n}^{\dagger}\Lambda_{E^q})+(\Lambda_{E^n}\Lambda_{E^q}^{\dagger})_{jl}Tr(\Lambda_{E^n}^{\dagger}h_{E^q})+(Y_E)_{jq}(\Lambda_{E^q}^{\dagger}h_{E^p}Y_E^*)_{lp}
\nonumber\\&&-(Y_E^{\dagger})_{ql}(3h_{E^q}\Lambda_{D^p}^*Y_D+h_{E^q}\Lambda_{E^p}^*Y_E)_{jp}+(\Lambda_{E^q}^{\dagger}Eh_{E^q}
\nonumber\\&&+3\Lambda_{E^q}^t\Lambda_{D^p}\Lambda_{D^p}^{\dagger}h_{E^q})_{lj}+(\Lambda_{E^q}^{\dagger}h_{E^p}\Lambda_{E^p}^{\dagger}\Lambda_{E^q}+\Lambda_{E^q}^{\dagger}\Lambda_{E^p}\Lambda_{E^p}^{\dagger}h_{E^q})_{lj}
\nonumber\\&&+6(\Lambda_{D^n}\Lambda_{D^q}^{\dagger})_{jl}(h_{U^m}\Lambda_{U^m}^{\dagger})_{qn}+3(Y_E)_{jq}(\Lambda_{E^q}^{\dagger}h_{D^p}Y_D^*)_{lp}+3(\Lambda_{E^q}^{\dagger}h_{D^p}\Lambda_{D^p}^{\dagger}\Lambda_{E^q})_{lj}
\nonumber\\&&+6(h_{D^n}\Lambda_{D^q}^{\dagger})_{jl}[(\tilde{D})_{nq}+Tr(\Lambda_{D^n}^{\dagger}\Lambda_{D^q})+(\Lambda_{U^m}\Lambda_{U^m}^{\dagger})_{qn}]
\nonumber\\&&+6(\Lambda_{D^n}\Lambda_{D^q}^{\dagger})_{jl}Tr(\Lambda_{D^n}^{\dagger}h_{D^q})+3(\Lambda_{D^q}^*Dh_{D^q}^t+\Lambda_{D^q}^*Th_{D^q}^t)_{lj}
\nonumber\\&&+3(\Lambda_{D^q}\Lambda_{D^p}^{\dagger}h_{D^p}\Lambda_{D^q}^{\dagger}+h_{D^q}\Lambda_{D^p}^{\dagger}\Lambda_{D^p}\Lambda_{D^q}^{\dagger})_{jl}\}
\nonumber\\&&-2(\Lambda_{E^k})_{lj}\{(16M_3g_3^2-\frac{2}{5}M_1g_1^2)(\Lambda_{D^q}\Lambda_{D^q}^{\dagger})_{il}+\frac{6}{5}M_1g_1^2(E+\Lambda_{E^q}\Lambda_{E^q}^{\dagger})_{il}
\nonumber\\&&+(\frac{2}{5}g_1^2-16g_3^2)(h_{D^q}\Lambda_{D^q}^{\dagger})_{il}-\frac{6}{5}g_1^2(h_EY_E^{\dagger}+h_{E^q}\Lambda_{E^q}^{\dagger})_{il}
\nonumber\\&&+2(h_EY_E^{\dagger}E+Eh_EY_E^{\dagger})_{il}+(Y_E^{\dagger})_{ql}(h_E)_{in}Tr(\Lambda_{E^n}^{\dagger}\Lambda_{E^q})
\nonumber\\&&+2(\Lambda_{E^n}\Lambda_{E^q}^{\dagger})_{il}(Y_E^{\dagger}h_E)_{nq}+(h_EY_E^{\dagger})_{il}Tr(E+3D)
\nonumber\\&&+E_{il}Tr(h_EY_E^{\dagger})+(h_E)_{iq}(3\Lambda_{E^q}^{\dagger}\Lambda_{D^p}Y_D^*+\Lambda_{E^q}^{\dagger}\Lambda_{E^p}Y_D^*)_{lp}
\nonumber\\&&-(Y_E^{\dagger})_{ql}(\Lambda_{E^q}\Lambda_{E^p}^*h_E)_{ip}+(\Lambda_{E^q}^{\dagger}h_EY_E^{\dagger}\Lambda_{E^q})_{li}+3(\Lambda_{D^q}^*h_UY_U^{\dagger}\Lambda_{D^q}^t)_{li}
\nonumber\\&&+3(E)_{il}Tr(h_DY_D^{\dagger})-3(Y_E^{\dagger})_{ql}(\Lambda_{E^q}\Lambda_{D^p}^*h_D)_{ip}+6(\Lambda_{D^n}\Lambda_{D^q}^{\dagger})_{il}(Y_D^{\dagger}h_D)_{nq}
\nonumber\\&&+3(\Lambda_{D^q}^*h_DY_D^{\dagger}\Lambda_{D^q}^t)_{li}+(Y_E^{\dagger})_{ql}(Y_E)_{in}Tr(\Lambda_{E^n}^{\dagger}h_{E^q})+2(h_{E^n}\Lambda_{E^q}^{\dagger})_{il}(\tilde{E})_{nq}
\nonumber\\&&+(h_{E^n}\Lambda_{E^q}^{\dagger})_{il}Tr(\Lambda_{E^n}^{\dagger}\Lambda_{E^q})+(\Lambda_{E^n}\Lambda_{E^q}^{\dagger})_{il}Tr(\Lambda_{E^n}^{\dagger}h_{E^q})+(Y_E)_{iq}(\Lambda_{E^q}^{\dagger}h_{E^p}Y_E^*)_{lp}
\nonumber\\&&-3(Y_E^{\dagger})_{ql}(h_{E^q}\Lambda_{D^p}^*Y_D)_{ip}-(Y_E^{\dagger})_{ql}(h_{E^q}\Lambda_{E^p}^*Y_E)_{ip}+(\Lambda_{E^q}^{\dagger}Eh_{E^q}
\nonumber\\&&+3\Lambda_{E^q}^{\dagger}\Lambda_{D^p}\Lambda_{D^p}^{\dagger}h_{E^q})_{li}+(\Lambda_{E^q}^{\dagger}h_{E^p}\Lambda_{E^p}^{\dagger}\Lambda_{E^q}+\Lambda_{E^q}^{\dagger}\Lambda_{E^p}\Lambda_{E^p}^{\dagger}h_{E^q})_{li}
\nonumber\\&&+6(\Lambda_{D^n}\Lambda_{D^q}^{\dagger})_{il}(h_{U^m}\Lambda_{U^m}^{\dagger})_{qn}+3(Y_E)_{iq}(\Lambda_{E^q}^{\dagger}h_{D^p}Y_D^*)_{lp}+3(\Lambda_{E^q}^{\dagger}h_{D^p}\Lambda_{D^p}^{\dagger}\Lambda_{E^q})_{li}
\nonumber\\&&+6(h_{D^n}\Lambda_{D^q}^{\dagger})_{il}[(\tilde{D})_{nq}+Tr(\Lambda_{D^n}^{\dagger}\Lambda_{D^q})+(\Lambda_{U^m}\Lambda_{U^m}^{\dagger})_{qn}]
\nonumber\\&&+6(\Lambda_{D^n}\Lambda_{D^q}^{\dagger})_{il}Tr(\Lambda_{D^n}^{\dagger}h_{D^q})+3(\Lambda_{D^q}^*Dh_{D^q}^t+\Lambda_{D^q}^*Th_{D^q}^t)_{li}
\nonumber\\&&+3(\Lambda_{D^q}\Lambda_{D^p}^{\dagger}h_{D^p}\Lambda_{D^q}^{\dagger}+h_{D^q}\Lambda_{D^p}^{\dagger}\Lambda_{D^p}\Lambda_{D^q}^{\dagger})_{il}\}
\nonumber\\&&+(h_{E^k})_{ij}\{\frac{234}{25}g_1^4+\frac{15}{2}g_2^4+\frac{207}{50}g_1^4+\frac{9}{5}g_1^2g_2^2\}
\nonumber\\&&-(\Lambda_{E^k})_{ij}\{30M_2g_2^4+\frac{18}{5}(M_1+M_2)g_2^2g_1^2+\frac{414}{25}M_1g_1^4+4(\frac{936}{25})g_1^4)\}
\end{eqnarray}
\begin{eqnarray}
{(16\pi^2)^2}\beta^{(2)}_{(h_{D^k})_{ij}}&=&(h_{D^l})_{ij}\{(\frac{16}{3}g_3^2+\frac{16}{15}g_1^2)(\Lambda_{U^q}\Lambda_{U^q}^{\dagger})_{kl}-2(D)_{lk}Tr(E+3D)
\nonumber\\&&+(6g_2^2+\frac{2}{5}g_1^2)[D_{kl}+Tr(\Lambda_{D^k}\Lambda_{D^l}^{\dagger})]-2(Y_D^{\dagger}DY_D)_{lk}-2(Y_D^{\dagger}TY_D)_{lk}
\nonumber\\&&-2(Y_D^t\Lambda_{D^p}^{\dagger}\Lambda_{D^p}Y_D^*)_{kl}-2Tr(\Lambda_{D^k}^t\Lambda_{D^l}^*[D+T]
\nonumber\\&&+\Lambda_{D^l}^{\dagger}\Lambda_{D^k}\Lambda_{D^q}^{\dagger}\Lambda_{D^q})-Tr(6\Lambda_{D^q}\Lambda_{D^q}^{\dagger}\Lambda_{D^k}\Lambda_{D^l}^{\dagger}+2\Lambda_{E^q}\Lambda_{E^q}^{\dagger}\Lambda_{D^k}\Lambda_{D^l}^{\dagger}
\nonumber\\&&+2E\Lambda_{D^k}\Lambda_{D^l}^{\dagger})-4(\Lambda_{U^m}\tilde{D}\Lambda_{U^m}^{\dagger})_{kl}-4(\Lambda_{U^m}^{\dagger}\Lambda_{U^p}\Lambda_{U^p}^{\dagger}\Lambda_{U^m})_{lk}
\nonumber\\&&+4(\Lambda_{U^m}^{\dagger})_{lq}(\Lambda_{U^m})_{kn}Tr(\Lambda_{D^n}\Lambda_{D^q}^{\dagger})-(6Y_D^{\dagger}\Lambda_{D^k}^t\Lambda_{D^p}^*Y_D)_{lp}
\nonumber\\&&-2(Y_D^{\dagger}\Lambda_{D^k}^t\Lambda_{E^p}^*Y_E)_{lp}-(6Y_D^t\Lambda_{D^l}^{\dagger}\Lambda_{D^p}Y_D^*+2Y_D^t\Lambda_{D^l}^{\dagger}\Lambda_{E^p}Y_E^*)_{kp}
\nonumber\\&&-2(\Lambda_{U^q}^{\dagger}\Lambda_{U^n})_{lk}[Tr(\Lambda_{U^n}^{\dagger}\Lambda_{U^q})+2(\tilde{T})_{nq}]\}
\nonumber\\&&+(h_{D^k})_{il}\{\frac{2}{5}g_1^2[(D+2T)_{jl}+(\Lambda_{D^q}^{\dagger}\Lambda_{D^q})_{lj}]-(\Lambda_{D^m}^{\dagger}E\Lambda_{D^m})_{lj}
\nonumber\\&&-2(D^2)_{jl}-(D)_{jl}Tr(E+3D)-2(T^2)_{jl}-3(T)_{jl}Tr(T)
\nonumber\\&&-2(\Lambda_{D^n}^{\dagger}\Lambda_{D^m})_{lj}[(\tilde{D})_{mn}+(\Lambda_{U^q}\Lambda_{U^q}^{\dagger})_{nm}+Tr(\Lambda_{D^m}^{\dagger}\Lambda_{D^n})]
\nonumber\\&&-3(\Lambda_{D^m}^{\dagger}\Lambda_{D^q}\Lambda_{D^q}^{\dagger}\Lambda_{D^m})_{lj}-(\Lambda_{D^m}^{\dagger}\Lambda_{E^q}\Lambda_{E^q}^{\dagger}\Lambda_{D^m})_{lj}-(Y_D)_{jn}(3\Lambda_{D^n}^{\dagger}\Lambda_{D^m}Y_D^*
\nonumber\\&&+\Lambda_{D^n}^{\dagger}\Lambda_{E^m}Y_E^*)_{lm}-2(Y_D\Lambda_{U^q}^{\dagger}\Lambda_{U^q}Y_D^{\dagger})_{jl}-2(Y_D)_{jm}(Y_D^{\dagger})_{nl}Tr(\Lambda_{D^m}^{\dagger}\Lambda_{D^n})
\nonumber\\&&-(Y_D^*)_{ln}(3\Lambda_{D^n}^t\Lambda_{D^m}^*Y_D+\Lambda_{D^n}^t\Lambda_{E^m}^*Y_E)_{jm}-(Y_U)_{jn}(Y_U^{\dagger})_{ql}Tr(\Lambda_{U^n}^{\dagger}\Lambda_{U^q})\}
\nonumber\\&&+(h_{D^k})_{lj}\{(16g_3^2-\frac{2}{5}g_1^2)(\Lambda_{D^q}\Lambda_{D^q}^{\dagger})_{il}+\frac{6}{5}g_1^2(E+\Lambda_{E^q}\Lambda_{E^q}^{\dagger})_{il}
\nonumber\\&&-2(E^2)_{il}-(Y_E^{\dagger})_{ql}(Y_E)_{in}Tr(\Lambda_{E^n}^{\dagger}\Lambda_{E^q})-2(\Lambda_{E^n}\Lambda_{E^q}^{\dagger})_{il}(\tilde{E})_{nq}
\nonumber\\&&-(\Lambda_{E^n}\Lambda_{E^q}^{\dagger})_{il}Tr(\Lambda_{E^n}^{\dagger}\Lambda_{E^q})-(E)_{il}Tr(E+3D)
\nonumber\\&&-(Y_E)_{iq}(3\Lambda_{E^q}^{\dagger}\Lambda_{D^p}Y_D^*+\Lambda_{E^q}^{\dagger}\Lambda_{E^p}Y_E^*)_{lp}+(Y_E^*)_{lq}(3\Lambda_{E^q}\Lambda_{D^p}^*Y_D
\nonumber\\&&+\Lambda_{E^q}\Lambda_{E^p}^*Y_E)_{ip}-(\Lambda_{E^q}^{\dagger}E\Lambda_{E^q}+3\Lambda_{E^q}^{\dagger}\Lambda_{D^p}\Lambda_{D^p}^{\dagger}\Lambda_{E^q}+\Lambda_{E^q}^{\dagger}\Lambda_{E^p}\Lambda_{E^p}^{\dagger}\Lambda_{E^q})_{li}
\nonumber\\&&-6(\Lambda_{D^n}\Lambda_{D^q}^{\dagger})_{il}[(\tilde{D})_{nq}+Tr(\Lambda_{D^n}^{\dagger}\Lambda_{D^q})+(\Lambda_{U^m}\Lambda_{U^m}^{\dagger})_{qn}]
\nonumber\\&&-3(\Lambda_{D^q}^*D\Lambda_{D^q}^t+\Lambda_{D^q}^*T\Lambda_{D^q}^t)_{li}-3(\Lambda_{D^q}\Lambda_{D^p}^{\dagger}\Lambda_{D^p}\Lambda_{D^q}^{\dagger})_{il}\}
\nonumber\\&&-(h_D)_{jk}\{(\frac{2}{5}g_1^2-16g_3^2)(\Lambda_{D^q}Y_D^*)_{iq}-\frac{6}{5}g_1^2(\Lambda_{E^q}Y_E^*)_{iq}
\nonumber\\&&-(3\Lambda_{E^n}^t\Lambda_{D^p}^*\Lambda_{D^p}^tY_E^*+\Lambda_{E^n}^t\Lambda_{E^p}^*\Lambda_{E^p}^tY_E^*+3\Lambda_{E^n}^tE^tY_E^*)_{in}
\nonumber\\&&-(\Lambda_{E^q}^tY_E^*)_{il}Tr(\Lambda_{E^l}^t\Lambda_{E^q}^*)-3(E\Lambda_{D^m}Y_D^*-E\Lambda_{E^m}Y_E^*)_{im}
\nonumber\\&&+9(\Lambda_{D^q}D^tY_D^*)_{iq}-6(\Lambda_{D^q}Y_D^*\Lambda_{U^m}^t\Lambda_{U^m}^*)_{iq}+6(\Lambda_{D^q}Y_D^*)_{il}Tr(\Lambda_{D^l}^*\Lambda_{D^q}^t)
\nonumber\\&&+3(\Lambda_{D^n}Y_U^*Y_U^tY_D^*+\Lambda_{D^n}\Lambda_{D^m}^{\dagger}\Lambda_{D^m}Y_D^*)_{in}\}
\nonumber\\&&-2(\Lambda_{D^l})_{ij}\{(\frac{16}{3}M_3g_3^2+\frac{16}{15}M_1g_1^2)(\Lambda_{U^q}\Lambda_{U^q}^{\dagger})_{kl}
\nonumber\\&&+(6M_2g_2^2+\frac{2}{5}M_1g_1^2)[(D)_{kl}+Tr(\Lambda_{D^k}\Lambda_{D^l}^{\dagger})]+2(h_D^t\Lambda_{D^p}^{\dagger}\Lambda_{D^p}Y_D^*)_{kl}
\nonumber\\&&-(\frac{16}{3}g_3^2+\frac{16}{15}g_1^2)(h_{U^q}\Lambda_{U^q}^{\dagger})_{kl}-(6g_2^2+\frac{2}{5}g_1^2)[(h_DY_D^{\dagger})_{kl}+Tr(h_{D^k}\Lambda_{D^l}^{\dagger})]
\nonumber\\&&+2(\tilde{D})_{lk}Tr(Y_E^{\dagger}h_E)+2Tr(h_EY_E^{\dagger}\Lambda_{D^k}\Lambda_{D^l}^{\dagger})+2(Y_D^{\dagger}\Lambda_{D^k}^t\Lambda_{E^p}^*h_E)_{lp}
\nonumber\\&&+2(Y_D^{\dagger}h_UY_U^{\dagger}Y_D)_{lk}+2Tr(\Lambda_{D^k}^t\Lambda_{D^l}^*h_UY_U^{\dagger})+4(\Lambda_{U^q}^{\dagger}\Lambda_{U^n})_{lk}(Y_U^{\dagger}h_U)_{nq}
\nonumber\\&&+2(Y_D^{\dagger}h_D\tilde{D})_{lk}+2(Y_D^{\dagger}Dh_D)_{lk}+2(Y_D^{\dagger}Th_D)_{lk}
\nonumber\\&&+2(Y_D^{\dagger}h_D)_{lk}Tr(E+3D)+6\tilde{D}_{lk}Tr(Y_D^{\dagger}h_D)
\nonumber\\&&+2Tr(\Lambda_{D^k}^t\Lambda_{D^l}^*h_DY_D^{\dagger})+4(\Lambda_{U^m}Y_D^{\dagger}h_D\Lambda_{U^m}^{\dagger})_{kl}+6(Y_D^{\dagger}\Lambda_{D^k}^t\Lambda_{D^p}^*h_D)_{lp}
\nonumber\\&&+6(h_D^t\Lambda_{D^l}^{\dagger}\Lambda_{D^p}Y_D^*+2h_D^t\Lambda_{D^l}^{\dagger}\Lambda_{E^p}Y_E^*)_{kp}+Tr(2h_{E^q}\Lambda_{E^q}^{\dagger}\Lambda_{D^k}\Lambda_{D^l}^{\dagger})
\nonumber\\&&+2(Y_D^t\Lambda_{D^l}^{\dagger}h_{E^p}Y_E^*)_{kp}+4(h_{U^m}\tilde{D}\Lambda_{U^m}^{\dagger})_{kl}+4(\Lambda_{U^m}^{\dagger}h_{U^p}\Lambda_{U^p}^{\dagger}\Lambda_{U^m})_{lk}
\nonumber\\&&+4(\Lambda_{U^m}^{\dagger}\Lambda_{U^p}\Lambda_{U^p}^{\dagger}h_{U^m})_{lk}-4(\Lambda_{U^m}^{\dagger})_{lq}(h_{U^m})_{kn}Tr(\Lambda_{D^n}\Lambda_{D^q}^{\dagger})+2(Y_D^t\Lambda_{D^p}^{\dagger}h_{D^p}Y_D^*)_{kl}
\nonumber\\&&+2(\Lambda_{U^q}^{\dagger}h_{U^n})_{lk}[Tr(\Lambda_{U^n}^{\dagger}\Lambda_{U^q})+2(\tilde{T})_{nq}]+2(\Lambda_{U^q}^{\dagger}\Lambda_{U^n})_{lk}Tr(\Lambda_{U^n}^{\dagger}h_{U^q})
\nonumber\\&&+2Tr(h_{D^k}^t\Lambda_{D^l}^*(D+T)+\Lambda_{D^l}^{\dagger}h_{D^k}\Lambda_{D^q}^{\dagger}\Lambda_{D^q}+\Lambda_{D^l}^{\dagger}\Lambda_{D^k}\Lambda_{D^q}^{\dagger}h_{D^q})
\nonumber\\&&+Tr(6h_{D^q}\Lambda_{D^q}^{\dagger}\Lambda_{D^k}\Lambda_{D^l}^{\dagger}+6\Lambda_{D^q}\Lambda_{D^q}^{\dagger}h_{D^k}\Lambda_{D^l}^{\dagger}+2\Lambda_{E^q}\Lambda_{E^q}^{\dagger}h_{D^k}\Lambda_{D^l}^{\dagger}+2Eh_{D^k}\Lambda_{D^l}^{\dagger})
\nonumber\\&&-4(\Lambda_{U^m}^{\dagger})_{lq}(\Lambda_{U^m})_{kn}Tr(h_{D^n}\Lambda_{D^q}^{\dagger})+(6Y_D^{\dagger}h_{D^k}^t\Lambda_{D^p}^*Y_D)_{lp}
\nonumber\\&&+2(Y_D^{\dagger}h_{D^k}^t\Lambda_{E^p}^*Y_E)_{lp}+(6Y_D^t\Lambda_{D^l}^{\dagger}h_{D^p}Y_D^*)_{kp}\}
\nonumber\\&&-2(\Lambda_{D^k})_{il}\{\frac{2}{5}M_1g_1^2[(D+2T)_{jl}+(\Lambda_{D^q}^{\dagger}\Lambda_{D^q})_{lj}]
\nonumber\\&&-\frac{2}{5}g_1^2[(h_DY_D^{\dagger}+2h_UY_U^{\dagger})_{jl}+(\Lambda_{D^q}^{\dagger}h_{D^q})_{lj}]+3(h_UY_U^{\dagger})_{jl}Tr(T)
\nonumber\\&&+(D)_{jl}Tr(Y_E^{\dagger}h_E)+(\Lambda_{D^m}^{\dagger}h_EY_E^{\dagger}\Lambda_{D^m})_{lj}+(Y_D^*)_{ln}(3\Lambda_{D^n}^t\Lambda_{E^m}^*h_E)_{jm}
\nonumber\\&&+2(h_UY_U^{\dagger}T)_{jl}+2(Th_UY_U^{\dagger})_{jl}+(Y_D)_{jn}(\Lambda_{D^n}^{\dagger}h_{E^m}Y_E^*)_{lm}
\nonumber\\&&+3(T)_{jl}Tr(Y_U^{\dagger}h_U)+(h_U)_{jn}(Y_U^{\dagger})_{ql}Tr(\Lambda_{U^n}^{\dagger}\Lambda_{U^q})+2(h_DY_D^{\dagger}D
\nonumber\\&&+Dh_DY_D^{\dagger})_{jl}+(h_DY_D^{\dagger})_{jl}Tr(E+3D)+(Y_U)_{jn}(Y_U^{\dagger})_{ql}Tr(\Lambda_{U^n}^{\dagger}h_{U^q})
\nonumber\\&&+(D)_{jl}Tr(3Y_D^{\dagger}h_D)+2(\Lambda_{D^n}^{\dagger}\Lambda_{D^m})_{lj}(Y_D^{\dagger}h_D)_{mn}+(h_D)_{jn}(3\Lambda_{D^n}^{\dagger}\Lambda_{D^m}Y_D^*
\nonumber\\&&+\Lambda_{D^n}^{\dagger}\Lambda_{E^m}Y_E^*)_{lm}+2(h_D\Lambda_{U^q}^{\dagger}\Lambda_{U^q}Y_D^{\dagger})_{jl}+2(h_D)_{jm}(Y_D^{\dagger})_{nl}Tr(\Lambda_{D^m}^{\dagger}\Lambda_{D^n})
\nonumber\\&&+3(Y_D^*)_{ln}(\Lambda_{D^n}^t\Lambda_{D^m}^*h_D)_{jm}+(\Lambda_{D^m}^{\dagger}h_{E^q}\Lambda_{E^q}^{\dagger}\Lambda_{D^m})_{lj}
\nonumber\\&&+2(\Lambda_{D^n}^{\dagger}\Lambda_{D^n})_{lj}(h_{U^q}\Lambda_{U^q}^{\dagger})_{nm}+2(Y_D\Lambda_{U^q}^{\dagger}h_{U^q}Y_D^{\dagger})_{jl}
\nonumber\\&&+2(\Lambda_{D^n}^{\dagger}h_{D^m})_{lj}[(\tilde{D})_{mn}+(\Lambda_{U^q}\Lambda_{U^q}^{\dagger})_{nm}+Tr(\Lambda_{D^m}^{\dagger}\Lambda_{D^n})]
\nonumber\\&&+2(\Lambda_{D^n}^{\dagger}\Lambda_{D^m})_{lj}Tr(\Lambda_{D^m}^{\dagger}h_{D^n})+(\Lambda_{D^m}^{\dagger}Eh_{D^m})_{lj}+3(\Lambda_{D^m}^{\dagger}h_{D^q}\Lambda_{D^q}^{\dagger}\Lambda_{D^m}
\nonumber\\&&+\Lambda_{D^m}^{\dagger}\Lambda_{D^q}\Lambda_{D^q}^{\dagger}h_{D^m})_{lj}+(\Lambda_{D^m}^{\dagger}\Lambda_{E^q}\Lambda_{E^q}^{\dagger}h_{D^m})_{lj}+3(Y_D)_{jn}(\Lambda_{D^n}^{\dagger}h_{D^m}Y_D^*)_{lm}
\nonumber\\&&+2(Y_D)_{jm}(Y_D^{\dagger})_{nl}Tr(\Lambda_{D^m}^{\dagger}h_{D^n})+(Y_D^*)_{ln}(3h_{D^n}^t\Lambda_{D^m}^*Y_D+h_{D^n}^t\Lambda_{E^m}^*Y_E)_{jm}\}
\nonumber\\&&-2(\Lambda_{D^k})_{lj}\{(16M_3g_3^2-\frac{2}{5}M_1g_1^2)(\Lambda_{D^q}\Lambda_{D^q}^{\dagger})_{il}+\frac{6}{5}M_1g_1^2(E+\Lambda_{E^q}\Lambda_{E^q}^{\dagger})_{il}
\nonumber\\&&-(16g_3^2-\frac{2}{5}g_1^2)(h_{D^q}\Lambda_{D^q}^{\dagger})_{il}-\frac{6}{5}g_1^2(h_EY_E^{\dagger}+h_{E^q}\Lambda_{E^q}^{\dagger})_{il}
\nonumber\\&&+2(h_EY_E^{\dagger}E+Eh_EY_E^{\dagger})_{il}+(Y_E^{\dagger})_{ql}(h_E)_{in}Tr(\Lambda_{E^n}^{\dagger}\Lambda_{E^q})
\nonumber\\&&+(h_EY_E^{\dagger})_{il}Tr(E+3D)+(E)_{il}Tr(h_EY_E^{\dagger}+3h_DY_D^{\dagger})
\nonumber\\&&+(h_E)_{iq}(3\Lambda_{E^q}^{\dagger}\Lambda_{D^p}Y_D^*+\Lambda_{E^q}^{\dagger}\Lambda_{E^p}Y_E^*)_{lp}-(Y_E^*)_{lq}(\Lambda_{E^q}\Lambda_{E^p}^*h_E)_{ip}
\nonumber\\&&+(\Lambda_{E^q}^{\dagger}h_EY_E^{\dagger}\Lambda_{E^q})_{li}+3(\Lambda_{D^q}^*h_UY_U^{\dagger}\Lambda_{D^q}^t)_{li}+2(\Lambda_{E^n}\Lambda_{E^q}^{\dagger})_{il}(Y_E^{\dagger}h_E)_{nq}
\nonumber\\&&-3(Y_E^*)_{lq}(\Lambda_{E^q}\Lambda_{D^p}^*h_D)_{ip}+6(\Lambda_{D^n}\Lambda_{D^q}^{\dagger})_{il}(Y_D^{\dagger}h_D)_{nq}+3(\Lambda_{D^q}^*h_DY_D^{\dagger}\Lambda_{D^q}^t)_{li}
\nonumber\\&&+(Y_E^{\dagger})_{ql}(Y_E)_{in}Tr(\Lambda_{E^n}^{\dagger}h_{E^q})+2(h_{E^n}\Lambda_{E^q}^{\dagger})_{il}(\tilde{E})_{nq}
\nonumber\\&&+(\Lambda_{E^n}\Lambda_{E^q}^{\dagger})_{il}Tr(\Lambda_{E^n}^{\dagger}h_{E^q})+(Y_E)_{iq}(\Lambda_{E^q}^{\dagger}h_{E^p}Y_E^*)_{lp}-(Y_E^*)_{lq}(3h_{E^q}\Lambda_{D^p}^*Y_D
\nonumber\\&&+h_{E^q}\Lambda_{E^p}^*Y_E)_{ip}+(\Lambda_{E^q}^{\dagger}Eh_{E^q}+3\Lambda_{E^q}^{\dagger}\Lambda_{D^p}\Lambda_{D^p}^{\dagger}h_{E^q}
\nonumber\\&&+\Lambda_{E^q}^{\dagger}h_{E^p}\Lambda_{E^p}^{\dagger}\Lambda_{E^q}+\Lambda_{E^q}^{\dagger}\Lambda_{E^p}\Lambda_{E^p}^{\dagger}h_{E^q})_{li}+6(\Lambda_{D^n}\Lambda_{D^q}^{\dagger})_{il}(h_{U^m}\Lambda_{U^m}^{\dagger})_{qn}
\nonumber\\&&+3(Y_E)_{iq}(\Lambda_{E^q}^{\dagger}h_{D^p}Y_D^*)_{lp}+3(\Lambda_{E^q}^{\dagger}h_{D^p}\Lambda_{D^p}^{\dagger}\Lambda_{E^q})_{li}+6(h_{D^n}\Lambda_{D^q}^{\dagger})_{il}[(\tilde{D})_{nq}
\nonumber\\&&+Tr(\Lambda_{D^n}^{\dagger}\Lambda_{D^q})+(\Lambda_{U^m}\Lambda_{U^m}^{\dagger})_{qn}]+6(\Lambda_{D^n}\Lambda_{D^q}^{\dagger})_{il}Tr(\Lambda_{D^n}^{\dagger}h_{D^q})
\nonumber\\&&+3(\Lambda_{D^q}^*Dh_{D^q}^t+\Lambda_{D^q}^*Th_{D^q}^t)_{li}+(h_{E^n}\Lambda_{E^q}^{\dagger})_{il}Tr(\Lambda_{E^n}^{\dagger}\Lambda_{E^q})
\nonumber\\&&+3(\Lambda_{D^q}\Lambda_{D^p}^{\dagger}h_{D^p}\Lambda_{D^q}^{\dagger}+h_{D^q}\Lambda_{D^p}^{\dagger}\Lambda_{D^p}\Lambda_{D^q}^{\dagger})_{il}\}
\nonumber\\&&+2(Y_D)_{jk}\{(\frac{2}{5}M_1g_1^2-16M_3g_3^2)(\Lambda_{D^q}Y_D^*)_{iq}-\frac{6}{5}M_1g_1^2(\Lambda_{E^q}Y_E^*)_{iq}
\nonumber\\&&-(\frac{2}{5}g_1^2-16g_3^2)(h_{D^q}Y_D^*)_{iq}+\frac{6}{5}g_1^2(h_{E^q}Y_E^*)_{iq}+(3\Lambda_{E^n}^tY_E^*h_E^tY_E^*)_{in}
\nonumber\\&&+3(h_EY_E^{\dagger}\Lambda_{D^m}Y_D^*-h_EY_E^{\dagger}\Lambda_{E^m}Y_E^*)_{im}-3(\Lambda_{D^n}Y_U^*h_U^tY_D^*)_{in}
\nonumber\\&&-9(\Lambda_{D^q}Y_D^*h_D^tY_D^*)_{iq}+(3h_{E^n}^t\Lambda_{D^p}^*\Lambda_{D^p}^tY_E^*+h_{E^n}^t\Lambda_{E^p}^*\Lambda_{E^p}^tY_E^*
\nonumber\\&&+\Lambda_{E^n}^t\Lambda_{E^p}^*h_{E^p}^tY_E^*+3h_{E^n}^tE^tY_E^*)_{in}+(h_{E^q}^tY_E^*)_{il}Tr(\Lambda_{E^l}^t\Lambda_{E^q}^*)
\nonumber\\&&+(\Lambda_{E^q}^tY_E^*)_{il}Tr(h_{E^l}^t\Lambda_{E^q}^*)-3(Eh_{E^m}Y_E^*)_{im}+6(\Lambda_{D^q}Y_D^*h_{U^m}^t\Lambda_{U^m}^*)_{iq}
\nonumber\\&&+3(\Lambda_{E^n}^t\Lambda_{D^p}^*h_{D^p}^tY_E^*)_{in}+3(Eh_{D^m}Y_D^*)_{im}-6(\Lambda_{D^q}Y_D^*)_{il}Tr(\Lambda_{D^l}^*h_{D^q}^t)
\nonumber\\&&+6(h_{D^q}Y_D^*\Lambda_{U^m}^t\Lambda_{U^m}^*)_{iq}-6(h_{D^q}Y_D^*)_{il}Tr(\Lambda_{D^l}^*\Lambda_{D^q}^t)-9(h_{D^q}D^tY_D^*)_{iq}
\nonumber\\&&-3(Y_D^{\dagger}Th_{D^n}^t)_{ni}-3(h_{D^n}\Lambda_{D^m}^{\dagger}\Lambda_{D^m}Y_D^*+\Lambda_{D^n}\Lambda_{D^m}^{\dagger}h_{D^m}Y_D^*)_{in}\}
\nonumber\\&&+(h_{D^k})_{ij}\{\frac{15}{2}g_2^4+\frac{287}{90}g_1^4-\frac{16}{9}g_3^4+8g_3^2g_2^2+\frac{8}{9}g_3^2g_1^2+g_2^2g_1^2\}
\nonumber\\&&+(\Lambda_{D^k})_{ij}\{-30M_2g_2^2-\frac{574}{45}M_1g_1^4+\frac{64}{9}M_3g_3^4-16(M_3+M_2)g_3^2g_2^2
\nonumber\\&&-\frac{16}{9}(M_3+M_1)g_3^2g_1^2-2(M_1+M_2)g_2^2g_1^2\}
\end{eqnarray}

\begin{eqnarray}
{(16\pi^2)^2}\beta^{(2)}_{(h_{U^i})_{jk}}&=&(h_{U^l})_{jk}\{(6g_2^2-\frac{2}{5}g_1^2)(\tilde{T})_{li}+(\frac{8}{3}g_3^2-\frac{4}{15}g_1^2)Tr(\Lambda_{U^i}\Lambda_{U^l}^{\dagger})
\nonumber\\&&-2(Y_U^{\dagger}DY_U)_{li}-6(\tilde{T})_{li}Tr(T)-4Tr(\Lambda_{U^i}\Lambda_{U^l}^{\dagger}\Lambda_{U^p}\Lambda_{U^p}^{\dagger})
\nonumber\\&&-2(Y_U^{\dagger}\Lambda_{D^m}^t\Lambda_{D^m}^*Y_U)_{li}-4(\Lambda_{U^l}^{\dagger}\Lambda_{U^i})_{qm}Tr(\Lambda_{D^m}^{\dagger}\Lambda_{D^q})-2(Y_U^{\dagger}TY_U)_{li}
\nonumber\\&&-4Tr(\Lambda_{U^l}^{\dagger}\Lambda_{U^i}\tilde{D})\}+(h_{U^i})_{lk}\{(\frac{16}{3}g_3^2+\frac{16}{15}g_1^2)(\Lambda_{U^q}\Lambda_{U^q}^{\dagger})_{jl}
\nonumber\\&&+(6g_2^2+\frac{2}{5}g_1^2)[(D)_{jl}+Tr(\Lambda_{D^j}\Lambda_{D^l}^{\dagger})]-2(Y_D^{\dagger}DY_D+Y_D^{\dagger}TY_D)_{lj}
\nonumber\\&&-2(Y_D^t\Lambda_{D^p}^{\dagger}\Lambda_{D^p}Y_D^*)_{jl}-6Tr(\Lambda_{D^q}\Lambda_{D^q}^{\dagger}\Lambda_{D^j}\Lambda_{D^l}^{\dagger})-(6Y_D^{\dagger}\Lambda_{D^j}^t\Lambda_{D^p}^*Y_D)_{lp}
\nonumber\\&&-2Tr[\Lambda_{D^j}^t\Lambda_{D^l}^*(D+T)+\Lambda_{D^l}^{\dagger}\Lambda_{D^j}\Lambda_{D^q}^{\dagger}\Lambda_{D^q}]-2\tilde{D}_{lj}Tr(E+3D)
\nonumber\\&&+2Tr(\Lambda_{E^q}\Lambda_{E^q}^{\dagger}\Lambda_{D^j}\Lambda_{D^l}^{\dagger}+E\Lambda_{D^j}\Lambda_{D^l}^{\dagger})-4(\Lambda_{U^m}\tilde{D}\Lambda_{U^m}^{\dagger})_{jl}
\nonumber\\&&-4(\Lambda_{U^m}^{\dagger}\Lambda_{U^p}\Lambda_{U^p}^{\dagger}\Lambda_{U^m})_{lj}+4(\Lambda_{U^m}^{\dagger})_{lq}(\Lambda_{U^m})_{jn}Tr(\Lambda_{D^n}\Lambda_{D^q}^{\dagger})
\nonumber\\&&-2(Y_D^{\dagger}\Lambda_{D^j}^t\Lambda_{E^p}^*Y_E)_{lp}-(6Y_D^t\Lambda_{D^l}^{\dagger}\Lambda_{D^p}Y_D^*+2Y_D^t\Lambda_{D^l}^{\dagger}\Lambda_{E^p}Y_E^*)_{jp}
\nonumber\\&&-2(\Lambda_{U^q}^{\dagger}\Lambda_{U^n})_{lj}[Tr(\Lambda_{U^n}^{\dagger}\Lambda_{U^q})+2(Y_U^{\dagger}Y_U)_{nq}]\}
\nonumber\\&&+(h_{U^i})_{jl}\{(\frac{16}{3}g_3^2+\frac{16}{15}g_1^2)(\Lambda_{U^q}\Lambda_{U^q}^{\dagger})_{kl}+(6g_2^2+\frac{2}{5}g_1^2)[D_{kl}
\nonumber\\&&+Tr(\Lambda_{D^k}\Lambda_{D^l}^{\dagger})]-2(Y_D^{\dagger}DY_D)_{lk}-2(Y_D^{\dagger}TY_D)_{lk}
\nonumber\\&&-2\tilde{D}_{lk}Tr(E+3D)-2(Y_D^t\Lambda_{D^p}^{\dagger}\Lambda_{D^p}Y_D^*)_{kl}
\nonumber\\&&-2Tr[\Lambda_{D^k}^t\Lambda_{D^l}^*(D+T)]-2Tr(\Lambda_{D^l}^{\dagger}\Lambda_{D^k}\Lambda_{D^q}^{\dagger}\Lambda_{D^q})
\nonumber\\&&-Tr(6\Lambda_{D^q}\Lambda_{D^q}^{\dagger}\Lambda_{D^k}\Lambda_{D^l}^{\dagger}+2\Lambda_{E^q}\Lambda_{E^q}^{\dagger}\Lambda_{D^k}\Lambda_{D^l}^{\dagger}+2E\Lambda_{D^k}\Lambda_{D^l}^{\dagger})
\nonumber\\&&-4(\Lambda_{U^m}\tilde{D}\Lambda_{U^m}^{\dagger})_{kl}-4(\Lambda_{U^m}^{\dagger}\Lambda_{U^p}\Lambda_{U^p}^{\dagger}\Lambda_{U^m})_{lk}-6(Y_D^{\dagger}\Lambda_{D^k}^t\Lambda_{D^p}^*Y_D)_{lp}
\nonumber\\&&+4(\Lambda_{U^m}^{\dagger})_{lq}(\Lambda_{U^m})_{kn}Tr(\Lambda_{D^n}\Lambda_{D^q}^{\dagger})-(6Y_D^t\Lambda_{D^l}^{\dagger}\Lambda_{D^p}Y_D^*+2Y_D^t\Lambda_{D^l}^{\dagger}\Lambda_{E^p}Y_E^*)_{kp}
\nonumber\\&&-2(Y_D^{\dagger}\Lambda_{D^k}^t\Lambda_{E^p}^*Y_E)_{lp}-2(\Lambda_{U^q}^{\dagger}\Lambda_{U^n})_{lk}[Tr(\Lambda_{U^n}^{\dagger}\Lambda_{U^q})+2(\tilde{T})_{nq}]\}
\nonumber\\&&-2(\Lambda_{U^l})_{jk}\{(6M_2g_2^2-\frac{2}{5}M_1g_1^2)(\tilde{T})_{li}-(6g_2^2-\frac{2}{5}g_1^2)(Y_U^{\dagger}h_U)_{li}
\nonumber\\&&-(\frac{8}{3}g_3^2-\frac{4}{15}g_1^2)Tr(h_{U^i}\Lambda_{U^l}^{\dagger})+(\frac{8}{3}M_3g_3^2-\frac{4}{15}M_1g_1^2)Tr(\Lambda_{U^i}\Lambda_{U^l}^{\dagger})
\nonumber\\&&+2(Y_U^{\dagger}h_U\tilde{T}+Y_U^{\dagger}Th_U+Y_U^{\dagger}Dh_U)_{li}+2(Y_U^{\dagger}\Lambda_{D^m}^t\Lambda_{D^m}^*h_U)_{li}
\nonumber\\&&+6(Y_U^{\dagger}h_U)_{li}Tr(T)+6(\tilde{T})_{li}Tr(h_UY_U^{\dagger})+2(Y_U^{\dagger}h_DY_D^{\dagger}Y_U)_{li}
\nonumber\\&&+4Tr(\Lambda_{U^l}^{\dagger}\Lambda_{U^i}Y_D^{\dagger}h_D)+4(\Lambda_{U^l}^{\dagger}h_{U^i})_{nm}Tr(\Lambda_{D^m}^{\dagger}\Lambda_{D^n})
\nonumber\\&&+4Tr(h_{U^i}\Lambda_{U^l}^{\dagger}\Lambda_{U^p}\Lambda_{U^p}^{\dagger})+4Tr(\Lambda_{U^i}\Lambda_{U^l}^{\dagger}h_{U^p}\Lambda_{U^p}^{\dagger})
\nonumber\\&&+4Tr(\Lambda_{U^l}^{\dagger}h_{U^i}\tilde{D})+2(Y_U^{\dagger}h_{D^m}^t\Lambda_{D^m}^*Y_U)_{li}+4(\Lambda_{U^l}^{\dagger}\Lambda_{U^i})_{nm}Tr(\Lambda_{D^m}^{\dagger}h_{D^n})\}
\nonumber\\&&-2(\Lambda_{U^i})_{lk}\{(\frac{16}{3}M_3g_3^2+\frac{16}{15}M_1g_1^2)(\Lambda_{U^q}\Lambda_{U^q}^{\dagger})_{jl}+2(Y_D^{\dagger}\Lambda_{D^j}^t\Lambda_{E^p}^*h_E)_{lp}
\nonumber\\&&-(6g_2^2+\frac{2}{5}g_1^2)[(h_DY_D^{\dagger})_{jl}+Tr(h_{D^j}\Lambda_{D^l}^{\dagger})]+2Tr(h_EY_E^{\dagger}\Lambda_{D^j}\Lambda_{D^l}^{\dagger})
\nonumber\\&&+(6M_2g_2^2+\frac{2}{5}M_1g_1^2)[(D)_{jl}+Tr(\Lambda_{D^j}\Lambda_{D^l}^{\dagger})]-(\frac{16}{3}g_3^2+\frac{16}{15}g_1^2)(h_{U^q}\Lambda_{U^q}^{\dagger})_{jl}
\nonumber\\&&+2\tilde{D}_{lj}Tr(Y_E^{\dagger}h_E)+2Tr(\Lambda_{D^j}^t\Lambda_{D^l}^*h_UY_U^{\dagger})+4(\Lambda_{U^q}^{\dagger}\Lambda_{U^n})_{lj}(Y_U^{\dagger}h_U)_{nq}
\nonumber\\&&+2(Y_D^{\dagger}h_D\tilde{D}+Y_D^{\dagger}h_UY_U^{\dagger}Y_D)_{lj}+2(Y_D^{\dagger}Dh_D+Y_D^{\dagger}Th_D)_{lj}
\nonumber\\&&+2(Y_D^{\dagger}h_D)_{lj}Tr(E+3D)+6\tilde{D}_{lj}Tr(Y_D^{\dagger}h_D)
\nonumber\\&&+2(h_D^t\Lambda_{D^p}^{\dagger}\Lambda_{D^p}Y_D^*)_{jl}+2Tr(\Lambda_{D^j}^t\Lambda_{D^l}^*h_DY_D^{\dagger})+4(\Lambda_{U^m}Y_D^{\dagger}h_D\Lambda_{U^m}^{\dagger})_{jl}
\nonumber\\&&+6(Y_D^{\dagger}\Lambda_{D^j}^t\Lambda_{D^p}^*h_D)_{lp}+(6h_D^t\Lambda_{D^l}^{\dagger}\Lambda_{D^p}Y_D^*+2h_D^t\Lambda_{D^l}^{\dagger}\Lambda_{E^p}Y_E^*)_{jp}
\nonumber\\&&+2Tr(h_{E^q}\Lambda_{E^q}^{\dagger}\Lambda_{D^j}\Lambda_{D^l}^{\dagger})+2(Y_D^t\Lambda_{D^l}^{\dagger}h_{E^p}Y_E^*)_{jp}+4(h_{U^m}\tilde{D}\Lambda_{U^m}^{\dagger})_{jl}
\nonumber\\&&+4(\Lambda_{U^m}^{\dagger}h_{U^p}\Lambda_{U^p}^{\dagger}\Lambda_{U^m})_{lj}+4(\Lambda_{U^m}^{\dagger}\Lambda_{U^p}\Lambda_{U^p}^{\dagger}h_{U^m})_{lj}-4(\Lambda_{U^m}^{\dagger})_{lq}(h_{U^m})_{jn}Tr(\Lambda_{D^n}\Lambda_{D^q}^{\dagger})
\nonumber\\&&+2(\Lambda_{U^q}^{\dagger}h_{U^n})_{lj}[Tr(\Lambda_{U^n}^{\dagger}\Lambda_{U^q})+2(\tilde{T})_{nq}]+2(\Lambda_{U^q}^{\dagger}\Lambda_{U^n})_{lj}Tr(\Lambda_{U^n}^{\dagger}h_{U^q})
\nonumber\\&&+2(Y_D^t\Lambda_{D^p}^{\dagger}h_{D^p}Y_D^*)_{jl}+2Tr[h_{D^j}^t\Lambda_{D^l}^*(D+T)+\Lambda_{D^l}^{\dagger}h_{D^j}\Lambda_{D^q}^{\dagger}\Lambda_{D^q}
\nonumber\\&&+\Lambda_{D^l}^{\dagger}\Lambda_{D^j}\Lambda_{D^q}^{\dagger}h_{D^q}]+6Tr(h_{D^q}\Lambda_{D^q}^{\dagger}\Lambda_{D^j}\Lambda_{D^l}^{\dagger}+\Lambda_{D^q}\Lambda_{D^q}^{\dagger}h_{D^j}\Lambda_{D^l}^{\dagger})
\nonumber\\&&+2Tr(\Lambda_{E^q}\Lambda_{E^q}^{\dagger}h_{D^j}\Lambda_{D^l}^{\dagger}+Eh_{D^j}\Lambda_{D^l}^{\dagger})-4(\Lambda_{U^m}^{\dagger})_{lq}(\Lambda_{U^m})_{jn}Tr(h_{D^n}\Lambda_{D^q}^{\dagger})
\nonumber\\&&+(6Y_D^{\dagger}h_{D^j}^t\Lambda_{D^p}^*Y_D)_{lp}+2(Y_D^{\dagger}h_{D^j}^t\Lambda_{E^p}^*Y_E)_{lp}+(6Y_D^t\Lambda_{D^l}^{\dagger}h_{D^p}Y_D^*)_{jp}\}
\nonumber\\&&-2(\Lambda_{U^i})_{jl}\{(\frac{16}{3}M_3g_3^2+\frac{16}{15}M_1g_1^2)(\Lambda_{U^q}\Lambda_{U^q}^{\dagger})_{kl}-(6g_2^2+\frac{2}{5}g_1^2)(h_DY_D^{\dagger})_{kl}
\nonumber\\&&+(6M_2g_2^2+\frac{2}{5}M_1g_1^2)[(D)_{kl}+Tr(\Lambda_{D^k}\Lambda_{D^l}^{\dagger})]-(\frac{16}{3}g_3^2+\frac{16}{15}g_1^2)(h_{U^q}\Lambda_{U^q}^{\dagger})_{kl}
\nonumber\\&&+2(\tilde{D})_{lk}Tr(Y_E^{\dagger}h_E)+Tr(2h_EY_E^{\dagger}\Lambda_{D^k}\Lambda_{D^l}^{\dagger})+2(Y_D^{\dagger}\Lambda_{D^k}^t\Lambda_{E^p}^*h_E)_{lp}
\nonumber\\&&+2(Y_D^{\dagger}h_UY_U^{\dagger}Y_D)_{lk}+2Tr(\Lambda_{D^k}^t\Lambda_{D^l}^*h_UY_U^{\dagger})+4(\Lambda_{U^q}^{\dagger}\Lambda_{U^n})_{lk}(Y_U^{\dagger}h_U)_{nq}
\nonumber\\&&+2(Y_D^{\dagger}h_D\tilde{D})_{lk}+2(Y_D^{\dagger}Dh_D)_{lk}+2(Y_D^{\dagger}Th_D)_{lk}
\nonumber\\&&+6(\tilde{D})_{lk}Tr(Y_D^{\dagger}h_D)+2(h_D^t\Lambda_{D^p}^{\dagger}\Lambda_{D^p}Y_D^*)_{kl}+2Tr(\Lambda_{D^k}^t\Lambda_{D^l}^*h_DY_D^{\dagger})
\nonumber\\&&+4(\Lambda_{U^m}Y_D^{\dagger}h_D\Lambda_{U^m}^{\dagger})_{kl}+6(Y_D^{\dagger}\Lambda_{D^k}^t\Lambda_{D^p}^*h_D)_{lp}+6(h_D^t\Lambda_{D^l}^{\dagger}\Lambda_{D^p}Y_D^*
\nonumber\\&&+2h_D^t\Lambda_{D^l}^{\dagger}\Lambda_{E^p}Y_E^*)_{kp}+2Tr(h_{E^q}\Lambda_{E^q}^{\dagger}\Lambda_{D^k}\Lambda_{D^l}^{\dagger})+2(Y_D^t\Lambda_{D^l}^{\dagger}h_{E^p}Y_E^*)_{kp}
\nonumber\\&&+4(h_{U^m}\tilde{D}\Lambda_{U^m}^{\dagger})_{kl}+4(\Lambda_{U^m}^{\dagger}h_{U^p}\Lambda_{U^p}^{\dagger}\Lambda_{U^m})_{lk}+4(\Lambda_{U^m}^{\dagger}\Lambda_{U^p}\Lambda_{U^p}^{\dagger}h_{U^m})_{lk}
\nonumber\\&&-4(\Lambda_{U^m}^{\dagger})_{lq}(h_{U^m})_{kn}Tr(\Lambda_{D^n}\Lambda_{D^q}^{\dagger})+2(\Lambda_{U^q}^{\dagger}h_{U^n})_{lk}[Tr(\Lambda_{U^n}^{\dagger}\Lambda_{U^q})+2(\tilde{T})_{nq}]
\nonumber\\&&+2(\Lambda_{U^q}^{\dagger}\Lambda_{U^n})_{lk}Tr(\Lambda_{U^n}^{\dagger}h_{U^q})+2(Y_D^t\Lambda_{D^p}^{\dagger}h_{D^p}Y_D^*)_{kl}+2Tr[h_{D^k}^t\Lambda_{D^l}^*(D+T)]
\nonumber\\&&+2Tr(\Lambda_{D^l}^{\dagger}h_{D^k}\Lambda_{D^q}^{\dagger}\Lambda_{D^q}+\Lambda_{D^l}^{\dagger}\Lambda_{D^k}\Lambda_{D^q}^{\dagger}h_{D^q})+Tr(6h_{D^q}\Lambda_{D^q}^{\dagger}\Lambda_{D^k}\Lambda_{D^l}^{\dagger}
\nonumber\\&&+6\Lambda_{D^q}\Lambda_{D^q}^{\dagger}h_{D^k}\Lambda_{D^l}^{\dagger}+2\Lambda_{E^q}\Lambda_{E^q}^{\dagger}h_{D^k}\Lambda_{D^l}^{\dagger}+2Eh_{D^k}\Lambda_{D^l}^{\dagger})
\nonumber\\&&-4(\Lambda_{U^m}^{\dagger})_{lq}(\Lambda_{U^m})_{kn}Tr(h_{D^n}\Lambda_{D^q}^{\dagger})+6(Y_D^{\dagger}h_{D^k}^t\Lambda_{D^p}^*Y_D)_{lp}+2(Y_D^{\dagger}h_{D^k}^t\Lambda_{E^p}^*Y_E)_{lp}
\nonumber\\&&+6(Y_D^t\Lambda_{D^l}^{\dagger}h_{D^p}Y_D^*)_{kp}+2(Y_D^{\dagger}h_D)_{lk}Tr(E+3D)\}
\nonumber\\&&+(\Lambda_{U^i})_{jk}(\frac{32}{3}M_3g_3^4-\frac{112}{5}M_1g_1^4-\frac{128}{15}(M_1+M_3)g_3^2g_1^2)
\nonumber\\&&+(h_{U^i})_{jk}(-\frac{8}{3}g_3^4+\frac{28}{5}g_1^4+\frac{64}{15}g_3^2g_1^2)
\end{eqnarray}
The two loop $\beta$ functions  for the gaugino masses and soft
breaking masses are given by:

\begin{eqnarray}
{(16\pi^2)^2}\beta^{(2)}_{M_1}&=&{(16\pi^2)^2} \beta^{(2)}_{{M_1}_{(RPC)}}
-2M_1g_1^2\{\frac{12}{5}Tr(\Lambda_{U^i}\Lambda_{U^i}^{\dagger})+\frac{14}{5}Tr(\Lambda_{D^i}h_{D^i}^{\dagger})+\frac{9}{5}Tr(\Lambda_{E^i}\Lambda_{E^i}^{\dagger})\}
\nonumber\\&&+2g_1^2[\frac{12}{5}Tr(h_{U^i}\Lambda_{U^i}^{\dagger})+\frac{14}{5}Tr(h_{D^i}\Lambda_{D^i}^{\dagger})+\frac{9}{5}Tr(h_{E^i}\Lambda_{E^i}^{\dagger})]
\end{eqnarray}
\begin{eqnarray}
{(16\pi^2)^2}\beta^{(2)}_{M_2}&=&{(16\pi^2)^2} \beta^{(2)}_{{M_2}_{(RPC)}}-2M_2g_2^2[6Tr(\Lambda_{D^i}\Lambda_{D^i}^{\dagger})+Tr(\Lambda_{E^i}\Lambda_{E^i}^{\dagger})]
\nonumber\\&&+2g_2^2[6Tr(h_{D^i}\Lambda_{D^i}^{\dagger})+Tr(h_{E^i}\Lambda_{E^i}^{\dagger})]
\end{eqnarray}
\begin{eqnarray}
{(16\pi^2)^2}\beta^{(2)}_{M_3}&=&{(16\pi^2)^2} \beta^{(2)}_{{M_3}_{(RPC)}}-2M_3g_3^2[3Tr(\Lambda_{U^i}\Lambda_{U^i}^{\dagger})+4Tr(\Lambda_{D^i}\Lambda_{D^i}^{\dagger})]
\nonumber\\&&+2g_3^2[3Tr(h_{U^i}\Lambda_{U^i}^{\dagger})+4Tr(h_{D^i}\Lambda_{D^i}^{\dagger})]
\end{eqnarray}

\begin{eqnarray}
{(16\pi^2)^2}\beta^{(2)}_{(M^2_{\tilde{E}})_{ij}}&=&{(16\pi^2)^2}\beta^{(2)}_{(M^2_{\tilde{E}})_{RPC}}+(12M_2^2g_2^2-\frac{12}{5}M_1^2g_1^2)Tr(\Lambda_{E^j}\Lambda_{E^i}^{\dagger})-(6M_2g_2^2-\frac{6}{5}M_1g_1^2)Tr(\Lambda_{E^j}h_{E^i}^{\dagger})
\nonumber\\&&-(6M_2^*g_2^2-\frac{6}{5}M_1^*g_1^2)Tr(h_{E^j}\Lambda_{E^i}^{\dagger})+(6g_2^2-\frac{6}{5}g_1^2)[Tr(h_{E^j}h_{E^i}^{\dagger})
\nonumber\\&&-4(h_E^{\dagger}h_{E^l}\Lambda_{E^l}^{\dagger}Y_E+h_E^{\dagger}\Lambda_{E^l}\Lambda_{E^l}^{\dagger}h_E+Y_E^{\dagger}h_{E^l}h_{E^l}^{\dagger}Y_E+Y_E^{\dagger}\Lambda_{E^l}h_{E^l}^{\dagger}h_E)_{ij}
\nonumber\\&&-12(h_E^{\dagger}h_{D^l}\Lambda_{D^l}^{\dagger}Y_E+h_E^{\dagger}\Lambda_{D^l}\Lambda_{D^l}^{\dagger}h_E+Y_E^{\dagger}h_{D^l}h_{D^l}^{\dagger}Y_E+Y_E^{\dagger}\Lambda_{D^l}h_{D^l}^{\dagger}h_E)_{ij}
\nonumber\\&&+12(h_E^{\dagger}h_{E^j}\Lambda_{D^m}^*Y_D+h_E^{\dagger}\Lambda_{E^j}\Lambda_{D^m}^*h_D+Y_E^{\dagger}h_{E^j}h_{D^m}^*Y_D+Y_E^{\dagger}\Lambda_{E^j}h_{D^m}^*h_D)_{im}
\nonumber\\&&-4(h_E^{\dagger}\Lambda_{E^j}\Lambda_{E^m}^{\dagger}h_E+h_E^{\dagger}h_{E^j}\Lambda_{E^m}^{\dagger}Y_E+Y_E^{\dagger}h_{E^j}h_{E^m}^{\dagger}Y_E+Y_E^{\dagger}\Lambda_{E^j}h_{E^m}^{\dagger}h_E)_{im}
\nonumber\\&&-4Tr[h_{E^j}h_{E^i}^{\dagger}(E+\Lambda_{E^l}\Lambda_{E^l}^{\dagger}+3\Lambda_{D^l}\Lambda_{D^l}^{\dagger})]+12(Y_D^{\dagger}h_{D^m}^th_{E^i}^{\dagger}Y_E
\nonumber\\&&+Y_D^{\dagger}\Lambda_{D^m}^th_{E^i}^{\dagger}h_E)_{mj}-4Tr[\Lambda_{E^j}h_{E^i}^{\dagger}(h_EY_E^{\dagger}+h_{E^l}\Lambda_{E^l}^{\dagger}+3h_{D^l}\Lambda_{D^l}^{\dagger})]
\nonumber\\&&-4Tr[h_{E^j}\Lambda_{E^i}^{\dagger}(Y_Eh_E^{\dagger}+\Lambda_{E^l}h_{E^l}^{\dagger}+3\Lambda_{D^l}h_{D^l}^{\dagger})]+12(h_D^{\dagger}h_{D^m}^t\Lambda_{E^i}^{\dagger}Y_E
\nonumber\\&&+h_D^{\dagger}\Lambda_{D^m}^t\Lambda_{E^i}^{\dagger}h_E)_{mj}-4Tr[\Lambda_{E^j}\Lambda_{E^i}^{\dagger}(h_Eh_E^{\dagger}+h_{E^l}h_{E^l}^{\dagger}+3h_{D^l}h_{D^l}^{\dagger})]
\nonumber\\&&-4(h_E^{\dagger}h_{E^m}\Lambda_{E^i}^{\dagger}Y_E+h_E^{\dagger}\Lambda_{E^m}\Lambda_{E^i}^{\dagger}h_E+Y_E^{\dagger}h_{E^m}h_{E^i}^{\dagger}Y_E+Y_E^{\dagger}\Lambda_{E^m}h_{E^i}^{\dagger}h_E)_{mj}
\nonumber\\&&+(3g_2^2-\frac{3}{5}g_1^2)[4Tr(m_{\tilde{L}}^2\Lambda_{E^j}\Lambda_{E^i}^{\dagger})+(m_{\tilde{E}}^2)_{lj}Tr(\Lambda_{E^l}\Lambda_{E^i}^{\dagger})+(m_{\tilde{E}}^2)_{il}Tr(\Lambda_{E^j}\Lambda_{E^l}^{\dagger})
\nonumber\\&&+4(\Lambda_{E^i}^{\dagger}Y_E)_{lj}m_{\tilde{L}_lH_1}^2]+(12g_2^2-\frac{12}{5}g_1^2)(Y_E^{\dagger}\Lambda_{E^j})_{il}m_{H_1\tilde{L}_l}^2-2(Y_E^{\dagger}\Lambda_{E^k})_{iq}(\tilde{E})_{kj}m_{H_1\tilde{L}_q}^2
\nonumber\\&&-4(Y_E^{\dagger}E\Lambda_{E^j})_{iq}m_{H_1\tilde{L}_q}^2-4(Y_E^{\dagger}m_{\tilde{L}}^2\Lambda_{E^l}\Lambda_{E^l}^{\dagger}Y_E)_{ij}-4(Y_E^{\dagger}\Lambda_{E^l}({m_{\tilde{L}}^2})^t\Lambda_{E^l}^{\dagger}Y_E)_{ij}
\nonumber\\&&-4(m_{\tilde{E}}^2)_{kl}(Y_E^{\dagger}\Lambda_{E^k}\Lambda_{E^l}^{\dagger}Y_E)_{ij}-2(\tilde{E})_{il}(\Lambda_{E^l}^{\dagger}Y_E)_{qj}m_{\tilde{L}_qH_1}^2+2(Y_E^t\Lambda_{E^l}^{\dagger}Y_E)_{lj}(Y_E^{\dagger})_{iq}m_{\tilde{L}_qH_1}^2
\nonumber\\&&-4(Y_E^{\dagger}\Lambda_{E^l}\Lambda_{E^l}^{\dagger}m_{\tilde{L}}^2Y_E)_{ij}-2(Y_E^{\dagger}\Lambda_{E^l}\Lambda_{E^l}^{\dagger}Y_Em_{\tilde{E}}^2)_{ij}-4(Y_E^{\dagger}\Lambda_{E^l}\Lambda_{E^l}^{\dagger}\Lambda_{E^j})_{il}m_{H_1\tilde{L}_l}^2
\nonumber\\&&-4m_{H_1}^2(Y_E^{\dagger}\Lambda_{E^l}\Lambda_{E^l}^{\dagger}Y_E)_{ij}-12(Y_E^{\dagger}m_{\tilde{L}}^2\Lambda_{D^l}\Lambda_{D^l}^{\dagger}Y_E+Y_E^{\dagger}\Lambda_{D^l}(m_{\tilde{Q}}^{2})^t\Lambda_{D^l}^{\dagger}Y_E)_{ij}
\nonumber\\&&-12(m_{\tilde{D}}^2)_{kl}(Y_E^{\dagger}\Lambda_{D^k}\Lambda_{D^l}^{\dagger}Y_E)_{ij}+6(Y_D^t\Lambda_{D^l}^{\dagger}Y_E)_{lj}(Y_E^{\dagger})_{iq}m_{\tilde{L}_qH_1}^2
\nonumber\\&&-12(Y_E^{\dagger}\Lambda_{D^l}\Lambda_{D^l}^{\dagger}m_{\tilde{L}}^2Y_E)_{ij}-12m_{H_1}^2(Y_E^{\dagger}\Lambda_{D^l}\Lambda_{D^l}^{\dagger}Y_E)_{ij}-6(Y_E^{\dagger}\Lambda_{D^l}\Lambda_{D^l}^{\dagger}Y_Em_{\tilde{E}}^2)_{ij}
\nonumber\\&&-4(Y_E^{\dagger}\Lambda_{E^j})_{il}m_{H_1\tilde{L}_l}^2Tr(E+3D)-4\tilde{E}_{ij}(Y_E^{\dagger}\Lambda_{E^k})_{kl}m_{H_1\tilde{L}_l}^2
\nonumber\\&&+12(\tilde{E})_{ij}(Y_D^{\dagger}\Lambda_{D^k}^t)_{kl}m_{H_1\tilde{L}_l}^2+12(Y_E^{\dagger}m_{\tilde{L}}^2\Lambda_{E^j}\Lambda_{D^m}^*Y_D)_{im}
\nonumber\\&&+12(Y_E^{\dagger}\Lambda_{E^j}{m_{\tilde{L}}^2}^t\Lambda_{D^m}^*Y_D)_{im}+6(m_{\tilde{E}}^2)_{kj}(Y_E^{\dagger}\Lambda_{E^k}\Lambda_{D^m}^*Y_D)_{im}
\nonumber\\&&+6(\tilde{E})_{ij}(\Lambda_{D^m}^*Y_D)_{qm}m_{\tilde{L}_qH_1}^2-6(Y_E^t\Lambda_{D^m}^*Y_D)_{jm}(Y_E^{\dagger})_{il}m_{\tilde{L}_lH_1}^2
\nonumber\\&&+12(Y_E^{\dagger}\Lambda_{E^j}\Lambda_{D^m}^*m_{\tilde{Q}}^2Y_D)_{im}+12(Y_E^{\dagger}\Lambda_{E^j}\Lambda_{D^m}^*Y_Dm_{\tilde{D}}^2)_{im}-4m_{H_1}^2Tr(\Lambda_{E^j}\Lambda_{E^i}^{\dagger}E)
\nonumber\\&&+12m_{H_1}^2(Y_E^{\dagger}\Lambda_{E^j}\Lambda_{D^m}^*Y_D)_{im}-12(Y_E^{\dagger}\Lambda_{E^j}\Lambda_{D^m}^*\Lambda_{D^m}^t)_{il}m_{H_1\tilde{L}_l}^2
\nonumber\\&&-4(Y_E^{\dagger}m_{\tilde{L}}^2\Lambda_{E^j}\Lambda_{E^m}^{\dagger}Y_E)_{im}-4(Y_E^{\dagger}\Lambda_{E^j}(m_{\tilde{L}}^{2})^t\Lambda_{E^m}^{\dagger}Y_E)_{im}
\nonumber\\&&-2(m_{\tilde{E}}^2)_{kj}(Y_E^{\dagger}\Lambda_{E^k}\Lambda_{E^m}^{\dagger}Y_E)_{im}-2(\tilde{E})_{ij}(\Lambda_{E^m}^{\dagger}Y_E)_{qm}m_{\tilde{L}_qH_1}^2
\nonumber\\&&+2(Y_E^t\Lambda_{E^m}^{\dagger}Y_E)_{jm}(Y_E^{\dagger})_{il}m_{\tilde{L}_lH_1}^2-4(Y_E^{\dagger}\Lambda_{E^j}\Lambda_{E^m}^{\dagger}m_{\tilde{L}}^2Y_E)_{im}
\nonumber\\&&-4(Y_E^{\dagger}\Lambda_{E^j}\Lambda_{E^m}^{\dagger}Y_Em_{\tilde{E}}^2)_{im}-4m_{H_1}^2(Y_E^{\dagger}\Lambda_{E^j}\Lambda_{E^m}^{\dagger}Y_E)_{im}-4Tr(Em_{\tilde{L}}^2\Lambda_{E^j}\Lambda_{E^i}^{\dagger})
\nonumber\\&&-4(Y_E^{\dagger}\Lambda_{E^j}\Lambda_{E^m}^{\dagger}\Lambda_{E^m})_{il}m_{H_1\tilde{L}_l}^2-12(Y_E^{\dagger}\Lambda_{D^l}\Lambda_{D^l}^{\dagger}\Lambda_{E^j})_{iq}m_{H_1\tilde{L}_q}^2
\nonumber\\&&-4Tr(\Lambda_{E^j}{m_{\tilde{L}}^2}^t\Lambda_{E^i}^{\dagger}E)-2(m_{\tilde{E}}^2)_{lj}Tr(\Lambda_{E^l}\Lambda_{E^i}^{\dagger}E)-4(\Lambda_{E^i}^{\dagger}EY_E)_{qj}m_{\tilde{L}_qH_1}^2
\nonumber\\&&+2(Y_E^t\Lambda_{E^i}^{\dagger}E)_{jq}m_{\tilde{L}_qH_1}^2-4Tr(m_{\tilde{L}}^2\Lambda_{E^j}\Lambda_{E^i}^{\dagger}\Lambda_{E^l}\Lambda_{E^l}^{\dagger})-4Tr(\Lambda_{E^i}^{\dagger}\Lambda_{E^l}\Lambda_{E^l}^{\dagger}\Lambda_{E^j}(m_{\tilde{L}}^{2})^t)
\nonumber\\&&-2(m_{\tilde{E}}^2)_{kj}Tr(\Lambda_{E^k}\Lambda_{E^i}^{\dagger}\Lambda_{E^l}\Lambda_{E^l}^{\dagger})-4(\Lambda_{E^i}^{\dagger}\Lambda_{E^l}\Lambda_{E^l}^{\dagger}Y_E)_{qj}m_{\tilde{L}_qH_1}^2
\nonumber\\&&+2(Y_E^t\Lambda_{E^i}^{\dagger}\Lambda_{E^l}\Lambda_{E^l}^{\dagger})_{jq}m_{\tilde{L}_qH_1}^2-12Tr(m_{\tilde{L}}^2\Lambda_{E^j}\Lambda_{E^i}^{\dagger}\Lambda_{D^l}\Lambda_{D^l}^{\dagger})-12Tr(\Lambda_{E^i}^{\dagger}\Lambda_{D^l}\Lambda_{D^l}^{\dagger}\Lambda_{E^j}(m_{\tilde{L}}^{2})^t)
\nonumber\\&&-6(m_{\tilde{E}}^2)_{kj}Tr(\Lambda_{E^k}\Lambda_{E^i}^{\dagger}\Lambda_{D^l}\Lambda_{D^l}^{\dagger})-12(\Lambda_{E^i}^{\dagger}\Lambda_{D^l}\Lambda_{D^l}^{\dagger}Y_E)_{qj}m_{\tilde{L}_qH_1}^2+6(Y_E^t\Lambda_{E^i}^{\dagger}\Lambda_{D^l}\Lambda_{D^l}^{\dagger})_{jq}m_{\tilde{L}_qH_1}^2
\nonumber\\&&-4Tr(\Lambda_{E^j}\Lambda_{E^i}^{\dagger}m_{\tilde{L}}^2E)-4Tr(\Lambda_{E^j}\Lambda_{E^i}^{\dagger}Y_Em_{\tilde{E}}^2Y_E^{\dagger})+2(Y_E^{\dagger}\Lambda_{E^m}Y_E^*)_{mi}(Y_E)_{qj}m_{H_1\tilde{L}_q}^2
\nonumber\\&&-4(Y_E^{\dagger}\Lambda_{E^j}\Lambda_{E^i}^{\dagger}\Lambda_{E^k})_{kl}m_{H_1\tilde{L}_l}^2-4Tr(\Lambda_{E^j}\Lambda_{E^i}^{\dagger}m_{\tilde{L}}^2\Lambda_{E^l}\Lambda_{E^l}^{\dagger})-4Tr(\Lambda_{E^j}\Lambda_{E^i}^{\dagger}\Lambda_{E^l}(m_{\tilde{L}}^{2})^t\Lambda_{E^l}^{\dagger})
\nonumber\\&&-4(m_{\tilde{E}}^2)_{kl}Tr(\Lambda_{E^j}\Lambda_{E^i}^{\dagger}\Lambda_{E^k}\Lambda_{E^l}^{\dagger})-4(\Lambda_{E^l}^{\dagger}\Lambda_{E^j}\Lambda_{E^i}^{\dagger}Y_E)_{ql}m_{\tilde{L}_qH_1}^2
\nonumber\\&&+2(Y_E^t\Lambda_{E^l}^{\dagger}\Lambda_{E^j}\Lambda_{E^i}^{\dagger})_{lq}m_{\tilde{L}_qH_1}^2-12Tr(\Lambda_{E^j}\Lambda_{E^i}^{\dagger}m_{\tilde{L}}^2\Lambda_{D^l}\Lambda_{D^l}^{\dagger})
\nonumber\\&&-12Tr(\Lambda_{E^j}\Lambda_{E^i}^{\dagger}\Lambda_{D^l}(m_{\tilde{Q}}^{2})^t\Lambda_{D^l}^{\dagger})-12(m_{\tilde{D}}^2)_{kl}Tr(\Lambda_{E^j}\Lambda_{E^i}^{\dagger}\Lambda_{D^k}\Lambda_{D^l}^{\dagger})
\nonumber\\&&+6(Y_D^t\Lambda_{D^l}^{\dagger}\Lambda_{E^j}\Lambda_{E^i}^{\dagger})_{lq}m_{\tilde{L}_qH_1}^2+12(Y_E^t\Lambda_{E^i}^*m_{\tilde{L}}^2\Lambda_{D^m}Y_D^*)_{jm}+12(Y_E^t\Lambda_{E^i}^*\Lambda_{D^m}(m_{\tilde{Q}}^{2})^tY_D^*)_{jm}
\nonumber\\&&-12Tr(\tilde{D})(Y_E^t\Lambda_{E^i}^*)_{jl}m_{\tilde{L}_lH_1}^2+12(Y_D^{\dagger}\Lambda_{D^m}^t\Lambda_{E^i}^{\dagger}m_{\tilde{L}}^2Y_E)_{mj}+6(Y_D^{\dagger}\Lambda_{D^m}^{t}\Lambda_{E^i}^{\dagger}Y_Em_{\tilde{E}}^2)_{mj}
\nonumber\\&&+6(Y_D^{\dagger}\Lambda_{D^m}^t\Lambda_{E^i}^{\dagger}\Lambda_{E^j})_{ml}m_{H_1\tilde{L}_l}^2+12m_{H_1}^2(Y_D^{\dagger}\Lambda_{D^m}^{t}\Lambda_{E^i}^{\dagger}Y_E)_{mj}-4(Y_E^{\dagger}m_{\tilde{L}}^2\Lambda_{E^m}\Lambda_{E^i}^{\dagger}Y_E)_{mj}
\nonumber\\&&-4(Y_E^{\dagger}\Lambda_{E^m}(m_{\tilde{L}}^{2})^t\Lambda_{E^i}^{\dagger}Y_E)_{mj}-4(m_{\tilde{E}}^2Y_E^{\dagger}\Lambda_{E^k}\Lambda_{E^i}^{\dagger}Y_E)_{kj}-4Tr(\tilde{E})(\Lambda_{E^i}^{\dagger}Y_E)_{qj}m_{\tilde{L}_qH_1}^2
\nonumber\\&&-4(Y_E^{\dagger}\Lambda_{E^m}\Lambda_{E^i}^{\dagger}m_{\tilde{L}}^2Y_E)_{mj}-2(Y_E^{\dagger}\Lambda_{E^m}\Lambda_{E^i}^{\dagger}Y_Em_{\tilde{E}}^2)_{mj}-4(m_{H_1}^2)(Y_E^{\dagger}\Lambda_{E^m}\Lambda_{E^i}^{\dagger}Y_E)_{mj}
\nonumber\\&&-2(Y_E^{\dagger}\Lambda_{E^m}\Lambda_{E^i}^{\dagger}\Lambda_{E^j})_{mk}m_{H_1\tilde{L}_k}^2+12(m_{\tilde{D}}^2Y_D^{\dagger}\Lambda_{D^k}^t\Lambda_{E^i}^{\dagger}Y_E)_{kj}+2(E^t\Lambda_{E^i}^{\dagger}Y_E)_{qj}m_{\tilde{L}_qH_1}^2
\nonumber\\&&-2(\tilde{E})_{il}(\Lambda_{E^l}^{\dagger}Y_E)_{qj}m_{\tilde{L}_qH_1}^2-2(m_{\tilde{E}}^2Y_E^{\dagger}\Lambda_{E^l}\Lambda_{E^l}^{\dagger}Y_E)_{ij}-2(\tilde{E})_{lj}(Y_E^{\dagger}\Lambda_{E^l})_{iq}m_{H_1\tilde{L}_q}^2
\nonumber\\&&-6(m_{\tilde{E}}^2Y_E^{\dagger}\Lambda_{D^l}\Lambda_{D^l}^{\dagger}Y_E)_{ij}+6(Y_E^{\dagger}\Lambda_{D^l}Y_D^*)_{il}(Y_E)_{qj}m_{H_1\tilde{L}_q}^2-2(\tilde{E})_{ij}(\Lambda_{E^k}^{\dagger}Y_E)_{lk}m_{\tilde{L}_lH_1}^2
\nonumber\\&&+6(\tilde{E})_{ij}(\Lambda_{D^k}^*Y_D)_{lk}m_{\tilde{L}_lH_1}^2+6(m_{\tilde{E}}^2Y_E^{\dagger}\Lambda_{E^j}\Lambda_{D^m}^*Y_D)_{im}+6(\Lambda_{E^i}^{\dagger}\Lambda_{E^j}\Lambda_{D^m}^*Y_D)_{lm}m_{\tilde{L}_lH_1}^2
\nonumber\\&&-2(m_{\tilde{E}}^2Y_E^{\dagger}\Lambda_{E^j}\Lambda_{E^m}^{\dagger}Y_E)_{im}-2(\Lambda_{E^i}^{\dagger}\Lambda_{E^j}\Lambda_{E^m}^{\dagger}Y_E)_{lm}m_{\tilde{L}_lH_1}^2+2(Y_E^{\dagger}\Lambda_{E^j}E^t)_{iq}m_{H_1\tilde{L}_q}^2
\nonumber\\&&-2(m_{\tilde{E}}^2)_{ik}Tr(\Lambda_{E^j}\Lambda_{E^k}^{\dagger}E)+2(E\Lambda_{E^j}Y_E^*)_{qi}m_{H_1\tilde{L}_q}^2-6(m_{\tilde{E}}^2)_{ik}Tr(\Lambda_{E^j}\Lambda_{E^k}^{\dagger}\Lambda_{D^l}\Lambda_{D^l}^{\dagger})
\nonumber\\&&-2(m_{\tilde{E}}^2)_{ik}Tr(\Lambda_{E^j}\Lambda_{E^k}^{\dagger}\Lambda_{E^l}\Lambda_{E^l}^{\dagger})+2(\Lambda_{E^j}\Lambda_{E^i}^{\dagger}\Lambda_{E^l}Y_E^*)_{ql}m_{H_1\tilde{L}_q}^2
\nonumber\\&&+6(\Lambda_{E^j}\Lambda_{E^i}^{\dagger}\Lambda_{D^l}Y_D^*)_{ql}m_{H_1\tilde{L}_q}^2-6(\Lambda_{D^m}^*\Lambda_{D^m}^t\Lambda_{E^i}^{\dagger}Y_E)_{lj}m_{\tilde{L}_lH_1}^2
\nonumber\\&&+6(m_{\tilde{E}}^2)_{ik}(Y_D^{\dagger}\Lambda_{D^m}^t\Lambda_{E^k}^{\dagger}Y_E)_{mj}-6(Y_D^{\dagger}\Lambda_{D^m}^tY_E^*)_{mi}(Y_E)_{qj}m_{H_1\tilde{L}_q}^2
\nonumber\\&&-2(\Lambda_{E^m}^{\dagger}\Lambda_{E^m}\Lambda_{E^i}^{\dagger}Y_E)_{lj}m_{\tilde{L}_lH_1}^2-2(Y_E^{\dagger}\Lambda_{E^m}\Lambda_{E^k}^{\dagger}Y_E)_{mj}(m_{\tilde{E}}^2)_{ik}-\frac{6}{5}s'\delta_{ij}
\end{eqnarray}

\begin{eqnarray}
{(16\pi^2)^2}\beta^{(2)}_{(M^2_{\tilde{L}})_{ij}}&=&{(16\pi^2)^2}\beta^{(2)}_{(M^2_{\tilde{L}})_{RPC}}+\nonumber\\&&
(64M_3^2g_3^2-\frac{8}{5}M_1^2g_1^2)(\Lambda_{D^q}\Lambda_{D^q}^{\dagger})_{ij}+\frac{24}{5}M_1^2g_1^2(\Lambda_{E^q}\Lambda_{E^q}^{\dagger})_{ij}
\nonumber\\&&+(-32M_3g_3^2+\frac{4}{5}M_1g_1^2)(\Lambda_{D^q}h_{D^q}^{\dagger})_{ij}-\frac{12}{5}M_1g_1^2(\Lambda_{E^q}h_{E^q}^{\dagger})_{ij}
\nonumber\\&&+(-32M_3^*g_3^2+\frac{4}{5}M_1^*g_1^2)(h_{D^q}\Lambda_{D^q}^{\dagger})_{ij}-\frac{12}{5}M_1^*g_1^2(h_{E^q}\Lambda_{E^q}^{\dagger})_{ij}
\nonumber\\&&+(32g_3^2-\frac{4}{5}g_1^2)(h_{D^q}h_{D^q}^{\dagger
})_{ij}+\frac{12}{5}g_1^2(h_{E^q}h_{E^q}^{\dagger})_{ij}
\nonumber\\&&-2(h_E^{\dagger})_{kj}(h_E)_{il}Tr(\Lambda_{E^l}^{\dagger}\Lambda_{E^k})-2(h_E^{\dagger})_{kj}(Y_E)_{il}Tr(\Lambda_{E^l}^{\dagger}h_{E^k})
\nonumber\\&&-2(Y_E^{\dagger})_{kj}(h_E)_{il}Tr(h_{E^l}^{\dagger}\Lambda_{E^k})-2(Y_E^{\dagger})_{kj}(Y_E)_{il}Tr(h_{E^l}^{\dagger}h_{E^k})
\nonumber\\&&-4(\Lambda_{E^l}\Lambda_{E^k}^{\dagger})_{ij}(h_E^{\dagger}h_E)_{lk}-4(h_{E^l}\Lambda_{E^k}^{\dagger})_{ij}(h_E^{\dagger}Y_E)_{lk}-4(\Lambda_{E^l}h_{E^k}^{\dagger})_{ij}(Y_E^{\dagger}h_E)_{lk}
\nonumber\\&&-4(h_{E^l}h_{E^k}^{\dagger})_{ij}\tilde{E}_{lk}-2(h_{E^l}\Lambda_{E^k}^{\dagger})_{ij}Tr(h_{E^l}^{\dagger}\Lambda_{E^k})-2(h_{E^l}h_{E^k}^{\dagger})_{ij}Tr(\Lambda_{E^l}^{\dagger}\Lambda_{E^k})
\nonumber\\&&-2(\Lambda_{E^l}h_{E^k}^{\dagger})_{ij}Tr(\Lambda_{E^l}^{\dagger}h_{E^k})-2(\Lambda_{E^l}\Lambda_{E^k}^{\dagger})_{ij}Tr(h_{E^l}^{\dagger}h_{E^k})
\nonumber\\&&-2(h_E)_{ik}(3h_{E^k}^{\dagger}\Lambda_{D^p}Y_D^*+3\Lambda_{E^k}^{\dagger}\Lambda_{D^p}h_D^*+h_{E^k}^{\dagger}\Lambda_{E^p}Y_E^*+\Lambda_{E^k}^{\dagger}\Lambda_{E^p}h_E^*)_{jp}
\nonumber\\&&-2(Y_E)_{ik}(3h_{E^k}^{\dagger}h_{D^p}Y_D^*+3\Lambda_{E^k}^{\dagger}h_{D^p}h_D^*+h_{E^k}^{\dagger}h_{E^p}Y_E^*+\Lambda_{E^k}^{\dagger}h_{E^p}h_E^*)_{jp}
\nonumber\\&&+2(h_E^*)_{jk}(3\Lambda_{E^k}\Lambda_{D^p}^*h_D+3h_{E^k}\Lambda_{D^p}^*Y_D+\Lambda_{E^k}\Lambda_{E^p}^*h_E+h_{E^k}\Lambda_{E^p}^*Y_E)_{ip}
\nonumber\\&&+2(Y_E^*)_{jk}(3\Lambda_{E^k}h_{D^p}^*h_D+3h_{E^k}h_{D^p}^*Y_D+\Lambda_{E^k}h_{E^p}^*h_E+h_{E^k}h_{E^p}^*Y_E)_{ip}
\nonumber\\&&-2(h_{E^k}^{\dagger}h_EY_E^{\dagger}\Lambda_{E^k}+h_{E^k}^{\dagger}Eh_{E^k}+\Lambda_{E^k}^{\dagger}h_Eh_E^{\dagger}\Lambda_{E^k}+\Lambda_{E^k}^{\dagger}Y_Eh_E^{\dagger}h_{E^k})_{ji}
\nonumber\\&&-6(h_{E^k}^{\dagger}h_{D^p}\Lambda_{D^p}^{\dagger}\Lambda_{E^k}+h_{E^k}^{\dagger}\Lambda_{D^p}\Lambda_{D^p}^{\dagger}h_{E^k}+\Lambda_{E^k}^{\dagger}h_{D^p}h_{D^p}^{\dagger}\Lambda_{E^k}+\Lambda_{E^k}^{\dagger}\Lambda_{D^p}h_{D^p}^{\dagger}h_{E^k})_{ji}
\nonumber\\&&-2(h_{E^k}^{\dagger}h_{E^p}\Lambda_{E^p}^{\dagger}\Lambda_{E^k}+h_{E^k}^{\dagger}\Lambda_{E^p}\Lambda_{E^p}^{\dagger}h_{E^k}+\Lambda_{E^k}^{\dagger}h_{E^p}h_{E^p}^{\dagger}\Lambda_{E^k}+\Lambda_{E^k}^{\dagger}\Lambda_{E^p}h_{E^p}^{\dagger}h_{E^k})_{ji}
\nonumber\\&&-12(h_{D^l}h_{D^k}^{\dagger})_{ij}[\tilde{D}_{lk}+Tr(\Lambda_{D^l}^{\dagger}\Lambda_{D^k})+(\Lambda_{U^q}\Lambda_{U^q}^{\dagger})_{kl}]
\nonumber\\&&-12(\Lambda_{D^l}h_{D^k}^{\dagger})_{ij}[(Y_D^{\dagger}h_D)_{lk}+Tr(\Lambda_{D^l}^{\dagger}h_{D^k})+(h_{U^q}\Lambda_{U^q}^{\dagger})_{kl}]
\nonumber\\&&-12(\Lambda_{D^l}\Lambda_{D^k}^{\dagger})_{ij}[(h_D^{\dagger}h_D)_{lk}+Tr(h_{D^l}^{\dagger}h_{D^k})+(h_{U^q}h_{U^q}^{\dagger})_{kl}]
\nonumber\\&&-12(h_{D^l}\Lambda_{D^k}^{\dagger})_{ij}[(h_D^{\dagger}Y_D)_{lk}+Tr(h_{D^l}^{\dagger}\Lambda_{D^k})+(\Lambda_{U^q}h_{U^q}^{\dagger})_{kl}]
\nonumber\\&&-6(h_{D^k}^*h_DY_D^{\dagger}\Lambda_{D^k}^t+h_{D^k}^*h_UY_U^{\dagger}\Lambda_{D^k}^t+h_{D^k}^*Y_DY_D^{\dagger}h_{D^k}^t+h_{D^k}^*Y_UY_U^{\dagger}h_{D^k}^t)_{ji}
\nonumber\\&&-6(\Lambda_{D^k}^*h_Dh_D^{\dagger}\Lambda_{D^k}^t+\Lambda_{D^k}^*Y_Dh_D^{\dagger}h_{D^k}^t+\Lambda_{D^k}^*h_Uh_U^{\dagger}\Lambda_{D^k}^t+\Lambda_{D^k}^*Y_Uh_U^{\dagger}h_{D^k}^t)_{ji}
\nonumber\\&&-6(h_{D^k}h_{D^p}^{\dagger}\Lambda_{D^p}\Lambda_{D^k}^{\dagger}+\Lambda_{D^k}h_{D^p}^{\dagger}h_{D^p}\Lambda_{D^k}^{\dagger}+\Lambda_{D^k}\Lambda_{D^p}^{\dagger}h_{D^p}h_{D^k}^{\dagger}+h_{D^k}\Lambda_{D^p}^{\dagger}\Lambda_{D^p}h_{D^k}^{\dagger})_{ij}
\nonumber\\&&+(16g_3^2-\frac{2}{5}g_1^2)[(m_{\tilde{L}}^2\Lambda_{D^q}\Lambda_{D^q}^{\dagger})_{ij}+2(\Lambda_{D^q}{m_{\tilde{Q}}^2}^t\Lambda_{D^q}^{\dagger})_{ij}+2(m^2_{\tilde{D}})_{lq}(\Lambda_{D^l}\Lambda_{D^q}^{\dagger})_{ij}
\nonumber\\&&-(Y_D^t\Lambda_{D^q}^{\dagger})_{qj}m_{\tilde{L}_iH_1}^2]-4(EY_E)_{ik}(\Lambda_{E^k}^{\dagger})_{qj}m_{\tilde{L}_qH_1}^2+3(\Lambda_{E^k}^*EY_E)_{jk}m_{\tilde{L}_iH_1}^2
\nonumber\\&&+\frac{6}{5}g_1^2(\Lambda_{E^k})_{il}(Y_E^{\dagger})_{kj}m_{H_1\tilde{L}_l}^2+\frac{6}{5}g_1^2[(m_{\tilde{L}}^2\Lambda_{E^q}\Lambda_{E^q}^{\dagger})_{ij}+2(\Lambda_{E^q}{m_{\tilde{L}}^2}^t\Lambda_{E^q}^{\dagger})_{ij}
\nonumber\\&&+2(m_{\tilde{E}}^2)_{lq}(\Lambda_{E^l}\Lambda_{E^q}^{\dagger})_{ij}+(Y_E)_{iq}(\Lambda_{E^q}^{\dagger})_{kj}m_{\tilde{L}_kH_1}^2-(Y_E^t\Lambda_{E^q}^{\dagger})_{qj}m^2_{\tilde{L}_iH_1}]
\nonumber\\&&-4(\Lambda_{E^k})_{il}(Y_E^{\dagger}E)_{kj}m^2_{H_1\tilde{L}_l}-3(Y_E^{\dagger})_{kj}(E\Lambda_{E^k})_{il}m^2_{H_1\tilde{L}_l}
\nonumber\\&&-(Y_E^{\dagger})_{kj}(m_{\tilde{L}}^2Y_E)_{il}Tr(\Lambda_{E^l}^{\dagger}\Lambda_{E^k})-2(Y_E^{\dagger})_{kj}(Y_Em_{\tilde{E}}^2)_{il}Tr(\Lambda_{E^l}^{\dagger}\Lambda_{E^k})
\nonumber\\&&-2m^2_{H_1}(Y_E^{\dagger})_{kj}(Y_E)_{il}Tr(\Lambda_{E^l}^{\dagger}\Lambda_{E^k})-(Y_E^{\dagger})_{kj}(\Lambda_{E^l})_{iq}Tr(\Lambda_{E^l}^{\dagger}\Lambda_{E^k})m_{H_1\tilde{L}_q}^2
\nonumber\\&&-2(m_{\tilde{E}}^2Y_E^{\dagger})_{qj}(Y_E)_{il}Tr(\Lambda_{E^l}^{\dagger}\Lambda_{E^q})-2(\Lambda_{E^l}^{\dagger}E)_{mj}(Y_E)_{il}m^2_{\tilde{L}_mH_1}
\nonumber\\&&-2(m_{\tilde{L}}^2\Lambda_{E^l}\Lambda_{E^k}^{\dagger})_{ij}\tilde{E}_{lk}-3(\Lambda_{D^k}^*T\Lambda_{D^k}^t(m_{\tilde{L}}^{2})^t)_{ji}
\nonumber\\&&-4(\Lambda_{E^l}{m_{\tilde{L}}^2}^t\Lambda_{E^k}^{\dagger})_{ij}\tilde{E}_{lk}-4(m_{\tilde{E}}^2\tilde{E})_{qk}(\Lambda_{E^q}\Lambda_{E^k}^{\dagger})_{ij}
\nonumber\\&&-4(\Lambda_{E^l}\Lambda_{E^k}^{\dagger})_{ij}(Y_E^{\dagger}m_{\tilde{L}}^2Y_E)_{lk}-4(\Lambda_{E^l}\Lambda_{E^k}^{\dagger})_{ij}[\tilde{E}m_{\tilde{E}}^2]_{lk}
\nonumber\\&&-4m^2_{H_1}(\Lambda_{E^l}\Lambda_{E^k}^{\dagger})_{ij}\tilde{E}_{lk}-2(\Lambda_{E^l}\Lambda_{E^k}^{\dagger})_{ij}(Y_E^{\dagger}\Lambda_{E^k})_{lq}m_{H_1\tilde{L}_q}^2
\nonumber\\&&-(m_{\tilde{L}}^2\Lambda_{E^l}\Lambda_{E^k}^{\dagger})_{ij}Tr(\Lambda_{E^l}^{\dagger}\Lambda_{E^k})-2(\Lambda_{E^l}{m_{\tilde{L}}^2}^t\Lambda_{E^k}^{\dagger})_{ij}Tr(\Lambda_{E^l}^{\dagger}\Lambda_{E^k})
\nonumber\\&&-2(m_{\tilde{E}}^2)_{ql}(\Lambda_{E^q}\Lambda_{E^k}^{\dagger})_{ij}Tr(\Lambda_{E^l}^{\dagger}\Lambda_{E^k})-Tr(\Lambda_{E^l}^{\dagger}\Lambda_{E^k})(Y_E)_{il}(\Lambda_{E^k}^{\dagger})_{qj}m_{\tilde{L}_qH_1}^2
\nonumber\\&&+Tr(\Lambda_{E^l}^{\dagger}\Lambda_{E^k})(Y_E^t\Lambda_{E^k}^{\dagger})_{lj}m_{\tilde{L}_iH_1}^2-4(\Lambda_{E^l}\Lambda_{E^k}^{\dagger})_{ij}Tr(\Lambda_{E^l}^{\dagger}m_{\tilde{L}}^2\Lambda_{E^k})
\nonumber\\&&-2(m_{\tilde{E}}^2)_{qk}(\Lambda_{E^l}\Lambda_{E^k}^{\dagger})_{ij}Tr(\Lambda_{E^l}^{\dagger}\Lambda_{E^q})-4(\Lambda_{E^l}\Lambda_{E^k}^{\dagger})_{ij}(\Lambda_{E^l}^{\dagger}Y_E)_{qk}m_{\tilde{L}_qH_1}^2
\nonumber\\&&-Tr(2E+6D)(\Lambda_{E^k})_{il}(Y_E^{\dagger})_{kj}m_{H_1\tilde{L}_l}^2+3(\Lambda_{D^k}^*Y_D)_{jl}(\Lambda_{D^l}^*\Lambda_{D^k}^t)_{qi}m_{\tilde{L}_qH_1}^2
\nonumber\\&&-E_{ij}(Y_E^{\dagger}\Lambda_{E^k})_{kl}m_{H_1\tilde{L}_l}^2+6E_{ij}(Y_D^{\dagger}\Lambda_{D^k}^t)_{kl}m_{H_1\tilde{L}_l}^2
\nonumber\\&&-3(m_{\tilde{L}}^2Y_E)_{ik}(\Lambda_{E^k}^{\dagger}\Lambda_{D^p}Y_D^*)_{jp}-(m_{\tilde{L}}^2Y_E)_{ik}(\Lambda_{E^k}^{\dagger}\Lambda_{E^p}Y_E^*)_{jp}
\nonumber\\&&-m_{H_1}^2(Y_E)_{ik}[6\Lambda_{E^k}^{\dagger}\Lambda_{D^p}Y_D^*+2\Lambda_{E^k}^{\dagger}\Lambda_{E^p}Y_E^*]_{jp}
\nonumber\\&&-6(Y_Em_{\tilde{E}}^2)_{ik}(\Lambda_{E^k}^{\dagger}\Lambda_{D^p}Y_D^*)_{jp}-2(Y_Em_{\tilde{E}}^2)_{ik}(\Lambda_{E^k}^{\dagger}\Lambda_{E^p}Y_E^*)_{jp}
\nonumber\\&&-3(\Lambda_{E^k})_{il}m_{H_1\tilde{L}_l}^2(\Lambda_{E^k}^{\dagger}\Lambda_{D^p}Y_D^*)_{jp}-(\Lambda_{E^k})_{il}m_{H_1\tilde{L}_l}^2(\Lambda_{E^k}^{\dagger}\Lambda_{E^p}Y_E^*)_{jp}
\nonumber\\&&-6(Y_E)_{ik}(\Lambda_{E^k}^{\dagger}m_{\tilde{L}}^2\Lambda_{D^p}Y_D^*)_{jp}-6(Y_E)_{ik}(\Lambda_{E^k}^{\dagger}\Lambda_{D^p}{m_{\tilde{Q}}^2}^tY_D^*)_{jp}
\nonumber\\&&-3(Y_E)_{ik}(\Lambda_{E^k}^{\dagger}\Lambda_{D^l}Y_D^*{m_{\tilde{D}}^2}^t)_{jl}+6(Y_E)_{ik}(\Lambda_{E^k}^{\dagger})_{jl}Tr(D)m_{\tilde{L}_lH_1}^2
\nonumber\\&&-2(Y_E)_{ik}(\Lambda_{E^k}^{\dagger}m_{\tilde{L}}^2\Lambda_{E^p}Y_E^*)_{jp}-2(Y_E)_{ik}(\Lambda_{E^k}^{\dagger}\Lambda_{E^p}{m_{\tilde{L}}^2}^tY_E^*)_{jp}
\nonumber\\&&-(Y_E)_{ik}(m_{\tilde{E}}^2Y_E^{\dagger}\Lambda_{E^l}^t\Lambda_{E^k}^*)_{lj}+2(Y_E)_{ik}(\Lambda_{E^k}^{\dagger})_{jl}Tr(E)m_{\tilde{L}_lH_1}^2
\nonumber\\&&+6m_{H_1}^2(Y_E^{\dagger})_{kj}(\Lambda_{E^k}\Lambda_{D^p}^*Y_D)_{ip}-3(Y_E^{\dagger})_{kj}(\Lambda_{E^k}\Lambda_{D^p}^*\Lambda_{D^p}^t)_{il}m_{H_1\tilde{L}_l}^2
\nonumber\\&&+2(Y_E^{\dagger})_{kj}(\Lambda_{E^k}\Lambda_{E^p}^*m_{\tilde{L}}^2Y_E+\Lambda_{E^k}\Lambda_{E^p}^*Y_Em_{\tilde{E}}^2)_{ip}+2m_{H_1}^2(Y_E^{\dagger})_{kj}(\Lambda_{E^k}\Lambda_{E^p}^*Y_E)_{ip}
\nonumber\\&&+m_{H_1\tilde{L}_l}^2(\Lambda_{E^k}\Lambda_{E^p}^*\Lambda_{E^p})_{il}(Y_E^{\dagger})_{kj}+3(Y_E^{\dagger})_{kj}(m_{\tilde{L}}^2\Lambda_{E^k}\Lambda_{D^p}^*Y_D+2\Lambda_{E^k}{m_{\tilde{L}}^2}^t\Lambda_{D^p}^*Y_D)_{ip}
\nonumber\\&&+6(m_{\tilde{E}}^2Y_E^{\dagger})_{lj}(\Lambda_{E^l}\Lambda_{D^p}^*Y_D)_{ip}+6E_{ij}(\Lambda_{D^p}^*Y_D)_{kp}m_{\tilde{L}_kH_1}^2
\nonumber\\&&-3(Y_D^t\Lambda_{D^p}^{\dagger}E)_{jp}m_{\tilde{L}_iH_1}^2+(Y_E^{\dagger})_{kj}(m_{\tilde{L}}^2\Lambda_{E^k}\Lambda_{E^p}^*Y_E+2\Lambda_{E^k}{m_{\tilde{L}}^2}^t\Lambda_{E^p}^*Y_E)_{ip}
\nonumber\\&&+2(m_{\tilde{E}}^2Y_E^{\dagger})_{lj}(\Lambda_{E^l}\Lambda_{E^p}^*Y_E)_{ip}+2E_{ij}(\Lambda_{E^p}^*Y_E)_{kp}m_{\tilde{L}_kH_1}^2
\nonumber\\&&-(Y_E^t\Lambda_{E^p}^{\dagger}E)_{pj}m_{\tilde{L}_iH_1}^2-2(\Lambda_{E^k}^{\dagger}m_{\tilde{L}}^2E\Lambda_{E^k})_{ji}
\nonumber\\&&-2(\Lambda_{E^k}^{\dagger}Y_Em_{\tilde{E}}^2Y_E^{\dagger}\Lambda_{E^k})_{ji}-(2\Lambda_{E^k}^{\dagger}Em_{\tilde{L}}^2\Lambda_{E^k}+\Lambda_{E^k}^{\dagger}E\Lambda_{E^k}{m_{\tilde{L}}^2}^t)_{ji}
\nonumber\\&&-(\Lambda_{E^k}^{\dagger}\Lambda_{E^m})_{jq}(Y_E^{\dagger}\Lambda_{E^k})_{mi}m_{H_1\tilde{L}_q}^2-2m_{H_1}^2(\Lambda_{E^k}^{\dagger}E\Lambda_{E^k})_{ji}
\nonumber\\&&-2(m_{\tilde{E}}^2)_{lk}(\Lambda_{E^k}^{\dagger}E\Lambda_{E^l})_{ji}-(\Lambda_{E^k}^{\dagger}EY_E)_{jk}m_{\tilde{L}_iH_1}^2+6(Y_E^{\dagger})_{kj}(\Lambda_{E^k}\Lambda_{D^p}^*m_{\tilde{Q}}^2Y_D
\nonumber\\&&+\Lambda_{E^k}\Lambda_{D^p}^*Y_Dm_{\tilde{D}}^2)_{ip}-6(\Lambda_{E^k}^{\dagger}m_{\tilde{L}}^2\Lambda_{D^p}\Lambda_{D^p}^{\dagger}\Lambda_{E^k}+\Lambda_{E^k}^{\dagger}\Lambda_{D^p}{m_{\tilde{Q}}^2}^t\Lambda_{D^p}^{\dagger}\Lambda_{E^k})_{ji}
\nonumber\\&&-6(m_{\tilde{D}}^2)_{lp}(\Lambda_{E^k}^{\dagger}\Lambda_{D^l}\Lambda_{D^p}^{\dagger}\Lambda_{E^k})_{ji}+3(Y_D^t\Lambda_{D^p}^{\dagger}\Lambda_{E^k})_{pi}(\Lambda_{E^k}^{\dagger})_{jl}m_{\tilde{L}_lH_1}^2
\nonumber\\&&+3(2\Lambda_{E^k}^{\dagger}\Lambda_{D^p}\Lambda_{D^p}^{\dagger}m_{\tilde{L}}^2\Lambda_{E^k}+\Lambda_{E^k}^{\dagger}\Lambda_{D^p}\Lambda_{D^p}^{\dagger}\Lambda_{E^k}{m_{\tilde{L}}^2}^t)_{ji}
\nonumber\\&&-6(m_{\tilde{E}}^2)_{lk}(\Lambda_{E^k}^{\dagger}\Lambda_{D^p}\Lambda_{D^p}^{\dagger}\Lambda_{E^l})_{ji}-(\Lambda_{E^k}^{\dagger}\Lambda_{E^p}\Lambda_{E^p}^{\dagger}Y_E)_{jk}m_{\tilde{L}_iH_1}^2
\nonumber\\&&-3(\Lambda_{E^k}^{\dagger}\Lambda_{D^p}\Lambda_{D^p}^{\dagger}Y_E)_{jk}m_{\tilde{L}_iH_1}^2+3(\Lambda_{E^k}^{\dagger}\Lambda_{D^p}\Lambda_{D^p}^{\dagger})_{jq}(Y_E)_{ik}m_{\tilde{L}_qH_1}^2
\nonumber\\&&+2(\Lambda_{E^k}^{\dagger}\Lambda_{E^p}\Lambda_{E^p}^{\dagger})_{jq}(Y_E)_{ik}m_{\tilde{L}_qH_1}^2-2(m_{\tilde{E}}^2)_{lk}(\Lambda_{E^k}^{\dagger}\Lambda_{E^p}\Lambda_{E^p}^{\dagger}\Lambda_{E^l})_{ji}
\nonumber\\&&-(2\Lambda_{E^k}^{\dagger}\Lambda_{E^p}\Lambda_{E^p}^{\dagger}m_{\tilde{L}}^2\Lambda_{E^k}+\Lambda_{E^k}^{\dagger}\Lambda_{E^p}\Lambda_{E^p}^{\dagger}\Lambda_{E^k}{m_{\tilde{L}}^2}^t)_{ji}
\nonumber\\&&-2(\Lambda_{E^k}^{\dagger}m_{\tilde{L}}^2\Lambda_{E^p}\Lambda_{E^p}^{\dagger}\Lambda_{E^k}+\Lambda_{E^k}^{\dagger}\Lambda_{E^p}{m_{\tilde{L}}^2}^t\Lambda_{E^p}^{\dagger}\Lambda_{E^k})_{ji}
\nonumber\\&&-2(m_{\tilde{E}}^2)_{lp}(\Lambda_{E^k}^{\dagger}\Lambda_{E^l}\Lambda_{E^p}^{\dagger}\Lambda_{E^k})_{ji}-(\Lambda_{E^k}^{\dagger}Y_E)_{jp}(\Lambda_{E^p}^{\dagger}\Lambda_{E^k})_{qi}m_{\tilde{L}_qH_1}^2
\nonumber\\&&+(Y_E^t\Lambda_{E^p}^{\dagger}\Lambda_{E^k})_{pi}(\Lambda_{E^k}^{\dagger})_{jl}m_{\tilde{L}_lH_1}^2+6(\Lambda_{U^q}\Lambda_{U^q}^{\dagger}Y_D^t\Lambda_{D^k}^{\dagger})_{kj}m_{\tilde{L}_iH_1}^2
\nonumber\\&&-12(\Lambda_{D^l}{m_{\tilde{Q}}^2}^t\Lambda_{D^k}^{\dagger})_{ij}[(\tilde{D})_{lk}+Tr(\Lambda_{D^l}^{\dagger}\Lambda_{D^k})+(\Lambda_{U^q}\Lambda_{U^q}^{\dagger})_{kl}]
\nonumber\\&&-6(m_{\tilde{L}}^2\Lambda_{D^l}\Lambda_{D^k}^{\dagger})_{ij}\tilde{D}_{lk}-6(m_{\tilde{L}}^2\Lambda_{D^l}\Lambda_{D^k}^{\dagger})_{ij}Tr(\Lambda_{D^l}^{\dagger}\Lambda_{D^k})
\nonumber\\&&-6(m_{\tilde{L}}^2\Lambda_{D^l}\Lambda_{D^k}^{\dagger})_{ij}(\Lambda_{U^q}\Lambda_{U^q}^{\dagger})_{kl}+9(\Lambda_{D^k}^*DY_D)_{jk}m_{\tilde{L}_iH_1}^2.
\nonumber\\&&+6(\Lambda_{D^k}^*Y_D)_{jl}Tr(\Lambda_{D^l}^{\dagger}\Lambda_{D^k})m_{\tilde{L}_iH_1}^2-12(\Lambda_{D^l}\Lambda_{D^k}^{\dagger})_{ij}(Y_D^{\dagger}m_{\tilde{Q}}^2Y_D)_{lk}
\nonumber\\&&-12(\Lambda_{D^l}\Lambda_{D^k}^{\dagger})_{ij}(\tilde{D}m_{\tilde{D}}^2)_{lk}-12m_{H_1}^2(\Lambda_{D^l}\Lambda_{D^k}^{\dagger})_{ij}(\tilde{D})_{lk}
\nonumber\\&&+6m_{H_1\tilde{L}_q}^2(Y_D^{\dagger}\Lambda_{D^k}^t)_{lq}(\Lambda_{D^l}\Lambda_{D^k}^{\dagger})_{ij}-12(\Lambda_{D^l}\Lambda_{D^k}^{\dagger})_{ij}[Tr(\Lambda_{D^l}^{\dagger}m_{\tilde{L}}^2\Lambda_{D^k})
\nonumber\\&&+Tr(\Lambda_{D^l}^{\dagger}\Lambda_{D^k}{m_{\tilde{Q}}^2}^t)]-4(Y_E^{\dagger})_{kj}(Y_E)_{il}Tr(\Lambda_{E^l}^{\dagger}m_{\tilde{L}}^2\Lambda_{E^k})
\nonumber\\&&-12(m_{\tilde{D}}^2)_{qk}(\Lambda_{D^l}\Lambda_{D^k}^{\dagger})_{ij}Tr(\Lambda_{D^l}^{\dagger}\Lambda_{D^q})+6(\Lambda_{D^l}\Lambda_{D^k}^{\dagger})_{ij}(Y_D^t\Lambda_{D^l}^{\dagger})_{kq}m_{\tilde{L}_qH_1}^2
\nonumber\\&&-12(\Lambda_{D^l}\Lambda_{D^k}^{\dagger})_{ij}(\Lambda_{U^q}m_{\tilde{D}}^2\Lambda_{U^q}^{\dagger})_{kl}-12(\Lambda_{D^l}\Lambda_{D^k}^{\dagger})_{ij}({m_{\tilde{D}}^2}^t\Lambda_{U^q}\Lambda_{U^q}^{\dagger})_{kl}
\nonumber\\&&-12(\Lambda_{D^l}\Lambda_{D^k}^{\dagger})_{ij}(\Lambda_{U^p}\Lambda_{U^q}^{\dagger})_{kl}(m_{\tilde{U}}^2)_{pq}-6(\Lambda_{D^k}^*m_{\tilde{Q}}^2D\Lambda_{D^k}^t)_{ji}
\nonumber\\&&-6(\Lambda_{D^k}^*Y_Dm_{\tilde{D}}^2Y_D^{\dagger}\Lambda_{D^k}^t)_{ji}-6m_{H_1}^2(\Lambda_{D^k}^*D\Lambda_{D^k}^t)_{ji}
\nonumber\\&&+3(\Lambda_{D^k}^*\Lambda_{D^m}^t)_{jq}(Y_D^{\dagger}\Lambda_{D^k}^t)_{mi}m_{H_1\tilde{L}_q}^2-3(m_{\tilde{L}}^2\Lambda_{D^k}D^t\Lambda_{D^k}^{\dagger})_{ij}
\nonumber\\&&-6(\Lambda_{D^k}{m_{\tilde{Q}}^2}^tD^t\Lambda_{D^k}^{\dagger})_{ij}-6(m_{\tilde{D}}^2)_{lk}(\Lambda_{D^l}D^t\Lambda_{D^k}^{\dagger})_{ij}-6(m_{\tilde{D}}^2)_{lk}(\Lambda_{D^k}^*T\Lambda_{D^l}^t)_{ji}
\nonumber\\&&+3(\Lambda_{D^k}^*TY_D)_{jk}m_{\tilde{L}_iH_1}^2-6(\Lambda_{D^k}^*m_{\tilde{Q}}^2T\Lambda_{D^k}^t)_{ji}-6(\Lambda_{D^k}^*Y_Um_{\tilde{U}}^2Y_U^{\dagger}\Lambda_{D^k}^t)_{ji}
\nonumber\\&&-3(m_{\tilde{L}}^2\Lambda_{D^k}\Lambda_{D^p}^{\dagger}\Lambda_{D^p}\Lambda_{D^k}^{\dagger})_{ij}-6(\Lambda_{D^k}{m_{\tilde{Q}}^2}^t\Lambda_{D^p}^{\dagger}\Lambda_{D^p}\Lambda_{D^k}^{\dagger})_{ij}
\nonumber\\&&+3(Y_D^t\Lambda_{D^p}^{\dagger}\Lambda_{D^p}\Lambda_{D^k}^{\dagger})_{kj}m_{\tilde{L}_iH_1}^2-6(\Lambda_{D^k}\Lambda_{D^p}^{\dagger}m_{\tilde{L}}^2\Lambda_{D^p}\Lambda_{D^k}^{\dagger})_{ij}
\nonumber\\&&-6(m_{\tilde{D}}^2)_{lp}(\Lambda_{D^k}\Lambda_{D^p}^{\dagger}\Lambda_{D^l}\Lambda_{D^k}^{\dagger})_{ij}+3(Y_D^t\Lambda_{D^k}^{\dagger})_{pj}(\Lambda_{D^k}\Lambda_{D^p}^{\dagger})_{il}m_{\tilde{L}_lH_1}^2
\nonumber\\&&-12(m_{\tilde{D}}^2)_{pl}Tr(\Lambda_{D^l}^{\dagger}\Lambda_{D^k})(\Lambda_{D^p}\Lambda_{D^k}^{\dagger})_{ij}-6(m_{\tilde{D}}^2)_{lk}(\Lambda_{D^l}\Lambda_{D^p}^{\dagger}\Lambda_{D^p}\Lambda_{D^k}^{\dagger})_{ij}
\nonumber\\&&+(16g_3^2-\frac{2}{5}g_1^2)[(\Lambda_{D^q}\Lambda_{D^q}^{\dagger}m_{\tilde{L}}^2)_{ij}-(\Lambda_{D^q}Y_D^*)_{iq}m_{H_1\tilde{L}_j}^2]
\nonumber\\&&+\frac{6}{5}g_1^2[(Y_E)_{ik}(\Lambda_{E^k}^{\dagger})_{lj}m_{\tilde{L}_lH_1}^2+(\Lambda_{E^q}\Lambda_{E^q}^{\dagger}m_{\tilde{L}}^2)_{ij}+(\Lambda_{E^q})_{ik}(Y_E^{\dagger})_{qj}m_{H_1\tilde{L}_k}^2
\nonumber\\&&-(\Lambda_{E^q}Y_E^*)_{iq}m_{H_1\tilde{L}_j}^2]-6m_{H_2}^2(\Lambda_{D^k}^*T\Lambda_{D^k}^t)_{ji}-6(\Lambda_{D^k}^*Tm^2_{\tilde{Q}}\Lambda_{D^k}^t)_{ij}
\nonumber\\&&-2(Y_E)_{il}(\Lambda_{E^l}^{\dagger}E)_{qj}m_{\tilde{L}_qH_1}^2-(Y_E^{\dagger}m_{\tilde{L}}^2)_{kj}(Y_E)_{il}Tr(\Lambda_{E^l}^{\dagger}\Lambda_{E^k})
\nonumber\\&&-m_{\tilde{L}_qH_1}^2(\Lambda_{E^k}^{\dagger})_{qj}(Y_E)_{il}Tr(\Lambda_{E^l}^{\dagger}\Lambda_{E^k})+(Y_E^{\dagger})_{kj}(E\Lambda_{E^k}^t)_{iq}m_{H_1\tilde{L}_q}^2
\nonumber\\&&-2\tilde{E}_{lk}(\Lambda_{E^l}\Lambda_{E^k}^{\dagger}m_{\tilde{L}}^2)_{ij}+3(Y_E^{\dagger}E\Lambda_{E^l}^t)_{li}m_{H_1\tilde{L}_j}^2-(Y_E^{\dagger}\Lambda_{E^p}^t\Lambda_{E^k}^*m_{\tilde{L}}^2)_{pj}(Y_E)_{ik}
\nonumber\\&&-(\Lambda_{E^l}\Lambda_{E^k}^{\dagger}m_{\tilde{L}}^2)_{ij}Tr(\Lambda_{E^l}^{\dagger}\Lambda_{E^k})-(\Lambda_{E^l})_{iq}(Y_E^{\dagger})_{kj}Tr(\Lambda_{E^l}^{\dagger}\Lambda_{E^k})m_{H_1\tilde{L}_q}^2
\nonumber\\&&+(\Lambda_{E^l}Y_E^*)_{ik}Tr(\Lambda_{E^l}^{\dagger}\Lambda_{E^k})m_{H_1\tilde{L}_j}^2-(Y_E^{\dagger}\Lambda_{E^k})_{lq}(\Lambda_{E^l}\Lambda_{E^k}^{\dagger})_{ij}m_{H_1\tilde{L}_q}^2
\nonumber\\&&+(\Lambda_{E^k}Y_E^*)_{ql}(\Lambda_{E^l}\Lambda_{E^k}^{\dagger})_{ij}m_{H_1\tilde{L}_q}^2-E_{ij}(\Lambda_{E^k}^{\dagger}Y_E)_{lk}m_{\tilde{L}_lH_1}^2
\nonumber\\&&-3(Y_E)_{ik}(Y_D^{\dagger}\Lambda_{D^p}^t\Lambda_{E^k}^*m_{\tilde{L}}^2)_{pj}-3(E\Lambda_{D^p}Y_D^*)_{ip}m_{H_1\tilde{L}_j}^2-(E\Lambda_{E^p}Y_E^*)_{ip}m_{H_1\tilde{L}_j}^2
\nonumber\\&&-3(Y_E)_{ik}(m_{\tilde{D}}^2Y_D^{\dagger}\Lambda_{D^p}^t\Lambda_{E^k}^*)_{pj}+3(Y_E)_{ik}(\Lambda_{E^k}^{\dagger}\Lambda_{D^p}\Lambda_{D^p}^{\dagger})_{jl}m_{\tilde{L}_lH_1}^2
\nonumber\\&&-(Y_E)_{ik}(m^2_{\tilde{E}}Y_E^{\dagger}\Lambda_{E^p}^t\Lambda_{E^k}^*)_{pj}+(Y_E^{\dagger})_{kj}(E\Lambda_{E^k})_{qi}m_{H_1\tilde{L}_q}^2
\nonumber\\&&+(Y_E^{\dagger}m_{\tilde{L}}^2)_{kj}(3\Lambda_{E^k}\Lambda_{D^p}^*Y_D+\Lambda_{E^k}\Lambda_{E^p}^*Y_E)_{ip}+(\Lambda_{E^k}^{\dagger})_{lj}(3\Lambda_{E^k}\Lambda_{D^p}^*Y_D
\nonumber\\&&+\Lambda_{E^k}\Lambda_{E^p}^*Y_E)_{ip}m_{\tilde{L}_lH_1}^2+(Y_E^{\dagger})_{kj}(E\Lambda_{E^k}^t)_{li}m_{H_1\tilde{L}_l}^2-({m_{\tilde{L}}^2}^t\Lambda_{E^k}^{\dagger}E\Lambda_{E^k})_{ji}
\nonumber\\&&-(\Lambda_{E^m}^{\dagger}\Lambda_{E^k})_{li}(\Lambda_{E^k}^{\dagger}Y_E)_{jm}m_{\tilde{L}_lH_1}^2-3({m_{\tilde{L}}^2}^t\Lambda_{E^k}^{\dagger}\Lambda_{D^p}\Lambda_{D^p}^{\dagger}\Lambda_{E^k})_{ji}
\nonumber\\&&-3(Y_E^{\dagger}\Lambda_{D^p}\Lambda_{D^p}^{\dagger}\Lambda_{E^k})_{ki}m_{H_1\tilde{L}_j}^2+3(Y_E^{\dagger})_{kj}(\Lambda_{D^p}\Lambda_{D^p}^{\dagger}\Lambda_{E^k})_{qi}m_{H_1\tilde{L}_q}^2
\nonumber\\&&+3(\Lambda_{E^k})_{qi}(\Lambda_{E^k}^{\dagger}\Lambda_{D^p}Y_D^*)_{jp}m_{H_1\tilde{L}_q}^2-({m_{\tilde{L}}^2}^t\Lambda_{E^k}^{\dagger}\Lambda_{E^p}\Lambda_{E^p}^{\dagger}\Lambda_{E^k})_{ji}
\nonumber\\&&-(Y_E^{\dagger}\Lambda_{E^p}\Lambda_{E^p}^{\dagger}\Lambda_{E^k})_{ki}m_{H_1\tilde{L}_j}^2+(Y_E^{\dagger})_{kj}(\Lambda_{E^p}\Lambda_{E^p}^{\dagger}\Lambda_{E^k})_{qi}m_{H_1\tilde{L}_q}^2
\nonumber\\&&-(Y_E^{\dagger}\Lambda_{E^k})_{pi}(\Lambda_{E^k}^{\dagger}\Lambda_{E^p})_{jl}m_{H_1\tilde{L}_l}^2+(\Lambda_{E^k}^{\dagger}\Lambda_{E^p}Y_E^*)_{jp}(\Lambda_{E^k})_{qi}m_{H_1\tilde{L}_q}^2
\nonumber\\&&-12(\Lambda_{U^q}\Lambda_{U^q}^{\dagger}{m_{\tilde{D}}^2}^t)_{kl}(\Lambda_{D^l}\Lambda_{D^k}^{\dagger})_{ij}-3(\Lambda_{D^k}D^t\Lambda_{D^k}^{\dagger}m_{\tilde{L}}^2)_{ij}-3((m_{\tilde{L}}^{2})^t\Lambda_{D^k}^*T\Lambda_{D^k}^t)_{ji}
\nonumber\\&&+3(Y_D^{\dagger}T\Lambda_{D^k}^t)_{ki}m_{H_1\tilde{L}_j}^2+3(\Lambda_{D^k}Y_D^*)_{ip}(\Lambda_{D^p}\Lambda_{D^k}^{\dagger})_{qj}m_{H_1\tilde{L}_q}^2-3(\Lambda_{D^k}\Lambda_{D^p}^{\dagger}\Lambda_{D^p}\Lambda_{D^k}^{\dagger}m_{\tilde{L}}^2)_{ij}
\nonumber\\&&+3(\Lambda_{D^k}\Lambda_{D^p}^{\dagger}\Lambda_{D^p}Y_D^*)_{ik}m_{H_1\tilde{L}_j}^2-6(\Lambda_{D^k}\Lambda_{D^p}^{\dagger}\Lambda_{D^p}{m_{\tilde{Q}}^2}^t\Lambda_{D^k}^{\dagger})_{ij}
\nonumber\\&&-6(\Lambda_{D^l}\Lambda_{D^k}^{\dagger}m_{\tilde{L}}^2)_{ij}[\tilde{D}_{lk}+Tr(\Lambda_{D^l}^{\dagger}\Lambda_{D^k})+(\Lambda_{U^q}\Lambda_{U^q}^{\dagger})_{kl}]
\nonumber\\&&+9(Y_D^{\dagger}D\Lambda_{D^l}^t)_{li}m_{H_1\tilde{L}_j}^2+6(\Lambda_{D^l}Y_D^*\Lambda_{U^q}\Lambda_{U^q}^{\dagger})_{il}m_{H_1\tilde{L}_j}^2
\nonumber\\&&+6(\Lambda_{D^l}Y_D^*)_{ik}Tr(\Lambda_{D^l}^{\dagger}\Lambda_{D^k})m_{H_1\tilde{L}_j}^2-12(m_{\tilde{D}}^2
\tilde{D})_{lk}(\Lambda_{D^l}\Lambda_{D^k}^{\dagger})_{ij}
\nonumber\\&&+6(\Lambda_{D^l}^*Y_D)_{qk}(\Lambda_{D^l}\Lambda_{D^k}^{\dagger})_{ij}m_{\tilde{L}_qH_1}^2+6(\Lambda_{D^k}Y_D^*)_{ql}(\Lambda_{D^l}\Lambda_{D^k}^{\dagger})_{ij}m_{H_1\tilde{L}_q}^2+\frac{3}{5}s'\delta_{ij}
\end{eqnarray}
\begin{eqnarray}
{(16\pi^2)^2}\beta^{(2)}_{(M^2_{\tilde{Q}})_{ij}}&=&{(16\pi^2)^2}\beta^{(2)}_{(M^2_{\tilde{Q}})_{RPC}}\nonumber\\&&+
\frac{4}{5}g_1^2[2M_1^2(\Lambda_{D^q}^{\dagger}\Lambda_{D^q})_{ji}-M_1(h_{D^q}^{\dagger}\Lambda_{D^q})_{ji}-M_1^*(\Lambda_{D^q}^{\dagger}h_{D^q})_{ji}+(h_{D^q}^{\dagger}h_{D^q})_{ji}]
\nonumber\\&&-4(h_{D^l}^{\dagger}h_{D^m})_{ji}[(\tilde{D})_{ml}+(\Lambda_{U^q}\Lambda_{U^q}^{\dagger})_{lm}+Tr(\Lambda_{D^m}^{\dagger}\Lambda_{D^l})]
\nonumber\\&&-4(h_{D^l}^{\dagger}\Lambda_{D^m})_{ji}[(Y_D^{\dagger}h_D)_{ml}+(h_{U^q}\Lambda_{U^q}^{\dagger})_{lm}+Tr(\Lambda_{D^m}^{\dagger}h_{D^l})]
\nonumber\\&&-4(\Lambda_{D^l}^{\dagger}h_{D^m})_{ji}[(h_D^{\dagger}Y_D)_{ml}+(\Lambda_{U^q}h_{U^q}^{\dagger})_{lm}+Tr(h_{D^m}^{\dagger}\Lambda_{D^l})]
\nonumber\\&&-4(\Lambda_{D^l}^{\dagger}\Lambda_{D^m})_{ji}[(h_D^{\dagger}h_D)_{ml}+(h_{U^q}h_{U^q}^{\dagger})_{lm}+Tr(h_{D^m}^{\dagger}h_{D^l})]
\nonumber\\&&-2(h_{D^m}^{\dagger}h_EY_E^{\dagger}\Lambda_{D^m}+h_{D^m}^{\dagger}Eh_{D^m}+\Lambda_{D^m}^{\dagger}h_Eh_E^{\dagger}\Lambda_{D^m}+\Lambda_{D^m}^{\dagger}Y_Eh_E^{\dagger}h_{D^m})_{ji}
\nonumber\\&&-6(h_{D^m}^{\dagger}h_{D^q}\Lambda_{D^q}^{\dagger}\Lambda_{D^m}+h_{D^m}^{\dagger}\Lambda_{D^q}\Lambda_{D^q}^{\dagger}h_{D^m})_{ji}-6(\Lambda_{D^m}^{\dagger}h_{D^q}h_{D^q}^{\dagger}\Lambda_{D^m}+\Lambda_{D^m}^{\dagger}\Lambda_{D^q}h_{D^q}^{\dagger}h_{D^m})_{ji}
\nonumber\\&&-2(h_{D^m}^{\dagger}h_{E^q}\Lambda_{E^q}^{\dagger}\Lambda_{D^m}+h_{D^m}^{\dagger}\Lambda_{E^q}\Lambda_{E^q}^{\dagger}h_{D^m}+\Lambda_{D^m}^{\dagger}h_{E^q}h_{E^q}^{\dagger}\Lambda_{D^m}+\Lambda_{D^m}^{\dagger}\Lambda_{E^q}h_{E^q}^{\dagger}h_{D^m})_{ji}
\nonumber\\&&-2(h_D)_{il}[3h_{D^l}^{\dagger}\Lambda_{D^m}Y_D^*+3\Lambda_{D^l}^{\dagger}\Lambda_{D^m}h_D^*+h_{D^l}^{\dagger}\Lambda_{E^m}Y_E^*+\Lambda_{D^l}^{\dagger}\Lambda_{E^m}h_E^*]_{jm}
\nonumber\\&&-6(Y_D)_{il}[h_{D^l}^{\dagger}h_{D^m}Y_D^*+\Lambda_{D^l}^{\dagger}h_{D^m}h_D^*]_{jm}-2(Y_U)_{il}(Y_U^{\dagger})_{kj}Tr(h_{U^l}^{\dagger}h_{U^k})
\nonumber\\&&-2(Y_D)_{il}(h_{D^l}^{\dagger}h_{E^m}Y_E^*+\Lambda_{D^l}^{\dagger}h_{E^m}h_E^*)_{jm}-4(h_Dh_{U^q}^{\dagger}\Lambda_{U^q}Y_D^{\dagger}
\nonumber\\&&+Y_Dh_{U^q}^{\dagger}h_{U^q}Y_D^{\dagger})_{ij}-4(h_D\Lambda_{U^q}^{\dagger}\Lambda_{U^q}h_D^{\dagger}+Y_D\Lambda_{U^q}^{\dagger}h_{U^q}h_D^{\dagger})_{ij}
\nonumber\\&&-4(h_D)_{im}(h_D^{\dagger})_{lj}Tr(\Lambda_{D^m}^{\dagger}\Lambda_{D^l})-4(Y_D)_{im}(h_D^{\dagger})_{lj}Tr(\Lambda_{D^m}^{\dagger}h_{D^l})
\nonumber\\&&-4(h_D)_{im}(Y_D^{\dagger})_{lj}Tr(h_{D^m}^{\dagger}\Lambda_{D^l})-4(Y_D)_{im}(Y_D^{\dagger})_{lj}Tr(h_{D^m}^{\dagger}h_{D^l})
\nonumber\\&&-6(h_D^*)_{jl}(h_{D^l}^t\Lambda_{D^m}^*Y_D+\Lambda_{D^l}^t\Lambda_{D^m}^*h_D)_{im}-2(h_D^*)_{jl}(h_{D^l}^t\Lambda_{E^m}^*Y_E
\nonumber\\&&+\Lambda_{D^l}^t\Lambda_{E^m}^*h_E)_{im}-6(Y_D^*)_{jl}(h_{D^l}^th_{D^m}^*Y_D+\Lambda_{D^l}^th_{D^m}^*h_D)_{im}
\nonumber\\&&-2(Y_D^*)_{jl}(h_{D^l}^th_{E^m}^*Y_E+\Lambda_{D^l}^th_{E^m}^*h_E)_{im}-2(h_U)_{il}(h_U^{\dagger})_{kj}Tr(\Lambda_{U^l}^{\dagger}\Lambda_{U^k})
\nonumber\\&&-2(Y_U)_{il}(h_U^{\dagger})_{kj}Tr(\Lambda_{U^l}^{\dagger}h_{U^k})-2(h_U)_{il}(Y_U^{\dagger})_{kj}Tr(h_{U^l}^{\dagger}\Lambda_{U^k})
\nonumber\\&&+\frac{4}{5}g_1^2[-(\Lambda_{D^k})_{li}(Y_D^{\dagger})_{kj}m^2_{H_1\tilde{L}_l}+(\Lambda_{D^q}^{\dagger}m_{\tilde{L}}^2\Lambda_{D^q})_{ji}
\nonumber\\&&+(\Lambda_{D^q}^{\dagger}\Lambda_{D^q}{m_{\tilde{Q}}^2}^t)_{ji}+2(m_{\tilde{D}}^2)_{lq}(\Lambda_{D^q}^{\dagger}\Lambda_{D^l})_{ji}-(\Lambda_{D^q}^{\dagger})_{jm}(Y_D)_{iq}m^2_{\tilde{L}_mH_1}]
\nonumber\\&&+4(\Lambda_{D^k})_{li}(Y_D^{\dagger}D)_{kj}m^2_{H_1\tilde{L}_l}+4(D\Lambda_{D^q}^t)_{il}(Y_D^{\dagger})_{qj}m^2_{H_1\tilde{L}_l}
\nonumber\\&&+2Tr(E+3D)(\Lambda_{D^k})_{li}(Y_D^{\dagger})_{kj}m^2_{H_1\tilde{L}_l}+6(D)_{ij}(Y_D^{\dagger}\Lambda_{D^k}^t)_{kl}m^2_{H_1\tilde{L}_l}
\nonumber\\&&-(D)_{ij}(Y_E^{\dagger}\Lambda_{E^k})_{kl}m^2_{H_1\tilde{L}_l}-2(\tilde{D})_{ml}[2(\Lambda_{D^l}^{\dagger}m_{\tilde{L}}^2\Lambda_{D^m})_{ji}+(\Lambda_{D^l}^{\dagger}\Lambda_{D^m}{m_{\tilde{Q}}^2}^t)_{ji}
\nonumber\\&&+2(m_{\tilde{D}}^2)_{qm}(\Lambda_{D^l}^{\dagger}\Lambda_{D^q})_{ji}-2(\Lambda_{D^l}^{\dagger})_{jq}(Y_D)_{im}m^2_{\tilde{L}_qH_1}]-6(\Lambda_{D^m}^{\dagger}\Lambda_{D^q}\Lambda_{D^q}^{\dagger}m_{\tilde{L}}^2\Lambda_{D^m})_{ji}
\nonumber\\&&-2(\Lambda_{U^q}\Lambda_{U^q}^{\dagger})_{lm}(2\Lambda_{D^l}^{\dagger}m_{\tilde{L}}^2\Lambda_{D^m}+\Lambda_{D^l}^{\dagger}\Lambda_{D^m}{m_{\tilde{Q}}^2}^t)_{ji}
\nonumber\\&&-4(m^2_{\tilde{D}}\Lambda_{U^q}^{\dagger}\Lambda_{U^q})_{ml}(\Lambda_{D^l}^{\dagger}\Lambda_{D^m})_{ji}-2Tr(\Lambda_{D^m}^{\dagger}\Lambda_{D^l})[2(\Lambda_{D^l}^{\dagger}m_{\tilde{L}}^2\Lambda_{D^m})_{ji}
\nonumber\\&&+(\Lambda_{D^l}^{\dagger}\Lambda_{D^m}{m_{\tilde{Q}}^2}^t)_{ji}+2(m_{\tilde{D}}^2)_{qm}(\Lambda_{D^l}^{\dagger}\Lambda_{D^q})_{ji}-2(\Lambda_{D^l}^{\dagger})_{jq}(Y_D)_{im}m^2_{\tilde{L}_qH_1}]
\nonumber\\&&+4(\Lambda_{D^l}^{\dagger}\Lambda_{D^m})_{ji}(Y_D^{\dagger}\Lambda_{D^l}^t)_{mk}m^2_{H_1\tilde{L}_k}-4(\Lambda_{D^l}^{\dagger}\Lambda_{D^m})_{ji}(Y_D^{\dagger}m^2_{\tilde{Q}}Y_D
\nonumber\\&&+Y_D^{\dagger}m^2_{H_1}Y_D+\tilde{D}m_{\tilde{D}}^2)_{ml}-(\Lambda_{D^m}^{\dagger}\Lambda_{E^q}\Lambda_{E^q}^{\dagger}\Lambda_{D^m}{m_{\tilde{Q}}^2}^t)_{ji}
\nonumber\\&&-4(\Lambda_{D^l}^{\dagger}\Lambda_{D^m})_{ji}[(\Lambda_{U^q}m_{\tilde{D}}^2\Lambda_{U^q}^{\dagger}+{m_{\tilde{D}}^2}^t\Lambda_{U^q}\Lambda_{U^q}^{\dagger})_{lm}+(\Lambda_{U^k}\Lambda_{U^q}^{\dagger})_{lm}(m_{\tilde{U}}^2)_{kq}]
\nonumber\\&&-4(\Lambda_{D^l}^{\dagger}\Lambda_{D^m})_{ji}Tr(\Lambda_{D^m}^{\dagger}m_{\tilde{L}}^2\Lambda_{D^l}+\Lambda_{D^m}^{\dagger}\Lambda_{D^l}{m_{\tilde{Q}}^2}^t)
\nonumber\\&&-4(\Lambda_{D^l}^{\dagger}\Lambda_{D^m})_{ji}Tr(\Lambda_{D^m}^{\dagger}\Lambda_{D^k})(m^2_{\tilde{D}})_{kl}+4(\Lambda_{D^l}^{\dagger}\Lambda_{D^m})_{ji}(Y_D^t\Lambda_{D^m}^{\dagger})_{lk}m^2_{\tilde{L}_kH_1}
\nonumber\\&&-2(\Lambda_{D^m}^{\dagger}m_{\tilde{L}}^2E\Lambda_{D^m}+\Lambda_{D^m}^{\dagger}Y_Em_{\tilde{E}}^2Y_E^{\dagger}\Lambda_{D^m})_{ji}+2(\Lambda_{U^q}\Lambda_{U^q}^{\dagger}Y_D^t)_{li}(\Lambda_{D^l}^{\dagger})_{jp}m^2_{\tilde{L}_pH_1}
\nonumber\\&&-2(\Lambda_{D^m}^{\dagger}E\Lambda_{D^m})_{ji}m^2_{H_1}-2(\Lambda_{D^m}^{\dagger}\Lambda_{E^k})_{jl}m^2_{H_1\tilde{L}_l}(Y_E^{\dagger}\Lambda_{D^m})_{ki}
\nonumber\\&&-2(\Lambda_{D^m}^{\dagger}Em_{\tilde{L}}^2\Lambda_{D^m})_{ji}-(\Lambda_{D^m}^{\dagger}E\Lambda_{D^m}{m_{\tilde{Q}}^2}^t)_{ji}-2(m_{\tilde{D}}^2)_{km}(\Lambda_{D^m}^{\dagger}E\Lambda_{D^k})_{ji}
\nonumber\\&&-3(\Lambda_{D^m}^{\dagger}\Lambda_{D^q}\Lambda_{D^q}^{\dagger}\Lambda_{D^m}{m_{\tilde{Q}}^2}^t)_{ji}-6(m_{\tilde{D}}^2)_{lq}(\Lambda_{D^m}^{\dagger}\Lambda_{D^l}\Lambda_{D^q}^{\dagger}\Lambda_{D^m})_{ji}
\nonumber\\&&-6(m_{\tilde{D}}^2)_{lm}(\Lambda_{D^m}^{\dagger}\Lambda_{D^q}\Lambda_{D^q}^{\dagger}\Lambda_{D^l})_{ji}+6(\Lambda_{D^m}^{\dagger}\Lambda_{D^q}\Lambda_{D^q}^{\dagger})_{jl}(Y_D)_{im}m^2_{\tilde{L}_lH_1}
\nonumber\\&&-6(\Lambda_{D^m}^{\dagger}m_{\tilde{L}}^2\Lambda_{D^q}\Lambda_{D^q}^{\dagger}\Lambda_{D^m}+\Lambda_{D^m}^{\dagger}\Lambda_{D^q}{m_{\tilde{Q}}^2}^t\Lambda_{D^q}^{\dagger}\Lambda_{D^m})_{ji}
\nonumber\\&&+3(\Lambda_{D^m}^{\dagger})_{jk}(Y_D^t\Lambda_{D^q}^{\dagger}\Lambda_{D^m})_{qi}m^2_{\tilde{L}_kH_1}-2(\Lambda_{D^m}^{\dagger}\Lambda_{E^q}\Lambda_{E^q}^{\dagger}m_{\tilde{L}}^2\Lambda_{D^m})_{ji}
\nonumber\\&&-2(m_{\tilde{D}}^2)_{lm}(\Lambda_{D^m}^{\dagger}\Lambda_{E^q}\Lambda_{E^q}^{\dagger}\Lambda_{D^l})_{ji}+2(\Lambda_{D^m}^{\dagger}\Lambda_{E^q}\Lambda_{E^q}^{\dagger})_{jl}(Y_D)_{im}m^2_{\tilde{L}_lH_1}
\nonumber\\&&-2(\Lambda_{D^m}^{\dagger}m_{\tilde{L}}^2\Lambda_{E^q}\Lambda_{E^q}^{\dagger}\Lambda_{D^m}+\Lambda_{D^m}^{\dagger}\Lambda_{E^q}{m_{\tilde{L}}^2}^t\Lambda_{E^q}^{\dagger}\Lambda_{D^m})_{ji}
\nonumber\\&&-2(m^2_{\tilde{E}})_{lq}(\Lambda_{D^m}^{\dagger}\Lambda_{E^l}\Lambda_{E^q}^{\dagger}\Lambda_{D^m})_{ji}-2(\Lambda_{D^m}^{\dagger}Y_E)_{jq}(\Lambda_{E^q}^{\dagger}\Lambda_{D^m})_{ki}m^2_{\tilde{L}_kH_1}
\nonumber\\&&+(\Lambda_{D^m}^{\dagger})_{jk}(Y_E^t\Lambda_{E^q}^{\dagger}\Lambda_{D^m})_{qi}m^2_{\tilde{L}_kH_1}-4(Y_D\Lambda_{U^q}^{\dagger}\Lambda_{U^l}Y_D^{\dagger})_{ij}(m_{\tilde{U}}^2)_{lq}
\nonumber\\&&-3(\Lambda_{D^l}^{\dagger}\Lambda_{D^m}Y_D^*)_{jm}[(m^2_{\tilde{Q}}Y_D+2Y_Dm_{\tilde{D}}^2)_{il}+2m^2_{H_1}(Y_D)_{il}]
\nonumber\\&&+6(\Lambda_{D^l}^{\dagger}\Lambda_{D^m}Y_D^*)_{jm}(\Lambda_{D^l})_{ki}m^2_{H_1\tilde{L}_k}-2(Y_D)_{il}(\Lambda_{D^l}^{\dagger}\Lambda_{E^k}Y_E^*m_{\tilde{E}}^{2t})_{jk}
\nonumber\\&&-(\Lambda_{D^l}^{\dagger}\Lambda_{E^m}Y_E^*)_{jm}[(m_{\tilde{Q}}^2Y_D+2Y_Dm_{\tilde{D}}^2)_{il}+2m^2_{H_1}(Y_D)_{il}
\nonumber\\&&-2(\Lambda_{D^l})_{ki}m^2_{H_1\tilde{L}_k}]-3(Y_D)_{il}(2\Lambda_{D^l}^{\dagger}m_{\tilde{L}}^2\Lambda_{D^m}Y_D^*+\Lambda_{D^l}^{\dagger}\Lambda_{D^m}{m_{\tilde{Q}}^2}^t)_{jm}
\nonumber\\&&-6(Y_D)_{il}(m_{\tilde{D}}^2)_{km}(\Lambda_{D^l}^{\dagger}\Lambda_{D^k}Y_D^*)_{jm}+6(Y_D)_{il}(\Lambda_{D^l}^{\dagger})_{jk}Tr(D)m^2_{\tilde{L}_kH_1}
\nonumber\\&&-2(Y_D)_{il}(\Lambda_{D^l}^{\dagger}m_{\tilde{L}}^2\Lambda_{E^m}Y_E^*+\Lambda_{D^l}^{\dagger}\Lambda_{E^m}{m_{\tilde{L}}^2}^tY_E^*)_{jm}
\nonumber\\&&-2(m_{\tilde{Q}}^2Y_D\Lambda_{U^q}^{\dagger}\Lambda_{U^q}Y_D^{\dagger}+2Y_Dm_{\tilde{D}}^2\Lambda_{U^q}^{\dagger}\Lambda_{U^q}Y_D^{\dagger})_{ij}-4m_{H_1}^2(Y_D\Lambda_{U^q}^{\dagger}\Lambda_{U^q}Y_D^{\dagger})_{ij}
\nonumber\\&&+2m_{H_1\tilde{L}_l}^2(\Lambda_{D^k})_{li}(\Lambda_{U^q}^{\dagger}\Lambda_{U^q}Y_D^{\dagger})_{kj}-4(Y_D\Lambda_{U^q}^{\dagger}\Lambda_{U^q}m_{\tilde{D}}^2Y_D^{\dagger}+Y_D\Lambda_{U^q}^{\dagger}{m_{\tilde{D}}^2}^t\Lambda_{U^q}Y_D^{\dagger})_{ij}
\nonumber\\&&-2(m_{\tilde{Q}}^2Y_D+2Y_Dm_{\tilde{D}}^2)_{im}(Y_D^{\dagger})_{lj}Tr(\Lambda_{D^m}^{\dagger}\Lambda_{D^l})+2(Y_D)_{il}Tr(\tilde{E})(\Lambda_{D^l}^{\dagger})_{jk}m^2_{\tilde{L}_kH_1}
\nonumber\\&&-4m^2_{H_1}(Y_D)_{im}(Y_D^{\dagger})_{lj}Tr(\Lambda_{D^m}^{\dagger}\Lambda_{D^l})+4(\Lambda_{D^m})_{ki}(Y_D^{\dagger})_{lj}Tr(\Lambda_{D^m}^{\dagger}\Lambda_{D^l})m_{H_1\tilde{L}_k}^2
\nonumber\\&&-4(Y_D)_{im}(Y_D^{\dagger})_{lj}Tr(\Lambda_{D^m}^{\dagger}m_{\tilde{L}}^2\Lambda_{D^l}+\Lambda_{D^m}^{\dagger}\Lambda_{D^l}{m_{\tilde{Q}}^2}^t)
\nonumber\\&&-4(m_{\tilde{D}}^2)_{kl}(Y_D)_{im}(Y_D^{\dagger})_{lj}Tr(\Lambda_{D^m}^{\dagger}\Lambda_{D^k})-6(m_{\tilde{D}}^2Y_D^{\dagger})_{kj}(Y_D^t\Lambda_{D^m}^{\dagger}\Lambda_{D^k})_{mi}
\nonumber\\&&-6(Y_D^{\dagger})_{lj}(\Lambda_{D^l}^t\Lambda_{D^m}^*m_{\tilde{Q}}^2Y_D+\Lambda_{D^l}^t\Lambda_{D^m}^*Y_Dm_{\tilde{D}}^2)_{im}
\nonumber\\&&-6(Y_D^*)_{jl}m_{H_1}^2(\Lambda_{D^l}^t\Lambda_{D^m}^*Y_D)_{im}+6m^2_{H_1\tilde{L}_k}(Y_D^*)_{jl}(\Lambda_{D^l}^t\Lambda_{D^m}^*\Lambda_{D^m}^t)_{ik}
\nonumber\\&&-3(Y_D^{\dagger})_{lj}(2Y_D^t\Lambda_{D^m}^{\dagger}m_{\tilde{L}}^2\Lambda_{D^l}+Y_D^t\Lambda_{D^m}^{\dagger}\Lambda_{D^l}{m_{\tilde{Q}}^2}^t)_{mi}
\nonumber\\&&+6(D)_{ij}m^2_{\tilde{L}_kH_1}(Y_D^t\Lambda_{D^m}^{\dagger})_{mk}-(Y_D^{\dagger})_{lj}(2Y_E^t\Lambda_{E^m}^{\dagger}m_{\tilde{L}}^2\Lambda_{D^l}
\nonumber\\&&+Y_E^t\Lambda_{E^m}^{\dagger}\Lambda_{D^l}{m^2_{\tilde{Q}}}^t)_{mi}+4(Y_D)_{im}(Y_D^{\dagger})_{lj}(Y_D^t\Lambda_{D^m}^{\dagger})_{lk}m^2_{\tilde{L}_kH_1}
\nonumber\\&&-2(m_{\tilde{D}}^2Y_D^{\dagger})_{kj}(Y_E^t\Lambda_{E^m}^{\dagger}\Lambda_{D^k})_{mi}+2(D)_{ij}m_{\tilde{L}_kH_1}^2(Y_E^t\Lambda_{E^m}^{\dagger})_{mk}
\nonumber\\&&-2(Y_D^{\dagger})_{lj}(\Lambda_{D^l}^t\Lambda_{E^m}^*m_{\tilde{L}}^2Y_E+\Lambda_{D^l}^t\Lambda_{E^m}^*Y_Em_{\tilde{E}}^2)_{im}-(Y_U^{\dagger}m_{\tilde{Q}}^2)_{kj}(Y_U)_{il}Tr(\Lambda_{U^l}^{\dagger}\Lambda_{U^k})
\nonumber\\&&-2m_{H_1}^2(Y_D^{\dagger})_{lj}(\Lambda_{D^l}^t\Lambda_{E^m}^*Y_E)_{im}-(Y_D^{\dagger})_{lj}(\Lambda_{D^l}^t\Lambda_{E^m}^*\Lambda_{E^m})_{ik}m^2_{H_1\tilde{L}_k}
\nonumber\\&&-[2m^2_{H_2}(Y_U)_{il}+(m_{\tilde{Q}}^2Y_U+2Y_Um_{\tilde{U}}^2)_{il}](Y_U^{\dagger})_{kj}Tr(\Lambda_{U^l}^{\dagger}\Lambda_{U^k})
\nonumber\\&&-2(Y_U)_{il}(Y_U^{\dagger})_{kj}[Tr(\Lambda_{U^l}^{\dagger}\Lambda_{U^k}m_{\tilde{D}}^2+{m_{\tilde{D}}^2}^t\Lambda_{U^k}\Lambda_{U^l}^{\dagger})+(m_{\tilde{U}}^2)_{qk}Tr(\Lambda_{U^l}^{\dagger}\Lambda_{U^q})]
\nonumber\\&&+\frac{2}{5}g_1^2[-(Y_D)_{ik}(\Lambda_{D^k}^{\dagger})_{jl}m_{\tilde{L}_lH_1}^2+({m_{\tilde{Q}}^2}^t\Lambda_{D^q}^{\dagger}\Lambda_{D^q})_{ji}]
\nonumber\\&&-2[(\tilde{D})_{ml}+(\Lambda_{U^q}\Lambda_{U^q}^{\dagger})_{lm}+Tr(\Lambda_{D^m}^{\dagger}\Lambda_{D^l})]({m^2_{\tilde{Q}}}^t\Lambda_{D^l}^{\dagger}\Lambda_{D^m})_{ji}
\nonumber\\&&+2(Y_D^*\Lambda_{U^q}\Lambda_{U^q}^{\dagger})_{jm}(\Lambda_{D^m})_{ki}m_{H_1\tilde{L}_k}^2-({m_{\tilde{Q}}^2}^t\Lambda_{D^m}^{\dagger}E\Lambda_{D^m})_{ji}
\nonumber\\&&+(Y_D^{\dagger})_{mj}(\Lambda_{E^q}\Lambda_{E^q}^{\dagger}\Lambda_{D^m})_{ki}m_{H_1\tilde{L}_k}^2-({m_{\tilde{Q}}^2}^t\Lambda_{D^m}^{\dagger}\Lambda_{E^q}\Lambda_{E^q}^{\dagger}\Lambda_{D^m})_{ji}
\nonumber\\&&-3(Y_D)_{il}(Y_D^{\dagger}m_{\tilde{Q}}^2\Lambda_{D^m}^t\Lambda_{D^l}^*)_{mj}-(Y_D)_{il}({m_{\tilde{Q}}^2}^t\Lambda_{D^l}^{\dagger}\Lambda_{E^m}Y_E^*)_{jm}
\nonumber\\&&+(D)_{ij}(\Lambda_{E^m}Y_E^*)_{km}m^2_{H_1\tilde{L}_k}-3(Y_D)_{il}({m_{\tilde{Q}}^2}^t\Lambda_{D^l}^{\dagger}\Lambda_{D^m}Y_D^*)_{jm}
\nonumber\\&&+2(\Lambda_{D^k}^{\dagger})_{jl}(Y_D\Lambda_{U^q}^{\dagger}\Lambda_{U^q})_{ik}m_{\tilde{L}_lH_1}^2-2(Y_D)_{im}(Y_D^{\dagger}m_{\tilde{Q}}^2)_{lj}Tr(\Lambda_{D^m}^{\dagger}\Lambda_{D^l})
\nonumber\\&&-(Y_D^{\dagger}m_{\tilde{Q}}^2)_{lj}[3\Lambda_{D^l}^t\Lambda_{D^m}^*Y_D+\Lambda_{D^l}^t\Lambda_{E^m}^*Y_E]_{im}-2(Y_D\Lambda_{U^q}^{\dagger}\Lambda_{U^q}Y_D^{\dagger}m_{\tilde{Q}}^2)_{ij}
\nonumber\\&&-3({m^2_{\tilde{Q}}}^t\Lambda_{D^m}^{\dagger}\Lambda_{D^q}\Lambda_{D^q}^{\dagger}\Lambda_{D^m})_{ji}+(\Lambda_{D^l}^{\dagger})_{jk}(3\Lambda_{D^l}^t\Lambda_{D^m}^*Y_D+\Lambda_{D^l}^t\Lambda_{E^m}^*Y_E)_{im}m^2_{\tilde{L}_kH_1}-\frac{1}{5}s'\delta_{ij}
\end{eqnarray}

\begin{eqnarray}
{(16\pi^2)^2}\beta^{(2)}_{(M^2_{\tilde{D}})_{ij}}&=&{(16\pi^2)^2}\beta^{(2)}_{(M^2_{\tilde{D}})_{RPC}}+\nonumber\\&&
(\frac{64}{3}M_3^2g_3^2+\frac{64}{15}M_1^2g_1^2)(\Lambda_{U^q}\Lambda_{U^q}^{\dagger})_{ji}+(24M_2^2g_2^2+\frac{8}{5}M_1^2g_1^2)Tr(\Lambda_{D^j}\Lambda_{D^i}^{\dagger})
\nonumber\\&&-(\frac{32}{3}M_3g_3^2+\frac{32}{15}M_1g_1^2)(\Lambda_{U^q}h_{U^q}^{\dagger})_{ji}-(12M_2g_2^2+\frac{4}{5}M_1g_1^2)Tr(\Lambda_{D^j}h_{D^i}^{\dagger})
\nonumber\\&&-(\frac{32}{3}M_3^*g_3^2+\frac{32}{15}M_1^*g_1^2)(h_{U^q}\Lambda_{U^q}^{\dagger})_{ji}-(12M_2^*g_2^2+\frac{4}{5}M_1^*g_1^2)Tr(h_{D^j}\Lambda_{D^i}^{\dagger})
\nonumber\\&&+(\frac{32}{3}g_3^2+\frac{32}{15}g_1^2)(h_{U^q}h_{U^q}^{\dagger})_{ji}+(12g_2^2+\frac{4}{5}g_1^2)Tr(h_{D^j}h_{D^i}^{\dagger})
\nonumber\\&&-4(h_D^th_{D^p}^{\dagger}\Lambda_{D^p}Y_D^*+Y_D^th_{D^p}^{\dagger}h_{D^p}Y_D^*+h_D^t\Lambda_{D^p}^{\dagger}\Lambda_{D^p}h_D^*+Y_D^t\Lambda_{D^p}^{\dagger}h_{D^p}h_D^*)_{ji}
\nonumber\\&&-4Tr(h_{D^j}^th_{D^i}^*D+\Lambda_{D^j}^th_{D^i}^*h_DY_D^{\dagger})-4Tr(h_{D^j}^t\Lambda_{D^i}^*Y_Dh_D^{\dagger}+\Lambda_{D^j}^t\Lambda_{D^i}^*h_Dh_D^{\dagger})
\nonumber\\&&-4Tr(h_{D^j}^th_{D^i}^*T+\Lambda_{D^j}^th_{D^i}^*h_UY_U^{\dagger})-4Tr(\Lambda_{D^i}^{\dagger}h_{D^j}h_{D^q}^{\dagger}\Lambda_{D^q}+\Lambda_{D^i}^{\dagger}\Lambda_{D^j}h_{D^q}^{\dagger}h_{D^q})
\nonumber\\&&-4Tr(h_{D^j}^t\Lambda_{D^i}^*Y_Uh_U^{\dagger}+\Lambda_{D^j}^t\Lambda_{D^i}^*h_Uh_U^{\dagger})-4Tr(h_{D^i}^{\dagger}h_{D^j}\Lambda_{D^q}^{\dagger}\Lambda_{D^q}+h_{D^i}^{\dagger}\Lambda_{D^j}\Lambda_{D^q}^{\dagger}h_{D^q})
\nonumber\\&&-12Tr(h_{D^q}h_{D^q}^{\dagger}\Lambda_{D^j}\Lambda_{D^i}^{\dagger}+\Lambda_{D^q}h_{D^q}^{\dagger}h_{D^j}\Lambda_{D^i}^{\dagger})-12Tr(h_{D^q}\Lambda_{D^q}^{\dagger}\Lambda_{D^j}h_{D^i}^{\dagger}
\nonumber\\&&+\Lambda_{D^q}\Lambda_{D^q}^{\dagger}h_{D^j}h_{D^i}^{\dagger})-4Tr(h_{E^q}h_{E^q}^{\dagger}\Lambda_{D^j}\Lambda_{D^i}^{\dagger}+\Lambda_{E^q}h_{E^q}^{\dagger}h_{D^j}\Lambda_{D^i}^{\dagger})
\nonumber\\&&-4Tr(h_{E^q}\Lambda_{E^q}^{\dagger}\Lambda_{D^j}h_{D^i}^{\dagger}+\Lambda_{E^q}\Lambda_{E^q}^{\dagger}h_{D^j}h_{D^i}^{\dagger})-4Tr(h_Eh_E^{\dagger}\Lambda_{D^j}\Lambda_{D^i}^{\dagger}
\nonumber\\&&+Y_Eh_E^{\dagger}h_{D^j}\Lambda_{D^i}^{\dagger})-4Tr(h_EY_E^{\dagger}\Lambda_{D^j}h_{D^i}^{\dagger}+Eh_{D^j}h_{D^i}^{\dagger})
\nonumber\\&&-8(h_{U^m}h_D^{\dagger}Y_D\Lambda_{U^m}^{\dagger}+\Lambda_{U^m}h_D^{\dagger}h_D\Lambda_{U^m}^{\dagger}+h_{U^m}\tilde{D}h_{U^m}^{\dagger}+\Lambda_{U^m}Y_D^{\dagger}h_Dh_{U^m}^{\dagger})_{ji}
\nonumber\\&&-8(h_{U^m}^{\dagger}h_{U^p}\Lambda_{U^p}^{\dagger}\Lambda_{U^m}+h_{U^m}^{\dagger}\Lambda_{U^p}\Lambda_{U^p}^{\dagger}h_{U^m}+\Lambda_{U^m}^{\dagger}h_{U^p}h_{U^p}^{\dagger}\Lambda_{U^m}+\Lambda_{U^m}^{\dagger}\Lambda_{U^p}h_{U^p}^{\dagger}h_{U^m})_{ij}
\nonumber\\&&+8(h_{U^m}^{\dagger})_{ik}(h_{U^m})_{jl}Tr(\Lambda_{D^l}\Lambda_{D^k}^{\dagger})+8(\Lambda_{U^m}^{\dagger})_{ik}(\Lambda_{U^m})_{jl}Tr(h_{D^l}h_{D^k}^{\dagger})
\nonumber\\&&+8(h_{U^m}^{\dagger})_{ik}(\Lambda_{U^m})_{jl}Tr(h_{D^l}\Lambda_{D^k}^{\dagger})+8(\Lambda_{U^m}^{\dagger})_{ik}(h_{U^m})_{jl}Tr(\Lambda_{D^l}h_{D^k}^{\dagger})
\nonumber\\&&-12(h_D^{\dagger}h_{D^j}^t\Lambda_{D^p}^*Y_D+h_D^{\dagger}\Lambda_{D^j}^t\Lambda_{D^p}^*h_D+Y_D^{\dagger}h_{D^j}^th_{D^p}^*Y_D+Y_D^{\dagger}\Lambda_{D^j}^th_{D^p}^*h_D)_{ip}
\nonumber\\&&-4(h_D^{\dagger}h_{D^j}^t\Lambda_{E^p}^*Y_E+h_D^{\dagger}\Lambda_{D^j}^t\Lambda_{E^p}^*h_E+Y_D^{\dagger}h_{D^j}^th_{E^p}^*Y_E+Y_D^{\dagger}\Lambda_{D^j}^th_{E^p}^*h_E)_{ip}
\nonumber\\&&-12(h_D^th_{D^i}^{\dagger}\Lambda_{D^p}Y_D^*+Y_D^th_{D^i}^{\dagger}h_{D^p}Y_D^*+h_D^t\Lambda_{D^i}^{\dagger}\Lambda_{D^p}h_D^*+Y_D^t\Lambda_{D^i}^{\dagger}h_{D^p}h_D^*)_{jp}
\nonumber\\&&-4(h_D^th_{D^i}^{\dagger}\Lambda_{E^p}Y_E^*+Y_D^th_{D^i}^{\dagger}h_{E^p}Y_E^*+h_D^t\Lambda_{D^i}^{\dagger}\Lambda_{E^p}h_E^*+Y_D^t\Lambda_{D^i}^{\dagger}h_{E^p}h_E^*)_{jp}
\nonumber\\&&-4(h_{U^k}^{\dagger}\Lambda_{U^l})_{ij}Tr(\Lambda_{U^l}^{\dagger}h_{U^k})-4(\Lambda_{U^k}^{\dagger}h_{U^l})_{ij}Tr(h_{U^l}^{\dagger}\Lambda_{U^k})
\nonumber\\&&-4(\Lambda_{U^k}^{\dagger}\Lambda_{U^l})_{ij}Tr(h_{U^l}^{\dagger}h_{U^k})-8(h_{U^k}^{\dagger}h_{U^l})_{ij}\tilde{T}_{lk}-4(h_{U^k}^{\dagger}h_{U^l})_{ij}Tr(\Lambda_{U^l}^{\dagger}\Lambda_{U^k})
\nonumber\\&&-8(h_{U^k}^{\dagger}\Lambda_{U^l})_{ij}(Y_U^{\dagger}h_U)_{lk}-8(\Lambda_{U^k}^{\dagger}h_{U^l})_{ij}(h_U^{\dagger}Y_U)_{lk}-8(\Lambda_{U^k}^{\dagger}\Lambda_{U^l})_{ij}(h_U^{\dagger}h_U)_{lk}
\nonumber\\&&+(\frac{16}{3}g_3^2+\frac{16}{15}g_1^2)[2(\Lambda_{U^q}m_{\tilde{D}}^2\Lambda_{U^q}^{\dagger})_{ji}+({m_{\tilde{D}}^2}^t\Lambda_{U^q}\Lambda_{U^q}^{\dagger})_{ji}+2(m_{\tilde{U}}^2)_{lq}(\Lambda_{U^l}\Lambda_{U^q}^{\dagger})_{ji}
\nonumber\\&&+(\Lambda_{U^q}\Lambda_{U^q}^{\dagger}{m_{\tilde{D}}^2}^t)_{ji}]-(6g_2^2+\frac{2}{5}g_1^2)[(\Lambda_{D^k})_{lj}(Y_D^{\dagger})_{ki}m_{H_1\tilde{L}_l}^2+(Y_D)_{jk}(\Lambda_{D^k}^{\dagger})_{il}m^2_{\tilde{L}_lH_1}]
\nonumber\\&&+(6g_2^2+\frac{2}{5}g_1^2)[2Tr(m^2_{\tilde{L}}\Lambda_{D^j}\Lambda_{D^i}^{\dagger}+\Lambda_{D^j}{m^2_{\tilde{Q}}}^t\Lambda_{D^i}^{\dagger})+(m_{\tilde{D}}^2)_{lj}Tr(\Lambda_{D^l}\Lambda_{D^i}^{\dagger})
\nonumber\\&&+(m_{\tilde{D}}^2)_{il}Tr(\Lambda_{D^j}\Lambda_{D^l}^{\dagger})]-(6g_2^2+\frac{2}{5}g_1^2)[(\Lambda_{D^i}^*Y_D)_{kj}m^2_{\tilde{L}_kH_1}
\nonumber\\&&+(\Lambda_{D^j}Y_D^*)_{ki}m^2_{H_1\tilde{L}_k}]+4(\Lambda_{D^k}Y_D^*)_{li}\tilde{D}_{kj}m^2_{H_1\tilde{L}_l}+4(Y_D^{\dagger}D\Lambda_{D^j}^t)_{il}m^2_{H_1\tilde{L}_l}
\nonumber\\&&+4(Y_D^{\dagger}T\Lambda_{D^j}^t)_{il}m^2_{H_1\tilde{L}_l}+4(Y_D^{\dagger}\Lambda_{D^j}^t)_{il}(m^2_{H_1\tilde{L}_l})Tr(E+3D)
\nonumber\\&&-4\tilde{D}_{ij}(Y_E^{\dagger}\Lambda_{E^k})_{kl}m^2_{H_1\tilde{L}_l}+12\tilde{D}_{ij}(Y_D^{\dagger}\Lambda_{D^k}^t)_{kl}m^2_{H_1\tilde{L}_l}-4(Y_D^{\dagger}\Lambda_{D^p}^t\Lambda_{D^p}^*m^2_{\tilde{Q}}Y_D)_{ij}
\nonumber\\&&-2(Y_D^{\dagger}\Lambda_{D^p}^t\Lambda_{D^p}^*Y_Dm_{\tilde{D}}^2)_{ij}-4m_{H_1}^2(Y_D^{\dagger}\Lambda_{D^p}^t\Lambda_{D^p}^*Y_D)_{ij}+4(Y_D^{\dagger}\Lambda_{D^p}^t\Lambda_{D^p}^*\Lambda_{D^j}^t)_{il}m_{H_1\tilde{L}_l}^2
\nonumber\\&&-4(Y_D^t\Lambda_{D^p}^{\dagger}m^2_{\tilde{L}}\Lambda_{D^p}Y_D^*)_{ji}-4(Y_D^t\Lambda_{D^p}^{\dagger}\Lambda_{D^p}{m^2_{\tilde{Q}}}^tY_D^*)_{ji}-2(m^2_{\tilde{D}})_{lp}(Y_D^t\Lambda_{D^p}^{\dagger}\Lambda_{D^l}Y_D^*)_{ji}
\nonumber\\&&+4\tilde{D}_{ip}(\Lambda_{D^p}^*Y_D)_{qj}m^2_{\tilde{L}_qH_1}-4Tr(D^t\Lambda_{D^i}^{\dagger}m^2_{\tilde{L}}\Lambda_{D^j})-4Tr(D^t\Lambda_{D^i}^{\dagger}\Lambda_{D^j}{m^2_{\tilde{Q}}}^t)
\nonumber\\&&-2(m^2_{\tilde{D}})_{lj}Tr(D^t\Lambda_{D^i}^{\dagger}\Lambda_{D^l})+4(\Lambda_{D^i}^*DY_D)_{kj}m^2_{\tilde{L}_kH_1}-4Tr(Y_U^*Y_U^t\Lambda_{D^i}^{\dagger}m^2_{\tilde{L}}\Lambda_{D^j})
\nonumber\\&&-4Tr(Y_U^*Y_U^t\Lambda_{D^i}^{\dagger}\Lambda_{D^j}{m^2_{\tilde{Q}}}^t)-2(m^2_{\tilde{D}})_{lj}Tr(Y_U^*Y_U^t\Lambda_{D^i}^{\dagger}\Lambda_{D^l})+4(\Lambda_{D^i}^*TY_D)_{kj}m^2_{\tilde{L}_kH_1}
\nonumber\\&&-4Tr(\Lambda_{D^j}^t\Lambda_{D^i}^*m^2_{\tilde{Q}}D)-4Tr(Y_Dm^2_{\tilde{D}}Y_D^{\dagger}\Lambda_{D^j}^t\Lambda_{D^i}^*)-4m^2_{H_1}Tr(D\Lambda_{D^j}^t\Lambda_{D^i}^*)
\nonumber\\&&+4(Y_D^{\dagger}\Lambda_{D^j}^t\Lambda_{D^i}^*\Lambda_{D^k}^t)_{kl}m^2_{H_1\tilde{L}_l}-4Tr(m^2_{\tilde{Q}}T\Lambda_{D^j}^t\Lambda_{D^i}^*)-4Tr(Y_Um^2_{\tilde{U}}Y_U^{\dagger}\Lambda_{D^j}^t\Lambda_{D^i}^*)
\nonumber\\&&-4m^2_{H_2}Tr(T\Lambda_{D^j}^t\Lambda_{D^i}^*)-4Tr(\Lambda_{D^q}^{\dagger}\Lambda_{D^q}\Lambda_{D^i}^{\dagger}m^2_{\tilde{L}}\Lambda_{D^j})-4Tr(\Lambda_{D^q}^{\dagger}\Lambda_{D^q}\Lambda_{D^i}^{\dagger}\Lambda_{D^j}{m^2_{\tilde{Q}}}^t)
\nonumber\\&&+4(Y_D^t\Lambda_{D^q}^{\dagger}\Lambda_{D^q}\Lambda_{D^i}^{\dagger})_{jk}m^2_{\tilde{L}_kH_1}-4Tr(\Lambda_{D^i}^{\dagger}\Lambda_{D^j}\Lambda_{D^q}^{\dagger}\Lambda_{D^q}{m^2_{\tilde{Q}}}^t)-4Tr(\Lambda_{D^i}^{\dagger}\Lambda_{D^j}\Lambda_{D^q}^{\dagger}m_{\tilde{L}}^2\Lambda_{D^q})
\nonumber\\&&-4(m^2_{\tilde{D}})_{lq}Tr(\Lambda_{D^i}^{\dagger}\Lambda_{D^j}\Lambda_{D^q}^{\dagger}\Lambda_{D^l})+4(Y_D^t\Lambda_{D^i}^{\dagger}\Lambda_{D^j}\Lambda_{D^q}^{\dagger})_{qk}m^2_{\tilde{L}_kH_1}-12Tr(\Lambda_{D^q}^{\dagger}\Lambda_{D^j}\Lambda_{D^i}^{\dagger}m^2_{\tilde{L}}\Lambda_{D^q})
\nonumber\\&&-12Tr(\Lambda_{D^q}^{\dagger}\Lambda_{D^j}\Lambda_{D^i}^{\dagger}\Lambda_{D^q}{m^2_{\tilde{Q}}}^t)-12(m_{\tilde{D}}^2)_{lq}Tr(\Lambda_{D^q}^{\dagger}\Lambda_{D^j}\Lambda_{D^i}^{\dagger}\Lambda_{D^l})
\nonumber\\&&+12(Y_D^t\Lambda_{D^q}^{\dagger}\Lambda_{D^j}\Lambda_{D^i}^{\dagger})_{qk}m^2_{\tilde{L}_kH_1}-12Tr(\Lambda_{D^i}^{\dagger}\Lambda_{D^q}\Lambda_{D^q}^{\dagger}m_{\tilde{L}}^2\Lambda_{D^j})\nonumber\\&&-12Tr(\Lambda_{D^i}^{\dagger}\Lambda_{D^q}\Lambda_{D^q}^{\dagger}\Lambda_{D^j}{m^2_{\tilde{Q}}}^t)-2(m_{\tilde{D}}^2)_{lj}Tr(\Lambda_{D^q}^{\dagger}\Lambda_{D^q}\Lambda_{D^i}^{\dagger}\Lambda_{D^l})
\nonumber\\&&-6(m_{\tilde{D}}^2)_{lj}Tr(\Lambda_{D^i}^{\dagger}\Lambda_{D^q}\Lambda_{D^q}^{\dagger}\Lambda_{D^l})+12(Y_D^t\Lambda_{D^i}^{\dagger}\Lambda_{D^q}\Lambda_{D^q}^{\dagger})_{jk}m_{\tilde{L}_kH_1}^2
\nonumber\\&&-4Tr(m_{\tilde{L}}^2\Lambda_{E^q}\Lambda_{E^q}^{\dagger}\Lambda_{D^j}\Lambda_{D^i}^{\dagger})-4Tr(\Lambda_{E^q}{m_{\tilde{L}}^2}^t\Lambda_{E^q}^{\dagger}\Lambda_{D^j}\Lambda_{D^i}^{\dagger})
\nonumber\\&&-4(m_{\tilde{E}}^2)_{lq}Tr(\Lambda_{E^l}\Lambda_{E^q}^{\dagger}\Lambda_{D^j}\Lambda_{D^i}^{\dagger})-4(\Lambda_{E^q}^{\dagger}\Lambda_{D^j}\Lambda_{D^i}^{\dagger}Y_E)_{kq}m^2_{\tilde{L}_kH_1}
\nonumber\\&&+4(Y_E^t\Lambda_{E^q}^{\dagger}\Lambda_{D^j}\Lambda_{D^i}^{\dagger})_{qk}m^2_{\tilde{L}_kH_1}-4Tr(\Lambda_{E^q}\Lambda_{E^q}^{\dagger}m_{\tilde{L}}^2\Lambda_{D^j}\Lambda_{D^i}^{\dagger})
\nonumber\\&&-4Tr(\Lambda_{E^q}\Lambda_{E^q}^{\dagger}\Lambda_{D^j}{m_{\tilde{Q}}^2}^t\Lambda_{D^i}^{\dagger})-2(m_{\tilde{D}})_{lj}Tr(\Lambda_{E^q}\Lambda_{E^q}^{\dagger}\Lambda_{D^l}\Lambda_{D^i}^{\dagger})
\nonumber\\&&+4(Y_D^t\Lambda_{D^i}^{\dagger}\Lambda_{E^q}\Lambda_{E^q}^{\dagger})_{jk}m_{\tilde{L}_kH_1}^2-4Tr(m_{\tilde{L}}^2E\Lambda_{D^j}\Lambda_{D^i}^{\dagger}+Y_Em_{\tilde{E}}^2Y_E^{\dagger}\Lambda_{D^j}\Lambda_{D^i}^{\dagger})
\nonumber\\&&-4m^2_{H_1}Tr(E\Lambda_{D^j}\Lambda_{D^i}^{\dagger})-4(Y_E^{\dagger}\Lambda_{D^j}\Lambda_{D^i}^{\dagger}\Lambda_{E^k})_{kl}m^2_{H_1\tilde{L}_l}-4Tr(\Lambda_{D^i}^{\dagger}Em_{\tilde{L}}^2\Lambda_{D^j})
\nonumber\\&&-4Tr(\Lambda_{D^i}^{\dagger}E\Lambda_{D^j}{m_{\tilde{Q}}^2}^t)-2(m_{\tilde{D}}^2)_{lj}Tr(\Lambda_{D^i}^{\dagger}E\Lambda_{D^l})-4(\Lambda_{U^m}Y_D^t\Lambda_{D^k}^{\dagger})_{jn}(\Lambda_{U^m}^{\dagger})_{ik}m^2_{\tilde{L}_nH_1}
\nonumber\\&&-8(\Lambda_{U^m}m_{\tilde{D}}^2\tilde{D}\Lambda_{U^m}^{\dagger})_{ji}-4({m_{\tilde{D}}^2}^t\Lambda_{U^m}\tilde{D}\Lambda_{U^m}^{\dagger})_{ji}-8(m_{\tilde{U}}^2)_{lm}(\Lambda_{U^l}\tilde{D}\Lambda_{U^m}^{\dagger})_{ji}
\nonumber\\&&-8(\Lambda_{U^m}Y_D^{\dagger}m_{\tilde{Q}}^2Y_D\Lambda_{U^m}^{\dagger})_{ji}-8(\Lambda_{U^m}\tilde{D}m_{\tilde{D}}^2\Lambda_{U^m}^{\dagger})_{ji}-8m_{H_1}^2(\Lambda_{U^m}\tilde{D}\Lambda_{U^m}^{\dagger})_{ji}
\nonumber\\&&+4(\Lambda_{U^m}^{\dagger})_{ki}(\Lambda_{U^m}Y_D^{\dagger}\Lambda_{D^k}^t)_{jl}m^2_{H_1\tilde{L}_l}-8(\Lambda_{U^m}^{\dagger}\Lambda_{U^p}m_{\tilde{D}}^2\Lambda_{U^p}^{\dagger}\Lambda_{U^m}+\Lambda_{U^m}^{\dagger}{m_{\tilde{D}}^2}^t\Lambda_{U^p}\Lambda_{U^p}^{\dagger}\Lambda_{U^m})_{ij}
\nonumber\\&&-8(\Lambda_{U^m}^{\dagger}\Lambda_{U^l}\Lambda_{U^p}^{\dagger}\Lambda_{U^m})_{ij}(m_{\tilde{U}}^2)_{lp}-4(\Lambda_{U^m}^{\dagger}\Lambda_{U^p}\Lambda_{U^p}^{\dagger}\Lambda_{U^m}m_{\tilde{D}}^2)_{ij}
\nonumber\\&&-8(\Lambda_{U^m}^{\dagger}\Lambda_{U^p}\Lambda_{U^p}^{\dagger}{m_{\tilde{D}}^2}^t\Lambda_{U^m})_{ij}-8(m_{\tilde{U}}^2)_{lm}(\Lambda_{U^m}^{\dagger}\Lambda_{U^p}\Lambda_{U^p}^{\dagger}\Lambda_{U^l})_{ij}
\nonumber\\&&+4(\Lambda_{U^m}^{\dagger})_{ik}(\Lambda_{U^m}m_{\tilde{D}}^2)_{jl}Tr(\Lambda_{D^l}\Lambda_{D^k}^{\dagger})+4(\Lambda_{U^m}^{\dagger})_{ik}({m_{\tilde{D}}^2}^t\Lambda_{U^m})_{jl}Tr(\Lambda_{D^l}\Lambda_{D^k}^{\dagger})
\nonumber\\&&+4(\Lambda_{U^m}^{\dagger})_{ik}(\Lambda_{U^q})_{jl}(m_{\tilde{U}}^2)_{qm}Tr(\Lambda_{D^l}\Lambda_{D^k}^{\dagger})+8(\Lambda_{U^m}^{\dagger})_{ik}(\Lambda_{U^m})_{jl}Tr(\Lambda_{D^l}{m_{\tilde{Q}}^2}^t\Lambda_{D^k}^{\dagger}
\nonumber\\&&+m_{\tilde{L}}^2\Lambda_{D^l}\Lambda_{D^k}^{\dagger})+4(\Lambda_{U^m}^{\dagger})_{ik}(\Lambda_{U^m}{m_{\tilde{D}}^2}^t)_{jq}Tr(\Lambda_{D^q}\Lambda_{D^k}^{\dagger})-12(Y_D^{\dagger}m_{\tilde{Q}}^2\Lambda_{D^j}^t\Lambda_{D^p}^*Y_D)_{ip}
\nonumber\\&&-12(Y_D^t\Lambda_{D^p}^{\dagger}m_{\tilde{L}}^2\Lambda_{D^j}Y_D^*)_{pi}-6(m_{\tilde{D}}^2)_{lj}(Y_D^t\Lambda_{D^p}^{\dagger}\Lambda_{D^l}Y_D^*)_{pi}
\nonumber\\&&+12\tilde{D}_{ij}(m^2_{\tilde{L}_kH_1})(Y_D^t\Lambda_{D^p}^{\dagger})_{pk}-12(Y_D^{\dagger}\Lambda_{D^j}^t\Lambda_{D^p}^*m_{\tilde{Q}}^2Y_D)_{ip}
\nonumber\\&&-6(Y_D^{\dagger}\Lambda_{D^j}^t\Lambda_{D^p}^*Y_Dm_{\tilde{D}}^2)_{ip}-12m_{H_1}^2(Y_D^{\dagger}\Lambda_{D^j}^t\Lambda_{D^p}^*Y_D)_{ip}+12(Y_D^{\dagger}\Lambda_{D^j}^t\Lambda_{D^p}^*\Lambda_{D^p}^t)_{il}m^2_{H_1\tilde{L}_l}
\nonumber\\&&-4(Y_E^t\Lambda_{E^p}^{\dagger}m_{\tilde{L}}^2\Lambda_{D^j}Y_D^*)_{pi}-2(m_{\tilde{D}}^2)_{lj}(Y_E^t\Lambda_{E^p}^{\dagger}\Lambda_{D^l}Y_D^*)_{pi}+4\tilde{D}_{ij}(Y_E^t\Lambda_{E^p}^{\dagger})_{pk}m^2_{\tilde{L}_kH_1}
\nonumber\\&&-4(Y_D^{\dagger}\Lambda_{D^j}^t\Lambda_{E^p}^*m_{\tilde{L}}^2Y_E)_{ip}-2(Y_D^{\dagger}\Lambda_{D^j}^t\Lambda_{E^p}^*Y_Em_{\tilde{E}}^2)_{ip}-4m_{H_1}^2(Y_D^{\dagger}\Lambda_{D^j}^t\Lambda_{E^p}^*Y_E)_{ip}
\nonumber\\&&-4(Y_D^{\dagger}\Lambda_{D^j}^t\Lambda_{E^p}^*\Lambda_{E^p})_{il}m^2_{H_1\tilde{L}_l}-12(Y_D^{\dagger}\Lambda_{D^p}^t\Lambda_{D^i}^*m_{\tilde{Q}}^2Y_D)_{pj}-4(Y_D^{\dagger}m_{\tilde{Q}}^2\Lambda_{D^j}^t\Lambda_{E^p}^*Y_E)_{ip}
\nonumber\\&&-6(Y_D^{\dagger}\Lambda_{D^p}^t\Lambda_{D^i}^*Y_Dm_{\tilde{D}}^2)_{pj}-12m^2_{H_1}(Y_D^{\dagger}\Lambda_{D^p}^t\Lambda_{D^i}^*Y_D)_{pj}+12(Y_D^{\dagger}\Lambda_{D^p}^t\Lambda_{D^i}^*\Lambda_{D^j}^t)_{pl}m^2_{H_1\tilde{L}_l}
\nonumber\\&&-12(Y_D^t\Lambda_{D^i}^{\dagger}m_{\tilde{L}}^2\Lambda_{D^p}Y_D^*)_{jp}-12(Y_D^{\dagger}m_{\tilde{Q}}^2\Lambda_{D^p}^t\Lambda_{D^i}^*Y_D)_{pj}
\nonumber\\&&-6(m_{\tilde{D}}^2)_{lp}(Y_D^t\Lambda_{D^i}^{\dagger}\Lambda_{D^l}Y_D^*)_{jp}+12Tr(D)(Y_D^t\Lambda_{D^i}^{\dagger})_{jk}m^2_{\tilde{L}_kH_1}
\nonumber\\&&-4(Y_E^{\dagger}\Lambda_{E^p}^t\Lambda_{D^i}^*m_{\tilde{Q}}^2Y_D)_{pj}-2(Y_E^{\dagger}\Lambda_{E^p}^t\Lambda_{D^i}^*Y_Dm_{\tilde{D}}^2)_{pj}
\nonumber\\&&-4m^2_{H_1}(Y_E^{\dagger}\Lambda_{E^p}^t\Lambda_{D^i}^*Y_D)_{pj}+4(Y_E^{\dagger}\Lambda_{E^p}^t\Lambda_{D^i}^*\Lambda_{D^j}^t)_{pl}m^2_{H_1\tilde{L}_l}
\nonumber\\&&-4(Y_D^t\Lambda_{D^i}^{\dagger}m_{\tilde{L}}^2\Lambda_{E^p}Y_E^*)_{jp}-4(Y_E^{\dagger}m_{\tilde{L}}^2\Lambda_{E^p}^t\Lambda_{D^i}^*Y_D)_{pj}
\nonumber\\&&-2(Y_D^t\Lambda_{D^i}^{\dagger}\Lambda_{E^l}Y_E^*(m_{\tilde{E}}^{2})^t)_{jl}+4(Y_D^t\Lambda_{D^i}^{\dagger})_{jk}Tr(E)m^2_{\tilde{L}_kH_1}
\nonumber\\&&-2(\Lambda_{U^k}^{\dagger}\Lambda_{U^l}m_{\tilde{D}}^2)_{ij}Tr(\Lambda_{U^l}^{\dagger}\Lambda_{U^k})-4(\Lambda_{U^k}^{\dagger}{m_{\tilde{D}}^2}^t\Lambda_{U^l})_{ij}Tr(\Lambda_{U^l}^{\dagger}\Lambda_{U^k})
\nonumber\\&&-4(m_{\tilde{U}}^2)_{ql}(\Lambda_{U^k}^{\dagger}\Lambda_{U^q})_{ij}Tr(\Lambda_{U^l}^{\dagger}\Lambda_{U^k})-8(\Lambda_{U^k}^{\dagger}\Lambda_{U^l})_{ij}Tr(\Lambda_{U^l}^{\dagger}\Lambda_{U^k}m_{\tilde{D}}^2)
\nonumber\\&&-4(m_{\tilde{U}}^2)_{qk}(\Lambda_{U^k}^{\dagger}\Lambda_{U^l})_{ij}Tr(\Lambda_{U^l}^{\dagger}\Lambda_{U^q})-4(\Lambda_{U^k}^{\dagger}\Lambda_{U^l}m_{\tilde{D}}^2)_{ij}\tilde{T}_{lk}
\nonumber\\&&-8(\Lambda_{U^k}^{\dagger}{m_{\tilde{D}}^2}^t\Lambda_{U^l})_{ij}\tilde{T}_{lk}-8(m_{\tilde{U}}^2\tilde{T})_{lk}(\Lambda_{U^k}^{\dagger}\Lambda_{U^l})_{ij}
\nonumber\\&&-8(\Lambda_{U^k}^{\dagger}\Lambda_{U^l})_{ij}(Y_U^{\dagger}m_{\tilde{Q}}^2Y_U)_{lk}-8(\Lambda_{U^k}^{\dagger}\Lambda_{U^l})_{ij}(\tilde{T}m_{\tilde{U}}^2)_{lk}
\nonumber\\&&-8m_{H_2}^2(\Lambda_{U^k}^{\dagger}\Lambda_{U^l})_{ij}\tilde{T}_{lk}+4(\Lambda_{U^m}^{\dagger}m_{\tilde{D}}^2)_{iq}(\Lambda_{U^m})_{jl}Tr(\Lambda_{D^l}\Lambda_{D^q}^{\dagger})
\nonumber\\&&-4(\Lambda_{U^m}^{\dagger}Y_D^{\dagger}\Lambda_{D^l}^t)_{iq}(\Lambda_{U^m})_{jl}m_{H_1\tilde{L}_q}^2-2(m_{\tilde{D}}^2Y_D^{\dagger}\Lambda_{D^j}^t\Lambda_{E^p}^*Y_E)_{ip}
\nonumber\\&&-2(m_{\tilde{D}}^2)_{il}Tr(E\Lambda_{D^j}\Lambda_{D^l}^{\dagger})+4(E\Lambda_{D^j}Y_D^*)_{ki}m^2_{H_1\tilde{L}_k}-4\tilde{T}_{lk}(m_{\tilde{D}}^2\Lambda_{U^k}^{\dagger}\Lambda_{U^l})_{ij}
\nonumber\\&&-2Tr(\Lambda_{U^l}^{\dagger}\Lambda_{U^k})(m_{\tilde{D}}^2\Lambda_{U^k}^{\dagger}\Lambda_{U^l})_{ij}-2(m_{\tilde{E}}^2Y_E^{\dagger}\Lambda_{E^p}^t\Lambda_{D^i}^*Y_D)_{pj}-2(m_{\tilde{D}}^2)_{pl}(Y_D^t\Lambda_{D^l}^{\dagger}\Lambda_{D^p}Y_D^*)_{ji}
\nonumber\\&&-2(m_{\tilde{D}}^2Y_D^{\dagger}\Lambda_{D^p}^t\Lambda_{D^p}^*Y_D)_{ij}-2(m_{\tilde{D}})_{il}Tr(\Lambda_{D^j}D^t\Lambda_{D^l}^{\dagger})
\nonumber\\&&-2(m_{\tilde{D}}^2)_{il}Tr(\Lambda_{D^j}Y_U^*Y_U^t\Lambda_{D^l}^{\dagger})-2(m_{\tilde{D}}^2)_{il}Tr(\Lambda_{D^l}^{\dagger}\Lambda_{D^j}\Lambda_{D^q}^{\dagger}\Lambda_{D^q})
\nonumber\\&&-6(m_{\tilde{D}}^2)_{il}Tr(\Lambda_{D^q}\Lambda_{D^q}^{\dagger}\Lambda_{D^j}\Lambda_{D^l}^{\dagger})-2(m_{\tilde{D}}^2)_{il}Tr(\Lambda_{E^q}\Lambda_{E^q}^{\dagger}\Lambda_{D^j}\Lambda_{D^l}^{\dagger})
\nonumber\\&&+4(\Lambda_{U^m})_{jk}(\Lambda_{D^k}^*Y_D\Lambda_{U^m}^{\dagger})_{li}m^2_{\tilde{L}_lH_1}-4(\Lambda_{U^m}\tilde{D}\Lambda_{U^m}^{\dagger}{m_{\tilde{D}}^2}^t)_{ji}-4(m_{\tilde{D}}^2\Lambda_{U^m}^{\dagger}\Lambda_{U^p}\Lambda_{U^p}^{\dagger}\Lambda_{U^m})_{ij}
\nonumber\\&&+4(\Lambda_{U^m})_{jl}Tr(\Lambda_{D^l}\Lambda_{D^k}^{\dagger})[(m_{\tilde{D}}^2\Lambda_{U^m}^{\dagger})_{ik}+(\Lambda_{U^m}^{\dagger}{m_{\tilde{D}}^2}^t)_{ik}+(m_{\tilde{U}}^2)_{mq}(\Lambda_{U^q}^{\dagger})_{ik}]
\nonumber\\&&-6(m_{\tilde{D}}^2Y_D^{\dagger}\Lambda_{D^j}^t\Lambda_{D^p}^*Y_D)_{ip}-6(Y_D^{\dagger}\Lambda_{D^j}^t\Lambda_{D^l}^*Y_Dm_{\tilde{D}}^2)_{il}
\nonumber\\&&-2(m_{\tilde{D}}^2)_{il}(Y_D^t\Lambda_{D^l}^{\dagger}\Lambda_{E^p}Y_E^*)_{jp}-6(m_{\tilde{D}}^2Y_D^{\dagger}\Lambda_{D^p}^t\Lambda_{D^i}^*Y_D)_{pj}
\nonumber\\&&-2((m_{\tilde{E}}^{2})^tY_E^t\Lambda_{E^l}^{\dagger}\Lambda_{D^j}Y_D^*)_{li}-6(m_{\tilde{D}}^2)_{il}(Y_D^t\Lambda_{D^l}^{\dagger}\Lambda_{D^p}Y_D^*)_{jp}-\frac{2}{5}s'\delta_{ij}
\end{eqnarray}

\begin{eqnarray}
{(16\pi^2)^2}\beta^{(2)}_{(M^2_{\tilde{U}})_{ij}}&=&{(16\pi^2)^2}\beta^{(2)}_{(M^2_{\tilde{U}})_{RPC}}+\nonumber\\&&(\frac{32}{3}M_3^2g_3^2-\frac{16}{15}M_1^2g_1^2)Tr(\Lambda_{U^j}\Lambda_{U^i}^{\dagger})-(\frac{16}{3}M_3^*g_3^2-\frac{8}{15}M_1^*g_1^2)Tr(h_{U^j}\Lambda_{U^i}^{\dagger})
\nonumber\\&&+(\frac{16}{3}g_3^2-\frac{8}{15}g_1^2)Tr(h_{U^j}h_{U^i}^{\dagger})-8(h_{U^i}^{\dagger}\Lambda_{U^j})_{lm}Tr(\Lambda_{D^m}^{\dagger}h_{D^l})
\nonumber\\&&-4(h_U^{\dagger}\Lambda_{D^m}^t\Lambda_{D^m}^*h_{U}+h_U^{\dagger}h_{D^m}^t\Lambda_{D^m}^*Y_U)_{ij}-4(Y_U^{\dagger}h_{D^m}^th_{D^m}^*Y_U+Y_U^{\dagger}\Lambda_{D^m}^th_{D^m}^*h_U)_{ij}
\nonumber\\&&-(\frac{16}{3}M_3g_3^2-\frac{8}{15}M_1g_1^2)Tr(\Lambda_{U^j}h_{U^i}^{\dagger})-8(h_{U^i}^{\dagger}h_{U^j})_{lm}Tr(\Lambda_{D^m}^{\dagger}\Lambda_{D^l})
\nonumber\\&&-8(\Lambda_{U^i}^{\dagger}h_{U^j})_{lm}Tr(h_{D^m}^{\dagger}\Lambda_{D^l})-8(\Lambda_{U^i}^{\dagger}\Lambda_{U^j})_{lm}Tr(h_{D^m}^{\dagger}h_{D^l})
\nonumber\\&&-8Tr(h_{U^j}h_{U^i}^{\dagger}\Lambda_{U^p}\Lambda_{U^p}^{\dagger}+\Lambda_{U^j}h_{U^i}^{\dagger}h_{U^p}\Lambda_{U^p}^{\dagger}+h_{U^j}\Lambda_{U^i}^{\dagger}\Lambda_{U^p}h_{U^p}^{\dagger}+\Lambda_{U^j}\Lambda_{U^i}^{\dagger}h_{U^p}h_{U^p}^{\dagger})
\nonumber\\&&-8Tr(h_{U^i}^{\dagger}h_{U^j}\tilde{D}+h_{U^i}^{\dagger}\Lambda_{U^j}Y_D^{\dagger}h_{D})-8Tr(\Lambda_{U^i}^{\dagger}h_{U^j}h_D^{\dagger}Y_D+\Lambda_{U^i}^{\dagger}\Lambda_{U^j}h_D^{\dagger}h_D)
\nonumber\\&&+(\frac{8}{3}g_3^2-\frac{4}{15}g_1^2)[4Tr(\Lambda_{U^j}m_{\tilde{D}}^2\Lambda_{U^i}^{\dagger})+(m_{\tilde{U}}^2)_{il}Tr(\Lambda_{U^j}\Lambda_{U^l}^{\dagger})+(m_{\tilde{U}}^2)_{lj}Tr(\Lambda_{U^l}\Lambda_{U^i}^{\dagger})]
\nonumber\\&&+2(Y_U^{\dagger}\Lambda_{D^k}^t)_{iq}(Y_D^{\dagger}Y_U)_{kj}m_{H_1\tilde{L}_q}^2-4(Y_U^t\Lambda_{D^m}^{\dagger}m_{\tilde{L}}^2\Lambda_{D^m}Y_U^*)_{ji}
\nonumber\\&&-4(Y_U^t\Lambda_{D^m}^{\dagger}\Lambda_{D^m}{m_{\tilde{Q}}^2}^tY_U^*)_{ji}-4(m_{\tilde{D}}^2)_{lm}(Y_U^t\Lambda_{D^m}^{\dagger}\Lambda_{D^l}Y_U^*)_{ji}
\nonumber\\&&+4(Y_U^{\dagger}Y_D)_{im}(\Lambda_{D^m}^*Y_U)_{lj}m_{\tilde{L}_lH_1}^2-4(Y_U^{\dagger}\Lambda_{D^m}^t\Lambda_{D^m}^*m_{\tilde{Q}}^2Y_U)_{ij}
\nonumber\\&&-2(Y_U^{\dagger}\Lambda_{D^m}^t\Lambda_{D^m}^*Y_Um_{\tilde{U}}^2)_{ij}-4m_{H_2}^2(Y_U^{\dagger}\Lambda_{D^m}^t\Lambda_{D^m}^*Y_U)_{ij}
\nonumber\\&&-8Tr(\Lambda_{D^m}^{\dagger}\Lambda_{D^l})(\Lambda_{U^i}^{\dagger}\Lambda_{U^j}m_{\tilde{D}}^2)_{lm}-8Tr(\Lambda_{D^m}^{\dagger}\Lambda_{D^l})(\Lambda_{U^i}^{\dagger}{m_{\tilde{D}}^2}^t\Lambda_{U^j})_{lm}
\nonumber\\&&-4(m_{\tilde{U}}^2)_{qj}(\Lambda_{U^i}^{\dagger}\Lambda_{U^q})_{lm}Tr(\Lambda_{D^m}^{\dagger}\Lambda_{D^l})-8(\Lambda_{U^i}^{\dagger}\Lambda_{U^j})_{lm}Tr(\Lambda_{D^m}^{\dagger}m_{\tilde{L}}^2\Lambda_{D^l})
\nonumber\\&&-8(\Lambda_{U^i}^{\dagger}\Lambda_{U^j})_{lm}Tr(\Lambda_{D^m}^{\dagger}\Lambda_{D^l}{m_{\tilde{Q}}^2}^t)-8(m_{\tilde{D}}^2\Lambda_{U^i}^{\dagger}\Lambda_{U^j})_{qm}Tr(\Lambda_{D^m}^{\dagger}\Lambda_{D^q})
\nonumber\\&&+8(\Lambda_{D^m}^*Y_D\Lambda_{U^i}^{\dagger}\Lambda_{U^j})_{qm}m_{\tilde{L}_qH_1}^2-8Tr(\Lambda_{U^j}m_{\tilde{D}}^2\Lambda_{U^i}^{\dagger}\Lambda_{U^p}\Lambda_{U^p}^{\dagger})
\nonumber\\&&-8Tr({m_{\tilde{D}}^2}^t\Lambda_{U^j}\Lambda_{U^i}^{\dagger}\Lambda_{U^p}\Lambda_{U^p}^{\dagger})-4(m_{\tilde{U}}^2)_{lj}Tr(\Lambda_{U^l}\Lambda_{U^i}^{\dagger}\Lambda_{U^p}\Lambda_{U^p}^{\dagger})
\nonumber\\&&-8Tr(\Lambda_{U^j}\Lambda_{U^i}^{\dagger}{m_{\tilde{D}}^2}^t\Lambda_{U^p}\Lambda_{U^p}^{\dagger})-8Tr(\Lambda_{U^j}\Lambda_{U^i}^{\dagger}\Lambda_{U^p}m_{\tilde{D}}^2\Lambda_{U^p}^{\dagger})
\nonumber\\&&-8(m_{\tilde{U}}^2)_{lp}Tr(\Lambda_{U^j}\Lambda_{U^i}^{\dagger}\Lambda_{U^l}\Lambda_{U^p}^{\dagger})-8Tr(\Lambda_{U^i}^{\dagger}\Lambda_{U^j}m_{\tilde{D}}^2\tilde{D})+8(\Lambda_{U^i}^{\dagger}\Lambda_{U^j}Y_D^{\dagger}\Lambda_{D^k}^t)_{kl}m_{H_1\tilde{L}_l}^2
\nonumber\\&&-8Tr(\Lambda_{U^i}^{\dagger}{m_{\tilde{D}}^2}^t\Lambda_{U^j}\tilde{D})-4Tr(\Lambda_{U^i}^{\dagger}\Lambda_{U^l}\tilde{D})(m_{\tilde{U}}^2)_{lj}-8m_{H_1}^2Tr(\Lambda_{U^i}^{\dagger}\Lambda_{U^j}\tilde{D})
\nonumber\\&&-2(m_{\tilde{U}}^2Y_U^{\dagger}\Lambda_{D^m}^t\Lambda_{D^m}^*Y_U)_{ij}+2(Y_D^{\dagger}Y_U)_{mj}(\Lambda_{D^m}Y_U^*)_{qi}m_{H_1\tilde{L}_q}^2
\nonumber\\&&-4(m_{\tilde{U}}^2)_{iq}(\Lambda_{U^q}^{\dagger}\Lambda_{U^j})_{lm}Tr(\Lambda_{D^m}^{\dagger}\Lambda_{D^l})-4(m_{\tilde{U}}^2)_{il}Tr(\Lambda_{U^l}^{\dagger}\Lambda_{U^p}\Lambda_{U^p}^{\dagger}\Lambda_{U^j})
\nonumber\\&&-8Tr(\Lambda_{U^i}^{\dagger}\Lambda_{U^j}Y_D^{\dagger}m_{\tilde{Q}}^2Y_D)-8Tr(\Lambda_{U^i}^{\dagger}\Lambda_{U^j}\tilde{D}m_{\tilde{D}}^2)-4(m_{\tilde{U}}^2)_{il}Tr(\Lambda_{U^l}^{\dagger}\Lambda_{U^j}\tilde{D})+\frac{4}{5}s'\delta_{ij}
\end{eqnarray}

\begin{eqnarray}
{(16\pi^2)^2}\beta^{(2)}_{(M^2_{{H_1}})}&=&{(16\pi^2)^2}\beta^{(2)}_{(M^2_{{H_1}})_{RPC}}+
-6Tr(h_Eh_E^{\dagger}\Lambda_{D^q}\Lambda_{D^q}^{\dagger}+Y_Eh_E^{\dagger}h_{D^q}\Lambda_{D^q}^{\dagger}+h_EY_E^{\dagger}\Lambda_{D^q}h_{D^q}^{\dagger}
\nonumber\\&&+Eh_{D^q}h_{D^q}^{\dagger})-2Tr(h_Eh_E^{\dagger}\Lambda_{E^q}\Lambda_{E^q}^{\dagger}+Y_Eh_E^{\dagger}h_{E^q}\Lambda_{E^q}^{\dagger}+h_EY_E^{\dagger}\Lambda_{E^q}h_{E^q}^{\dagger}+Eh_{E^q}h_{E^q}^{\dagger})
\nonumber\\&&-12Tr(h_D^{\dagger}h_D\Lambda_{U^q}^{\dagger}\Lambda_{U^q}+h_D^{\dagger}Y_D\Lambda_{U^q}^{\dagger}h_{U^q}+Y_D^{\dagger}h_Dh_{U^q}^{\dagger}\Lambda_{U^q}+\tilde{D}h_{U^q}^{\dagger}h_{U^q})
\nonumber\\&&-6Tr(h_Dh_D^{\dagger}\Lambda_{D^q}^t\Lambda_{D^q}^*+Y_Dh_D^{\dagger}h_{D^q}^t\Lambda_{D^q}^*+h_DY_D^{\dagger}\Lambda_{D^q}^th_{D^q}^*
\nonumber\\&&+Dh_{D^q}^th_{D^q}^*)-2(h_E^{\dagger}h_E)_{ij}Tr(\Lambda_{E^j}^{\dagger}\Lambda_{E^i})-2(h_E^{\dagger}Y_E)_{ij}Tr(\Lambda_{E^j}^{\dagger}h_{E^i})
\nonumber\\&&-2(Y_E^{\dagger}h_E)_{ij}Tr(h_{E^j}^{\dagger}\Lambda_{E^i})-2(\tilde{E})_{ij}Tr(h_{E^j}^{\dagger}h_{E^i})-12(h_D^{\dagger}h_{D})_{ik}Tr(\Lambda_{D^k}^{\dagger}\Lambda_{D^i})
\nonumber\\&&-12(h_D^{\dagger}Y_D)_{ik}Tr(\Lambda_{D^k}^{\dagger}h_{D^i})-12(Y_D^{\dagger}h_{D})_{ik}Tr(h_{D^k}^{\dagger}\Lambda_{D^i})
\nonumber\\&&-12(\tilde{D})_{ik}Tr(h_{D^k}^{\dagger}h_{D^i})-12Tr(\Lambda_{D^k}^{\dagger}\Lambda_{D^i})(\tilde{D}m_{\tilde{D}}^2+m_{\tilde{D}}^2\tilde{D})_{ik}
\nonumber\\&&-(16g_3^2-\frac{2}{5}g_1^2)[(\Lambda_{D^k}^*Y_D)_{lk}m_{\tilde{L}_lH_1}^2+(\Lambda_{D^k}Y_D^*)_{lk}m_{H_1\tilde{L}_l}^2]
\nonumber\\&&+\frac{6}{5}g_1^2[(Y_E^{\dagger}\Lambda_{E^k})_{kl}m_{H_1\tilde{L}_l}^2+(\Lambda_{E^k}^{\dagger}Y_E)_{lk}m_{\tilde{L}_lH_1}^2]-6Tr(Y_Em_{\tilde{E}}^2Y_E^{\dagger}\Lambda_{D^q}\Lambda_{D^q}^{\dagger})
\nonumber\\&&-6Tr(Em_{\tilde{L}}^2\Lambda_{D^q}\Lambda_{D^q}^{\dagger})-6Tr(E\Lambda_{D^q}(m_{\tilde{Q}}^{2})^t\Lambda_{D^q}^{\dagger})
\nonumber\\&&-6(m_{\tilde{D}}^2)_{lq}Tr(E\Lambda_{D^l}\Lambda_{D^q}^{\dagger})-6Tr(D^t\Lambda_{D^q}^{\dagger}m_{\tilde{L}}^2\Lambda_{D^q})
\nonumber\\&&-6(m_{\tilde{D}}^2)_{lq}Tr(D^t\Lambda_{D^q}^{\dagger}\Lambda_{D^l})-12(\tilde{D})_{ik}[Tr(\Lambda_{D^k}^{\dagger}m_{\tilde{L}}^2\Lambda_{D^i})
\nonumber\\&&+Tr(\Lambda_{D^k}^{\dagger}\Lambda_{D^i}(m_{\tilde{Q}}^{2})^t)]-9(Y_EE\Lambda_{E^k})_{kl}m_{H_1\tilde{L}_l}^2-2Tr[m_{\tilde{L}}^2E\Lambda_{E^q}\Lambda_{E^q}^{\dagger}]
\nonumber\\&&-2m_{H_1}^2Tr(E\Lambda_{E^q}\Lambda_{E^q}^{\dagger})-12Tr(Y_D\Lambda_{U^q}^{\dagger}\Lambda_{U^q}Y_D^{\dagger}m_{\tilde{Q}}^2)
\nonumber\\&&-12m_{H_1}^2Tr(\tilde{D}\Lambda_{U^q}^{\dagger}\Lambda_{U^q})-6m_{H_1}^2Tr(D\Lambda_{D^q}^t\Lambda_{D^q}^*)
\nonumber\\&&-6Tr(Y_Dm_{\tilde{D}}^2Y_D^{\dagger}\Lambda_{D^q}^t\Lambda_{D^q}^*)-12Tr(\Lambda_{D^k}^{\dagger}\Lambda_{D^i})[(Y_D^{\dagger}m_{\tilde{Q}}^2Y_D)_{ik}+m_{H_1}^2(\tilde{D})_{ik}]
\nonumber\\&&-2Tr(\Lambda_{E^j}^{\dagger}\Lambda_{E^i})[(Y_E^{\dagger}m_{\tilde{L}}^2Y_E)_{ij}+m_{H_1}^2(\tilde{E})_{ij}]-6Tr(m_{\tilde{Q}}^2\Lambda_{D^q}^t\Lambda_{D^q}^*D)
\nonumber\\&&-6Tr(m_{\tilde{Q}}^2D\Lambda_{D^q}^t\Lambda_{D^q}^*)-12Tr(m_{\tilde{D}}^2\Lambda_{U^q}^{\dagger}\Lambda_{U^q}\tilde{D})-6Tr(E\Lambda_{D^q}\Lambda_{D^q}^{\dagger}m_{\tilde{L}}^2)
\nonumber\\&&-2Tr(\Lambda_{E^q}^{\dagger}E\Lambda_{E^q}(m_{\tilde{L}}^{2})^t)-2(m_{\tilde{E}}^2)_{lq}Tr(\Lambda_{E^q}^{\dagger}E\Lambda_{E^l})
\nonumber\\&&-4(\tilde{E})_{ij}Tr(\Lambda_{E^j}^{\dagger}m_{\tilde{L}}^2\Lambda_{E^i})-12Tr(\tilde{D}\Lambda_{U^q}^{\dagger}{m_{\tilde{D}}^2}^t\Lambda_{U^q})
\nonumber\\&&-12(m_{\tilde{U}}^2)_{lq}Tr(\tilde{D}\Lambda_{U^q}^{\dagger}\Lambda_{U^l})-2Tr(\Lambda_{E^q}^{\dagger}Em_{\tilde{L}}^2\Lambda_{E^q})
\nonumber\\&&-12Tr(\tilde{D}\Lambda_{U^q}^{\dagger}\Lambda_{U^q}m_{\tilde{D}}^2)-2Tr(\Lambda_{E^q}\Lambda_{E^q}^{\dagger}Y_Em_{\tilde{E}}^2Y_E^{\dagger})+27m_{H_1\tilde{L}_l}^2(Y_D^{\dagger}D\Lambda_{D^q}^t)_{ql}
\nonumber\\&&-Tr(\Lambda_{E^j}^{\dagger}\Lambda_{E^i})(\tilde{E}m_{\tilde{E}}^2+m_{\tilde{E}}^2\tilde{E})_{ij}+27m_{\tilde{L}_lH_1}^2(\Lambda_{D^q}^*DY_D)_{lq}
\nonumber\\&&+6(\Lambda_{U^q}^{\dagger}\Lambda_{U^q}Y_D^{\dagger}\Lambda_{D^k}^t)_{kl}m_{H_1\tilde{L}_l}^2+3(Y_D^{\dagger}\Lambda_{D^q}^t\Lambda_{D^q}^*\Lambda_{D^k}^t)_{kl}m_{H_1\tilde{L}_l}^2
\nonumber\\&&-3(Y_E^{\dagger}\Lambda_{D^q}\Lambda_{D^q}^{\dagger}\Lambda_{E^k})_{kl}m_{H_1\tilde{L}_l}^2-(Y_E^{\dagger}\Lambda_{E^q}\Lambda_{E^q}^{\dagger}\Lambda_{E^k})_{kl}m_{H_1\tilde{L}_l}^2
\nonumber\\&&+3(Y_D^t\Lambda_{D^q}^{\dagger}E)_{ql}m_{\tilde{L}_lH_1}^2+(Y_E^t\Lambda_{E^q}^{\dagger}E)_{ql}m_{\tilde{L}_lH_1}^2-(m_{\tilde{E}}^2\tilde{E})_{lj}Tr(\Lambda_{E^j}^{\dagger}\Lambda_{E^l})
\nonumber\\&&-9(\Lambda_{E^k}^{\dagger}EY_E)_{lk}m_{\tilde{L}_lH_1}^2+3(\Lambda_{D^k}^*TY_D)_{lk}m_{\tilde{L}_lH_1}^2+3(E\Lambda_{D^q}Y_D^*)_{iq}m_{H_1\tilde{L}_i}^2
\nonumber\\&&-3(\Lambda_{E^k}^{\dagger}\Lambda_{D^q}\Lambda_{D^q}^{\dagger}Y_E)_{lk}m_{\tilde{L}_lH_1}^2-(\Lambda_{E^k}^{\dagger}\Lambda_{E^q}\Lambda_{E^q}^{\dagger}Y_E)_{lk}m_{\tilde{L}_lH_1}^2
\nonumber\\&&+6(\Lambda_{D^k}^*Y_D\Lambda_{U^q}^{\dagger}\Lambda_{U^q})_{lk}m_{\tilde{L}_lH_1}^2+3(\Lambda_{D^k}^*\Lambda_{D^q}^t\Lambda_{D^q}^*Y_D)_{lk}m_{\tilde{L}_lH_1}^2
\nonumber\\&&+6(Y_D^{\dagger}\Lambda_{D^k}^t)_{il}Tr(\Lambda_{D^k}^{\dagger}\Lambda_{D^i})m_{H_1\tilde{L}_l}^2-(Y_E^{\dagger}\Lambda_{E^j})_{il}Tr(\Lambda_{E^j}^{\dagger}\Lambda_{E^i})m_{H_1\tilde{L}_l}^2
\nonumber\\&&-Tr(\Lambda_{E^j}^{\dagger}\Lambda_{E^i})(\Lambda_{E^i}^{\dagger}Y_E)_{lj}m_{\tilde{L}_lH_1}^2+6Tr(\Lambda_{D^k}^{\dagger}\Lambda_{D^i})(\Lambda_{D^i}^*Y_D)_{lk}m_{\tilde{L}_lH_1}^2
\nonumber\\&&+(E\Lambda_{E^q}Y_E^*)_{lq}m_{H_1\tilde{L}_l}^2-(\tilde{E}m_{\tilde{E}}^2)_{il}Tr(\Lambda_{E^i}\Lambda_{E^l}^{\dagger})-6m^2_{H_1}Tr(\Lambda_{D^q}\Lambda_{D^q}^{\dagger}E)+\frac{3}{5}s'
\end{eqnarray}

\begin{eqnarray}
{(16\pi^2)^2}\beta^{(2)}_{(M^2_{{H_2}})}&=&{(16\pi^2)^2}\beta^{(2)}_{(M^2_{{H_2}})_{RPC}}+
-6Tr(h_Uh_U^{\dagger}\Lambda_{D^q}^{t}\Lambda_{D^q}^*+Y_Uh_U^{\dagger}h_{D^q}^{t}\Lambda_{D^q}^*+h_UY_U^{\dagger}\Lambda_{D^q}^{t}h_{D^q}^*+Th_{D^q}^{t}h_{D^q}^*)
\nonumber\\&&-6(h_U^{\dagger}h_U)_{jk}Tr(\Lambda_{U^k}^{\dagger}\Lambda_{U^j})-6(h_U^{\dagger}Y_U)_{jk}Tr(\Lambda_{U^k}^{\dagger}h_{U^j})-6(Y_U^{\dagger}h_U)_{jk}Tr(h_{U^k}^{\dagger}\Lambda_{U^j})
\nonumber\\&&-6Tr(m_{\tilde{Q}}^2T\Lambda_{D^q}^{t}\Lambda_{D^q}^*+Y_Um_{\tilde{U}}^2Y_U\Lambda_{D^q}^{t}\Lambda_{D^q}^*)-6m_{H_2}^2Tr(T\Lambda_{D^q}^{t}\Lambda_{D^q}^*)
\nonumber\\&&-6m_{H_2}^2\tilde{T}_{jk}Tr(\Lambda_{U^k}^{\dagger}\Lambda_{U^j})-6(Y_U^{\dagger}m_{\tilde{Q}}^2Y_U+\tilde{T}m_{\tilde{U}}^2)_{jk}Tr(\Lambda_{U^k}^{\dagger}\Lambda_{U^j})
\nonumber\\&&-6Tr(m_{\tilde{Q}}^2\Lambda_{D^q}^{t}\Lambda_{D^q}^*T)-12\tilde{T}_{jk}Tr(\Lambda_{U^k}^{\dagger}\Lambda_{U^j}m_{\tilde{D}}^2)-6Tr(T\Lambda_{D^q}^t(m_{\tilde{L}}^{2})^t\Lambda_{D^q}^*)
\nonumber\\&&-6Tr(Y_U^*Y_U^t\Lambda_{D^q}^{\dagger}\Lambda_{D^l})(m_{\tilde{D}}^2)_{lq}-6(m^2_{\tilde{U}}T)_{lk}Tr(\Lambda_{U^k}^{\dagger}\Lambda_{U^l})
\nonumber\\&&+6(Y_D^{\dagger}T\Lambda_{D^k}^{t})_{kl}m_{H_1\tilde{L}_l}^2+6(\Lambda_{D^q}^*TY_D)_{iq}m_{\tilde{L}_iH_1}^2-6\tilde{T}_{jk}Tr(h_{U^k}^{\dagger}h_{U^j})-\frac{3}{5}s'
\end{eqnarray}

\begin{eqnarray}
{(16\pi^2)^2}\beta^{(2)}_{(M^2_{{H_1\tilde{L_i}}})}&=&(\frac{8}{5}M_1^2g_1^2-64M_3^2g_3^2)(\Lambda_{D^q}^*Y_D)_{iq}-\frac{24}{5}M_1^2g_1^2(\Lambda_{E^q}^*Y_E)_{iq}
\nonumber\\&&+(-\frac{4}{5}M_1g_1^2+32M_3g_3^2)(h_{D^q}^*Y_D)_{iq}+\frac{12}{5}M_1g_1^2(h_{E^q}^*Y_E)_{iq}
\nonumber\\&&+(-\frac{4}{5}M_1^*g_1^2+32M_3^*g_3^2)(\Lambda_{D^q}^*h_D)_{iq}+\frac{12}{5}M_1^*g_1^2(\Lambda_{E^q}^*h_E)_{iq}
\nonumber\\&&+(\frac{4}{5}g_1^2-32g_3^2)(h_{D^q}^*h_D)_{iq}-\frac{12}{5}g_1^2(h_{E^q}^*h_E)_{iq}-6(h_{E^j}^{\dagger}h_{D^p}\Lambda_{D^p}^{\dagger}Y_E
\nonumber\\&&+h_{E^j}^{\dagger}\Lambda_{D^p}\Lambda_{D^p}^{\dagger}h_E+\Lambda_{E^j}^{\dagger}h_{D^p}h_{D^p}^{\dagger}Y_E+\Lambda_{E^j}^{\dagger}\Lambda_{D^p}h_{D^p}^{\dagger}h_E)_{ij}
\nonumber\\&&-2(h_{E^j}^{\dagger}h_{E^p}\Lambda_{E^p}^{\dagger}Y_E+h_{E^j}^{\dagger}\Lambda_{E^p}\Lambda_{E^p}^{\dagger}h_E+\Lambda_{E^j}^{\dagger}h_{E^p}h_{E^p}^{\dagger}Y_E
\nonumber\\&&+\Lambda_{E^j}^{\dagger}\Lambda_{E^p}h_{E^p}^{\dagger}h_E)_{ij}-6(h_{E^j}^{\dagger}h_E\tilde{E}+h_{E^j}^{\dagger}Eh_E
\nonumber\\&&+\Lambda_{E^j}^{\dagger}h_Eh_E^{\dagger}Y_E+\Lambda_{E^j}^{\dagger}Y_Eh_E^{\dagger}h_E)_{ij}-2(h_{E^k}^{\dagger}h_E)_{il}Tr(\Lambda_{E^l}^{\dagger}\Lambda_{E^k})
\nonumber\\&&-2(h_{E^k}^{\dagger}Y_E)_{il}Tr(\Lambda_{E^l}^{\dagger}h_{E^k})-2(\Lambda_{E^k}^{\dagger}h_E)_{il}Tr(h_{E^l}^{\dagger}\Lambda_{E^k})
\nonumber\\&&-6(h_E^*h_E^t\Lambda_{D^m}^*Y_D+h_E^*Y_E^t\Lambda_{D^m}^*h_D+Y_E^*h_E^th_{D^m}^*Y_D
\nonumber\\&&+E^th_{D^m}^*h_D)_{im}+6(h_E^*h_E^t\Lambda_{E^m}^*Y_E+h_E^*Y_E^t\Lambda_{E^m}^*h_E
\nonumber\\&&+Y_E^*h_E^th_{E^m}^*Y_E+E^th_{E^m}^*h_E)_{im}+18(h_{D^k}^*h_D\tilde{D}
\nonumber\\&&+h_{D^k}^*Dh_D+\Lambda_{D^k}^*h_Dh_D^{\dagger}Y_D+\Lambda_{D^k}^*Y_Dh_D^{\dagger}h_D)_{ik}
\nonumber\\&&-12(h_{D^k}^*h_D\Lambda_{U^m}^{\dagger}\Lambda_{U^m}+h_{D^k}^*Y_D\Lambda_{U^m}^{\dagger}h_{U^m}+\Lambda_{D^k}^*h_Dh_{U^m}^{\dagger}\Lambda_{U^m}
\nonumber\\&&+\Lambda_{D^k}^*Y_Dh_{U^m}^{\dagger}h_{U^m})_{ik}+12(h_{D^k}^*h_D)_{il}Tr(\Lambda_{D^l}\Lambda_{D^k}^{\dagger})
\nonumber\\&&+12(\Lambda_{D^k}^*h_D)_{il}Tr(\Lambda_{D^l}h_{D^k}^{\dagger})+12(\Lambda_{D^k}^*Y_D)_{il}Tr(h_{D^l}h_{D^k}^{\dagger})
\nonumber\\&&+6(h_{D^j}^*h_UY_U^{\dagger}Y_D+h_{D^j}^*Th_D+\Lambda_{D^j}^*h_Uh_U^{\dagger}Y_D+\Lambda_{D^j}^*Y_Uh_U^{\dagger}h_D)_{ij}
\nonumber\\&&+6(h_{D^j}^*h_{D^m}^t\Lambda_{D^m}^*Y_D+h_{D^j}^*\Lambda_{D^m}^t\Lambda_{D^m}^*h_D+\Lambda_{D^j}^*h_{D^m}^th_{D^m}^*Y_D
\nonumber\\&&+\Lambda_{D^j}^*\Lambda_{D^m}^th_{D^m}^*h_D)_{ij}-2(\Lambda_{E^k}^{\dagger}Y_E)_{il}Tr(h_{E^l}^{\dagger}h_{E^k})
\nonumber\\&&+12(h_{D^k}^*Y_D)_{il}Tr(h_{D^l}\Lambda_{D^k}^{\dagger})-3Tr(Y_D^t\Lambda_{D^m}^{\dagger}\Lambda_{D^m}Y_D^*)m_{H_1\tilde{L}_i}^2
\nonumber\\&&+(\frac{2}{5}g_1^2-16g_3^2)[2(\Lambda_{D^q}^*m_{\tilde{Q}}^2Y_D)_{iq}+2(\Lambda_{D^q}^*Y_Dm_{\tilde{D}}^2)_{iq}+m_{H_1}^2(\Lambda_{D^q}^*Y_D)_{iq}
\nonumber\\&&-(\Lambda_{D^q}^*\Lambda_{D^q}^t)_{il}m_{H_1\tilde{L}_l}^2]-\frac{6}{5}g_1^2[2(\Lambda_{E^q}^*m_{\tilde{L}}^2Y_E)_{iq}+2(\Lambda_{E^q}^*Y_Em_{\tilde{E}}^2)_{iq}
\nonumber\\&&+m_{H_1}^2(\Lambda_{E^q}^*Y_E)_{iq}+(\Lambda_{E^q}^*\Lambda_{E^q})_{il}m_{H_1\tilde{L}_l}^2]-6(\Lambda_{E^j}^{\dagger}m_{\tilde{L}}^2\Lambda_{D^p}\Lambda_{D^p}^{\dagger}Y_E)_{ij}
\nonumber\\&&-6(\Lambda_{E^j}^{\dagger}\Lambda_{D^p}{m_{\tilde{Q}}^2}^t\Lambda_{D^p}^{\dagger}Y_E)_{ij}-6(m_{\tilde{D}}^2)_{lp}(\Lambda_{E^j}^{\dagger}\Lambda_{D^l}\Lambda_{D^p}^{\dagger}Y_E)_{ij}
\nonumber\\&&+3(\Lambda_{E^j}^{\dagger})_{iq}(Y_D^t\Lambda_{D^p}^{\dagger}Y_E)_{pj}m^2_{\tilde{L}_qH_1}-6(\Lambda_{E^j}^{\dagger}\Lambda_{D^p}\Lambda_{D^p}^{\dagger}m_{\tilde{L}}^2Y_E)_{ij}
\nonumber\\&&-3m_{H_1}^2(\Lambda_{E^j}^{\dagger}\Lambda_{D^p}\Lambda_{D^p}^{\dagger}Y_E)_{ij}-6(\Lambda_{E^j}^{\dagger}\Lambda_{D^p}\Lambda_{D^p}^{\dagger}Y_Em_{\tilde{E}}^2)_{ij}
\nonumber\\&&-3(\Lambda_{E^j}^{\dagger}\Lambda_{D^p}\Lambda_{D^p}^{\dagger}\Lambda_{E^j})_{il}m_{H_1\tilde{L}_l}^2-2(\Lambda_{E^j}^{\dagger}m_{\tilde{L}}^2\Lambda_{E^p}\Lambda_{E^p}^{\dagger}Y_E)_{ij}
\nonumber\\&&-2(\Lambda_{E^j}^{\dagger}\Lambda_{E^p}(m_{\tilde{L}}^{2})^t\Lambda_{E^p}^{\dagger}Y_E)_{ij}-2(m_{\tilde{E}}^2)_{lp}(\Lambda_{E^j}^{\dagger}\Lambda_{E^l}\Lambda_{E^p}^{\dagger}Y_E)_{ij}
\nonumber\\&&-2(\Lambda_{E^j}^{\dagger}\Lambda_{E^p}\Lambda_{E^p}^{\dagger}m_{\tilde{L}}^2Y_E)_{ij}-2(\Lambda_{E^j}^{\dagger}\Lambda_{E^p}\Lambda_{E^p}^{\dagger}Y_Em_{\tilde{E}}^2)_{ij}-6(\Lambda_{E^j}^{\dagger}m_{\tilde{L}}^2EY_E)_{ij}
\nonumber\\&&-m_{H_1}^2(\Lambda_{E^j}^{\dagger}\Lambda_{E^p}\Lambda_{E^p}^{\dagger}Y_E)_{ij}-(\Lambda_{E^j}^{\dagger}\Lambda_{E^p}\Lambda_{E^p}^{\dagger}\Lambda_{E^j})_{il}m_{H_1\tilde{L}_l}^2-9m_{H_1}^2(\Lambda_{E^j}^{\dagger}EY_E)_{ij}
\nonumber\\&&-6(\Lambda_{E^j}^{\dagger}Y_Em_{\tilde{E}}^2\tilde{E})_{ij}-4(\Lambda_{E^j}^{\dagger}\Lambda_{E^k})_{iq}\tilde{E}_{kj}m_{H_1\tilde{L}_q}^2
-6(\Lambda_{E^j}^{\dagger}Em_{\tilde{L}}^2Y_E)_{ij}
\nonumber\\&&-6(\Lambda_{E^j}^{\dagger}EY_Em_{\tilde{E}}^2)_{ij}-5(\Lambda_{E^j}^{\dagger}E\Lambda_{E^j})_{il}m_{H_1\tilde{L}_l}^2-2(\Lambda_{E^k}^{\dagger}m_{\tilde{L}}^2Y_E)_{il}Tr(\Lambda_{E^l}^{\dagger}\Lambda_{E^k})
\nonumber\\&&-2(\Lambda_{E^k}^{\dagger}Y_Em_{\tilde{E}}^2)_{il}Tr(\Lambda_{E^l}^{\dagger}\Lambda_{E^k})-m_{H_1}^2(\Lambda_{E^k}^{\dagger}Y_E)_{il}Tr(\Lambda_{E^l}^{\dagger}\Lambda_{E^k})
\nonumber\\&&-(\Lambda_{E^k}^{\dagger}\Lambda_{E^l})_{iq}Tr(\Lambda_{E^l}^{\dagger}\Lambda_{E^k})m_{H_1\tilde{L}_q}^2-4(\Lambda_{E^k}^{\dagger}Y_E)_{il}Tr(\Lambda_{E^l}^{\dagger}m_{\tilde{L}}^2\Lambda_{E^k})
\nonumber\\&&-2(m_{\tilde{E}}^2)_{qk}(\Lambda_{E^k}^{\dagger}Y_E)_{il}Tr(\Lambda_{E^l}^{\dagger}\Lambda_{E^q})-6(\Lambda_{E^k}^{\dagger}Y_E)_{il}(\Lambda_{E^l}^{\dagger}Y_E)_{qk}m_{\tilde{L}_qH_1}^2
\nonumber\\&&-6(Y_D^t\Lambda_{D^m}^{\dagger}m_{\tilde{L}}^2E)_{mi}-6(E^t\Lambda_{D^m}^*Y_Dm_{\tilde{D}}^2)_{im}-6(Y_D^t\Lambda_{D^m}^{\dagger}Y_Em_{\tilde{E}}^2Y_E^{\dagger})_{mi}
\nonumber\\&&-9m_{H_1}^2(Y_D^t\Lambda_{D^m}^{\dagger}E)_{mi}-3(Y_D^t\Lambda_{D^m}^{\dagger}\Lambda_{E^k})_{mq}(Y_E^{\dagger})_{ki}m_{H_1\tilde{L}_q}^2-6(E^t\Lambda_{D^m}^*m_{\tilde{Q}}^2Y_D)_{im}
\nonumber\\&&+6(Y_E^t\Lambda_{E^m}^{\dagger}m_{\tilde{L}}^2E)_{mi}+6(Y_E^t\Lambda_{E^m}^{\dagger}Y_Em_{\tilde{E}}^2Y_E^{\dagger})_{mi}+4(Y_E^t\Lambda_{E^p}^{\dagger}Y_E)_{pj}(\Lambda_{E^j}^{\dagger})_{il}m_{\tilde{L}_lH_1}^2
\nonumber\\&&+3(Y_E^t\Lambda_{E^m}^{\dagger}\Lambda_{E^k})_{mq}(Y_E^{\dagger})_{ki}m_{H_1\tilde{L}_q}^2+9m_{H_1}^2(Y_E^t\Lambda_{E^m}^{\dagger}E)_{mi}+6(E^t\Lambda_{E^m}^*m_{\tilde{L}}^2Y_E)_{im}
\nonumber\\&&+6(E^t\Lambda_{E^m}^*Y_Em_{\tilde{E}}^2)_{im}+2(E^t\Lambda_{E^m}^*\Lambda_{E^m})_{il}m_{H_1\tilde{L}_l}^2+27m_{H_1}^2(\Lambda_{D^k}^*DY_D)_{ik}
\nonumber\\&&+18(\Lambda_{D^k}^*m_{\tilde{Q}}^2DY_D)_{ik}+18(\Lambda_{D^k}^*Y_Dm_{\tilde{D}}^2\tilde{D})_{ik}-12(\Lambda_{D^k}^*\Lambda_{D^m}^t)_{iq}\tilde{D}_{mk}m_{H_1\tilde{L}_q}^2
\nonumber\\&&+18(\Lambda_{D^k}^*Dm_{\tilde{Q}}^2Y_D)_{ik}+18(\Lambda_{D^k}^*DY_Dm_{\tilde{D}}^2)_{ik}-9(\Lambda_{D^k}^*D\Lambda_{D^k}^t)_{iq}m_{H_1\tilde{L}_q}^2
\nonumber\\&&-12(\Lambda_{D^k}^*m_{\tilde{Q}}^2Y_D\Lambda_{U^m}^{\dagger}\Lambda_{U^m})_{ik}-12(\Lambda_{D^k}^*Y_Dm_{\tilde{D}}^2\Lambda_{U^m}^{\dagger}\Lambda_{U^m})_{ik}
\nonumber\\&&-6m_{H_1}^2(\Lambda_{D^k}^*Y_D\Lambda_{U^m}^{\dagger}\Lambda_{U^m})_{ik}+6(\Lambda_{D^k}^*\Lambda_{D^q}^t)_{ij}(\Lambda_{U^m}^{\dagger}\Lambda_{U^m})_{qk}m_{H_1\tilde{L}_j}^2
\nonumber\\&&-12(\Lambda_{D^k}^*Y_D\Lambda_{U^m}^{\dagger}\Lambda_{U^m}m_{\tilde{D}}^2)_{ik}-12(\Lambda_{D^k}^*Y_D\Lambda_{U^m}^{\dagger}{m_{\tilde{D}}^2}^t\Lambda_{U^m})_{ik}
\nonumber\\&&-12(m_{\tilde{U}}^2)_{lm}(\Lambda_{D^k}^*Y_D\Lambda_{U^m}^{\dagger}\Lambda_{U^l})_{ik}+12(\Lambda_{D^k}^*m_{\tilde{Q}}^2Y_D)_{il}Tr(\Lambda_{D^l}\Lambda_{D^k}^{\dagger})
\nonumber\\&&+6(\Lambda_{D^k}^*Y_Dm_{\tilde{D}}^2)_{il}Tr(\Lambda_{D^l}\Lambda_{D^k}^{\dagger})+6m_{H_1}^2(\Lambda_{D^k}^*Y_D)_{il}Tr(\Lambda_{D^l}\Lambda_{D^k}^{\dagger})
\nonumber\\&&-6(\Lambda_{D^k}^*\Lambda_{D^l}^t)_{iq}Tr(\Lambda_{D^l}\Lambda_{D^k}^{\dagger})m_{H_1\tilde{L}_q}^2+12(\Lambda_{D^k}^*Y_D)_{il}Tr(m_{\tilde{L}}^2\Lambda_{D^l}\Lambda_{D^k}^{\dagger})
\nonumber\\&&+12(\Lambda_{D^k}^*Y_D)_{il}Tr(\Lambda_{D^l}(m_{\tilde{Q}}^{2})^t\Lambda_{D^k}^{\dagger})+6(m_{\tilde{D}}^2Y_D^t\Lambda_{D^k}^{\dagger})_{qi}Tr(\Lambda_{D^q}\Lambda_{D^k}^{\dagger})
\nonumber\\&&-6(\Lambda_{D^k}^*Y_DY_D^t\Lambda_{D^k}^{\dagger})_{iq}m_{\tilde{L}_qH_1}^2+6(\Lambda_{D^j}^*m_{\tilde{Q}}^2TY_D)_{ij}+6(\Lambda_{D^j}^*TY_Dm_{\tilde{D}}^2)_{ij}
\nonumber\\&&+6(\Lambda_{D^j}^*Y_Um_{\tilde{U}}^2Y_U^{\dagger}Y_D)_{ij}+6m_{H_2}^2(\Lambda_{D^j}^*TY_D)_{ij}+6(\Lambda_{D^j}^*Tm_{\tilde{Q}}^2Y_D)_{ij}
\nonumber\\&&+3m_{H_1}^2(\Lambda_{D^j}^*TY_D)_{ij}-3(\Lambda_{D^j}^*T\Lambda_{D^j}^t)_{iq}m_{H_1\tilde{L}_q}^2
\nonumber\\&&+6(\Lambda_{D^j}^*\Lambda_{D^m}^t\Lambda_{D^m}^*m_{\tilde{Q}}^2Y_D)_{ij}+6(\Lambda_{D^j}^*\Lambda_{D^m}^t\Lambda_{D^m}^*Y_Dm_{\tilde{D}}^2)_{ij}
\nonumber\\&&+3m_{H_1}^2(\Lambda_{D^j}^*\Lambda_{D^m}^t\Lambda_{D^m}^*Y_D)_{ij}-3(\Lambda_{D^j}^*\Lambda_{D^m}^t\Lambda_{D^m}^*\Lambda_{D^j}^t)_{iq}m_{H_1\tilde{L}_q}^2
\nonumber\\&&+6(Y_D^t\Lambda_{D^m}^{\dagger}m_{\tilde{L}}^2\Lambda_{D^m}\Lambda_{D^j}^{\dagger})_{ji}+6(Y_D^t\Lambda_{D^m}^{\dagger}\Lambda_{D^m}(m_{\tilde{Q}}^{2})^t\Lambda_{D^j}^{\dagger})_{ji}
\nonumber\\&&+6(m_{\tilde{D}}^2)_{lm}(Y_D^t\Lambda_{D^m}^{\dagger}\Lambda_{D^l}\Lambda_{D^j}^{\dagger})_{ji}+(\frac{2}{5}g_1^2-16g_3^2)[(Y_D^t\Lambda_{D^q}^{\dagger}m_{\tilde{L}}^2)_{qi}
\nonumber\\&&-m_{H_1\tilde{L}_i}^2Tr(D)]-\frac{6}{5}g_1^2[(Y_E^t\Lambda_{E^q}^{\dagger}m_{\tilde{L}}^2)_{qi}+(E)_{ki}m_{H_1\tilde{L}_k}^2-m_{H_1\tilde{L}_i}^2Tr(E)]
\nonumber\\&&-3({m_{\tilde{L}}^2}^t\Lambda_{E^j}^{\dagger}\Lambda_{D^p}\Lambda_{D^p}^{\dagger}Y_E)_{ij}-3Tr(Y_E^{\dagger}\Lambda_{D^p}\Lambda_{D^p}^{\dagger}Y_E)m_{H_1\tilde{L}_i}^2
\nonumber\\&&+6(\Lambda_{D^p}\Lambda_{D^p}^{\dagger}E)_{qi}m_{H_1\tilde{L}_q}^2+3(\Lambda_{E^j}^{\dagger}\Lambda_{D^p}Y_D^*)_{ip}(Y_E)_{qj}m_{H_1\tilde{L}_q}^2
\nonumber\\&&-((m_{\tilde{L}}^{2})^t\Lambda_{E^j}^{\dagger}\Lambda_{E^p}\Lambda_{E^p}^{\dagger}Y_E)_{ij}-Tr(Y_E^{\dagger}\Lambda_{E^p}\Lambda_{E^p}^{\dagger}Y_E)m_{H_1\tilde{L}_i}^2
\nonumber\\&&+(\Lambda_{E^j}^{\dagger}\Lambda_{E^p}Y_E^*)_{ip}(Y_E)_{qj}m_{H_1\tilde{L}_q}^2-3({m_{\tilde{L}}^2}^t\Lambda_{E^j}^{\dagger}EY_E)_{ij}-3Tr(E^2)m_{H_1\tilde{L}_i}^2
\nonumber\\&&+6(E^2)_{qi}m_{H_1\tilde{L}_q}^2-((m_{\tilde{L}}^{2})^t\Lambda_{E^k}^{\dagger}Y_E)_{il}Tr(\Lambda_{E^l}^{\dagger}\Lambda_{E^k})+3(Y_E^t\Lambda_{E^m}^{\dagger}Em_{\tilde{L}}^2)_{mi}
\nonumber\\&&-\tilde{E}_{lk}Tr(\Lambda_{E^l}^{\dagger}\Lambda_{E^k})m_{H_1\tilde{L}_i}^2+(Y_E^{\dagger})_{ki}(Y_E)_{ql}Tr(\Lambda_{E^l}^{\dagger}\Lambda_{E^k})m_{H_1\tilde{L}_q}^2
\nonumber\\&&-3(Y_D^t\Lambda_{D^m}^{\dagger}Em_{\tilde{L}}^2)_{mi}-3(Y_D^t\Lambda_{D^m}^{\dagger}Y_E)_{mk}(\Lambda_{E^k}^{\dagger})_{li}m_{\tilde{L}_lH_1}^2
\nonumber\\&&-3Tr(E)(E)_{qi}m_{H_1\tilde{L}_q}^2+9(Y_D^tD^t\Lambda_{D^k}^{\dagger}m_{\tilde{L}}^2)_{ki}-9Tr(D^2)m_{H_1\tilde{L}_i}^2
\nonumber\\&&-12(\Lambda_{D^k}^*Y_D)_{il}(\Lambda_{D^l}^*Y_D)_{qk}m_{\tilde{L}_qH_1}^2-6(\Lambda_{U^m}^t\Lambda_{U^m}^*Y_D^t\Lambda_{D^k}^{\dagger}m_{\tilde{L}}^2)_{ki}
\nonumber\\&&+6Tr(\tilde{D}\Lambda_{U^m}^{\dagger}\Lambda_{U^m})m_{H_1\tilde{L}_i}^2+6(Y_D^t\Lambda_{D^k}^{\dagger}m_{\tilde{L}}^2)_{li}Tr(\Lambda_{D^l}\Lambda_{D^k}^{\dagger})
\nonumber\\&&+12(m_{\tilde{D}}^2)_{kq}(\Lambda_{D^q}^*Y_D)_{il}Tr(\Lambda_{D^l}\Lambda_{D^k}^{\dagger})-6\tilde{D}_{kl}Tr(\Lambda_{D^l}\Lambda_{D^k}^{\dagger})m_{H_1\tilde{L}_i}^2
\nonumber\\&&-6m_{H_1\tilde{L}_j}^2(\Lambda_{D^l}Y_D^*)_{jk}(\Lambda_{D^k}^*Y_D)_{il}+3(Y_D^tT^t\Lambda_{D^j}^{\dagger}m_{\tilde{L}}^2)_{ji}
\nonumber\\&&-3Tr(DT)m_{H_1\tilde{L}_i}^2+3(Y_D^t\Lambda_{D^m}^{\dagger}\Lambda_{D^m}\Lambda_{D^j}^{\dagger}m_{\tilde{L}}^2)_{ji}+3Tr(D)(E)_{qi}m_{H_1\tilde{L}_q}^2,
\end{eqnarray}
where
\begin{eqnarray}
(m^2_{{H_1\tilde{L_i}}})=(m^2_{{\tilde{L_i}H_1}})^{*}, \  D=Y_DY_D^{\dagger}, \ T=Y_UY_U^{\dagger},\ E=Y_EY_E^{\dagger}, \ \tilde{D}=Y_D^{\dagger}Y_D, \ \tilde{T}=Y_U^{\dagger}Y_U, \ \tilde{E}=Y_E^{\dagger}Y_E,
\end{eqnarray}
and
\begin{eqnarray}
s'&=&2g^2_1[Tr(\Lambda_{D^q}^{\dagger}\Lambda_{D^q}{m^2_{\tilde{Q}}}^t)-Tr(\Lambda_{E^q}^{\dagger}\Lambda_{E^q}m^2_{\tilde{L}})+2Tr(\Lambda_{U^q}^{\dagger}\Lambda_{U^q}m^2_{\tilde{D}})
\nonumber\\&&-2(m^2_{\tilde{U}})_{kl}Tr(\Lambda_{U^l}^{\dagger}\Lambda_{U^k})+2(m^2_{\tilde{D}})_{kl}Tr(\Lambda_{D^l}^{\dagger}\Lambda_{D^k})+(m^2_{\tilde{E}})_{kl}Tr(\Lambda_{E^l}^{\dagger}\Lambda_{E^k})]
\end{eqnarray}

for the soft SUSY breaking masses, the $s'$-term comes from exclusion the U(1) Fayet-Iliopoulos (FI) D-term  by using its
equation of motion then this renormalization gives rise to extra contributions proportional to the U(1) gauge coupling~\cite{a7}.
To check  our results,first we have calculated the full one loop R-parity violating
RGEs, and have compared with ref~\cite{a7,a22}. Our results are the same as theirs. Moreover;
to test the above  method  we  have obtained the full two loop
R-parity conserving RGEs which are consistent with results in ref~\cite{a21}. Finally, when we set all lepton number violating couplings to be zero we have obtained the full two loop
the baryon number violating RGEs which we could compare with  results in ref~\cite{a14}. We have realized that we have made some mistakes at previous paper; here we have generalised
the latter results to include lepton number violating terms, and take the opportunity to correct some errors in Ref~\cite{a14}.
\section{Conclusions}
In this work, we have presented the full two-loop $\beta$-functions for all couplings for the RPV MSSM. In previous papers~\cite{a14,a23,a24}, we have studied the effects of some RPV couplings up two-loop RGEs corrections; there we have made a Form program to calculate some of two loop RPV $\beta$-functions as well as RPC $\beta$-functions. however we have not calculated explicitly the full two-loop $\beta$ functions for $R$
parity violating MSSM couplings.

Typically, we expect effects of the two loop $\beta$-functions make a difference of several percent in compare with effects of one loop on the standard running analysis such as Higgs physics and the scalar quark sector of the MSSM; however it is quite difficult to make consequential estimates of the size of the two-loop corrections without committing to a specific model. Moreover it is desirable from the point of view of consistency to use the full set of $\beta$-functions. In any event, we hope that the full set of $\beta$-functions for the most general R-parity violating version of the MSSM could be used in the future analysts.

\end{document}